**On the variability of regression shrinkage methods for clinical prediction models: simulation study on predictive performance**

Ben VAN CALSTER[1,2], Maarten VAN SMEDEN[2,3], Ewout W STEYERBERG[2]

1 KU Leuven, Department of Development and Regeneration, Herestraat 49 box 805, 3000 Leuven, Belgium; 2 Department of Biomedical Data Sciences, Leiden University Medical Center, PO Box 9600, 2300 RC Leiden, Netherlands; 3 Department of Clinical Epidemiology, Leiden University Medical Center, Albinusdreef 2, 2333 ZA Leiden, Netherlands

Running title: On the variability of shrinkage for prediction models

Corresponding author: Ben Van Calster, Ben.vancalster@kuleuven.be, +32 16 377788

Abstract:

When developing risk prediction models, shrinkage methods are recommended, especially when the sample size is limited. Several earlier studies have shown that the shrinkage of model coefficients can reduce overfitting of the prediction model and subsequently result in better predictive performance on average. In this simulation study, we aimed to investigate the variability of regression shrinkage on predictive performance for a binary outcome, with focus on the calibration slope. The slope indicates whether risk predictions are too extreme (slope<1) or not extreme enough (slope>1). We investigated the following shrinkage methods in comparison to standard maximum likelihood estimation: uniform shrinkage (likelihood-based and bootstrap-based), ridge regression, penalized maximum likelihood, LASSO regression, adaptive LASSO, non-negative garrote, and Firth's correction. There were three main findings. First, shrinkage improved calibration slopes on average. Second, the between-sample variability of calibration slopes was often increased relative to maximum likelihood. Among the shrinkage methods, the bootstrap-based uniform shrinkage worked well overall. In contrast to other shrinkage approaches, Firth's correction had only a small shrinkage effect but did so with low variability. Third, the correlation between the estimated shrinkage and the optimal shrinkage to remove overfitting was typically negative. Hence, although shrinkage improved predictions on average, it often worked poorly in individual datasets, in particular when shrinkage was most needed. The observed variability of shrinkage methods implies that these methods do not solve problems associated with small sample size or low number of events per variable.

Key findings

- On average, shrinkage results in calibration slopes that are closer to 1 than maximum likelihood estimation.
- The estimated amount of shrinkage is highly variable, in particular at low sample size; this often leads to more variability in calibration slopes than maximum likelihood estimation.
- Taken together, the mean squared deviation from the target value of the calibration slopes is lower when shrinkage is used.
- Typically, the estimated shrinkage is negatively correlated with the optimal shrinkage. Therefore, shrinkage does not guarantee improved risk predictions for any given dataset.
- Shrinkage does not lead to more reliable prediction models at low sample size, and hence does not justify using lower sample sizes.
- Overall, uniform shrinkage using the bootstrap works well. Firth's correction shrunk the least, but is attractive because of low variability. The Garrote showed erratic behavior.

1. Introduction

When developing clinical prediction models, the ultimate aim is to obtain risk estimates that work well on patients that were not used to develop the model.[1] To do so, we have to keep statistical overfitting under control. Assuming that data collection was done carefully, and according to standardized procedures and definitions, the values in a dataset reflect (1) true underlying distributions of and associations between variables, and (2) some amount of random variability. Overfitting occurs when a prediction model also captures these random idiosyncrasies of the development dataset, which by definition do not generalize to new data from the same population.[2] The risk of an overfitted model increases when the model building strategy is too ambitious for the available data, for example when the number of variables that are tested as potential model predictors is large given the available sample size.

A well-known rule of thumb for sample size for prediction models is to have at least 10 events per variable (EPV).[3-6] For binary outcomes, the number of events is the number of cases in the smallest of the two outcome levels. 'Variables' actually refers to the number of parameters that are considered for inclusion in the model (excluding intercepts). Some parameters may be checked but not included in the final model, and variables may be modeled using more than one parameter. Recent research has indicated that the EPV≥10 rule is too simplistic, and highlights that there are no good rules of thumb regarding sample size.[7-11] Therefore, the use of shrinkage methods is recommended when sample size is small.[5,6] Several studies have suggested that model performance improves on average when shrinkage methods are applied.[5,9,12-17] Some have suggested that shrinkage may be needed for EPV values up to 20 if the model is prespecified.[1] When variable selection has to be performed to develop the model, the required EPV for reliable selection may increase to 50.[1]

Most regression shrinkage methods deliberately induce bias in the coefficient estimates, by shrinking them towards 0, in order to reduce the expected variance in the predictions. As a consequence, for models with a binary outcome, these methods aim to prevent predicted risks that are too extreme, i.e. where small risks are underestimated, and high risks overestimated. This leads to better expected mean squared error of the predictions.[18] Since prediction focuses on reliable predictions, inducing bias in the model coefficients is not a key concern. Therefore, it seems that the use of shrinkage methods is always good when sample size is limited.[6] However, some observations are puzzling. Hans van Houwelingen already noted in 2001 that 'it is surprising to observe that the estimated shrinkage factors can be quite off the mark and are negatively correlated with optimal shrinkage factor'.[19] This would imply that shrinkage methods shrink too little when it is really needed, and vice versa. However, van Houwelingen's paper included only small simulation study focusing on uniform shrinkage factors. It is of interest to see whether this also occurs with more modern approaches to regression shrinkage, such as LASSO, Ridge, and Firth's correction.[20-22] Other studies suggest that some methods result in too much shrinkage on average, as indicated by an average calibration slope larger than 1.[9,14,16,23]

The aim of this simulation study was to investigate the performance of various modern shrinkage approaches for the validity of clinical prediction models that are developed with small number of predictors relative to the total sample size (low dimensional). This implies a situation in which some preselection of potentially important predictors has been done before the modeling (e.g. by expert opinion or based on previous studies). We address the performance on average, as well as performance for individual simulation runs. The latter is done by evaluating the between-sample variability in the amount of shrinkage provided by various methods, and the correlation between estimated shrinkage and optimal shrinkage.

2. Methods

2.1. Standard logistic regression

Standard logistic regression is the reference method, in which coefficients are determined by maximum likelihood (ML). Hence, no shrinkage is applied here. When the outcome variably $Y$ equals 1 for an event and 0 for a non-event, ML estimates the probability of an event ($Y = 1$) for patient $i$ ($\pi_i$) using a weighted combination of $p$ predictor variables $X_j$. We define $\pi_i$ as $P(Y = 1|\mathbf{x_i})$, with $i = 1, \ldots, n$, and $\mathbf{x_i}$ a vector of size $p$ with the values of the predictors for patient $i$. Assuming only linear effects and no interactions between the predictors, the logistic regression has the following form:

$$log\left(\frac{\pi_i}{1-\pi_i}\right) = \alpha + \sum_{j=1}^{p} \beta_j x_{ij} = \boldsymbol{\beta}'\mathbf{x_i},$$

where $\pi_i = \frac{\boldsymbol{\beta}'\mathbf{x_i}}{1+\boldsymbol{\beta}'\mathbf{x_i}}$, and $\boldsymbol{\beta}$ the vector containing the intercept $\alpha$ and the coefficients $\beta_j$. Coefficient estimates $\hat{\alpha}$ and $\hat{\beta}_j$ are obtained by finding the maximum of the log-likelihood function:

$$\ell(\boldsymbol{\beta}) = \sum_{i=1}^{n}\{y_i log(\pi_i(\boldsymbol{\beta})) + (1 - y_i)log(1 - \pi_i(\boldsymbol{\beta}))\}.$$

2.2. Shrinkage methods

*Likelihood-based uniform shrinkage (LU)*. Likelihood-Uniform shrinkage uses the likelihood-ratio statistic to compute a uniform shrinkage factor

$$s_{LU} = \frac{\chi^2_{model} - df}{\chi^2_{model}},$$

where $\chi^2_{model}$ is the likelihood-ratio statistic of the fitted model and *df* is the degrees of freedom for the number of candidate predictors considered for the model.[24] The shrunk model coefficients are then calculated as $\beta_{j,LU} = s_{LU}\beta_j$. After adjusting the coefficients, we re-estimated the intercept to guarantee that the average predicted risk equaled the event rate.

*Bootstrap-based uniform shrinkage (BU)*. The uniform shrinkage factor *s* can also be computed using a bootstrap procedure:[12]

1. A bootstrap sample is taken from the original data sample, that is, a random sample with replacement of the same size as the original sample.
2. If a selection procedure was used to select variables this is also applied in the bootstrap samples. The regression coefficients are estimated on the bootstrap sample, $\widehat{\boldsymbol{\beta}}_{\mathbf{bt}}$.
3. The linear predictor for each of the observations in the original sample is calculated using $\widehat{\boldsymbol{\beta}}_{\mathbf{bt}}$.
4. In the original sample, the linear predictor is used to predict the outcome. Retain coefficient for the regression of the linear predictor.
5. Repeat the procedure, steps 1 to 4, and the average coefficient from step 4 provides the shrinkage factor $s_{BU}$. We used 200 repetitions.
6. The shrunk coefficients are calculated as $\beta_{j,BU} = s_{BU}\beta_j$.
7. Re-estimat the intercept using maximum likelihood while keeping $\beta_{j,BU}$ fixed.

*Ridge regression*. Regression shrinkage is implemented via the ridge penalty, also known as the L2-penalty.[20] Ridge regression was extended to logistic regression initially by Schaefer and colleagues, and later by Le Cessie and Van Houwelingen.[25,26] The following penalized version of the log-likelihood function is maximized:

$$\ell(\boldsymbol{\beta}) \quad \lambda \sum_{j=1}^{p} \beta_j^2.$$

The tuning parameter, $\lambda$, controls the amount of shrinkage. The optimal value for this parameter can be estimated by, for example, generalized cross validation (GCV). Ridge regression shrinks the estimated coefficients towards zero (on average), with higher values of $\lambda$ leading to more shrinkage. Note that coefficients will not be shrunk to 0.

*Penalized maximum likelihood*. Similar to ridge, penalized maximum likelihood estimation (PMLE) maximizes a penalized version of $\ell(\boldsymbol{\beta})$.[27] The following function is maximized:

$$\ell(\boldsymbol{\beta}) \quad 0.5\lambda \sum_{j=1}^{p} (s_j \beta_j)^2,$$

where $s_j$ are scaling factors set to be the standard deviation of the predictor.

*LASSO regression*. LASSO is similar to ridge, but uses the L1-penalty that poses a constraint on the sum of the absolute value of the estimated coefficients.[21] For logistic regression, the LASSO optimizes to following function:

$$\ell(\boldsymbol{\beta}) \quad \lambda \sum_{j=1}^{p} |\beta_j|.$$

The L1-penalty allows coefficients to be shrunk to 0, and hence LASSO performs variables selection as well.

*Adaptive LASSO (AL).* The Adaptive LASSO is a variant of the LASSO where a weight is given for each parameter in the penalty term, in order to improve variable selection.[28] The optimized function is:

$$\ell(\boldsymbol{\beta}) \quad \lambda \sum_{j=1}^{p} w_j |\beta_j|,$$

where $w_j = \frac{1}{\left|\hat{\beta}_j^{init}\right|^\gamma}$ contains adaptive weights. The $\hat{\beta}_j^{init}$ are initial coefficient estimates for the predictors. We used the maximum likelihood estimate $\hat{\beta}_j$ as $\hat{\beta}_j^{init}$, and fixed $\gamma$ at 1.[15,28] Adaptive LASSO shrinks higher absolute values of $\hat{\beta}_j^{init}$ less than lower values.

*Non-negative garrote.* The non-negative garrote, developed by Breiman, selects variables and shrinks them at the same time.[29] Contrary to the LASSO, the non-negative garrote estimates do not change signs. This method was later extended to logistic regression.[30] The following function is maximized:

$$\sum_{i=1}^{n}\left\{y_i \log\left(\pi_i\left(\alpha + \sum_{j=1}^{p} x_j \tilde{\beta}_j\right)\right) + (1 - y_i) \log\left(1 - \pi_i\left(\alpha + \sum_{j=1}^{p} x_j \tilde{\beta}_j\right)\right)\right\} - \lambda \sum_{j=1}^{p} c_j,$$

where $\tilde{\beta}_j = c_j \hat{\beta}_j^{init}$, and $c_j \geq 0$. We used the maximum likelihood estimates for the initial coefficient estimates $\hat{\beta}_j^{init}$.

*Firth's penalized likelihood.* Firth developed a procedure to reduce the first order bias in the regression coefficients.[31] To do so, modified score functions are used to estimate model coefficients This avoids problems with separation, but also shrinks the coefficients.[32] In terms of the log-likelihood, Firth's correction optimizes

$$\ell(\boldsymbol{\beta}) + 0.5 \log|I(\boldsymbol{\beta})|,$$

where $I(\boldsymbol{\beta})$ is the Fisher information matrix evaluated at $\boldsymbol{\beta}$. For making predictions based on Firth's correction, we re-estimated the intercept.[16]

### 2.3. Simulation setup

We simulated data to predict a binary outcome. We used a full factorial simulation setup varying the following factors: EPV, the number and strength of predictors, the correlation between predictors, and the outcome event rate (Table 1). In total, this gave us 60 simulation scenarios. In the setting with 5 true predictors, the true coefficients of the predictors were 0.2, 0.2, 0.2, 0.5, and 0.8. These values were based on the Cohen's d measure of effect size, and would correspond to having three weak predictors (odds ratio 1.22), one moderate predictor (odds ratio 1.65), and one strong predictor (odds ratio 2.23).[33] In the setting with 10 true predictors, 6 had a coefficient of 0.2, 2 had a coefficient of 0.5 and 2 had a coefficient of 0.8. Noise predictors had coefficients of 0. The chosen values of the simulation factors had an impact on the true c-statistic (i.e. area under the receiver operating characteristic curve) of the model, the sample size of the simulated datasets, and the number of cases with an event (Table 2).

For every scenario, the simulations were performed as follows. First, for each of 1 million individuals the predictor values were generated by draws from a standard multivariate normal distribution, with equal pairwise correlations. The true model formula (linear predictor) was applied to each patient, with the intercept chosen to obtain the target event rate (Table 2). The inverse logit of the linear predictor was the true risk for that individual. Then, the outcome for each patient was generated through a Bernoulli trial using the true risk. A different dataset, but also with N=1,000,000, was generated for model validation. Predictors and outcomes were generated analogous to the development population, which means that our out-of-sample performance corresponds to a large sample internal validation setting.

We executed 1,000 simulation runs per simulation condition. For each run, we generated a development dataset of the appropriate size (Table 2) by randomly drawing without replacement from the development population. The event rate was fixed at the target value in each development dataset by applying stratified sampling. Next, the predictor variables were standardized, and all types of models were fitted. Ridge, LASSO and adaptive LASSO models models were estimated using the glmnet R package.[34] Penalized maximum likelihood models were developed using the rms R package.[27] Non-negative garrote models were developed using custom-made scripts in R.[30] For ridge, (adaptive) LASSO, and garrote, the tuning parameter was selected from a grid of 251 values between 0 (no shrinkage) and 64 (very large shrinkage). The 250 non-null values were equidistant on logarithmic scale. We used 10-fold cross-validation that minimized the cross-validated deviance. For penalized maximum likelihood, the tuning parameter was chosen using the corrected Akaike Information Criterion using a similar grid.[35] Firth's penalized models were developed using the logistf R package. Simulation samples suggestive of separation when using standard maximum likelihood were omitted from the results: when R warned that there were fitted probabilities of 0 or 1, or when the model did not converge. For logistic regression with bootstrap uniform shrinkage, bootstrap models suggestive of separation were replaced by other bootstrap replicates without separation.

The resulting models were validated on the accompanying full validation dataset. We calculated the the c-statistic and the calibration slope. Because the development and validation data are based on identical populations, the calibration intercept was of little interest and therefore not calculated.[36] At internal validation (i.e. when the underlying population is the same), the calibration slope measures bias of risk predictions in terms of spread.[36,37] A slope<1 suggests that predictions are too extreme: low risks are underestimated, high risks are overestimated. A slope>1 suggests the opposite. We calculated median slopes to assess the deviation from the target value 1. To investigate the variability in the slope, we calculated the median absolute deviation (MAD) of the log(slope). To combine bias (deviation of slope from 1 on average), and variability, we calculated root mean squared distance from the target value (RMSD) of the log(slope) over the 1,000 runs. We used the logarithm of the slope to acknowledge its asymmetry. A slope of 0.5 (half the target) corresponds to a similar quantitative deviation to a slope of 2 (double the target), but in opposite directions. The RMSD was calculated as the square root of the mean of $(log(1) - log(slope))^2$ over the 1,000 runs. Finally, we calculated the Spearman correlation between the estimated shrinkage and the optimal shrinkage over the 1,000 simulation runs. The optimal shrinkage was defined as $log(1) - log(\text{slope}_{ML})$, with $\text{slope}_{ML}$ the slope for the standard maximum likelihood model. The estimated shrinkage for a specific shrinkage approach was defined as $log(\text{slope}_{\text{shrinkage}}) - log(\text{slope}_{ML})$,. To calculate MAD, RMSD, and correlations, we winsorized slopes at 0.01 to avoid problems with rare instances of negative calibration slopes. When no variables were selected by (adaptive) LASSO or garrote, the calibration slope was arbitrarily set at 1000 to reflect the extreme amount of underfitting.

3. Results

There were few runs where separation was suggested (Table S1), except in the scenario with 3 EPV, 10 true predictors, 0.5 correlation and 0.5 event rate. Generally, results differed little between the 5 predictor and 10 prediction scenarios, therefore we focus here on the scenarios with 5 true predictors for the main document. Detailed results for all scenarios are provided in the appendix.

### 3.1. Performance on average

The median calibration slope approached 1 for all methods as EPV increased (Figure 1, Figure S1, Table S2). The standard maximum likelihood model yielded the lowest median calibration slopes. For ridge regression, the median slope at lower EPV values was consistently above 1, suggesting too much shrinkage on average. Penalized maximum likelihood and LASSO were slightly better, but in many scenarios showed median slopes above 1 as well. Other methods generally had median slopes below 1, with bootstrap uniform shrinkage usually having median slopes closest to 1. The use of Firth's correction was slightly better than maximum likelihood. Non-negative garrote had variable performance, but typically resulted in median slopes that were not much better than maximum likelihood.

The average c-statistics also converged to their respective true values as EPV increased (Figure S2). By design, uniform shrinkage had the same c-statistics as regular maximum likelihood. When predictors were correlated, ridge and penalized maximum likelihood had highest c-statistics. When predictors were uncorrelated and no noise predictors were present, LASSO had lower c-statistics than the maximum likelihood model. Adaptive LASSO only had better discrimination than maximum likelihood when noise predictors were present. Firth's correction did not improve the c-statistic. Non-negative garrote had poor c-statistics when event rate was high, and hence overall sample size was low.

### 3.2. Variability in the applied shrinkage

For the scenarios with 5 true predictors, pairwise correlations of 0.5 between predictors, and an event rate of 50%, box plots of the calibration slopes over the 1,000 simulation runs are shown in Figure 2. For all scenarios, box plots are given in Figure S3, and MAD in Figure S4. The variability of the calibration slope after shrinkage was larger than the variability based on maximum likelihood, except when Firth's correction was used. This increased variability was particularly strong when EPV is low, and correlations between predictors were low. Only when there were 10 true predictors with high intercorrelations, most shrinkage methods had lower variability than maximum likelihood.

Generally, shrinkage methods (except the Garrote) improved the RMSD relative to the maximum likelihood model (Table S3, Figure 3, Figure S5). However, LASSO, adaptive LASSO, ridge and penalized maximum likelihood methods often had higher RMSDs than maximum likelihood when predictors were uncorrelated and EPV or sample size was low. Ridge and penalized maximum likelihood often showed higher RMSD than other methods when predictors were correlated and EPV was high. The bootstrap uniform shrinkage method always had lower RMSD than maximum likelihood. This was also the case for Firth's correction, which shrunk little but with little variability.

Box plots of the c-statistics also showed high between-sample variability for all methods (Figure S6).

### 3.3. Correlation between estimated and optimal shrinkage

The Spearman correlation between estimated and optimal shrinkage was typically negative(Figures 4-5, Table S4, Figures S7-8). Firth's correction was the exception with consistently positive correlations. LASSO-based methods typically had the lowest negative correlations (closest to 0). For these methods, correlations where highest, and in particular cases even positive, in settings with more highly correlated predictors. The highest positive correlations between estimated and optimal shrinkage were found when there were 10 true predictors, there was non-zero true correlation between the predictors and the EPV was low.

### 3.4. Results for coefficient estimates and variable selection

Coefficient estimates of true predictors were positively biased (bias away from 0) when the maximum likelihood model was used (Figure S9). The bias decreased with increasing EPV. Using Firth's correction removed the bias. All other shrinkage methods induced negative bias (bias towards 0). At higher EPV, the garrote often did not apply shrinkage at all, such that average bias was about the same as that for maximum likelihood. With respect to noise predictors, ridge, penalized maximum likelihood, LASSO, adaptive LASSO, and garrote had bias in the estimated coefficients when there was correlation between predictors (Figure S10).

Regarding variable selection, adaptive LASSO selected less predictors than standard LASSO implementations (Figure S11). The garrote selected nearly all predictors when event rate was 10%. In simulation scenarios with noise predictors, these predictors were selected more often with increasing EPV, except when adaptive LASSO was used (Figure S12). When event rate was 10%, the garrote selected nearly all noise predictors.

### 4. Discussion

In this paper, we assessed the performance of various shrinkage methods for clinical risk prediction models using simulations. The specific focus was on variation in performance by assessing variability and the correlation between estimated and optimal shrinkage. Our key results were the following. First, shrinkage led to calibration slopes that were on average closer to the ideal value of unity than maximum likelihood. Firth's correction reduced the bias least among the methods that were considered. Ridge, and to a lesser extent penalized maximum likelihood and LASSO, tended to shrink too much overall. Second, despite improved performance on average, the performance of the shrinkage methods was highly variable, especially when sample size was relatively low. The exception was Firth's penalized likelihood, which showed remarkably stable performance. The garrote was notably unpredictable, with a preference for shrinking very strongly or hardly at all. Despite the increased variance, the RMSD of the calibration slopes was usually lower for shrinkage methods compared to standard maximum likelihood model. This was notably the case for Firth's correction, due to its limited variability, but also for bootstrap uniform shrinkage. Third, we commonly observed that the estimated shrinkage was inversely correlated with the optimal shrinkage. This confirmed and corroborated the early observation by van Houwelingen.[19] The negative correlation implies that shrinkage often does least when it is needed most. In addition, shrinkage was often too strong when it was less urgent. Firth's correction is a notable exception, which showed consistently positive correlations. Other exceptions were seen for LASSO methods when predictors were correlated. Fourth, there were differences between the shrinkage methods. Shrinkage using the bootstrap uniform shrinkage factor performed remarkably well, perhaps because this method explicitly uses the calibration slope for shrinkage estimation. Firth's penalized likelihood is interesting in that it almost surely improves the situation over maximum likelihood, with low variability and with more shrinkage when it is needed more. However, the magnitude of shrinkage is small on average.

These results have implications. Although shrinkage works on average by bringing the calibration slope closer to 1, it may not work as anticipated for any given dataset. The variability of uniform shrinkage, ridge, penalized maximum likelihood, LASSO and garrote methods is high, and the estimated shrinkage is often small when it is strongly needed, or large when it is hardly needed. The variability in the estimated shrinkage when sample size was low, suggests that shrinkage methods do not solve the problem of low sample size. Thus, the use of shrinkage does not justify using lower sample size for the development of prediction models. When sample size is low, it may even be advisable not to build a prediction model. Alternatively, a less complicated model can be considered, for example by discarding many predictors a priori. Of course, if clinically important predictors have to be discarded in order to improve the EPV, the relevance of the resulting model should be re-evaluated. Our results advise against a suggestion made in a previous study in the context of survival prediction models.[14] In that study, authors suggested that it may be possible to develop an acceptable model with EPV of 2.5 if methods like ridge or LASSO are used, although acknowledged that more work was required.[14] We cannot defend this suggestion based on our results.

One key issue that contributes to the high variability is that shrinkage parameters have to be estimated as well, such as global shrinkage factors or lambda values. Shrinkage methods are often recommended for small sample size or EPV situations, yet in these very situations it is notably hard to estimate a reasonable value for these parameters. Our results suggest that using shrinkage methods will not solve the problem of low sample size or too many predictors. In a recent systematic review of studies comparing logistic regression with modern machine learning methods (e.g. neural networks or random forests) for low-dimensional risk prediction modeling, no benefit of machine learning on the c-statistic was observed.[38] The median EPV in this review was 8, which is low even for standard logistic regression. When using more flexible algorithms, it becomes even harder to reliably determine the amount of shrinkage due to the large flexibility. This may explain the data hungriness of such flexible algorithms, or that these algorithms mainly thrive under high signal-to-noise settings.[39,40]

Our study focused on low-dimensional settings for which predictors were largely pre-specified. It would be relevant to investigate the issues of high variability and negative correlation in high-dimensional settings, settings where both sample size and the number of potential predictors are large (such as in some electronic health record studies), or in settings with categorical predictors. In addition, we investigated many well-known shrinkage methods. Nevertheless, it may be interesting to investigate whether our findings can be confirmed in other approaches, such as elastic net, smoothly clipped absolute deviation (SCAD), weighted fusion, or machine learning methods.[41-43] We recognize that some methods, e.g. support vector machines with linear kernel, can be seen as a form of penalized regression.[44]

Our work did not intend to lead to updated recommendations regarding appropriate sample sizes or EPV values for prediction modeling studies. However, our results are in line with previous work. In line with a large recent simulation study, model performance in our study was clearly related to event rate even when EPV was fixed.[9] Further, the results are consistent with the recent recommendation to base sample size on a maximal expected level of shrinkage.[10] In accordance with earlier work, we observed that methods like ridge or LASSO may have the tendency to shrink too much on average.[8,14-16] Perhaps the use of 10-fold cross-validation may contribute to this, because this implies that the tuning parameter is searched by training models on only 90% of the available sample size. However, in contrast with earlier statements,[6,14] the bootstrap uniform shrinkage method often performed relatively well in our simulations. These earlier statements seem to be based on simulations with 2.5 EPV, which is lower than the values considered in our study. Our results do not support the development of prediction models with such low EPV with any method, although more work on settings with very low event rates may be of interest.

In conclusion, shrinkage improves performance on average. The larger variability in calibration slope with the use of shrinkage methods, and the negative correlation between estimated and optimal shrinkage, suggest that shrinkage may not work well for any given dataset. The use of shrinkage is not a solution to the problem of low sample size or low EPV. In such cases more fundamental changes are needed, such as refraining from the development of a model, increasing sample size (e.g. using multicenter collaborations), or reducing a priori the number of predictors if this is clinically acceptable.

6. Funding

This research was funded by the Research Foundation – Flanders (FWO; grant G0B4716N), and by Internal Funds KU Leuven (grant C24/15/037).

7. Acknowledgements

We are indebted and grateful to Bavo De Cock for his help in programming the simulations.

Table 1. Overview of the simulation factors in the full factorial simulation design.

| Simulation factor | Factor levels |
| --- | --- |
| Events per variable | 3, 5, 10, 20, 50 |
| Predictors | 5 true predictors; 10 true predictors; 5 true and 5 noise predictors |
| Correlation between predictors | 0, 0.5 |
| Outcome event rate | 0.1, 0.5 |

Table 2. Overview of the characteristics of the 60 simulation scenarios

| Predictors | Correlation | Event rate | EPV | Events | Sample size | True c statistic | Model intercept |
|---|---|---|---|---|---|---|---|
| 5 true predictors, or 5 true + 5 noise predictors | 0 | 0.1 | 3 | 15 | 150 | 0.75 | -2.57 |
| | | | 5 | 25 | 250 | | |
| | | | 10 | 50 | 500 | | |
| | | | 20 | 100 | 1000 | | |
| | | | 50 | 250 | 2500 | | |
| | | 0.5 | 3 | 15 | 30 | 0.74 | 0 |
| | | | 5 | 25 | 50 | | |
| | | | 10 | 50 | 100 | | |
| | | | 20 | 100 | 200 | | |
| | | | 50 | 250 | 500 | | |
| | 0.5 | 0.1 | 3 | 15 | 150 | 0.83 | -2.98 |
| | | | 5 | 25 | 250 | | |
| | | | 10 | 50 | 500 | | |
| | | | 20 | 100 | 1000 | | |
| | | | 50 | 250 | 2500 | | |
| | | 0.5 | 3 | 15 | 30 | 0.81 | 0 |
| | | | 5 | 25 | 50 | | |
| | | | 10 | 50 | 100 | | |
| | | | 20 | 100 | 200 | | |
| | | | 50 | 250 | 500 | | |
| 10 true predictors | 0 | 0.1 | 3 | 30 | 300 | 0.82 | -2.88 |
| | | | 5 | 50 | 500 | | |
| | | | 10 | 100 | 1000 | | |
| | | | 20 | 200 | 2000 | | |
| | | | 50 | 500 | 5000 | | |
| | | 0.5 | 3 | 30 | 60 | 0.80 | 0 |
| | | | 5 | 50 | 100 | | |
| | | | 10 | 100 | 200 | | |
| | | | 20 | 200 | 400 | | |
| | | | 50 | 500 | 1000 | | |
| | 0.5 | 0.1 | 3 | 30 | 300 | 0.93 | -4.34 |
| | | | 5 | 50 | 500 | | |
| | | | 10 | 100 | 1000 | | |
| | | | 20 | 200 | 2000 | | |
| | | | 50 | 500 | 5000 | | |
| | | 0.5 | 3 | 30 | 60 | 0.91 | 0 |
| | | | 5 | 50 | 100 | | |
| | | | 10 | 100 | 200 | | |
| | | | 20 | 200 | 400 | | |
| | | | 50 | 500 | 1000 | | |

Figure 1. Median calibration slopes for the scenarios with 5 true predictors.

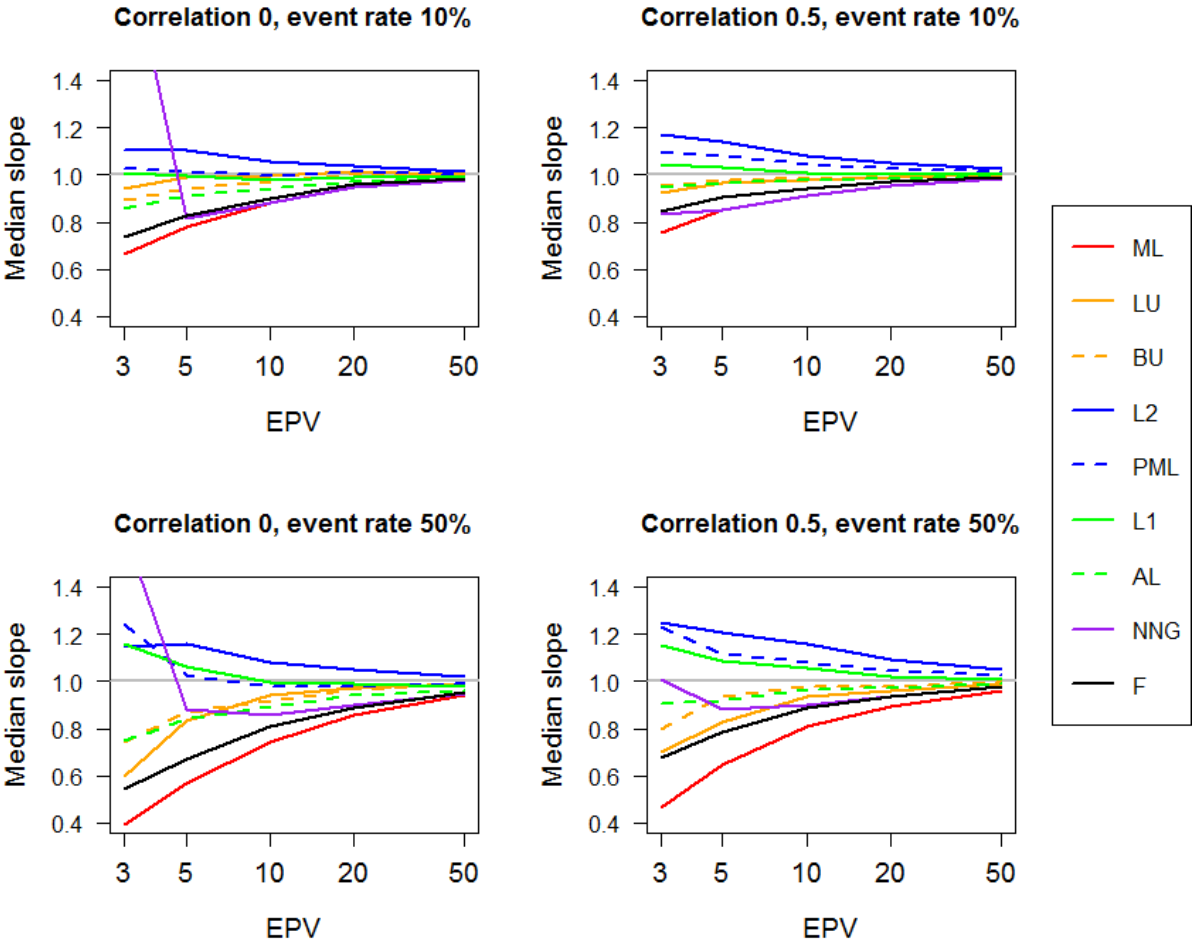

ML, maximum likelihood; LU, uniform shrinkage based on likelihood; BU, uniform shrinkage based on bootstrapping; L2, ridge (L2) regression; PML, penalized maximum likelihood; L1, LASSO (L1) regression; AL, adaptive LASSO; NNG, non-negative garrote; F, logistic regression with Firth's correction.

Figure 2. Box plots of the calibration slope over the 1,000 simulation runs for scenarios with 5 true predictors, no correlation between predictors, and 50% event rate. The events per variable (EPV) is indicated in the top left. The numbers at the bottom are the root mean squared distances (RMSD) of the log of the calibration slopes. The length of the whiskers is at most 1.5 times the interquartile range. Calibration slopes are winsorized at 0.1 and 10 for visualization purposes.

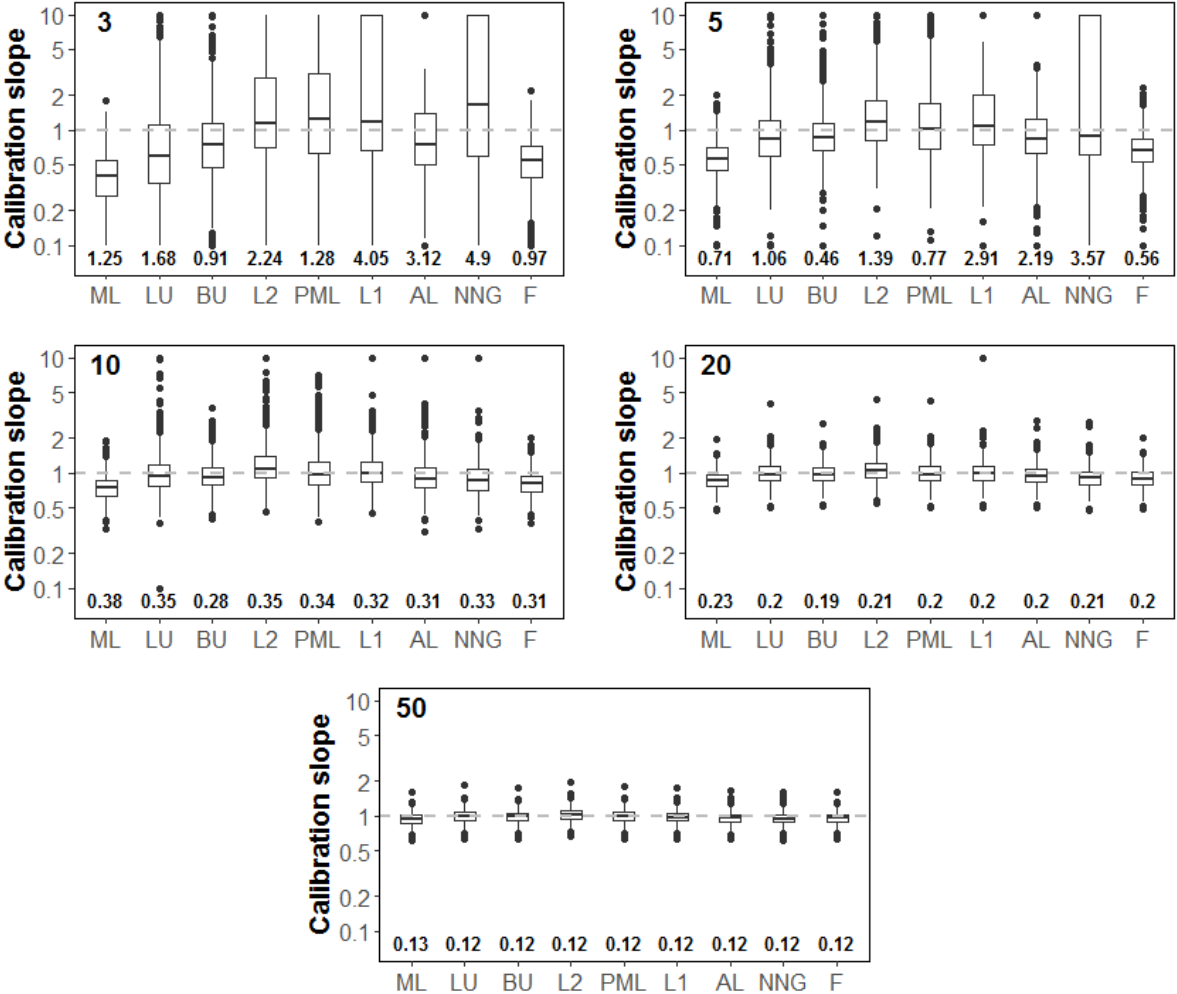

ML, maximum likelihood; LU, uniform shrinkage based on likelihood; BU, uniform shrinkage based on bootstrapping; L2, ridge (L2) regression; PML, penalized maximum likelihood; L1, LASSO (L1) regression; AL, adaptive LASSO; NNG, non-negative garrote; F, logistic regression with Firth's correction.

Figure 3. Root mean squared distance (RMSD) of the logarithm of the calibration slope over 1,000 simulation runs for scenarios with 5 true predictors.

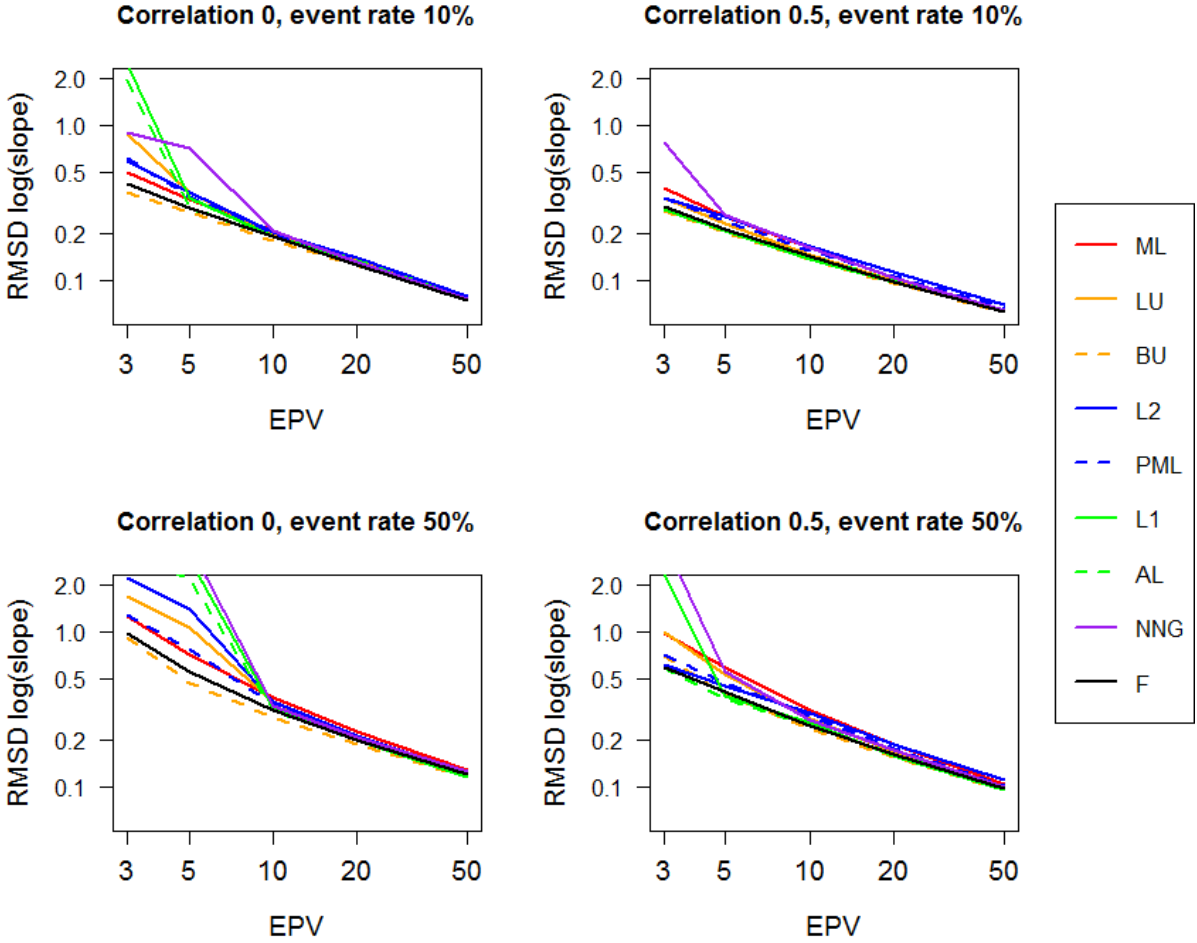

ML, maximum likelihood; LU, uniform shrinkage based on likelihood; BU, uniform shrinkage based on bootstrapping; L2, ridge (L2) regression; PML, penalized maximum likelihood; L1, LASSO (L1) regression; AL, adaptive LASSO; NNG, non-negative garrote; F, logistic regression with Firth's correction.

Figure 4. Scatter plots of the slope after shrinkage versus the slope based on maximum likelihood (no shrinkage) for the scenario with 5 true predictors, 0 correlation between predictors, 50% event rate, and 3 EPV. Each point represents one of the 1,000 simulation runs. The blue line is the diagonal, where both slopes are the same. The green lines show the ideal slope (unity). Red circles refer to simulation runs where maximum likelihood resulted in a slope>1.

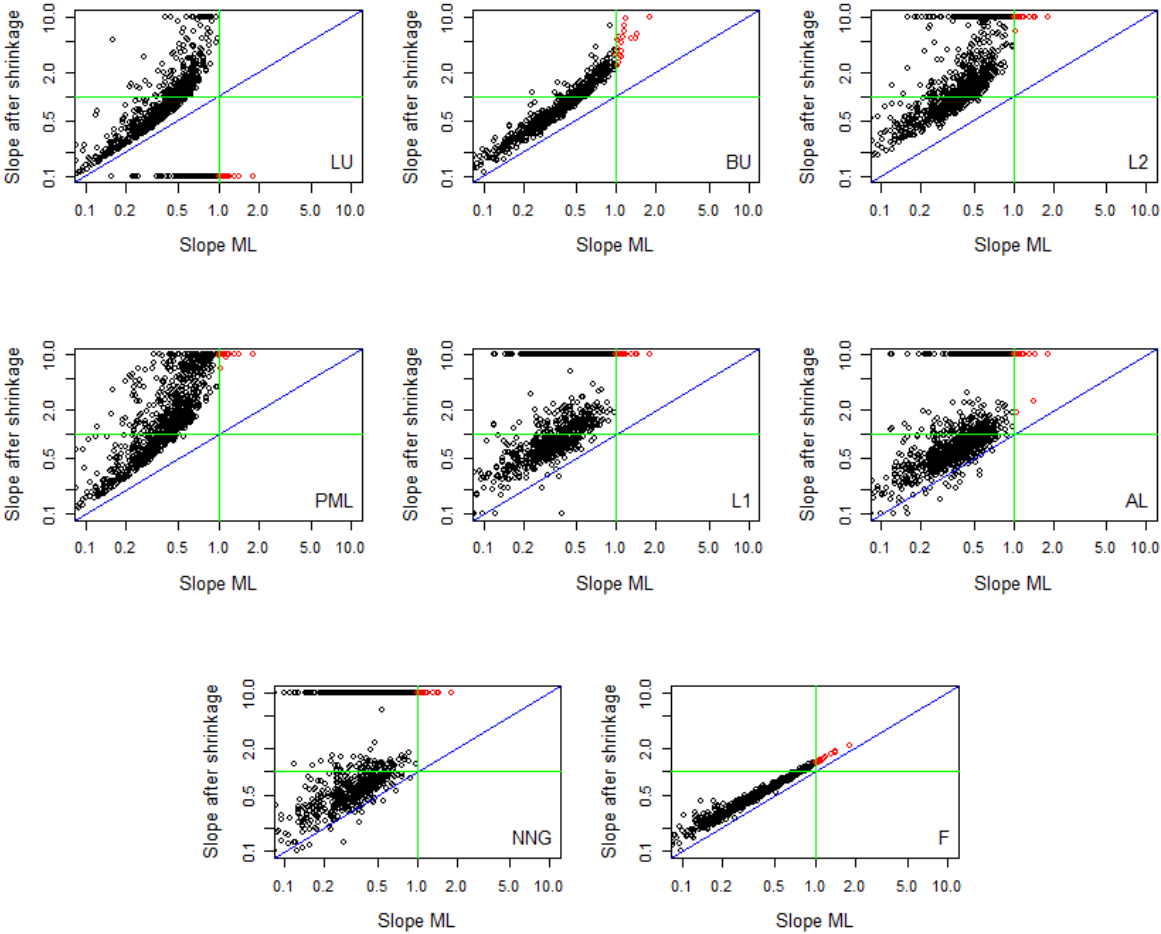

LU, uniform shrinkage based on likelihood; BU, uniform shrinkage based on bootstrapping; L2, ridge (L2) regression; PML, penalized maximum likelihood; L1, LASSO (L1) regression; AL, adaptive LASSO; NNG, non-negative garrote; F, logistic regression with Firth's correction.

Figure 5. Scatter plots of the slope after shrinkage versus the slope without shrinkage (maximum likelihood) for the scenario with 5 true predictors, 0 correlation between predictors, 50% event rate, and 10 EPV. Each point represents one of the 1,000 simulation runs. The blue line is the diagonal, where both slopes are the same. The green lines show the ideal slope (unity). Red circles refer to simulation runs where maximum likelihood resulted in a slope>1.

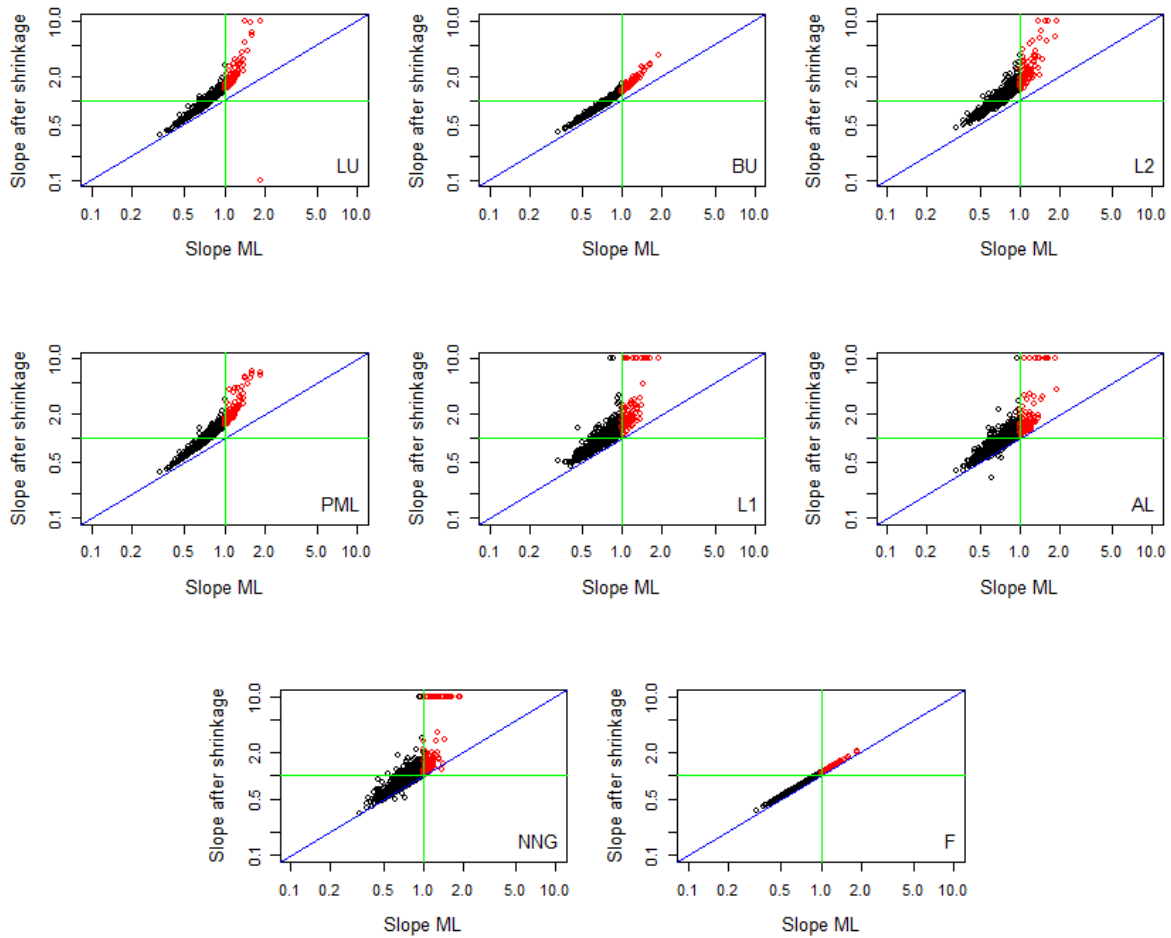

LU, uniform shrinkage based on likelihood; BU, uniform shrinkage based on bootstrapping; L2, ridge (L2) regression; PML, penalized maximum likelihood; L1, LASSO (L1) regression; AL, adaptive LASSO; NNG, non-negative garrote; F, logistic regression with Firth's correction.

**SUPPLEMENTARY MATERIAL**

Table S1. Summary of simulation runs suggestive of separation.

Table S2. Median calibration slope (with $5^{th}$ and $95^{th}$ percentile) by scenario and method.

Table S3. Root mean squared distance (RMSD) of the logarithm of the calibration slope by scenario and method.

Table S4. Correlation between estimated and optimal shrinkage by scenario and method.

Figure S1. Median calibration slopes per scenario and method.

Figure S2. Median c-statistic per scenario and method.

Figure S3. Box plots of calibration slopes over the 1,000 simulation runs for each scenario.

Figure S4. Median absolute deviation (MAD) of the logarithm of the calibration slope.

Figure S5. Root mean squared distance of the target value (RMSD) of the calibration slope.

Figure S6. Box plots of c-statistics over the 1,000 simulation runs for each scenario.

Figure S7. Scatter plots of calibration slopes with shrinkage vs calibration slopes without shrinkage (maximum likelihood).

Figure S8. Correlation between estimated and optimal shrinkage by scenario and method.

Figure S9. Mean bias in true coefficients per scenario and method.

Figure S10. Mean bias in noise coefficients per method, in scenarios with 5 true and 5 noise coefficients.

Figure S11. Mean number of selected variables per scenario, for methods based on LASSO procedures.

Figure S12. Mean number of selected noise coefficients per method, in scenarios with 5 true and 5 noise coefficients.

**On the variability of regression shrinkage methods for clinical prediction models: simulation study on predictive performance**

Ben VAN CALSTER, Maarten VAN SMEDEN, Ewout W STEYERBERG

**SUPPLEMENTARY MATERIAL**

Table S1. Summary of simulation runs suggestive of separation.

| Simulation scenario | Affected runs, n (%) | Events / N |
|---|---|---|
| 3 EPV, 10 true predictors, 0.5 correlation, 0.5 event rate | 124 (12%) | 30 / 60 |
| 3 EPV, 5 true predictors, 0.5 correlation, 0.5 event rate | 37 (4%) | 15 / 30 |
| 3 EPV, 5 true predictors, 0 correlation, 0.5 event rate | 8 (1%) | 15 / 30 |
| 5 EPV, 10 true predictors, 0.5 correlation, 0.5 event rate | 6 (1%) | 50 / 100 |
| 3 EPV, 5 true + 5 noise predictors, 0.5 correlation, 0.5 event rate | 2 (<1%) | 30 / 60 |
| 3 EPV, 5 true + 5 noise predictors, 0 correlation, 0.5 event rate | 1 (<1%) | 30 / 60 |
| 3 EPV, 10 true predictors, 0.5 correlation, 0.1 event rate | 1 (<1%) | 30 / 300 |
| All 53 other scenarios | 0 | NA |

Affected runs are runs where R warned that there were fitted probabilities of 0 or 1, or where the model did not converge. EPV, events per variable.

Table S2. Median calibration slope (with 5th and 95th percentile) by scenario and method.

| Simulation scenario | ML | LU | BU | L2 | PML | L1 | AL | NNG | Firth |
|---|---|---|---|---|---|---|---|---|---|
| 5 true predictors | | | | | | | | | |
| Corr 0, ER 10% | | | | | | | | | |
| EPV 3 | 0.67 (0.41-1.10) | 0.94 (0.45-3.24) | 0.89 (0.53-1.83) | 1.11 (0.60-4.79) | 1.03 (0.52-4.80) | 1.01 (0.54->10) | 0.86 (0.48->10) | 2.07 (0.44-6.86) | 0.74 (0.47-1.21) |
| EPV 5 | 0.78 (0.54-1.18) | 0.99 (0.62-2.00) | 0.94 (0.62-1.58) | 1.10 (0.68-2.33) | 1.01 (0.63-2.07) | 0.99 (0.65-2.19) | 0.91 (0.60-1.62) | 0.82 (0.54-3.64) | 0.82 (0.58-1.23) |
| EPV 10 | 0.88 (0.68-1.17) | 0.99 (0.73-1.43) | 0.97 (0.72-1.34) | 1.05 (0.78-1.56) | 1.00 (0.73-1.45) | 0.98 (0.73-1.44) | 0.94 (0.70-1.32) | 0.88 (0.68-1.18) | 0.90 (0.69-1.20) |
| EPV 20 | 0.95 (0.78-1.17) | 1.01 (0.82-1.28) | 1.00 (0.81-1.25) | 1.04 (0.84-1.32) | 1.01 (0.82-1.28) | 0.99 (0.80-1.25) | 0.97 (0.79-1.24) | 0.95 (0.78-1.17) | 0.96 (0.79-1.18) |
| EPV 50 | 0.98 (0.87-1.12) | 1.00 (0.89-1.16) | 1.00 (0.89-1.15) | 1.02 (0.90-1.17) | 1.00 (0.89-1.16) | 0.99 (0.88-1.13) | 0.99 (0.88-1.13) | 0.98 (0.87-1.12) | 0.98 (0.87-1.13) |
| Corr 0.5, ER 10% | | | | | | | | | |
| EPV 3 | 0.75 (0.47-1.13) | 0.92 (0.54-1.68) | 0.96 (0.59-1.53) | 1.17 (0.70-2.01) | 1.10 (0.62-1.94) | 1.04 (0.65-1.76) | 0.95 (0.60-1.62) | 0.84 (0.48-3.81) | 0.85 (0.55-1.24) |
| EPV 5 | 0.85 (0.62-1.19) | 0.97 (0.67-1.47) | 0.98 (0.71-1.41) | 1.14 (0.80-1.73) | 1.08 (0.75-1.63) | 1.03 (0.73-1.50) | 0.97 (0.69-1.38) | 0.85 (0.62-1.23) | 0.91 (0.67-1.25) |
| EPV 10 | 0.91 (0.73-1.16) | 0.98 (0.77-1.27) | 0.98 (0.79-1.26) | 1.08 (0.84-1.43) | 1.04 (0.81-1.38) | 1.01 (0.80-1.30) | 0.98 (0.77-1.26) | 0.91 (0.73-1.16) | 0.94 (0.76-1.19) |
| EPV 20 | 0.96 (0.81-1.13) | 0.99 (0.83-1.18) | 0.99 (0.84-1.18) | 1.05 (0.88-1.26) | 1.03 (0.86-1.23) | 1.00 (0.85-1.19) | 0.98 (0.83-1.18) | 0.96 (0.81-1.13) | 0.97 (0.83-1.15) |
| EPV 50 | 0.99 (0.89-1.10) | 1.00 (0.90-1.11) | 1.00 (0.90-1.11) | 1.03 (0.92-1.15) | 1.02 (0.91-1.14) | 1.00 (0.90-1.12) | 0.99 (0.89-1.11) | 0.99 (0.89-1.10) | 0.99 (0.89-1.10) |
| Corr 0, ER 50% | | | | | | | | | |
| EPV 3 | 0.40 (0.12-0.82) | 0.60 (-2.02-3.56) | 0.74 (0.18-2.22) | 1.15 (0.32->10) | 1.24 (0.20->10) | 1.16 (0.31->10) | 0.75 (0.24->10) | 1.64 (0.25->10) | 0.54 (0.19-1.07) |
| EPV 5 | 0.57 (0.29-1.00) | 0.83 (0.29-2.90) | 0.87 (0.42-1.92) | 1.16 (0.51->10) | 1.03 (0.40-6.31) | 1.06 (0.46->10) | 0.84 (0.38->10) | 0.88 (0.36->10) | 0.67 (0.36-1.17) |
| EPV 10 | 0.74 (0.50-1.12) | 0.94 (0.57-1.80) | 0.92 (0.60-1.56) | 1.08 (0.67-2.22) | 0.97 (0.59-1.95) | 0.99 (0.63-1.96) | 0.89 (0.57-1.69) | 0.86 (0.55-1.71) | 0.81 (0.54-1.21) |
| EPV 20 | 0.85 (0.65-1.17) | 0.97 (0.71-1.47) | 0.96 (0.72-1.38) | 1.05 (0.77-1.59) | 0.98 (0.71-1.47) | 0.99 (0.72-1.46) | 0.94 (0.70-1.35) | 0.90 (0.67-1.26) | 0.89 (0.68-1.22) |
| EPV 50 | 0.94 (0.77-1.13) | 0.99 (0.81-1.21) | 0.99 (0.81-1.20) | 1.02 (0.84-1.27) | 0.99 (0.81-1.21) | 0.98 (0.81-1.20) | 0.96 (0.79-1.16) | 0.95 (0.78-1.14) | 0.95 (0.79-1.15) |
| Corr 0.5, ER 50% | | | | | | | | | |
| EPV 3 | 0.47 (0.17-1.02) | 0.70 (0.19-2.96) | 0.79 (0.24-2.15) | 1.25 (0.56-3.98) | 1.23 (0.33-3.92) | 1.15 (0.54->10) | 0.90 (0.39-4.68) | 1.01 (0.35->10) | 0.68 (0.32-1.34) |
| EPV 5 | 0.64 (0.33-1.11) | 0.82 (0.39-1.96) | 0.94 (0.47-1.78) | 1.21 (0.64-2.66) | 1.12 (0.51-2.44) | 1.08 (0.59-2.20) | 0.92 (0.51-1.77) | 0.88 (0.47-5.48) | 0.78 (0.43-1.31) |
| EPV 10 | 0.81 (0.55-1.19) | 0.94 (0.60-1.51) | 0.98 (0.66-1.49) | 1.16 (0.75-1.86) | 1.08 (0.68-1.77) | 1.06 (0.70-1.66) | 0.96 (0.64-1.53) | 0.90 (0.60-1.41) | 0.89 (0.61-1.30) |
| EPV 20 | 0.89 (0.69-1.16) | 0.96 (0.73-1.28) | 0.98 (0.76-1.28) | 1.09 (0.83-1.46) | 1.04 (0.78-1.41) | 1.02 (0.78-1.34) | 0.97 (0.74-1.31) | 0.94 (0.72-1.23) | 0.93 (0.73-1.21) |
| EPV 50 | 0.96 (0.82-1.12) | 0.99 (0.84-1.17) | 1.00 (0.85-1.17) | 1.05 (0.89-1.25) | 1.03 (0.87-1.22) | 1.01 (0.86-1.19) | 0.99 (0.84-1.17) | 0.98 (0.83-1.14) | 0.98 (0.84-1.14) |
| 5 true and 5 noise predictors | | | | | | | | | |
| Corr 0, ER 10% | | | | | | | | | |
| EPV 3 | 0.65 (0.46-0.93) | 0.95 (0.57-2.02) | 0.91 (0.60-1.46) | 1.10 (0.67-2.27) | 1.00 (0.59-2.20) | 1.07 (0.64-2.47) | 0.88 (0.56-1.55) | 2.06 (0.48-3.45) | 0.71 (0.50-0.99) |
| EPV 5 | 0.77 (0.59-1.02) | 0.98 (0.69-1.48) | 0.95 (0.71-1.35) | 1.07 (0.75-1.65) | 0.99 (0.70-1.54) | 1.07 (0.73-1.67) | 0.93 (0.68-1.37) | 0.78 (0.59-3.10) | 0.81 (0.62-1.06) |
| EPV 10 | 0.88 (0.73-1.07) | 0.99 (0.79-1.27) | 0.99 (0.80-1.24) | 1.04 (0.84-1.34) | 0.99 (0.79-1.28) | 1.04 (0.82-1.40) | 0.96 (0.78-1.24) | 0.88 (0.73-1.07) | 0.90 (0.74-1.09) |
| EPV 20 | 0.94 (0.83-1.09) | 1.00 (0.87-1.18) | 1.00 (0.87-1.17) | 1.03 (0.89-1.21) | 1.01 (0.87-1.18) | 1.04 (0.89-1.24) | 0.99 (0.86-1.16) | 0.94 (0.83-1.09) | 0.95 (0.84-1.10) |
| EPV 50 | 0.98 (0.89-1.08) | 1.00 (0.91-1.11) | 1.00 (0.91-1.11) | 1.01 (0.92-1.12) | 1.00 (0.91-1.11) | 1.03 (0.93-1.16) | 1.00 (0.91-1.11) | 0.98 (0.89-1.08) | 0.98 (0.90-1.08) |
| Corr 0.5, ER 10% | | | | | | | | | |
| EPV 3 | 0.54 (0.32-0.85) | 0.79 (0.40-1.77) | 0.91 (0.50-1.57) | 1.14 (0.66-2.11) | 1.11 (0.55-2.08) | 1.05 (0.64-1.82) | 0.89 (0.52-1.55) | 2.06 (0.53-3.77) | 0.67 (0.42-1.00) |
| EPV 5 | 0.70 (0.49-0.99) | 0.89 (0.58-1.50) | 0.95 (0.67-1.43) | 1.11 (0.76-1.78) | 1.08 (0.70-1.72) | 1.05 (0.70-1.61) | 0.93 (0.64-1.42) | 0.77 (0.49-3.41) | 0.78 (0.56-1.09) |
| EPV 10 | 0.83 (0.65-1.06) | 0.94 (0.71-1.27) | 0.98 (0.76-1.28) | 1.07 (0.82-1.43) | 1.04 (0.78-1.38) | 1.04 (0.78-1.37) | 0.97 (0.74-1.28) | 0.83 (0.65-1.06) | 0.88 (0.69-1.11) |
| EPV 20 | 0.91 (0.77-1.08) | 0.98 (0.81-1.17) | 0.99 (0.83-1.18) | 1.04 (0.87-1.26) | 1.02 (0.85-1.23) | 1.04 (0.86-1.25) | 0.99 (0.82-1.20) | 0.91 (0.77-1.08) | 0.94 (0.79-1.10) |

| | ML | LU | BU | L2 | PML | L1 | AL | NNG | |
|---|---|---|---|---|---|---|---|---|---|
| EPV 50 | 0.97 (0.87-1.08) | 1.00 (0.89-1.11) | 1.00 (0.90-1.11) | 1.02 (0.92-1.14) | 1.01 (0.91-1.13) | 1.03 (0.92-1.15) | 1.00 (0.89-1.11) | 0.97 (0.87-1.08) | 0.98 (0.88-1.08) |
| Corr 0, ER 50% | | | | | | | | | |
| EPV 3 | 0.38 (0.19-0.66) | 0.68 (0.22-2.92) | 0.81 (0.39-1.66) | 1.14 (0.50->10) | 1.20 (0.37-4.35) | 1.12 (0.46->10) | 0.76 (0.33->10) | 0.97 (0.33->10) | 0.50 (0.26-0.82) |
| EPV 5 | 0.55 (0.35-0.83) | 0.84 (0.46-1.90) | 0.88 (0.54-1.49) | 1.10 (0.63-3.24) | 1.01 (0.52-2.75) | 1.09 (0.60-3.03) | 0.86 (0.50-1.63) | 0.90 (0.46->10) | 0.63 (0.42-0.94) |
| EPV 10 | 0.73 (0.56-0.98) | 0.93 (0.66-1.48) | 0.94 (0.70-1.38) | 1.06 (0.75-1.76) | 0.95 (0.67-1.58) | 1.07 (0.74-1.68) | 0.92 (0.66-1.39) | 0.91 (0.63-1.45) | 0.79 (0.61-1.05) |
| EPV 20 | 0.85 (0.70-1.05) | 0.97 (0.77-1.26) | 0.97 (0.78-1.25) | 1.04 (0.82-1.36) | 0.97 (0.77-1.27) | 1.03 (0.81-1.39) | 0.96 (0.77-1.24) | 0.94 (0.75-1.25) | 0.88 (0.72-1.09) |
| EPV 50 | 0.93 (0.82-1.07) | 0.98 (0.85-1.14) | 0.99 (0.86-1.15) | 1.01 (0.88-1.18) | 0.98 (0.85-1.14) | 1.03 (0.88-1.23) | 0.98 (0.84-1.14) | 0.97 (0.84-1.13) | 0.95 (0.83-1.08) |
| Corr 0.5, ER 50% | | | | | | | | | |
| EPV 3 | 0.45 (0.23-0.76) | 0.67 (0.29-1.59) | 0.91 (0.42-1.67) | 1.17 (0.63-2.26) | 1.15 (0.51-2.28) | 1.07 (0.59-2.17) | 0.87 (0.47-1.65) | 0.95 (0.45->10) | 0.61 (0.36-0.97) |
| EPV 5 | 0.64 (0.42-0.95) | 0.82 (0.49-1.46) | 0.96 (0.63-1.53) | 1.13 (0.72-1.90) | 1.09 (0.65-1.84) | 1.09 (0.69-1.74) | 0.94 (0.59-1.52) | 0.95 (0.56-1.60) | 0.74 (0.50-1.10) |
| EPV 10 | 0.80 (0.62-1.04) | 0.92 (0.69-1.26) | 0.99 (0.75-1.30) | 1.09 (0.82-1.47) | 1.05 (0.78-1.42) | 1.06 (0.77-1.42) | 0.97 (0.71-1.30) | 0.97 (0.69-1.31) | 0.87 (0.67-1.10) |
| EPV 20 | 0.90 (0.75-1.08) | 0.96 (0.80-1.18) | 0.99 (0.83-1.21) | 1.05 (0.87-1.29) | 1.02 (0.84-1.26) | 1.05 (0.86-1.28) | 0.99 (0.81-1.21) | 0.97 (0.79-1.22) | 0.93 (0.78-1.11) |
| EPV 50 | 0.96 (0.85-1.07) | 0.99 (0.87-1.10) | 1.00 (0.89-1.11) | 1.03 (0.91-1.15) | 1.01 (0.89-1.13) | 1.03 (0.91-1.17) | 0.99 (0.88-1.12) | 0.99 (0.88-1.11) | 0.97 (0.86-1.08) |
| 10 true predictors | | | | | | | | | |
| Corr 0, ER 10% | | | | | | | | | |
| EPV 3 | 0.73 (0.54-0.98) | 0.91 (0.62-1.41) | 0.96 (0.70-1.34) | 1.06 (0.75-1.62) | 0.94 (0.63-1.48) | 0.97 (0.70-1.47) | 0.88 (0.63-1.26) | 0.75 (0.54-3.05) | 0.80 (0.59-1.06) |
| EPV 5 | 0.83 (0.66-1.04) | 0.95 (0.73-1.29) | 0.98 (0.77-1.27) | 1.05 (0.81-1.39) | 0.96 (0.73-1.31) | 0.98 (0.75-1.28) | 0.92 (0.72-1.22) | 0.83 (0.66-1.04) | 0.87 (0.70-1.08) |
| EPV 10 | 0.91 (0.77-1.07) | 0.98 (0.81-1.17) | 0.99 (0.83-1.17) | 1.03 (0.86-1.23) | 0.98 (0.81-1.17) | 0.97 (0.82-1.17) | 0.95 (0.80-1.12) | 0.91 (0.77-1.07) | 0.93 (0.79-1.09) |
| EPV 20 | 0.95 (0.85-1.07) | 0.99 (0.88-1.12) | 0.99 (0.89-1.12) | 1.01 (0.90-1.14) | 0.99 (0.88-1.12) | 0.98 (0.87-1.10) | 0.96 (0.86-1.09) | 0.95 (0.85-1.07) | 0.96 (0.86-1.08) |
| EPV 50 | 0.98 (0.91-1.06) | 1.00 (0.93-1.07) | 1.00 (0.93-1.08) | 1.01 (0.94-1.09) | 1.00 (0.92-1.07) | 0.99 (0.92-1.07) | 0.99 (0.92-1.06) | 0.98 (0.91-1.06) | 0.99 (0.92-1.06) |
| Corr 0.5, ER 10% | | | | | | | | | |
| EPV 3 | 0.77 (0.52-1.05) | 0.85 (0.57-1.21) | 1.01 (0.68-1.36) | 1.16 (0.84-1.58) | 1.08 (0.71-1.50) | 1.06 (0.77-1.39) | 0.96 (0.68-1.28) | 0.77 (0.52-1.06) | 0.89 (0.63-1.17) |
| EPV 5 | 0.86 (0.68-1.08) | 0.91 (0.71-1.16) | 1.01 (0.79-1.25) | 1.12 (0.89-1.40) | 1.05 (0.81-1.34) | 1.04 (0.83-1.30) | 0.99 (0.76-1.23) | 0.86 (0.68-1.08) | 0.93 (0.74-1.15) |
| EPV 10 | 0.93 (0.79-1.07) | 0.96 (0.81-1.12) | 1.01 (0.85-1.16) | 1.08 (0.92-1.25) | 1.04 (0.86-1.21) | 1.02 (0.87-1.18) | 1.00 (0.85-1.16) | 0.93 (0.79-1.07) | 0.97 (0.82-1.11) |
| EPV 20 | 0.97 (0.86-1.08) | 0.98 (0.87-1.10) | 1.01 (0.90-1.12) | 1.05 (0.93-1.17) | 1.02 (0.91-1.15) | 1.01 (0.90-1.14) | 1.00 (0.88-1.12) | 0.97 (0.86-1.08) | 0.99 (0.88-1.10) |
| EPV 50 | 0.99 (0.92-1.06) | 0.99 (0.93-1.07) | 1.00 (0.94-1.07) | 1.02 (0.96-1.09) | 1.01 (0.94-1.08) | 1.00 (0.94-1.07) | 0.99 (0.93-1.07) | 0.99 (0.92-1.06) | 0.99 (0.93-1.07) |
| Corr 0, ER 50% | | | | | | | | | |
| EPV 3 | 0.44 (0.24-0.75) | 0.67 (0.32-1.77) | 0.90 (0.45-1.66) | 1.09 (0.58-3.13) | 0.96 (0.37-3.34) | 0.94 (0.51->10) | 0.73 (0.40-1.54) | 0.83 (0.40->10) | 0.59 (0.36-0.95) |
| EPV 5 | 0.62 (0.41-0.91) | 0.82 (0.50-1.51) | 0.94 (0.63-1.49) | 1.08 (0.68-1.94) | 0.90 (0.52-1.83) | 0.96 (0.63-1.85) | 0.83 (0.55-1.39) | 0.82 (0.51-1.48) | 0.73 (0.50-1.05) |
| EPV 10 | 0.79 (0.62-1.03) | 0.92 (0.69-1.29) | 0.97 (0.75-1.32) | 1.05 (0.79-1.47) | 0.93 (0.69-1.32) | 0.97 (0.75-1.39) | 0.91 (0.69-1.25) | 0.86 (0.66-1.22) | 0.85 (0.67-1.11) |
| EPV 20 | 0.89 (0.74-1.07) | 0.96 (0.78-1.18) | 0.99 (0.81-1.20) | 1.04 (0.85-1.29) | 0.96 (0.79-1.18) | 0.98 (0.81-1.20) | 0.94 (0.78-1.16) | 0.91 (0.75-1.10) | 0.92 (0.77-1.10) |
| EPV 50 | 0.95 (0.85-1.08) | 0.98 (0.87-1.12) | 0.99 (0.88-1.13) | 1.01 (0.90-1.15) | 0.98 (0.87-1.12) | 0.98 (0.87-1.10) | 0.96 (0.86-1.09) | 0.95 (0.85-1.08) | 0.97 (0.86-1.09) |
| Corr 0.5, ER 50% | | | | | | | | | |
| EPV 3 | 0.47 (0.23-0.85) | 0.59 (0.28-1.18) | 0.86 (0.36-1.67) | 1.27 (0.83-1.97) | 1.18 (0.60-1.93) | 1.13 (0.71-1.75) | 0.90 (0.56-1.42) | 0.96 (0.47-1.76) | 0.75 (0.48-1.15) |
| EPV 5 | 0.66 (0.39-0.97) | 0.76 (0.43-1.17) | 1.03 (0.59-1.46) | 1.19 (0.80-1.67) | 1.08 (0.61-1.61) | 1.07 (0.72-1.49) | 0.92 (0.62-1.34) | 0.90 (0.60-1.27) | 0.83 (0.54-1.15) |
| EPV 10 | 0.82 (0.62-1.06) | 0.89 (0.66-1.16) | 1.01 (0.77-1.27) | 1.14 (0.86-1.45) | 1.05 (0.77-1.39) | 1.05 (0.82-1.35) | 0.97 (0.74-1.25) | 0.91 (0.69-1.19) | 0.91 (0.70-1.16) |
| EPV 20 | 0.91 (0.74-1.09) | 0.94 (0.77-1.14) | 1.00 (0.82-1.20) | 1.08 (0.89-1.29) | 1.03 (0.84-1.25) | 1.03 (0.85-1.23) | 0.99 (0.82-1.18) | 0.94 (0.77-1.12) | 0.96 (0.79-1.14) |
| EPV 50 | 0.97 (0.86-1.06) | 0.98 (0.87-1.08) | 1.00 (0.89-1.11) | 1.04 (0.92-1.16) | 1.02 (0.90-1.12) | 1.01 (0.90-1.12) | 0.99 (0.88-1.10) | 0.97 (0.86-1.07) | 0.98 (0.87-1.08) |

Corr, correlation; ER, event rate; EPV, events per variable; ML, maximum likelihood; LU, uniform shrinkage based on likelihood; BU, uniform shrinkage based on bootstrap; L2, ridge (L2 penalty); PML, penalized maximum likelihood; L1, LASSO (L1 penalty); AL, adaptive LASSO; NNG, non-negative garrote.

Table S3. Root mean squared distance (RMSD) of the logarithm of the calibration slope by scenario and method.

| Simulation scenario | ML | LU | BU | L2 | PML | L1 | AL | NNG | Firth |
|---|---|---|---|---|---|---|---|---|---|
| 5 true predictors | | | | | | | | | |
| Corr 0, ER 10% | | | | | | | | | |
| EPV 3 | 0.50 | 0.87 | 0.37 | 0.58 | 0.61 | 2.47 | 1.94 | 0.90 | 0.41 |
| EPV 5 | 0.33 | 0.37 | 0.27 | 0.37 | 0.35 | 0.34 | 0.29 | 0.71 | 0.29 |
| EPV 10 | 0.21 | 0.20 | 0.18 | 0.21 | 0.20 | 0.19 | 0.19 | 0.21 | 0.19 |
| EPV 20 | 0.13 | 0.13 | 0.13 | 0.14 | 0.13 | 0.13 | 0.13 | 0.13 | 0.13 |
| EPV 50 | 0.08 | 0.08 | 0.08 | 0.08 | 0.08 | 0.07 | 0.07 | 0.08 | 0.07 |
| Corr 0.5, ER 10% | | | | | | | | | |
| EPV 3 | 0.39 | 0.34 | 0.28 | 0.34 | 0.34 | 0.29 | 0.29 | 0.77 | 0.30 |
| EPV 5 | 0.26 | 0.23 | 0.20 | 0.26 | 0.24 | 0.21 | 0.21 | 0.26 | 0.21 |
| EPV 10 | 0.16 | 0.15 | 0.14 | 0.17 | 0.15 | 0.14 | 0.14 | 0.16 | 0.14 |
| EPV 20 | 0.10 | 0.10 | 0.10 | 0.11 | 0.11 | 0.10 | 0.10 | 0.10 | 0.10 |
| EPV 50 | 0.06 | 0.06 | 0.06 | 0.07 | 0.07 | 0.06 | 0.06 | 0.06 | 0.06 |
| Corr 0, ER 50% | | | | | | | | | |
| EPV 3 | 1.25 | 1.68 | 0.91 | 2.24 | 1.28 | 4.05 | 3.12 | 4.90 | 0.97 |
| EPV 5 | 0.71 | 1.06 | 0.46 | 1.39 | 0.77 | 2.91 | 2.19 | 3.57 | 0.56 |
| EPV 10 | 0.38 | 0.35 | 0.28 | 0.35 | 0.34 | 0.32 | 0.31 | 0.33 | 0.31 |
| EPV 20 | 0.23 | 0.20 | 0.19 | 0.21 | 0.20 | 0.20 | 0.20 | 0.21 | 0.20 |
| EPV 50 | 0.13 | 0.12 | 0.12 | 0.12 | 0.12 | 0.12 | 0.12 | 0.12 | 0.12 |
| Corr 0.5, ER 50% | | | | | | | | | |
| EPV 3 | 0.96 | 0.99 | 0.70 | 0.61 | 0.70 | 2.33 | 0.58 | 3.77 | 0.59 |
| EPV 5 | 0.58 | 0.53 | 0.39 | 0.45 | 0.47 | 0.39 | 0.37 | 0.55 | 0.41 |
| EPV 10 | 0.31 | 0.27 | 0.24 | 0.30 | 0.28 | 0.25 | 0.25 | 0.27 | 0.25 |
| EPV 20 | 0.19 | 0.17 | 0.16 | 0.19 | 0.18 | 0.16 | 0.16 | 0.17 | 0.16 |
| EPV 50 | 0.10 | 0.10 | 0.10 | 0.11 | 0.10 | 0.09 | 0.10 | 0.10 | 0.10 |
| 5 true and 5 noise predictors | | | | | | | | | |
| Corr 0, ER 10% | | | | | | | | | |
| EPV 3 | 0.48 | 0.38 | 0.27 | 0.37 | 0.37 | 0.38 | 0.31 | 0.81 | 0.41 |
| EPV 5 | 0.30 | 0.22 | 0.19 | 0.24 | 0.23 | 0.24 | 0.21 | 0.56 | 0.26 |
| EPV 10 | 0.17 | 0.14 | 0.13 | 0.14 | 0.14 | 0.16 | 0.14 | 0.17 | 0.16 |
| EPV 20 | 0.10 | 0.09 | 0.09 | 0.10 | 0.09 | 0.11 | 0.09 | 0.10 | 0.09 |
| EPV 50 | 0.06 | 0.06 | 0.06 | 0.06 | 0.06 | 0.07 | 0.06 | 0.06 | 0.06 |
| Corr 0.5, ER 10% | | | | | | | | | |
| EPV 3 | 0.69 | 0.50 | 0.35 | 0.36 | 0.38 | 0.31 | 0.34 | 0.81 | 0.48 |
| EPV 5 | 0.42 | 0.29 | 0.22 | 0.26 | 0.27 | 0.24 | 0.24 | 0.73 | 0.32 |
| EPV 10 | 0.23 | 0.17 | 0.15 | 0.17 | 0.17 | 0.16 | 0.16 | 0.23 | 0.19 |
| EPV 20 | 0.13 | 0.11 | 0.10 | 0.11 | 0.11 | 0.11 | 0.11 | 0.13 | 0.12 |
| EPV 50 | 0.07 | 0.06 | 0.06 | 0.07 | 0.06 | 0.07 | 0.06 | 0.07 | 0.06 |
| Corr 0, ER 50% | | | | | | | | | |
| EPV 3 | 1.08 | 1.08 | 0.52 | 1.49 | 0.76 | 2.87 | 1.85 | 3.80 | 0.84 |
| EPV 5 | 0.65 | 0.55 | 0.32 | 0.47 | 0.48 | 0.48 | 0.39 | 1.87 | 0.52 |
| EPV 10 | 0.35 | 0.24 | 0.20 | 0.24 | 0.24 | 0.25 | 0.22 | 0.25 | 0.29 |
| EPV 20 | 0.20 | 0.15 | 0.13 | 0.15 | 0.15 | 0.16 | 0.15 | 0.16 | 0.17 |
| EPV 50 | 0.11 | 0.09 | 0.09 | 0.09 | 0.09 | 0.10 | 0.09 | 0.09 | 0.10 |
| Corr 0.5, ER 50% | | | | | | | | | |
| EPV 3 | 0.90 | 0.67 | 0.43 | 0.39 | 0.46 | 0.37 | 0.38 | 1.94 | 0.59 |
| EPV 5 | 0.52 | 0.37 | 0.26 | 0.31 | 0.32 | 0.28 | 0.29 | 0.33 | 0.38 |
| EPV 10 | 0.27 | 0.20 | 0.16 | 0.19 | 0.19 | 0.18 | 0.18 | 0.20 | 0.21 |
| EPV 20 | 0.15 | 0.12 | 0.11 | 0.13 | 0.12 | 0.13 | 0.12 | 0.13 | 0.13 |
| EPV 50 | 0.08 | 0.07 | 0.07 | 0.07 | 0.07 | 0.08 | 0.07 | 0.07 | 0.07 |
| 10 true predictors | | | | | | | | | |
| Corr 0, ER 10% | | | | | | | | | |
| EPV 3 | 0.36 | 0.24 | 0.20 | 0.23 | 0.24 | 0.21 | 0.24 | 0.61 | 0.28 |
| EPV 5 | 0.23 | 0.17 | 0.15 | 0.16 | 0.17 | 0.16 | 0.17 | 0.23 | 0.19 |
| EPV 10 | 0.14 | 0.11 | 0.10 | 0.11 | 0.11 | 0.11 | 0.11 | 0.14 | 0.12 |
| EPV 20 | 0.08 | 0.07 | 0.07 | 0.07 | 0.07 | 0.07 | 0.08 | 0.08 | 0.08 |
| EPV 50 | 0.05 | 0.04 | 0.04 | 0.04 | 0.04 | 0.04 | 0.04 | 0.05 | 0.04 |
| Corr 0.5, ER 10% | | | | | | | | | |
| EPV 3 | 0.34 | 0.28 | 0.20 | 0.23 | 0.22 | 0.18 | 0.19 | 0.34 | 0.22 |
| EPV 5 | 0.21 | 0.18 | 0.14 | 0.18 | 0.16 | 0.14 | 0.14 | 0.21 | 0.15 |
| EPV 10 | 0.12 | 0.10 | 0.09 | 0.12 | 0.10 | 0.09 | 0.09 | 0.12 | 0.10 |
| EPV 20 | 0.08 | 0.07 | 0.07 | 0.08 | 0.07 | 0.07 | 0.07 | 0.08 | 0.07 |
| EPV 50 | 0.04 | 0.04 | 0.04 | 0.04 | 0.04 | 0.04 | 0.04 | 0.04 | 0.04 |
| Corr 0, ER 50% | | | | | | | | | |
| EPV 3 | 0.90 | 0.65 | 0.41 | 0.49 | 0.63 | 1.86 | 0.49 | 2.88 | 0.61 |
| EPV 5 | 0.54 | 0.37 | 0.26 | 0.32 | 0.37 | 0.30 | 0.32 | 0.36 | 0.39 |
| EPV 10 | 0.28 | 0.20 | 0.17 | 0.19 | 0.20 | 0.18 | 0.19 | 0.23 | 0.22 |
| EPV 20 | 0.16 | 0.13 | 0.12 | 0.13 | 0.13 | 0.12 | 0.13 | 0.15 | 0.14 |
| EPV 50 | 0.09 | 0.07 | 0.07 | 0.07 | 0.07 | 0.07 | 0.08 | 0.09 | 0.08 |
| Corr 0.5, ER 50% | | | | | | | | | |
| EPV 3 | 0.88 | 0.70 | 0.51 | 0.35 | 0.38 | 0.28 | 0.29 | 0.40 | 0.39 |
| EPV 5 | 0.52 | 0.42 | 0.27 | 0.27 | 0.29 | 0.22 | 0.25 | 0.26 | 0.30 |
| EPV 10 | 0.26 | 0.21 | 0.15 | 0.19 | 0.18 | 0.15 | 0.16 | 0.19 | 0.18 |
| EPV 20 | 0.15 | 0.13 | 0.11 | 0.13 | 0.12 | 0.11 | 0.11 | 0.13 | 0.12 |
| EPV 50 | 0.08 | 0.07 | 0.06 | 0.08 | 0.07 | 0.06 | 0.07 | 0.07 | 0.07 |

Corr, correlation; ER, event rate; EPV, events per variable; ML, maximum likelihood; LU, uniform shrinkage based on likelihood; BU, uniform shrinkage based on bootstrap; L2, ridge (L2 penalty); PML, penalized maximum likelihood; L1, LASSO (L1 penalty); AL, adaptive LASSO; NNG, non-negative garrote.

Table S4. Spearman correlation between estimated and optimal shrinkage by scenario and method. Correlations lower than 0 are marked in red.

| Simulation scenario | LU | BU | L2 | PML | L1 | AL | NNG | Firth |
|---|---|---|---|---|---|---|---|---|
| 5 true predictors | | | | | | | | |
| Corr 0, ER 10% | | | | | | | | |
| EPV 3 | -0.71 | -0.64 | -0.58 | -0.83 | -0.43 | -0.29 | -0.31 | 0.64 |
| EPV 5 | -0.90 | -0.75 | -0.68 | -0.91 | -0.39 | -0.22 | -0.33 | 0.68 |
| EPV 10 | -0.94 | -0.72 | -0.58 | -0.96 | -0.17 | -0.15 | 0.33 | 0.76 |
| EPV 20 | -0.97 | -0.56 | -0.46 | -0.98 | -0.05 | -0.16 | 0.67 | 0.79 |
| EPV 50 | -0.97 | -0.31 | -0.31 | -0.99 | 0.06 | 0.02 | 0.72 | 0.80 |
| Corr 0.5, ER 10% | | | | | | | | |
| EPV 3 | -0.89 | -0.33 | -0.26 | -0.74 | -0.02 | 0.00 | -0.40 | 0.80 |
| EPV 5 | -0.92 | -0.33 | -0.36 | -0.82 | -0.01 | -0.02 | -0.15 | 0.83 |
| EPV 10 | -0.95 | -0.33 | -0.37 | -0.80 | 0.01 | -0.06 | -0.19 | 0.85 |
| EPV 20 | -0.95 | -0.19 | -0.30 | -0.71 | 0.01 | -0.02 | -0.44 | 0.87 |
| EPV 50 | -0.97 | -0.12 | -0.23 | -0.65 | -0.05 | 0.00 | -0.80 | 0.88 |
| Corr 0, ER 50% | | | | | | | | |
| EPV 3 | -0.35 | -0.64 | -0.42 | -0.61 | -0.15 | -0.09 | -0.05 | 0.59 |
| EPV 5 | -0.60 | -0.62 | -0.54 | -0.74 | -0.36 | -0.19 | -0.31 | 0.62 |
| EPV 10 | -0.85 | -0.71 | -0.58 | -0.86 | -0.34 | -0.21 | -0.21 | 0.58 |
| EPV 20 | -0.93 | -0.73 | -0.52 | -0.95 | -0.23 | -0.15 | -0.11 | 0.51 |
| EPV 50 | -0.96 | -0.49 | -0.43 | -0.98 | 0.00 | -0.04 | -0.01 | 0.50 |
| Corr 0.5, ER 50% | | | | | | | | |
| EPV 3 | -0.74 | -0.69 | 0.20 | -0.28 | 0.11 | 0.29 | -0.11 | 0.77 |
| EPV 5 | -0.88 | -0.41 | -0.11 | -0.59 | 0.00 | 0.15 | -0.05 | 0.80 |
| EPV 10 | -0.93 | -0.42 | -0.37 | -0.79 | -0.10 | 0.02 | -0.01 | 0.81 |
| EPV 20 | -0.96 | -0.35 | -0.38 | -0.82 | -0.03 | -0.01 | -0.01 | 0.82 |
| EPV 50 | -0.97 | -0.18 | -0.25 | -0.74 | 0.05 | 0.02 | 0.06 | 0.81 |
| 5 true and 5 noise predictors | | | | | | | | |
| Corr 0, ER 10% | | | | | | | | |
| EPV 3 | -0.81 | -0.71 | -0.65 | -0.83 | -0.44 | -0.23 | -0.37 | 0.68 |
| EPV 5 | -0.90 | -0.80 | -0.68 | -0.91 | -0.41 | -0.22 | -0.11 | 0.71 |
| EPV 10 | -0.94 | -0.78 | -0.61 | -0.95 | -0.40 | -0.28 | 0.62 | 0.76 |
| EPV 20 | -0.96 | -0.66 | -0.53 | -0.97 | -0.25 | -0.13 | 0.66 | 0.83 |
| EPV 50 | -0.98 | -0.41 | -0.37 | -0.98 | -0.18 | -0.09 | 0.62 | 0.84 |
| Corr 0.5, ER 10% | | | | | | | | |
| EPV 3 | -0.83 | -0.44 | 0.07 | -0.28 | 0.23 | 0.18 | 0.21 | 0.81 |
| EPV 5 | -0.88 | -0.32 | -0.16 | -0.53 | -0.01 | 0.01 | -0.45 | 0.82 |
| EPV 10 | -0.92 | -0.49 | -0.37 | -0.74 | -0.09 | -0.08 | -0.43 | 0.85 |
| EPV 20 | -0.95 | -0.50 | -0.37 | -0.80 | -0.12 | -0.11 | -0.60 | 0.87 |
| EPV 50 | -0.96 | -0.26 | -0.21 | -0.83 | -0.05 | -0.06 | -0.79 | 0.87 |
| Corr 0, ER 50% | | | | | | | | |
| EPV 3 | -0.54 | -0.52 | -0.34 | -0.52 | -0.13 | -0.02 | -0.08 | 0.64 |
| EPV 5 | -0.74 | -0.67 | -0.55 | -0.74 | -0.32 | -0.13 | -0.30 | 0.63 |
| EPV 10 | -0.88 | -0.78 | -0.61 | -0.90 | -0.44 | -0.16 | -0.28 | 0.66 |
| EPV 20 | -0.93 | -0.80 | -0.57 | -0.95 | -0.34 | -0.16 | -0.21 | 0.63 |
| EPV 50 | -0.97 | -0.59 | -0.44 | -0.98 | -0.26 | -0.15 | -0.16 | 0.65 |
| Corr 0.5, ER 50% | | | | | | | | |
| EPV 3 | -0.81 | -0.50 | 0.25 | -0.03 | 0.29 | 0.33 | 0.04 | 0.80 |
| EPV 5 | -0.88 | -0.32 | -0.10 | -0.39 | 0.08 | 0.08 | -0.08 | 0.82 |
| EPV 10 | -0.93 | -0.52 | -0.30 | -0.68 | -0.02 | 0.00 | -0.14 | 0.83 |
| EPV 20 | -0.95 | -0.49 | -0.33 | -0.79 | -0.02 | -0.06 | -0.12 | 0.84 |
| EPV 50 | -0.97 | -0.26 | -0.23 | -0.80 | -0.01 | -0.07 | -0.04 | 0.84 |
| 10 true predictors | | | | | | | | |
| Corr 0, ER 10% | | | | | | | | |
| EPV 3 | -0.86 | -0.49 | -0.43 | -0.87 | -0.21 | -0.06 | -0.50 | 0.79 |
| EPV 5 | -0.91 | -0.58 | -0.47 | -0.93 | -0.15 | -0.11 | -0.15 | 0.84 |
| EPV 10 | -0.94 | -0.52 | -0.38 | -0.96 | 0.02 | -0.03 | -0.06 | 0.86 |
| EPV 20 | -0.94 | -0.37 | -0.26 | -0.97 | 0.04 | 0.00 | -0.20 | 0.84 |
| EPV 50 | -0.95 | -0.16 | -0.17 | -0.97 | 0.09 | 0.07 | -0.55 | 0.84 |
| Corr 0.5, ER 10% | | | | | | | | |
| EPV 3 | -0.91 | 0.18 | 0.43 | -0.30 | 0.53 | 0.43 | -0.06 | 0.91 |
| EPV 5 | -0.92 | 0.43 | 0.18 | -0.52 | 0.34 | 0.19 | -0.30 | 0.91 |

|  | ML | LU | BU | L2 | PML | L1 | AL | NNG |
|---|---|---|---|---|---|---|---|---|
| EPV 10 | -0.92 | 0.27 | 0.08 | -0.56 | 0.25 | 0.14 | -0.54 | 0.90 |
| EPV 20 | -0.93 | 0.11 | -0.07 | -0.49 | 0.14 | 0.01 | -0.80 | 0.92 |
| EPV 50 | -0.93 | 0.04 | -0.03 | -0.42 | 0.14 | -0.01 | -0.92 | 0.91 |
| Corr 0, ER 50% | | | | | | | | |
| EPV 3 | -0.79 | -0.55 | -0.28 | -0.73 | -0.08 | 0.13 | -0.09 | 0.75 |
| EPV 5 | -0.85 | -0.38 | -0.42 | -0.85 | -0.15 | 0.02 | -0.05 | 0.78 |
| EPV 10 | -0.89 | -0.57 | -0.42 | -0.91 | -0.19 | -0.09 | -0.05 | 0.78 |
| EPV 20 | -0.93 | -0.53 | -0.34 | -0.96 | -0.08 | -0.11 | -0.01 | 0.81 |
| EPV 50 | -0.94 | -0.34 | -0.23 | -0.98 | 0.10 | 0.05 | 0.04 | 0.81 |
| Corr 0.5, ER 50% | | | | | | | | |
| EPV 3 | -0.87 | -0.61 | 0.72 | 0.43 | 0.70 | 0.67 | 0.39 | 0.89 |
| EPV 5 | -0.90 | 0.19 | 0.54 | -0.11 | 0.59 | 0.46 | 0.46 | 0.89 |
| EPV 10 | -0.92 | 0.47 | 0.16 | -0.53 | 0.34 | 0.25 | 0.18 | 0.92 |
| EPV 20 | -0.94 | 0.22 | 0.05 | -0.59 | 0.26 | 0.17 | 0.12 | 0.92 |
| EPV 50 | -0.95 | 0.14 | -0.01 | -0.55 | 0.14 | 0.06 | 0.06 | 0.92 |

Corr, correlation; ER, event rate; EPV, events per variable; ML, maximum likelihood; LU, uniform shrinkage based on likelihood; BU, uniform shrinkage based on bootstrap; L2, ridge (L2 penalty); PML, penalized maximum likelihood; L1, LASSO (L1 penalty); AL, adaptive LASSO; NNG, non-negative garrote.

Figure S1. Median calibration slopes per scenario and method. ML, maximum likelihood; LU, uniform shrinkage based on likelihood; BU, uniform shrinkage based on bootstrap; L2, ridge (L2 penalty); PML, penalized maximum likelihood; L1, LASSO (L1 penalty); AL, adaptive LASSO; NNG, non-negative garrote; F, Firth's correction.

A. Scenarios with 5 true predictors

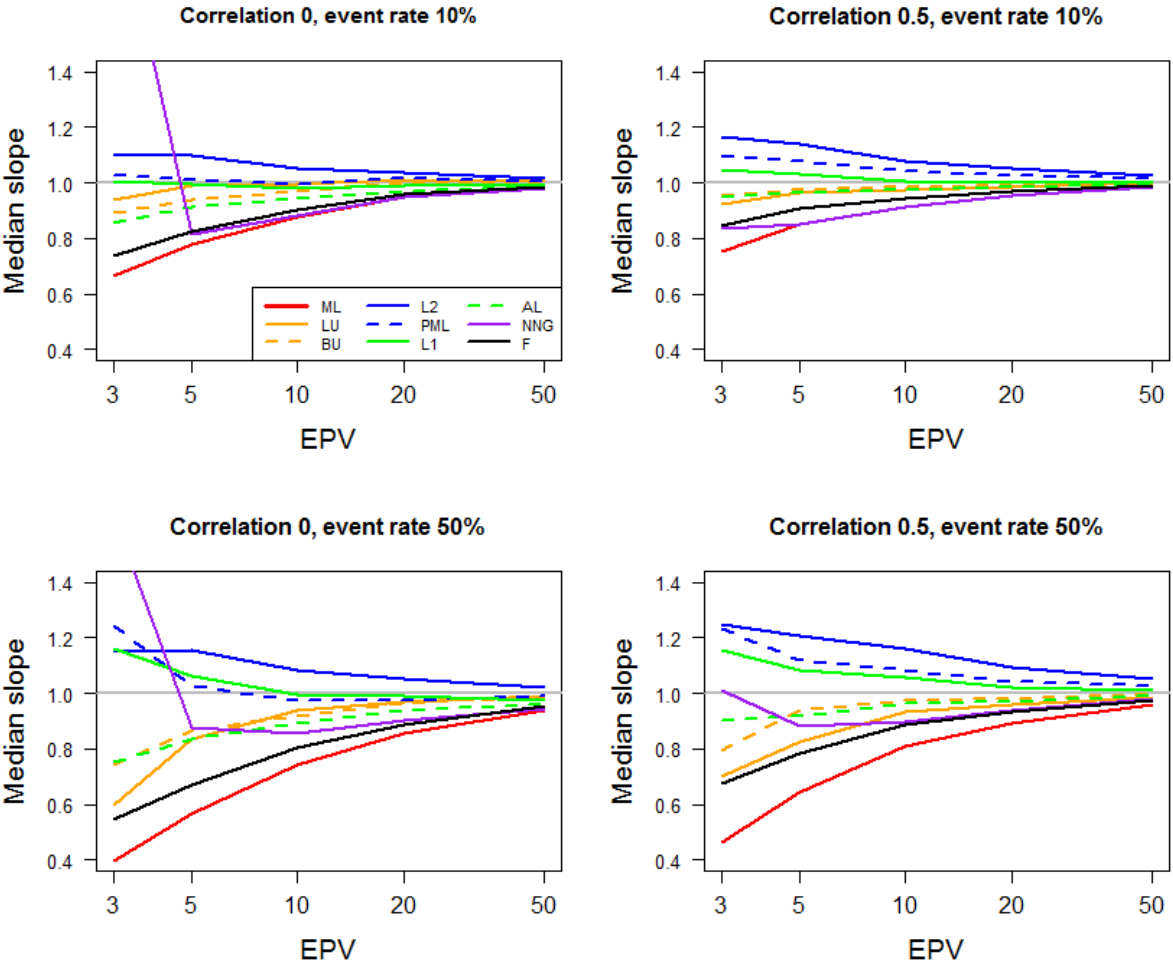

B. Scenarios with 5 true and 5 noise predictors

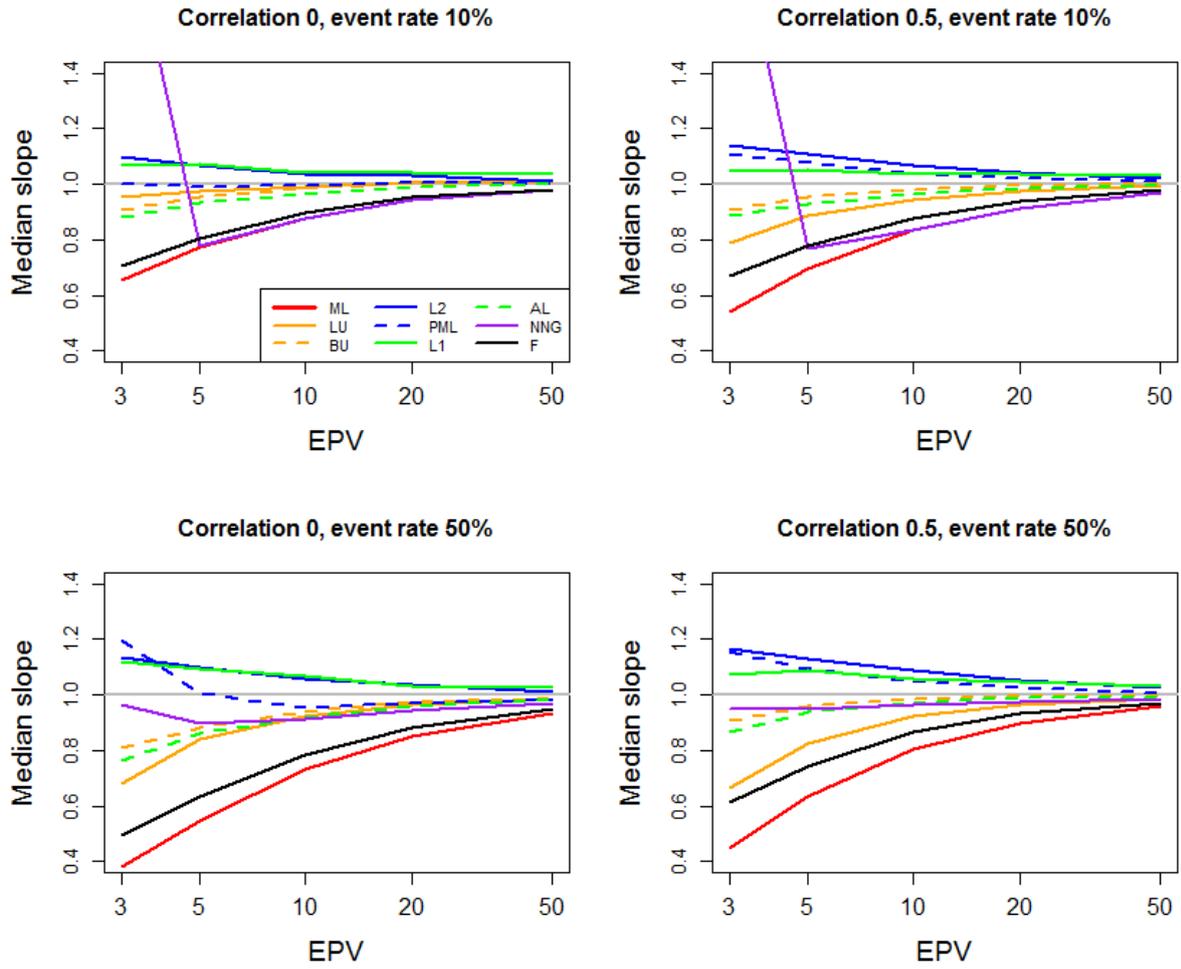

C. Scenarios with 10 true predictors

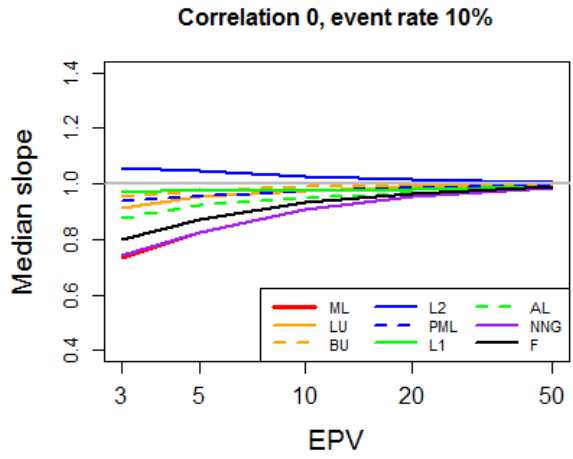
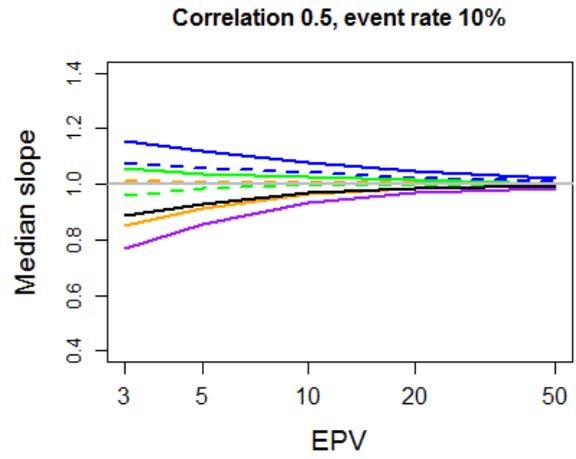
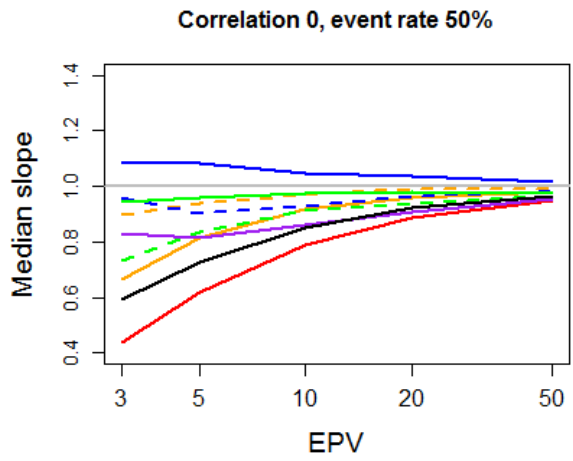
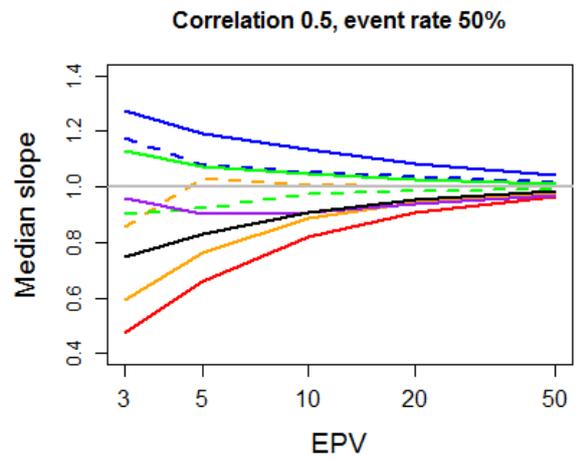

Figure S2. Median c-statistic per scenario and method. ML, maximum likelihood; LU, uniform shrinkage based on likelihood; BU, uniform shrinkage based on bootstrap; L2, ridge (L2 penalty); PML, penalized maximum likelihood; L1, LASSO (L1 penalty); AL, adaptive LASSO; NNG, non-negative garrote; F, Firth's correction.

A. Scenarios with 5 true predictors

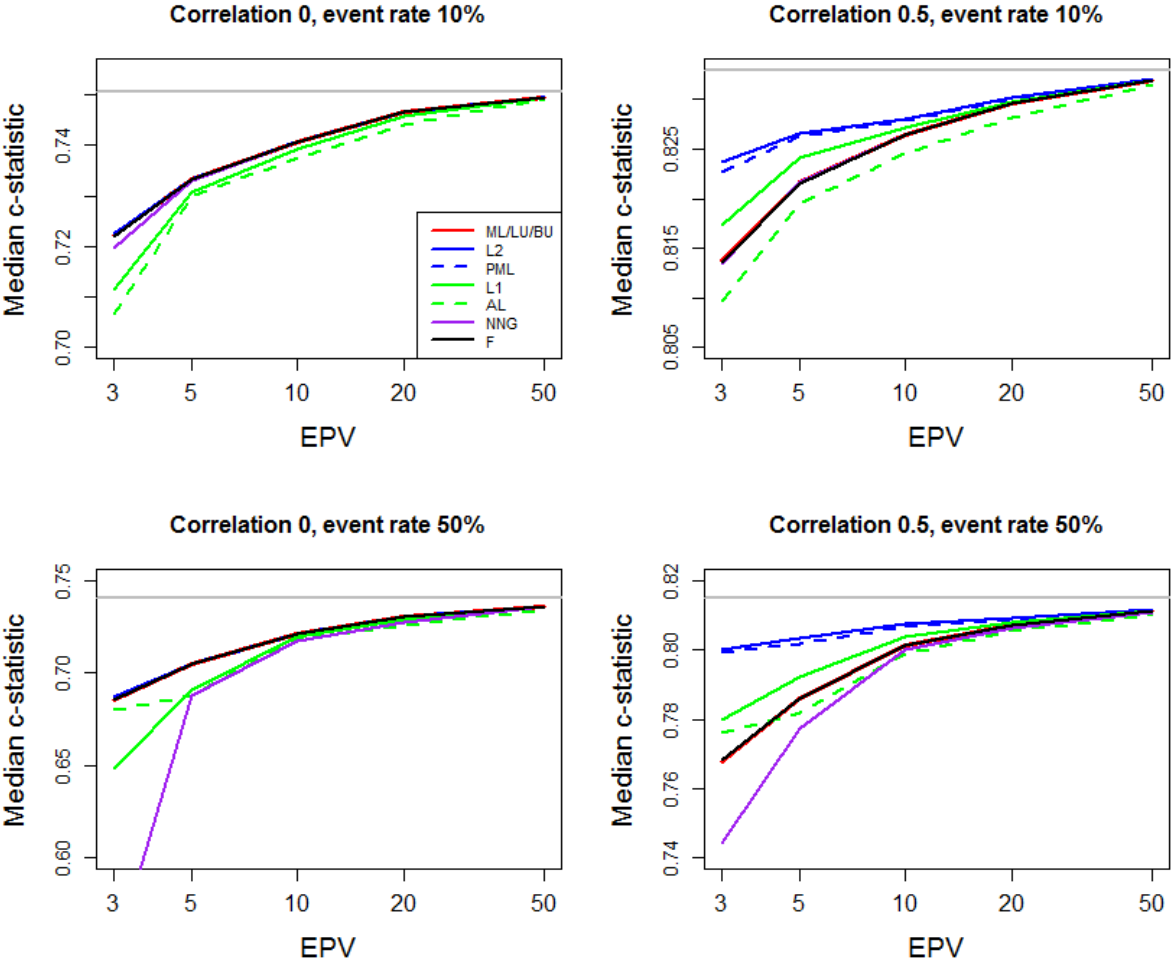

B. Scenarios with 5 true and 5 noise predictors

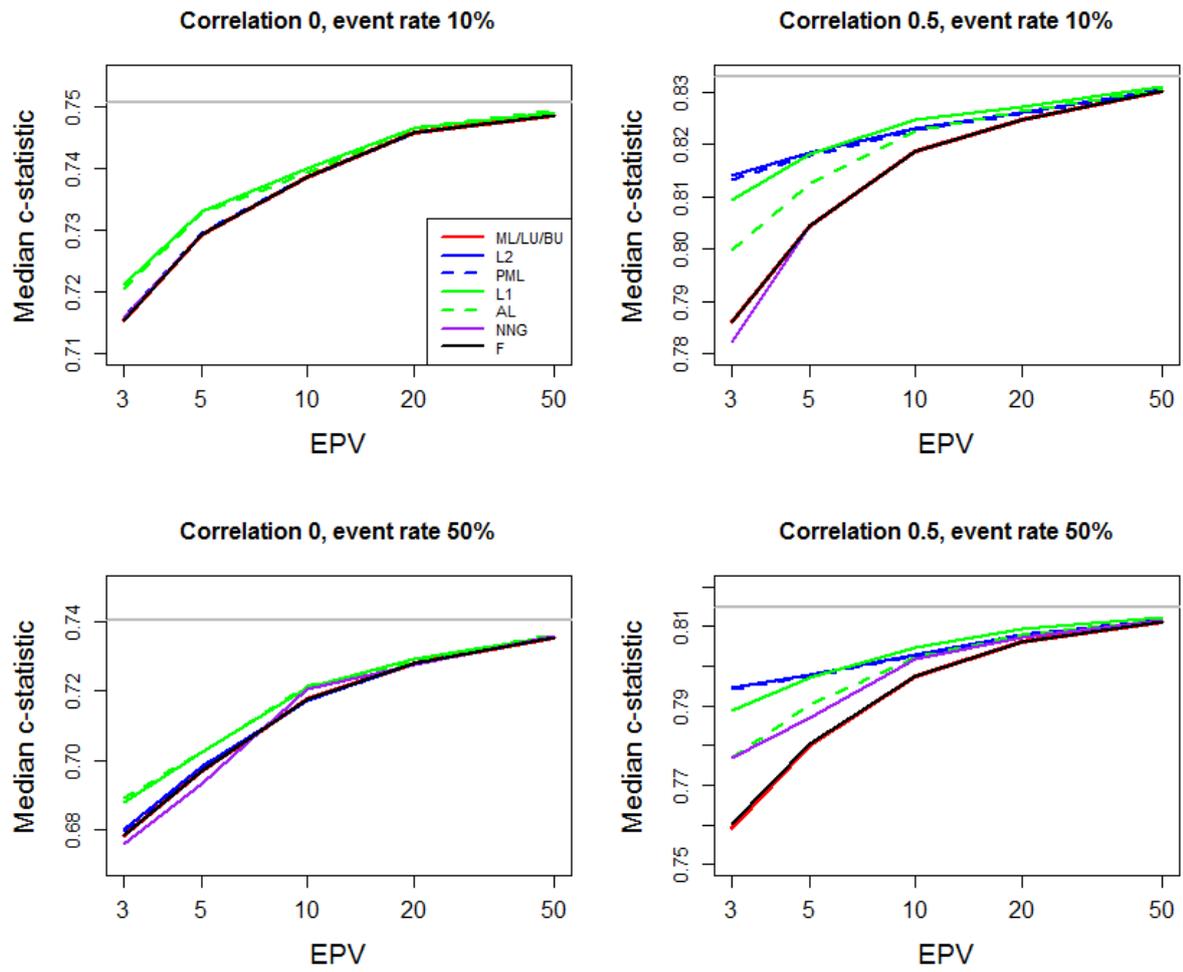

C. Scenarios with 10 true predictors

Figure S3. Box plots of calibration slopes over the 1,000 simulation runs for each scenario. The events per variable (EPV) is indicated in the top left. The numbers at the bottom are the root mean squared distances (RMSD) of the log of the calibration slopes. The length of the whiskers is at most 1.5 times the interquartile range. Calibration slopes are winsorized at 0.1 and 10 for visualization purposes. ML, maximum likelihood; LU, uniform shrinkage based on likelihood; BU, uniform shrinkage based on bootstrap; L2, ridge (L2 penalty); PML, penalized maximum likelihood; L1, LASSO (L1 penalty); AL, adaptive LASSO; NNG, non-negative garrote; F, Firth's correction.

A. 5 true predictors, correlation 0, event rate 10%

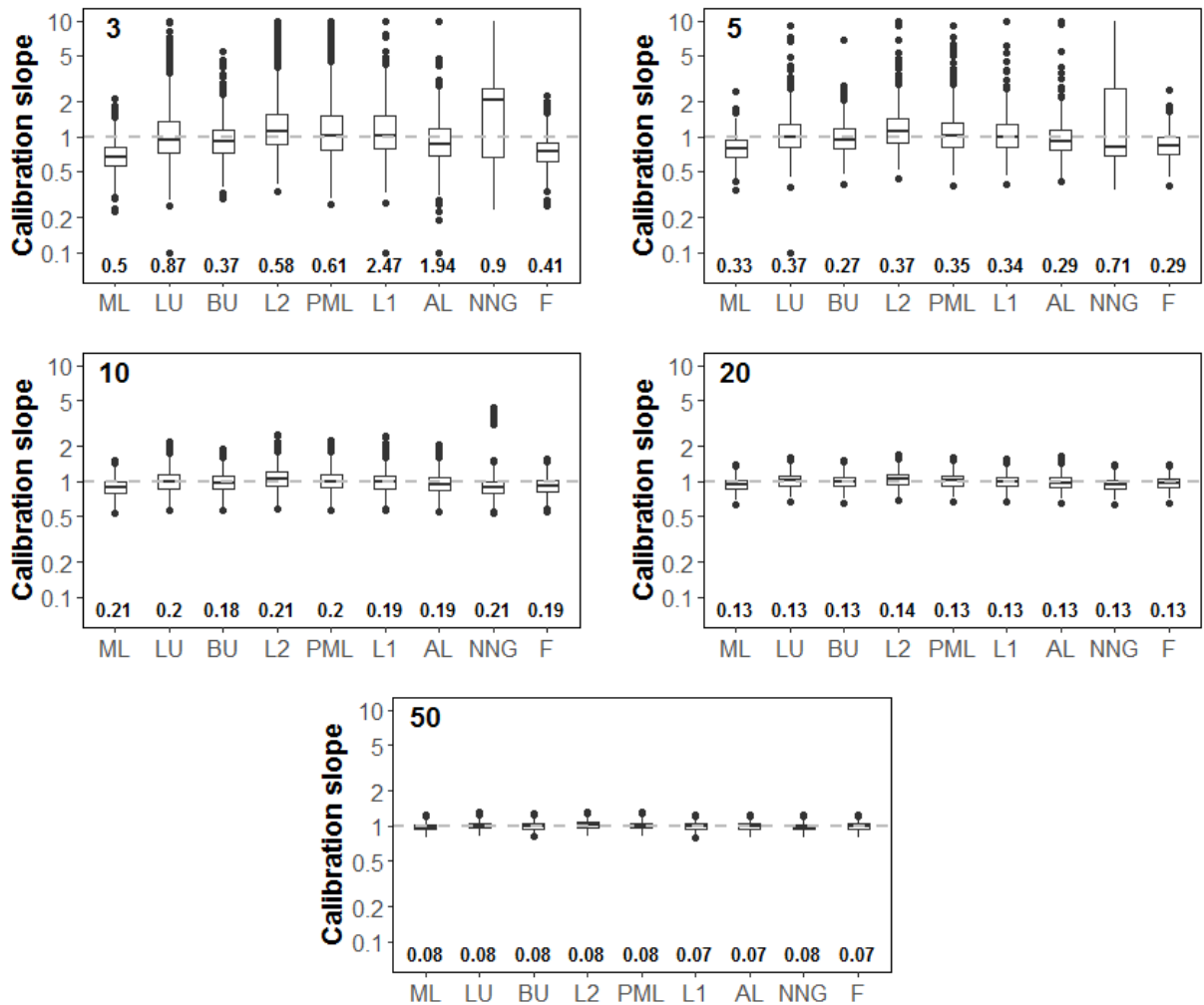

B. 5 true predictors, correlation 0.5, event rate 10%

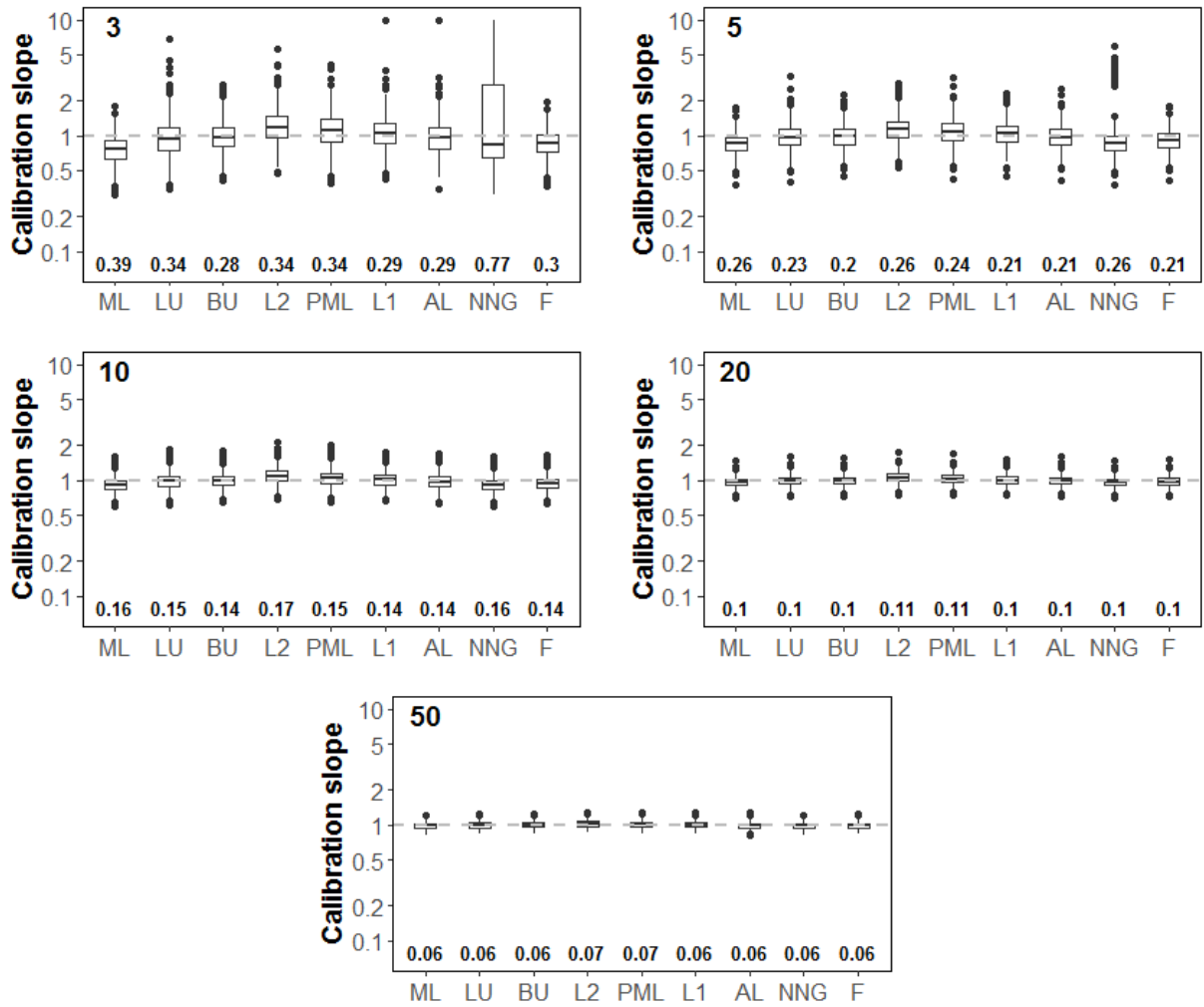

C. 5 true predictors, correlation 0, event rate 50%

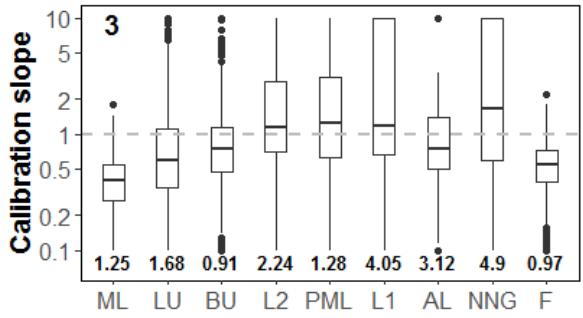
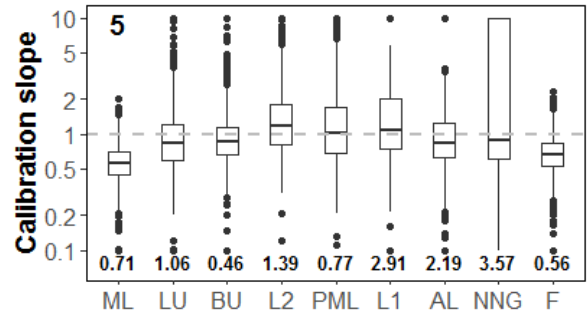
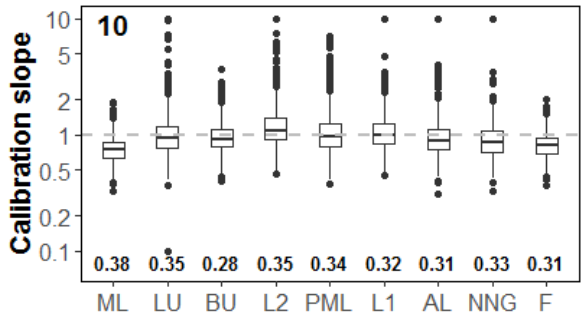
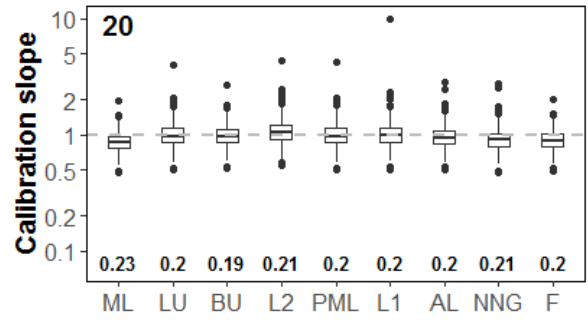
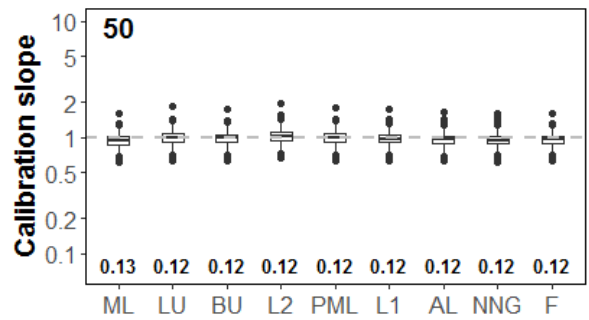

D. 5 true predictors, correlation 0.5, event rate 50%

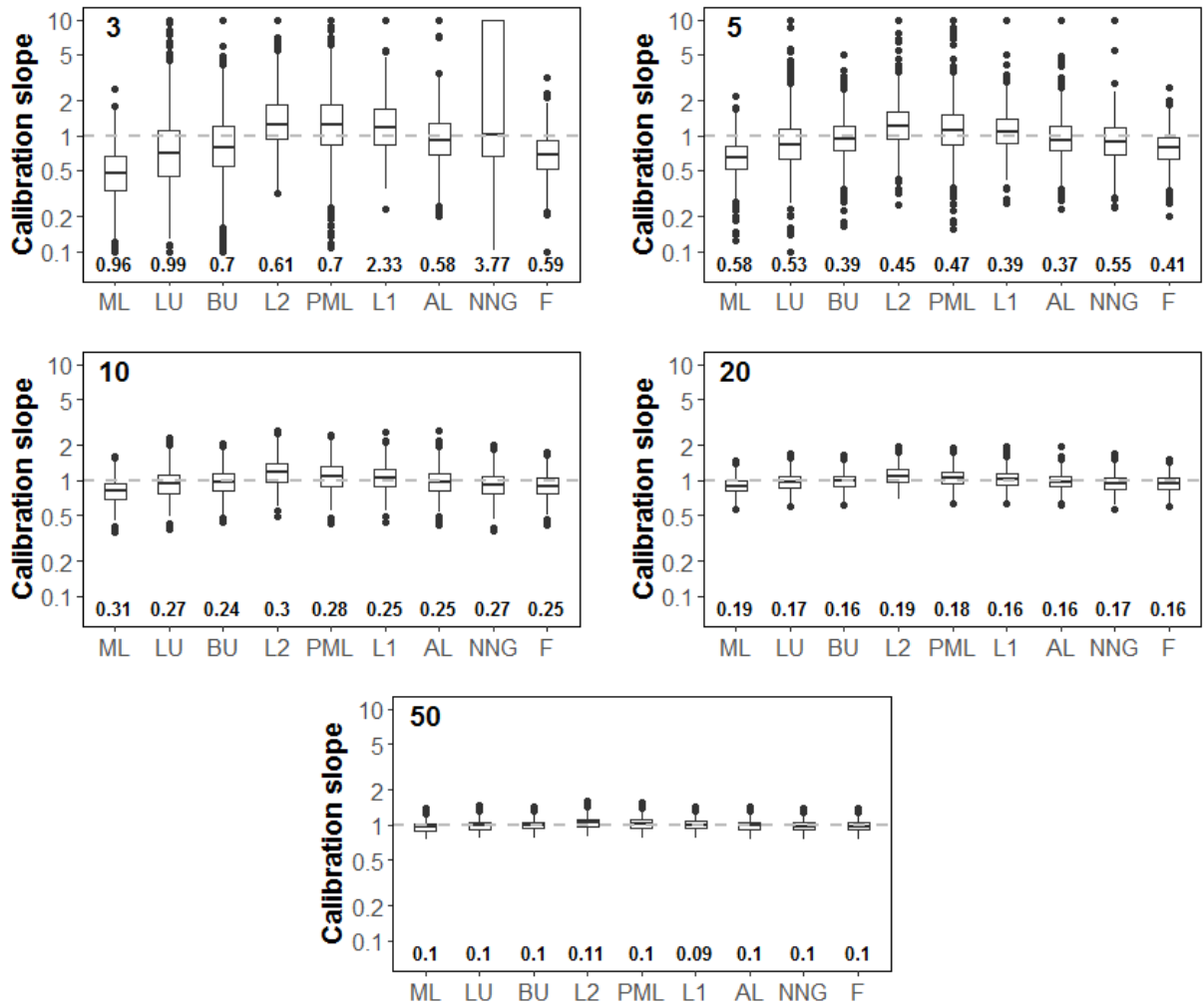

E. 5 true and 5 noise predictors, correlation 0, event rate 10%

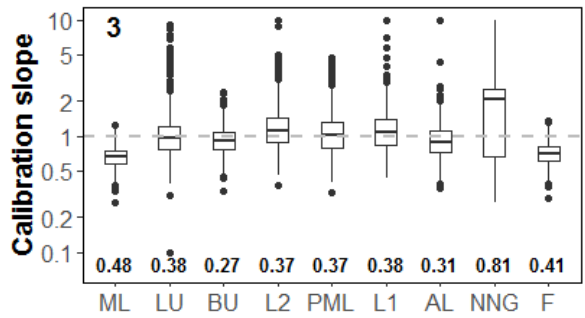
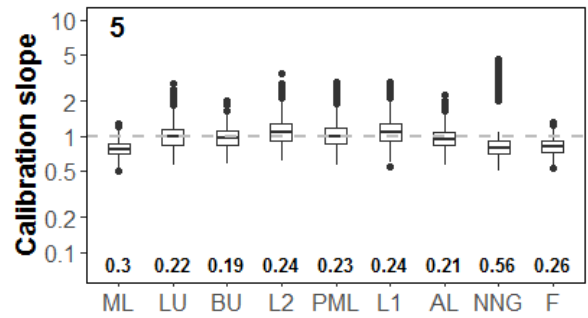
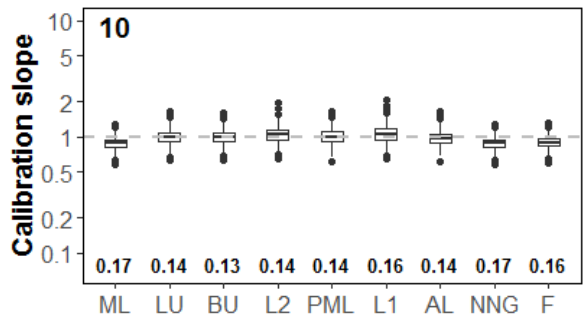
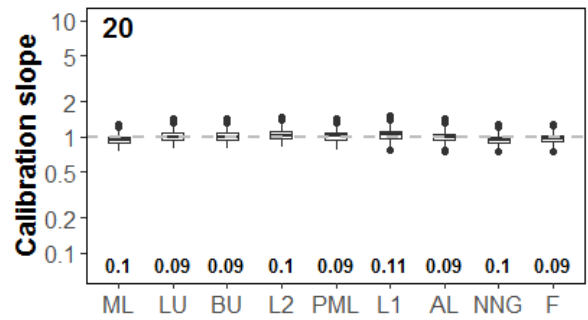
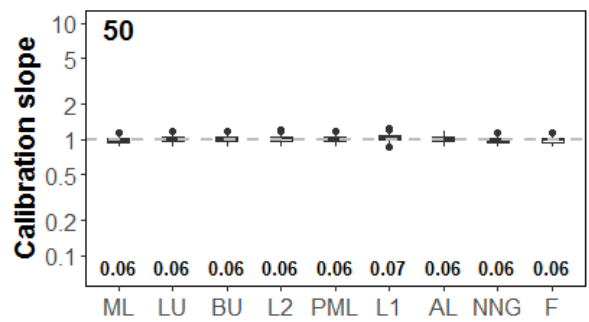

F. 5 true and 5 noise predictors, correlation 0.5, event rate 10%

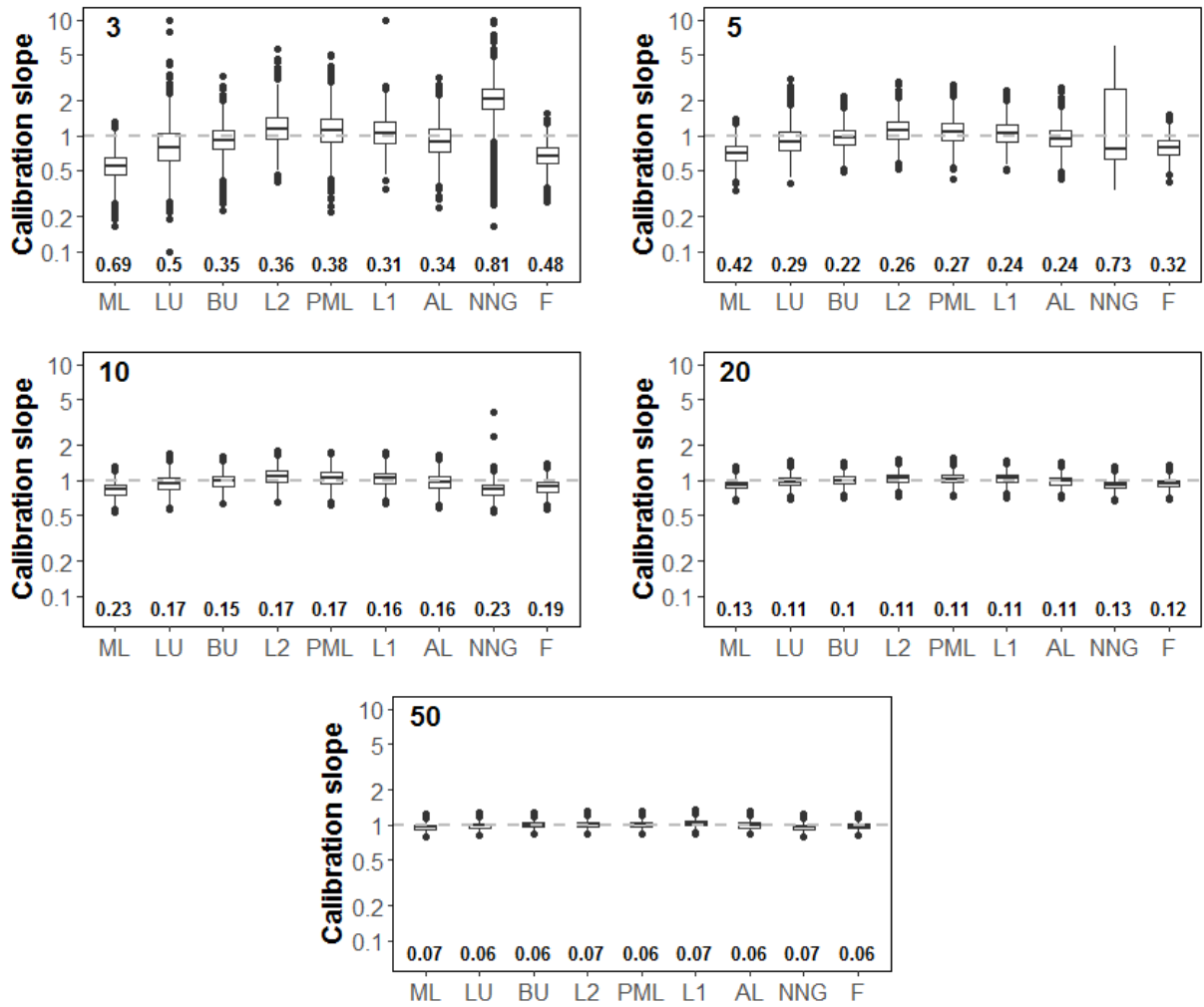

G. 5 true and 5 noise predictors, correlation 0, event rate 50%

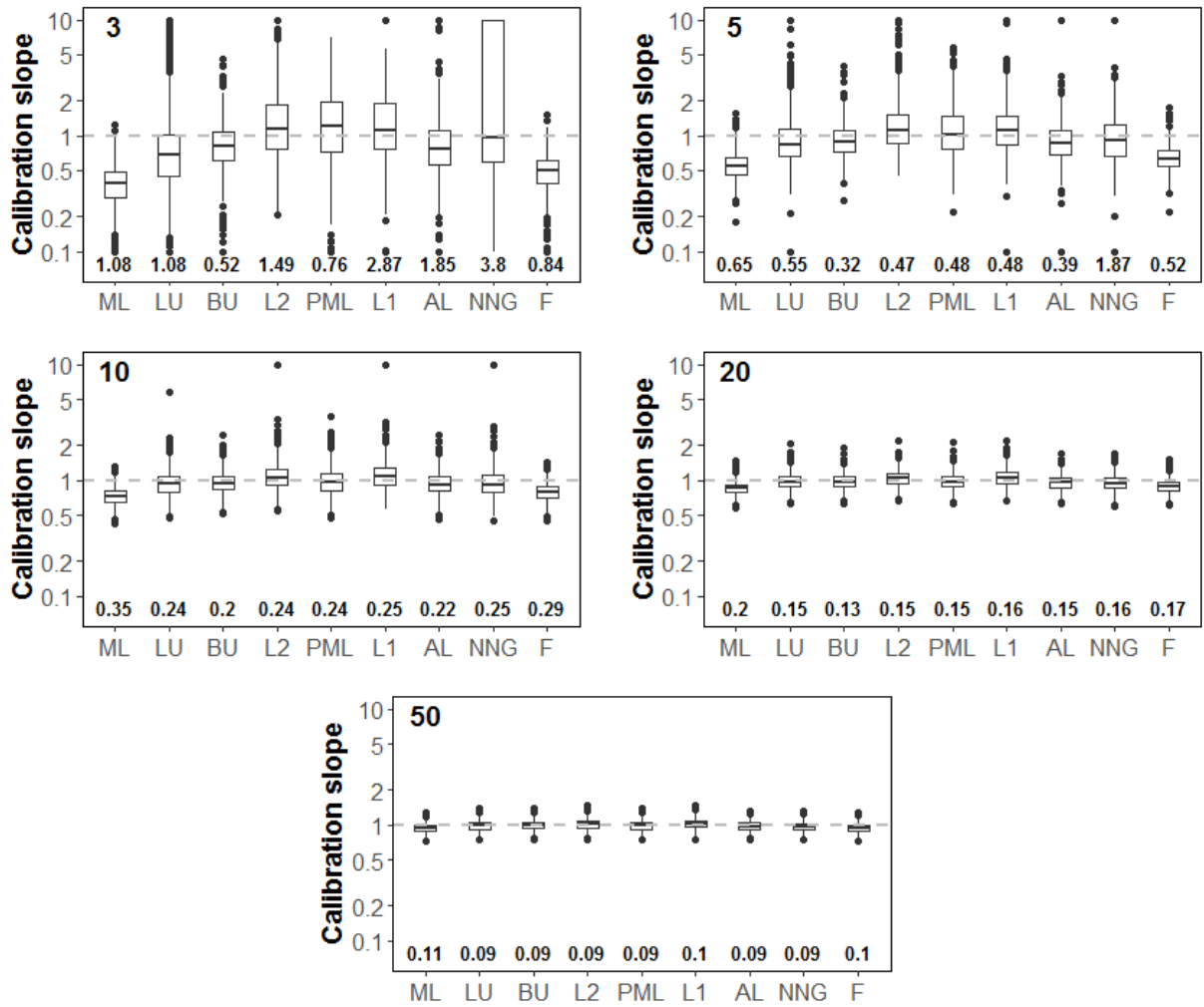

H. 5 true and 5 noise predictors, correlation 0.5, event rate 50%

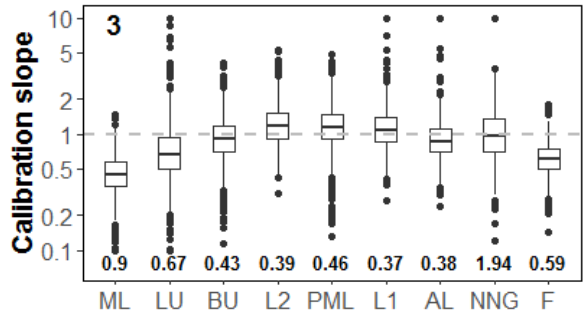
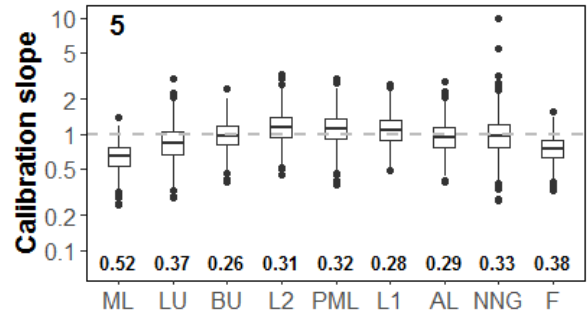
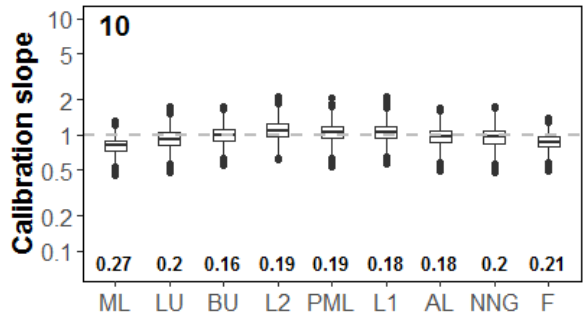
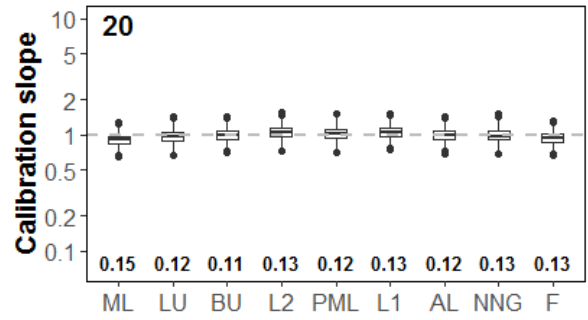
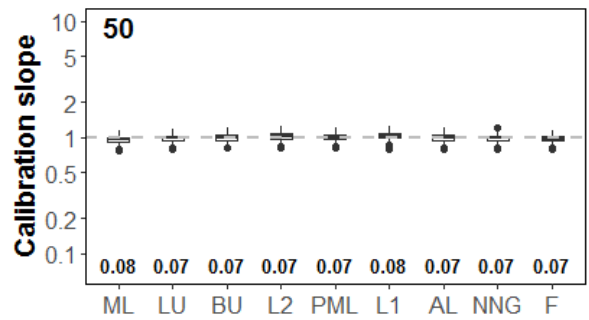

I. 10 true predictors, correlation 0, event rate 10%

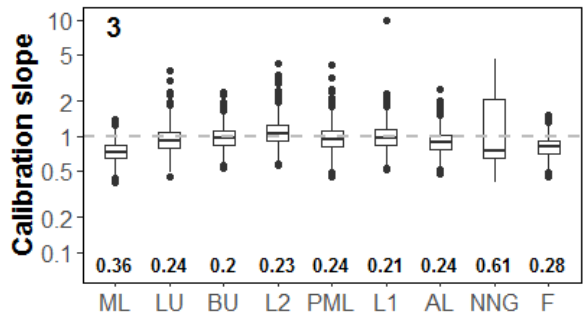
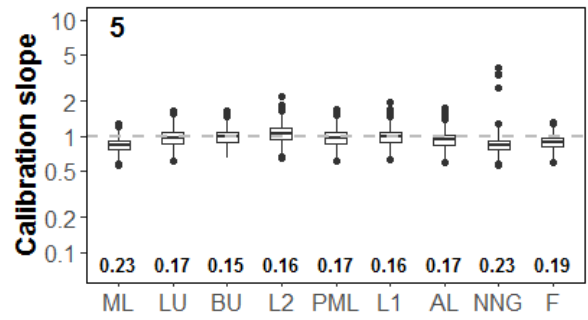
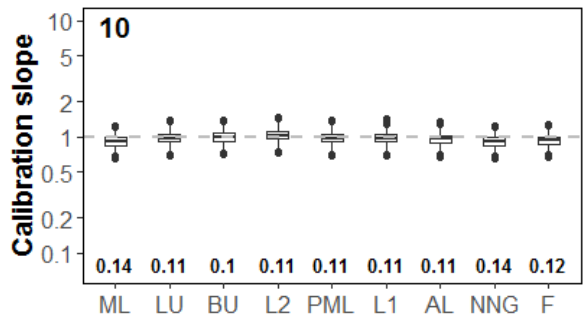
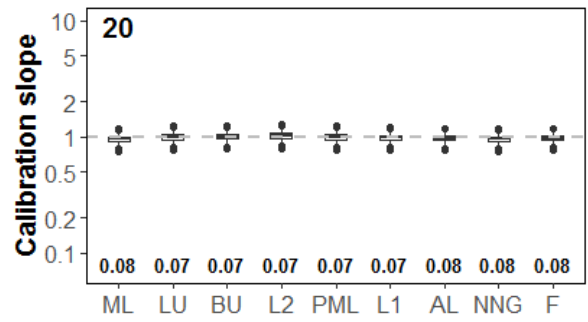
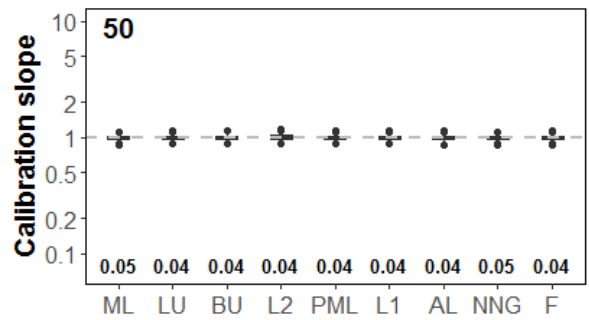

J. 10 true predictors, correlation 0.5, event rate 10%

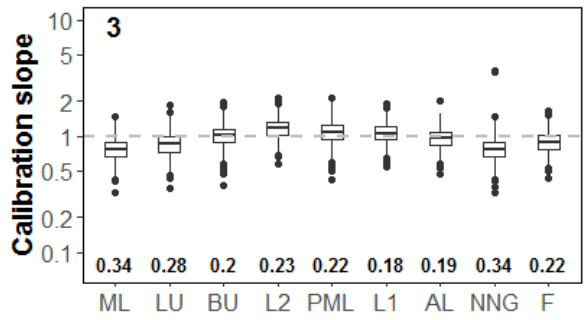
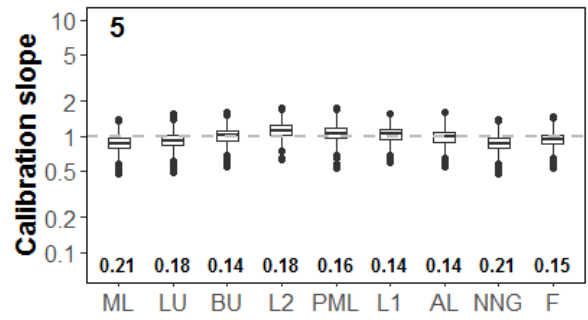
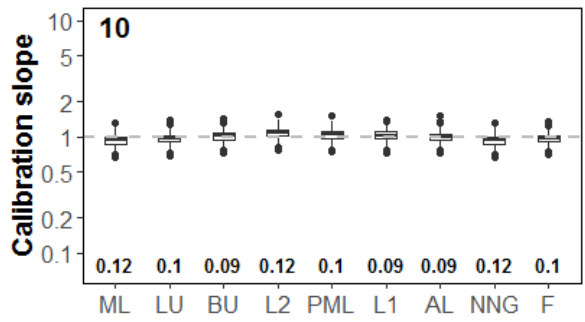
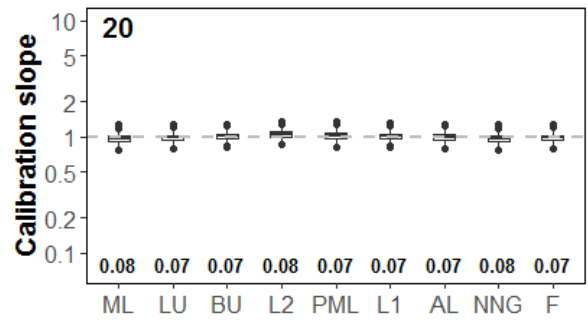
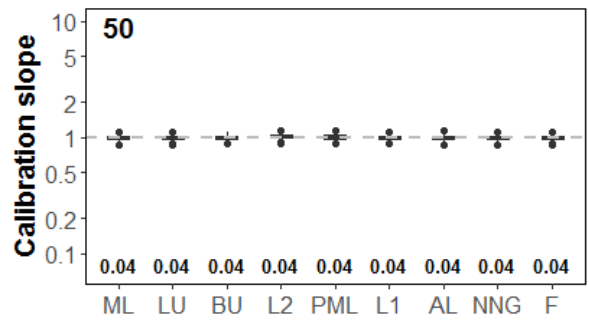

K. 10 true predictors, correlation 0, event rate 50%

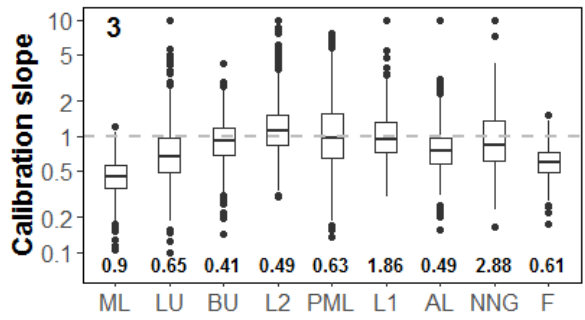
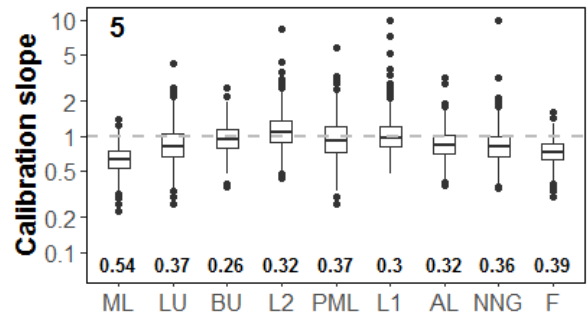
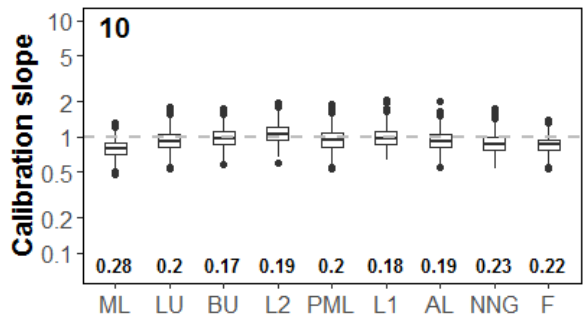
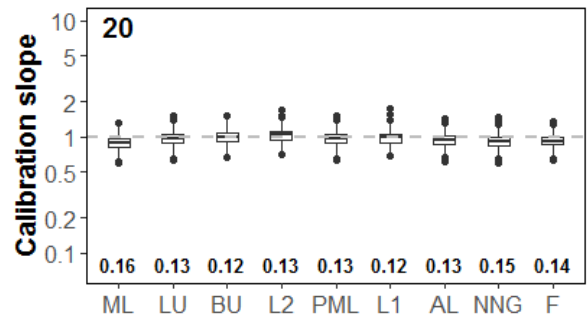
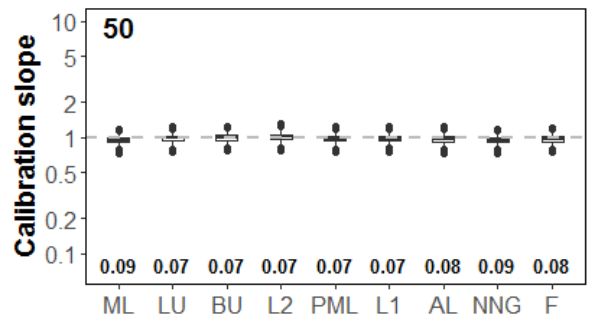

L. 10 true predictors, correlation 0.5, event rate 50%

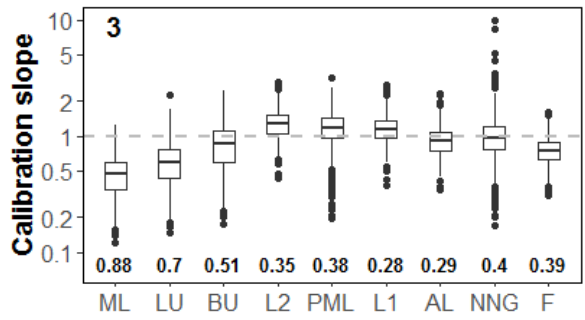
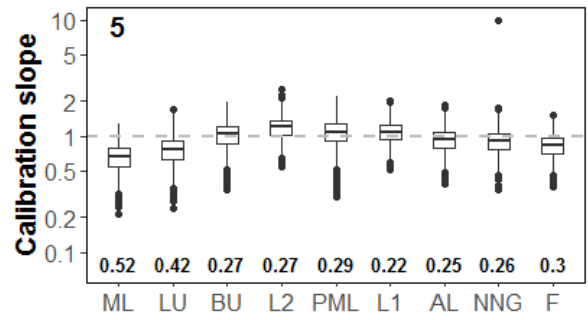
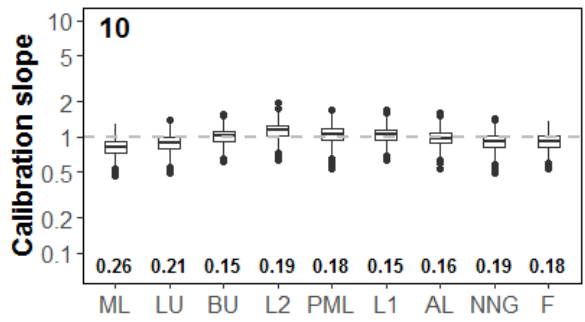
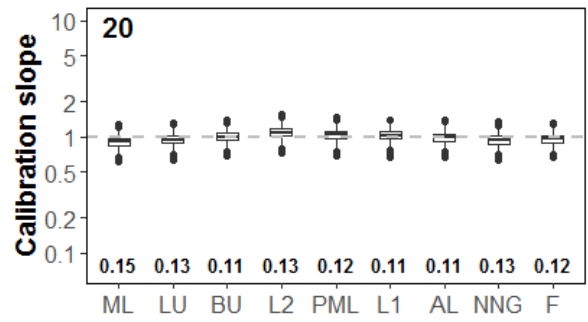
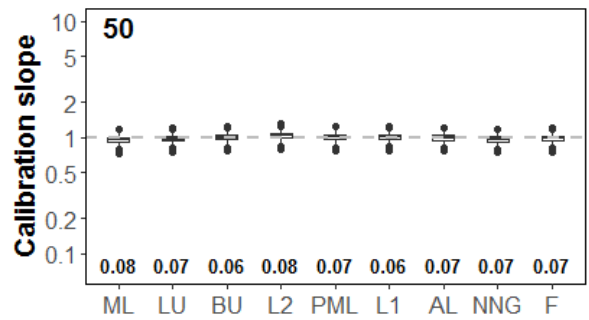

Figure S4. Median absolute deviation (MAD) of the logarithm of the calibration slope. ML, maximum likelihood; LU, uniform shrinkage based on likelihood; BU, uniform shrinkage based on bootstrap; L2, ridge (L2 penalty); PML, penalized maximum likelihood; L1, LASSO (L1 penalty); AL, adaptive LASSO; NNG, non-negative garrote; F, Firth's correction.

A. Scenarios with 5 true predictors

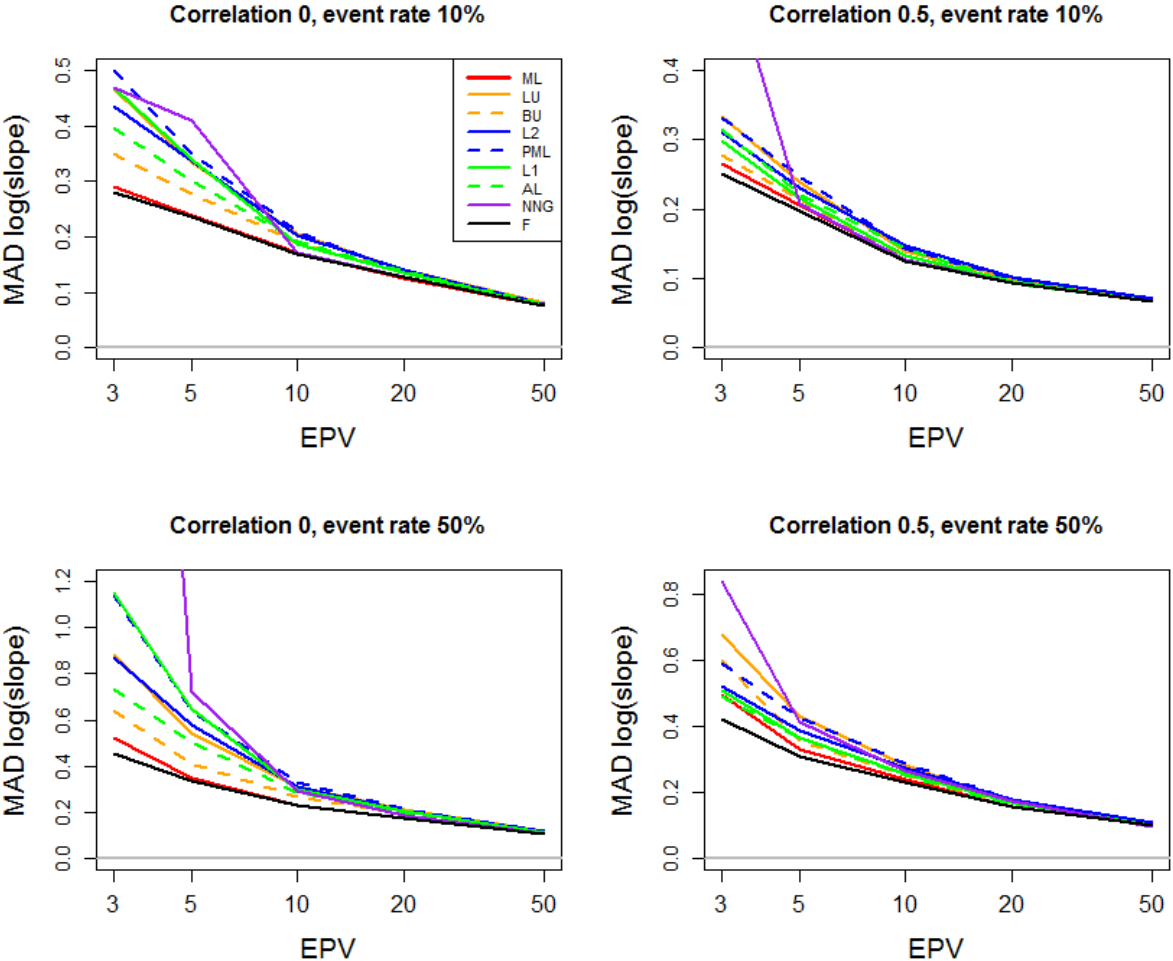

B. Scenarios with 5 true and 5 noise predictors

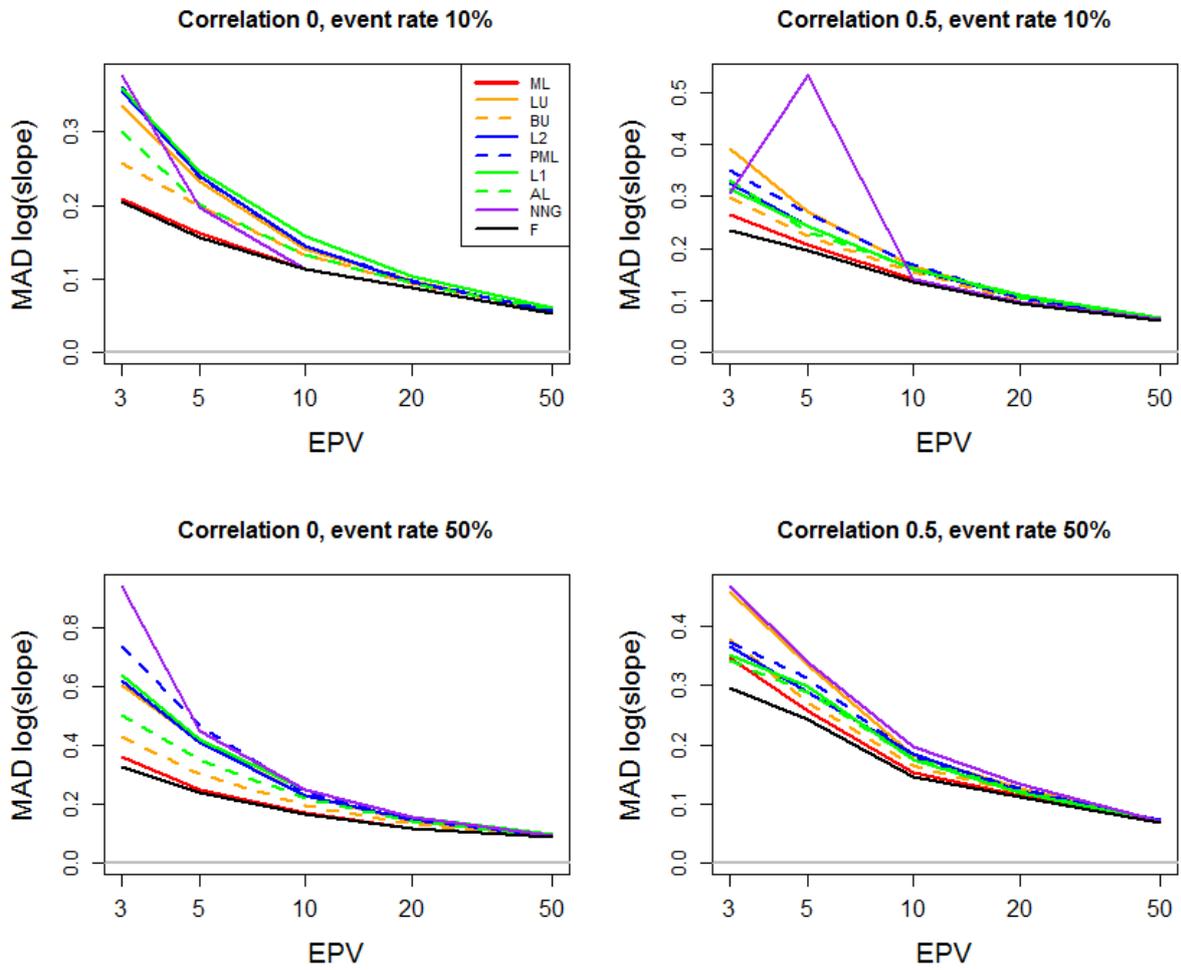

C. Scenarios with 10 true predictors

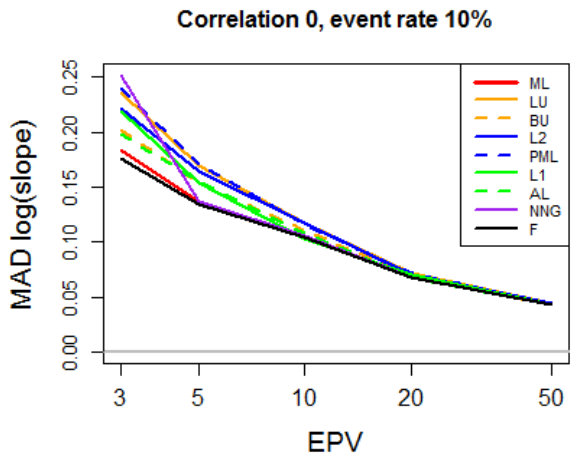
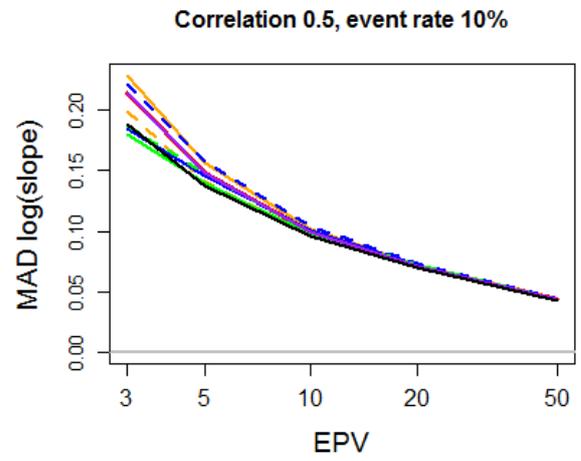
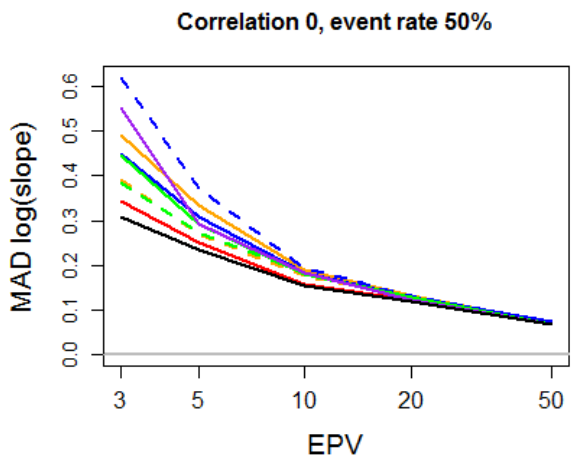
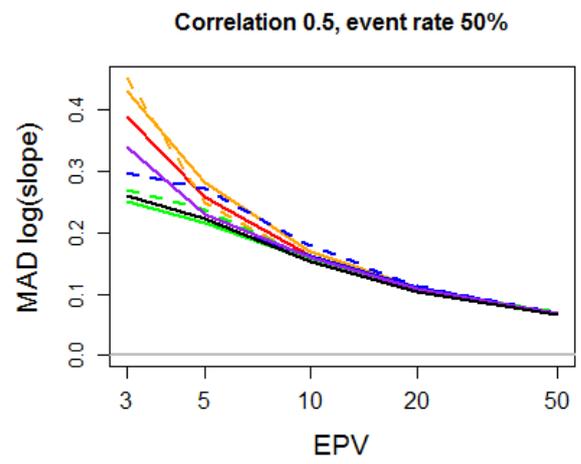

Figure S5. Root mean squared distance of the target value (RMSD) of the calibration slope. ML, maximum likelihood; LU, uniform shrinkage based on likelihood; BU, uniform shrinkage based on bootstrap; L2, ridge (L2 penalty); PML, penalized maximum likelihood; L1, LASSO (L1 penalty); AL, adaptive LASSO; NNG, non-negative garrote; F, Firth's correction.

A. Scenarios with 5 true predictors

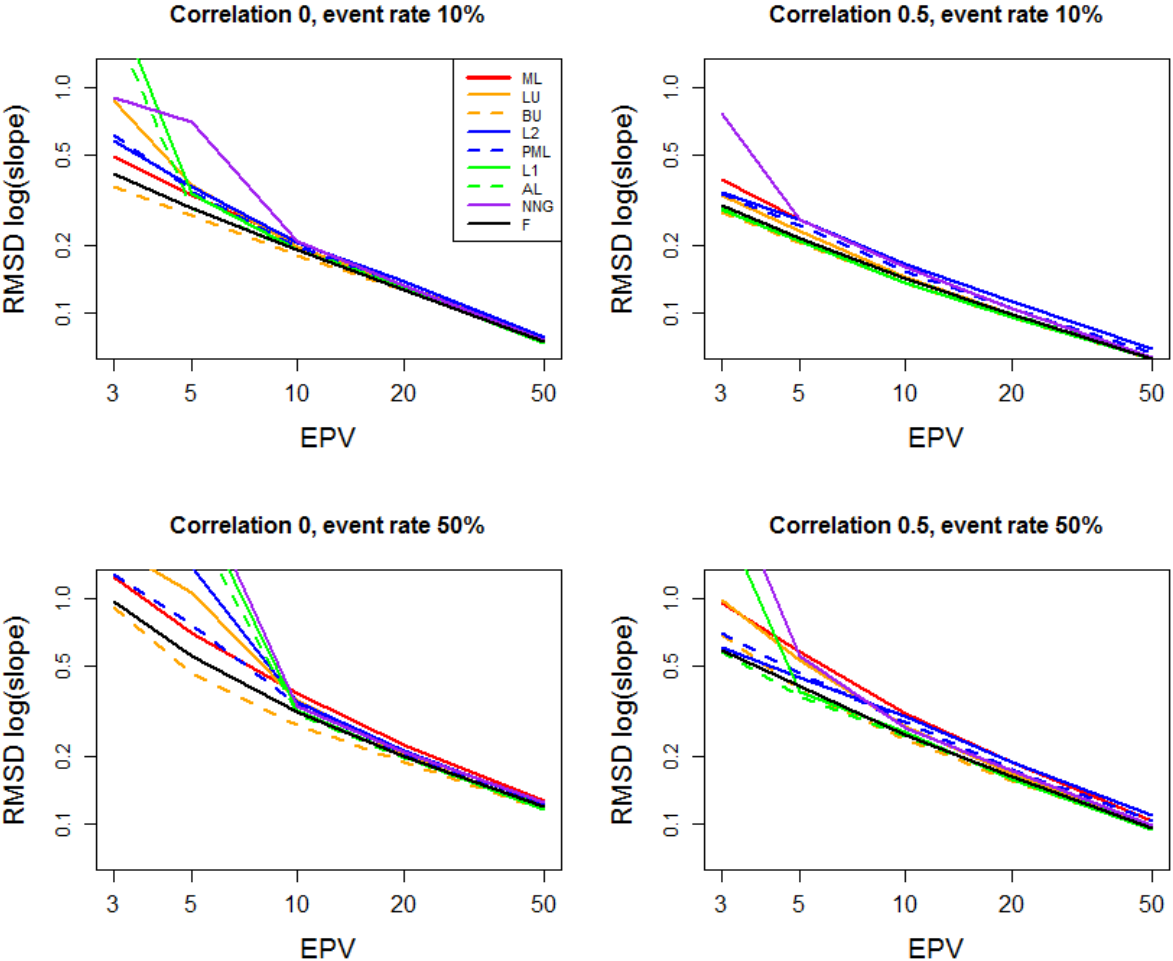

B. Scenarios with 5 true and 5 noise predictors

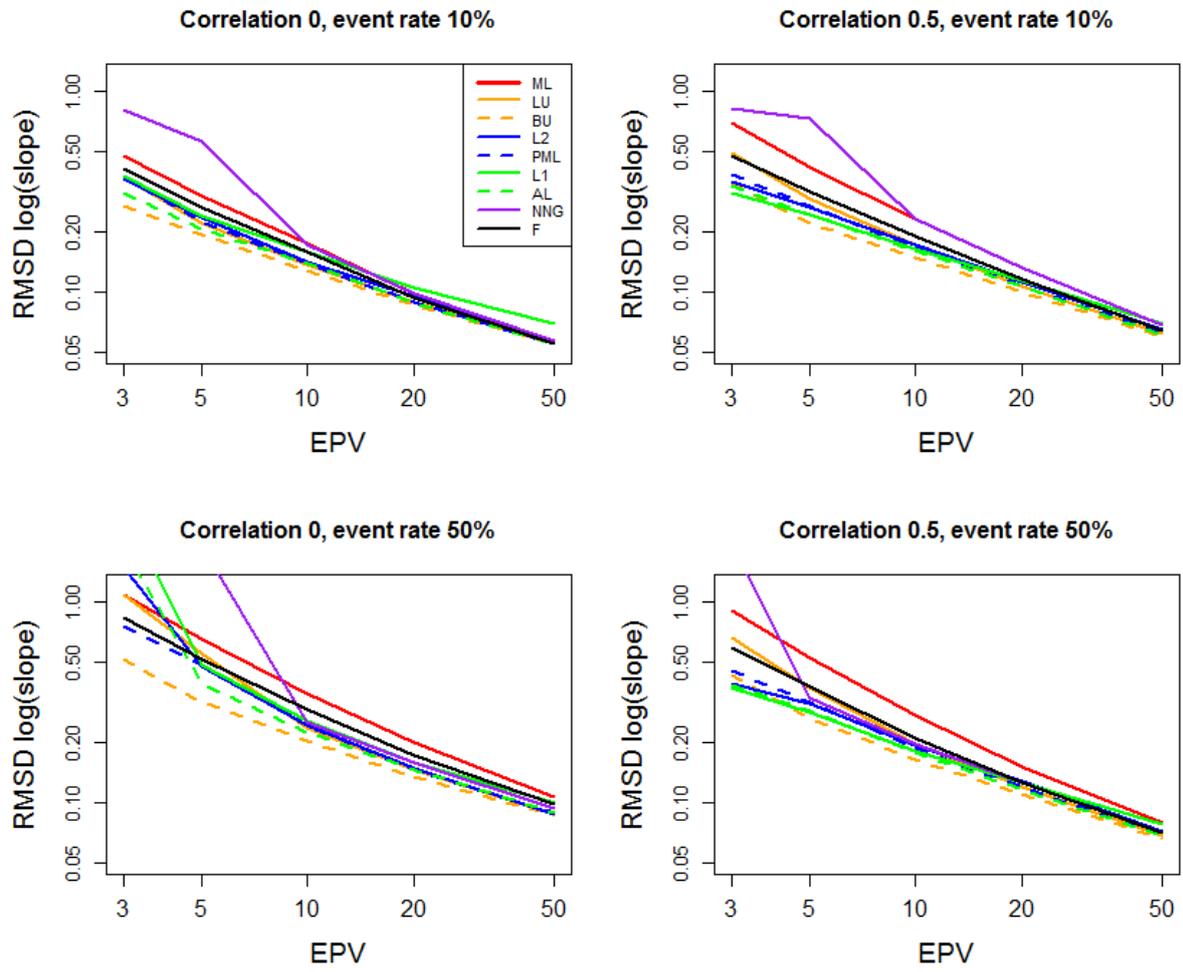

C. Scenarios with 10 true predictors

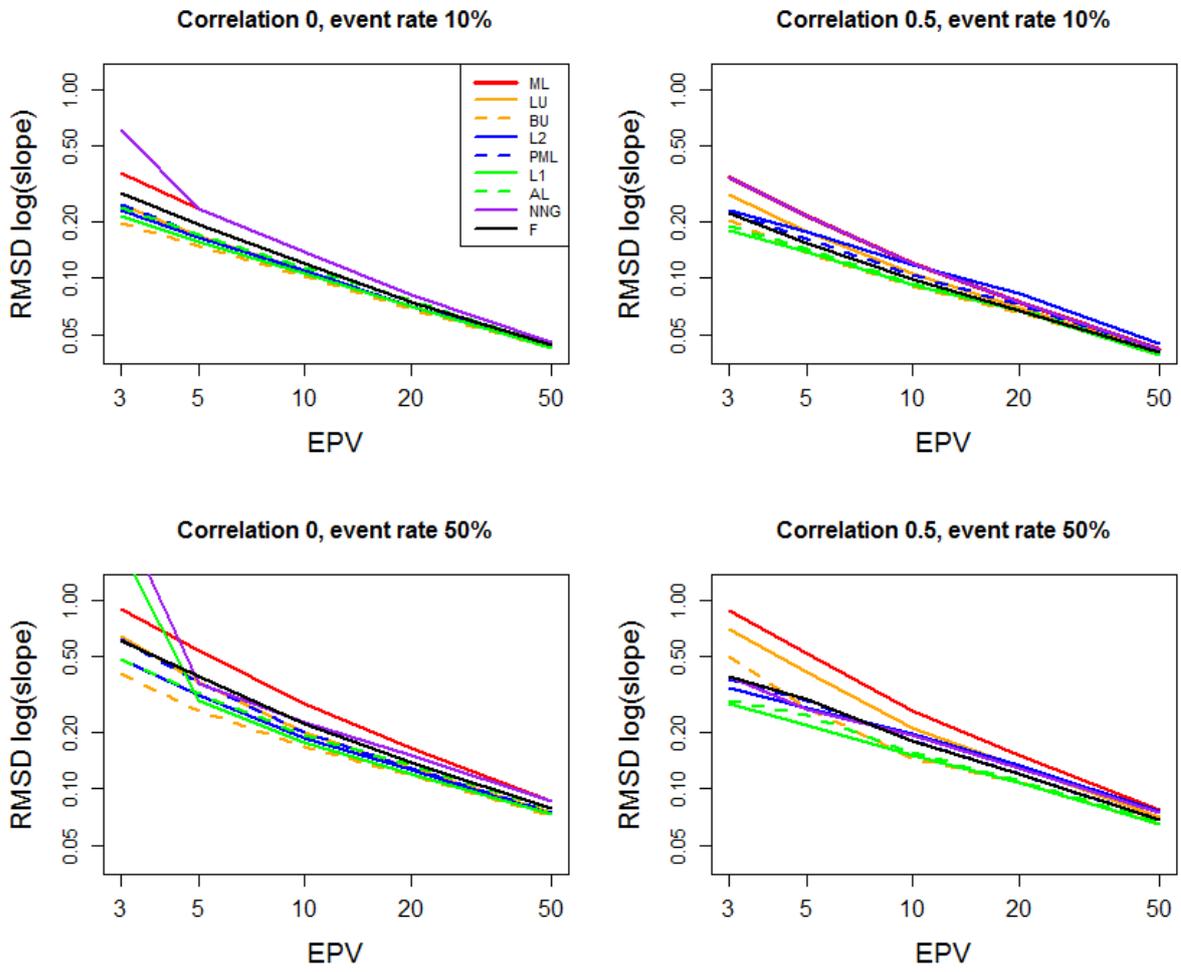

Figure S6. Box plots of c-statistics over the 1,000 simulation runs for each scenario. The events per variable (EPV) is indicated in the top left. The length of the whiskers is at most 1.5 times the interquartile range. C-statistics are winsorized at 0.5 for visualization purposes. ML, maximum likelihood; LU, uniform shrinkage based on likelihood; BU, uniform shrinkage based on bootstrap; L2, ridge (L2 penalty); PML, penalized maximum likelihood; L1, LASSO (L1 penalty); AL, adaptive LASSO; NNG, non-negative garrote; F, Firth's correction.

A. 5 true predictors, correlation 0, event rate 10%

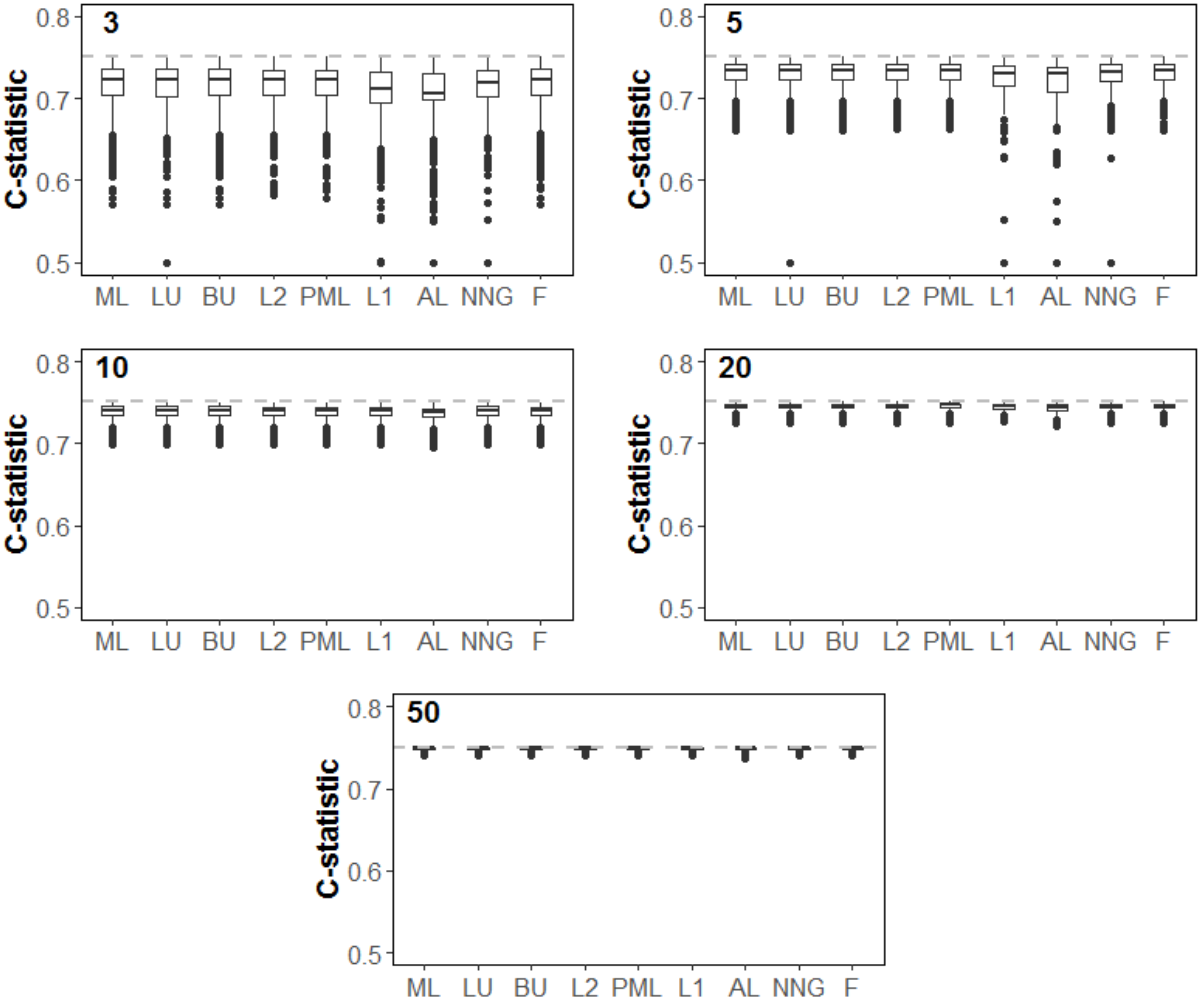

B. 5 true predictors, correlation 0.5, event rate 10%

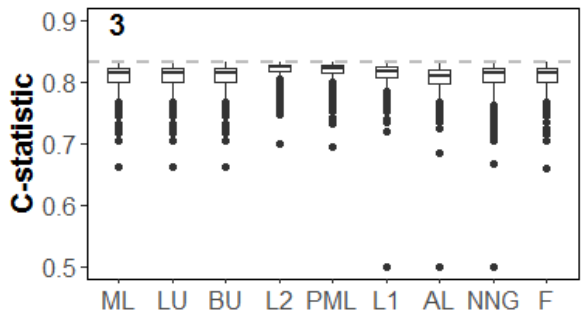
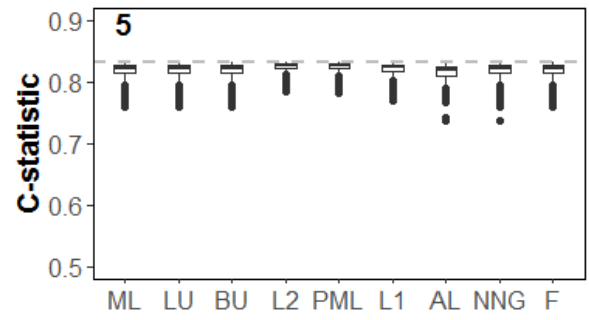
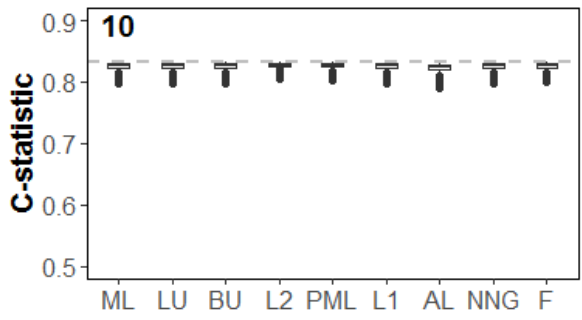
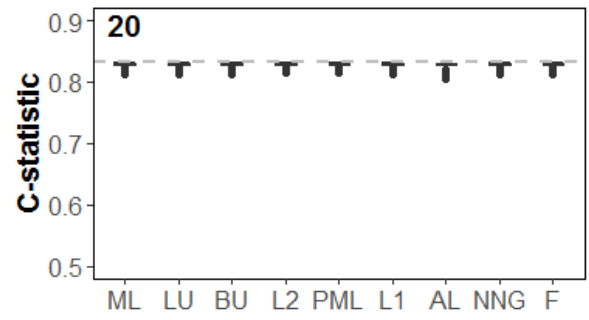
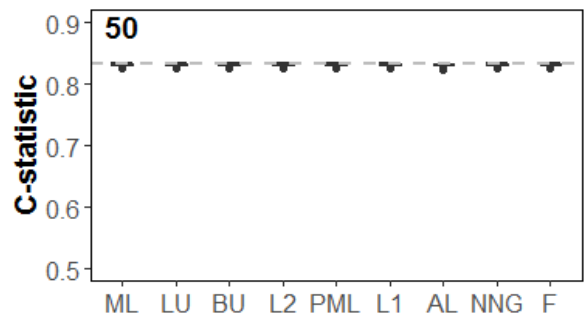

C. 5 true predictors, correlation 0, event rate 50%

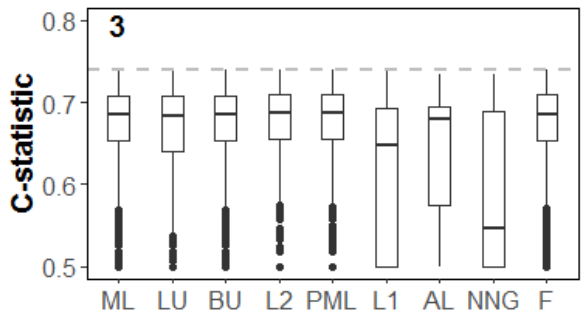
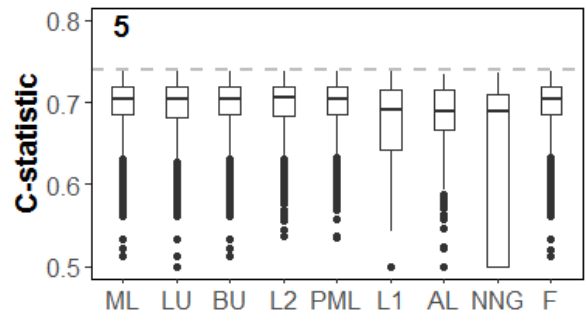
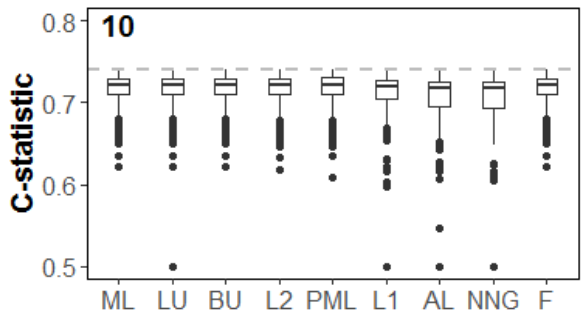
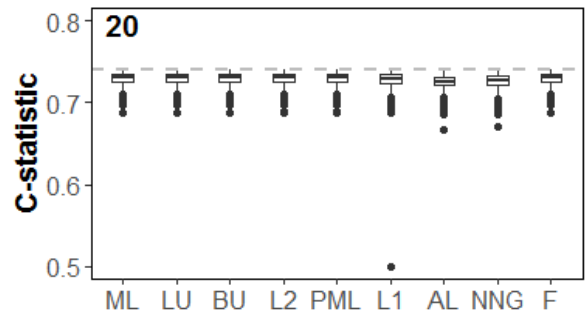
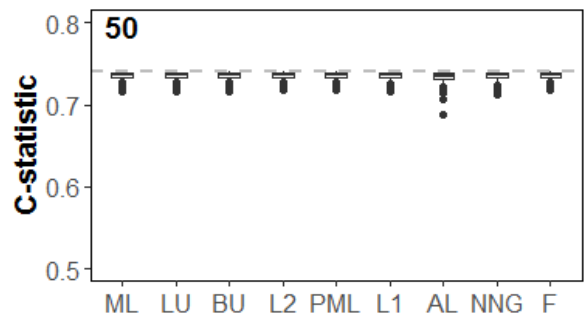

D. 5 true predictors, correlation 0.5, event rate 50%

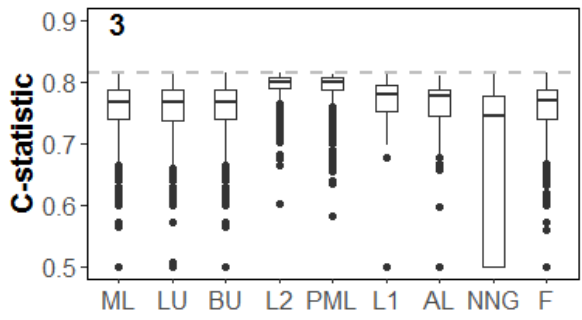
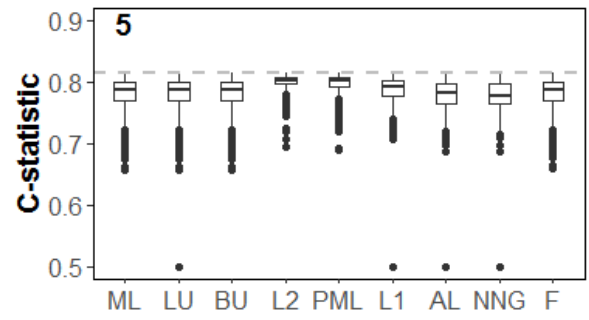
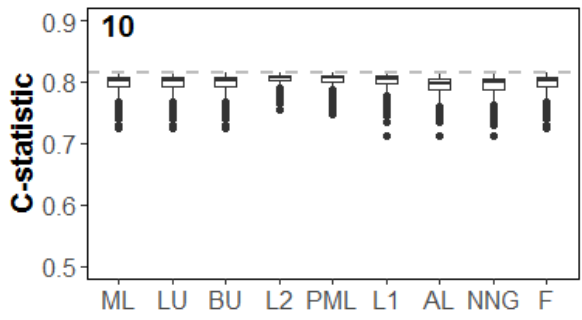
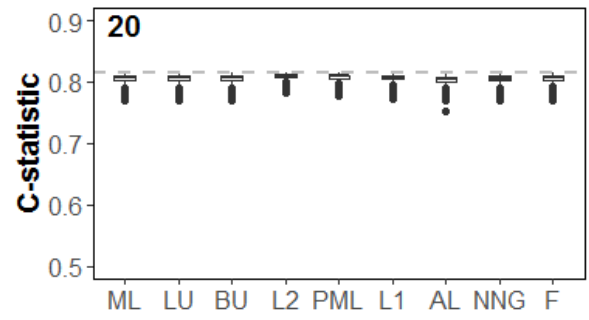
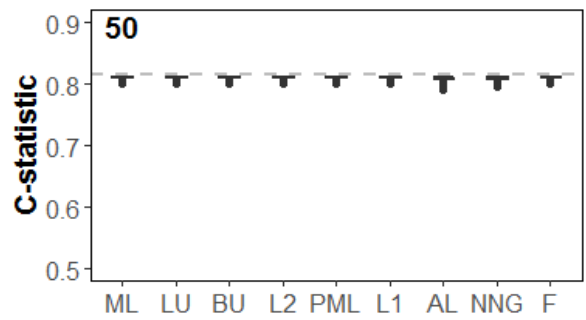

E. 5 true and 5 noise predictors, correlation 0, event rate 10%

F. 5 true and 5 noise predictors, correlation 0.5, event rate 10%

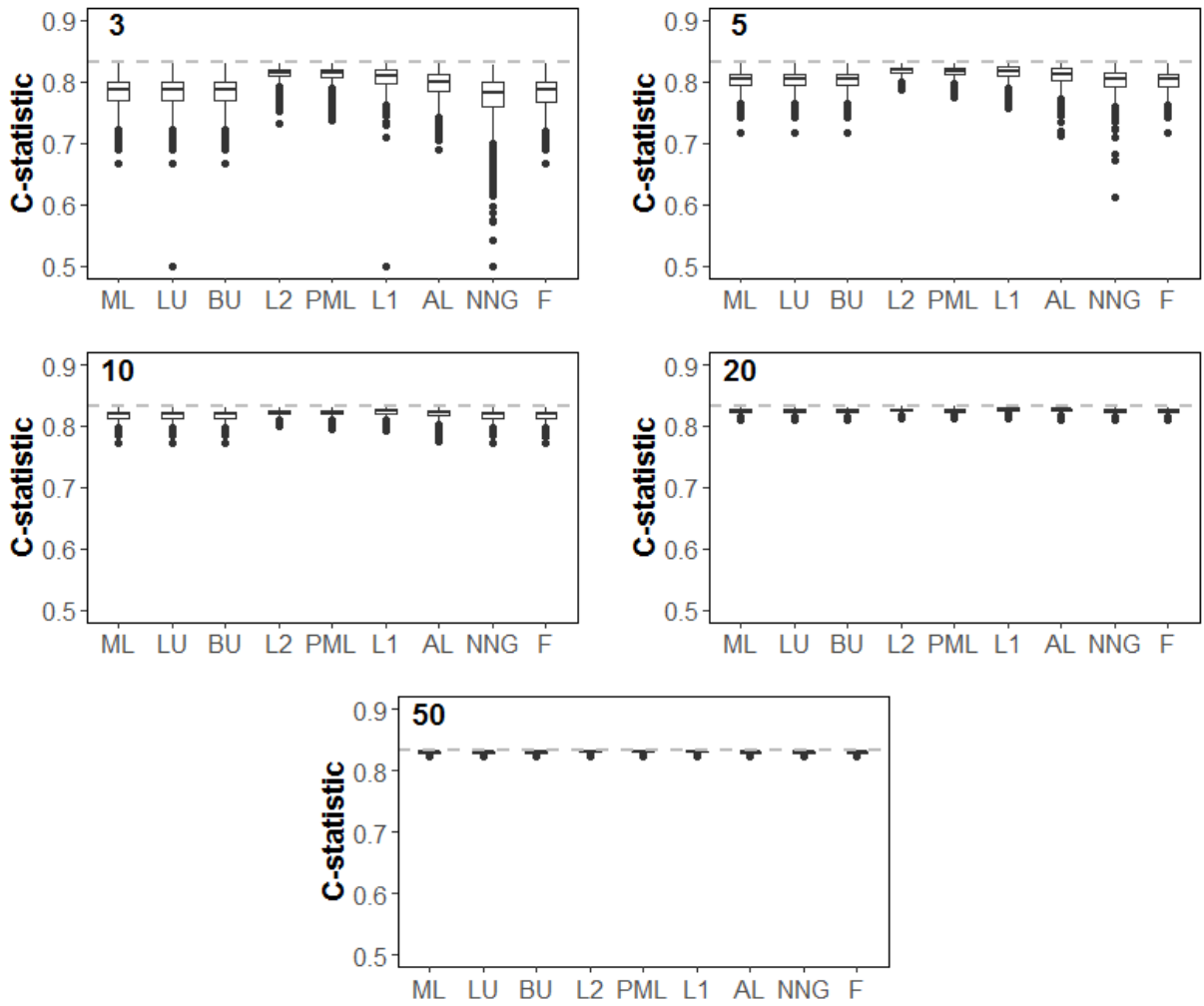

G. 5 true and 5 noise predictors, correlation 0, event rate 50%

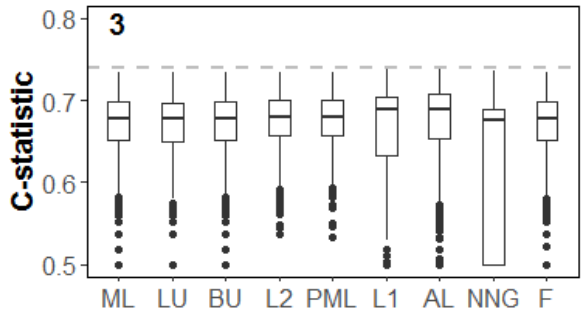
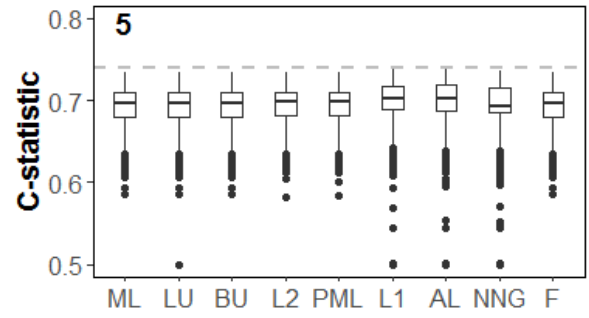
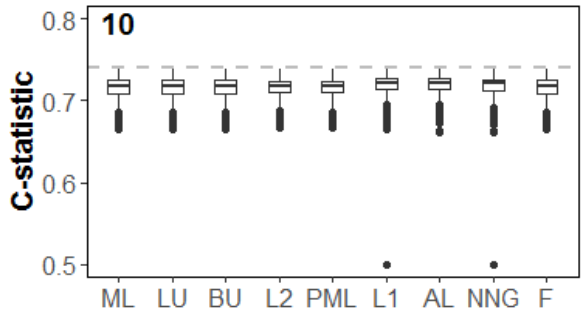
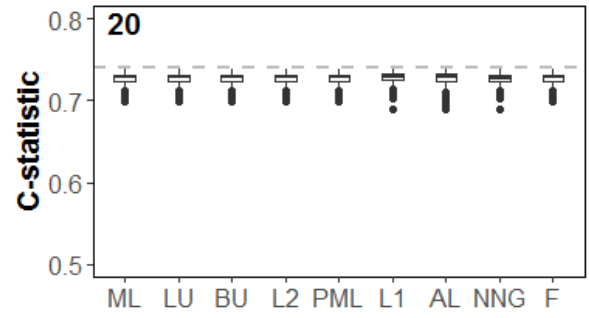
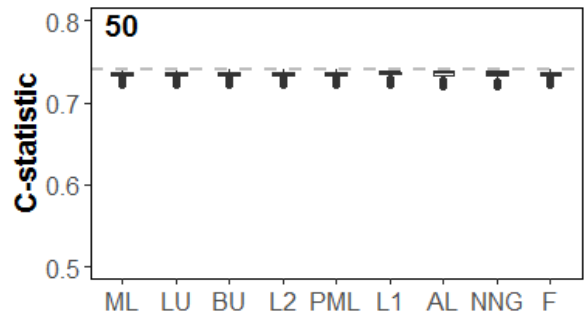

H. 5 true and 5 noise predictors, correlation 0.5, event rate 50%

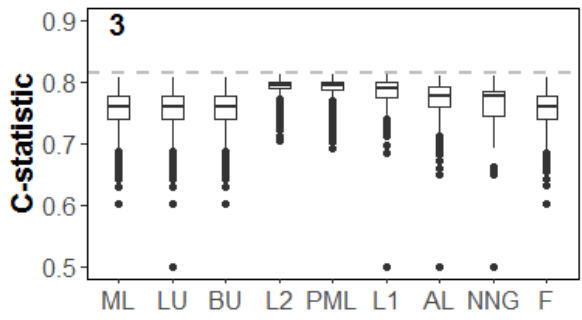
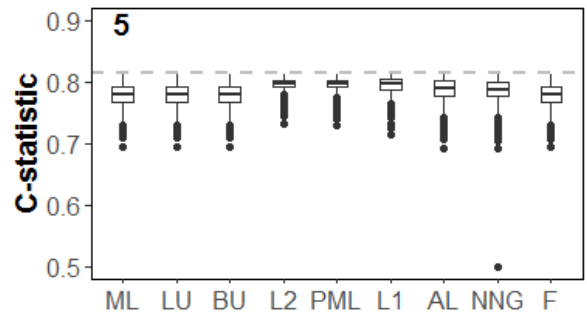
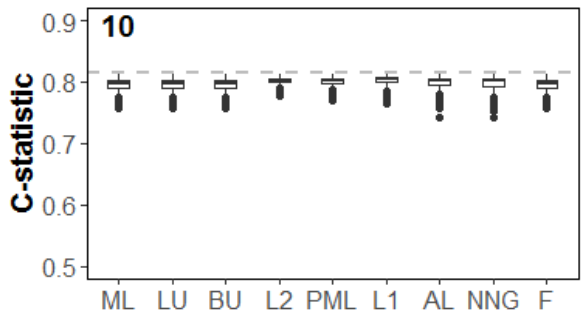
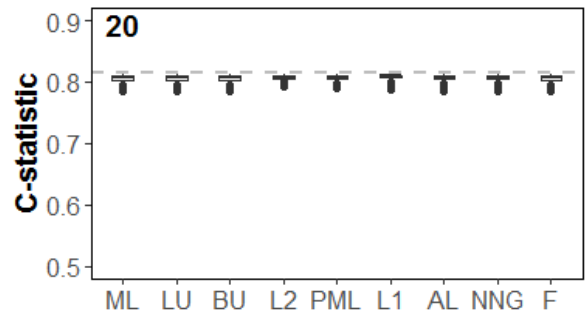
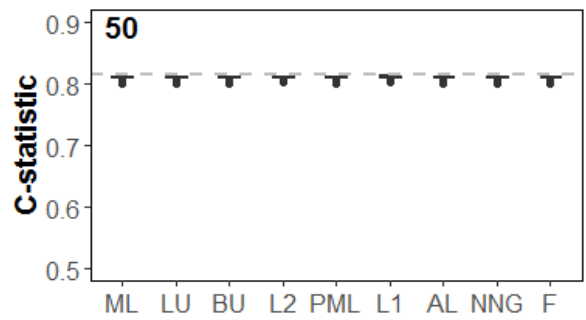

I. 10 true predictors, correlation 0, event rate 10%

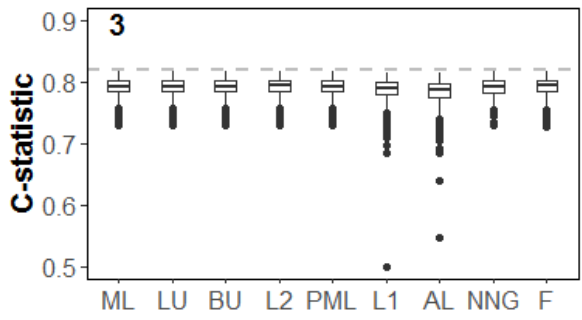
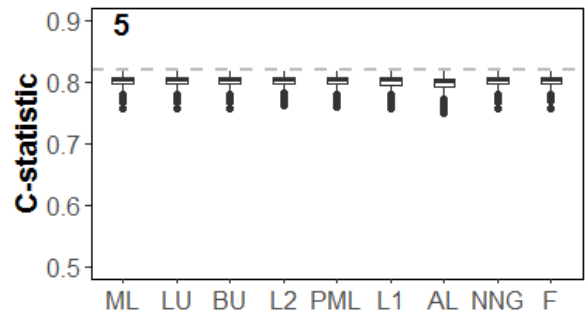
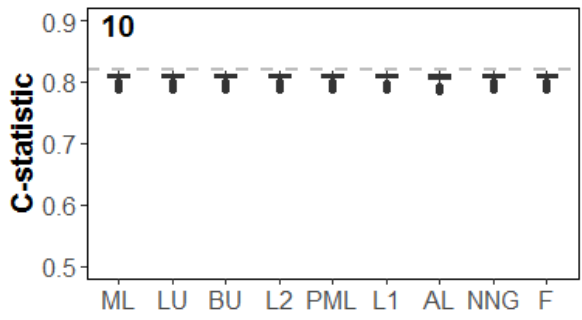
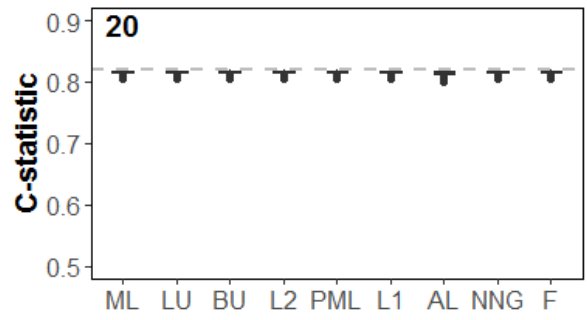
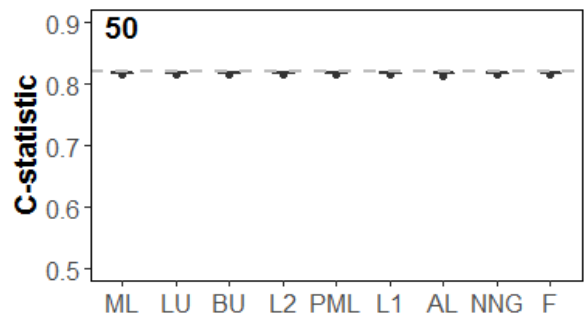

J. 10 true predictors, correlation 0.5, event rate 10%

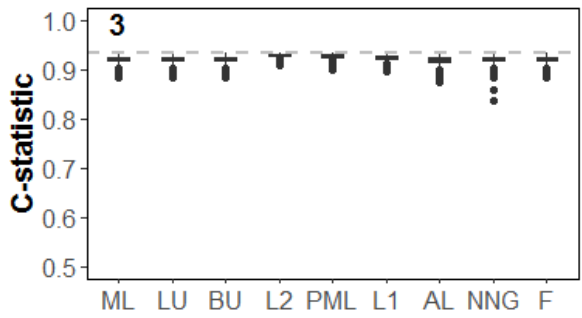
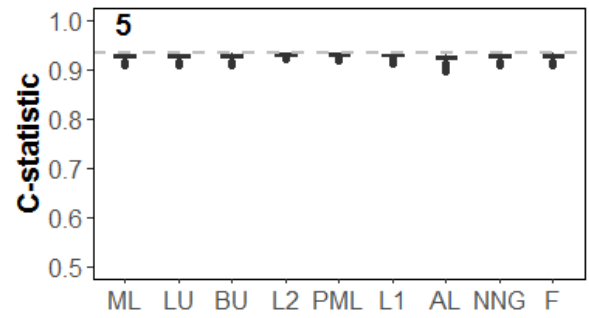
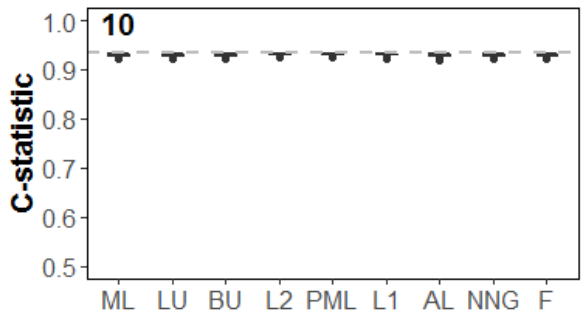
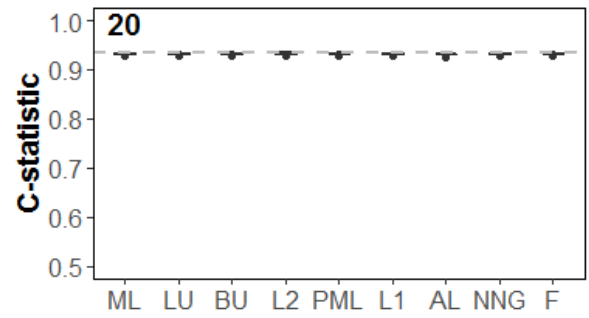
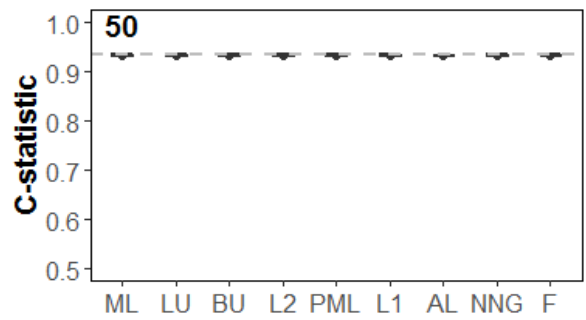

K. 10 true predictors, correlation 0, event rate 50%

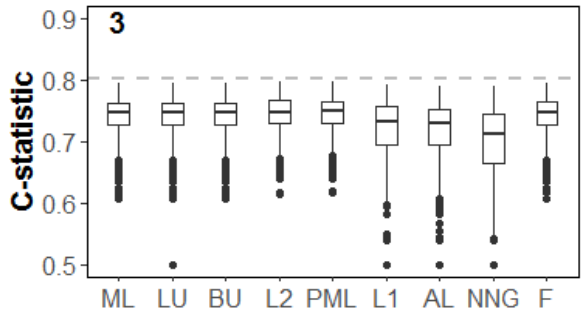
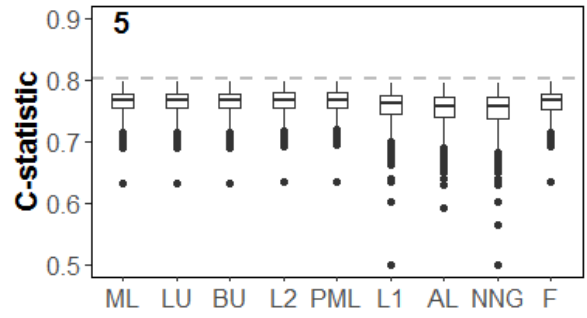
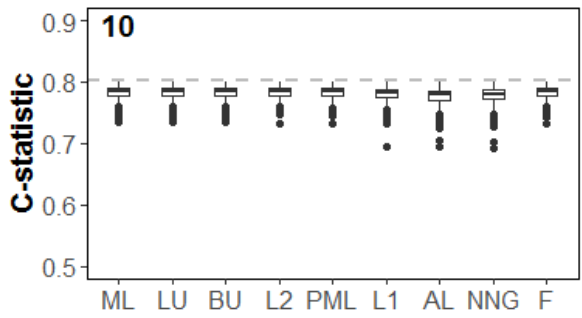
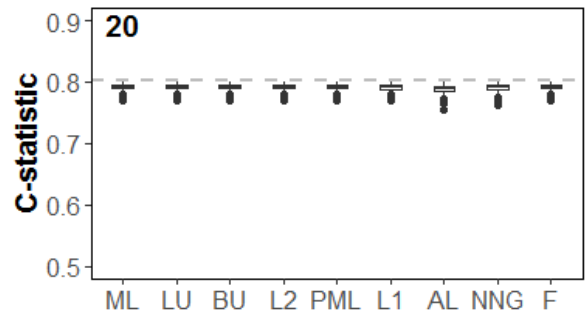
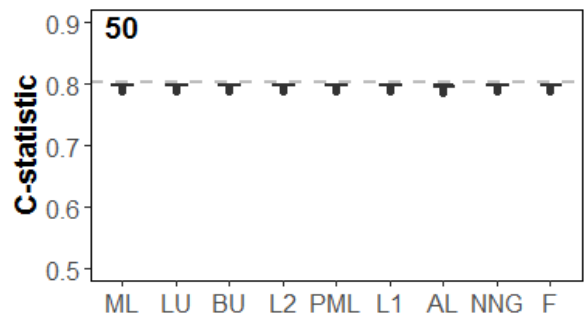

L. 10 true predictors, correlation 0.5, event rate 50%

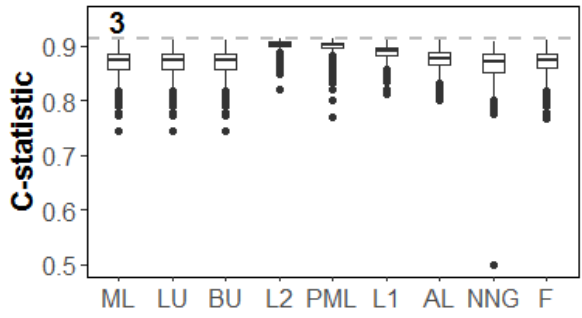
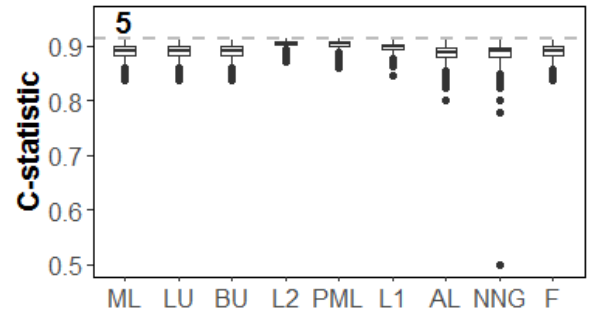
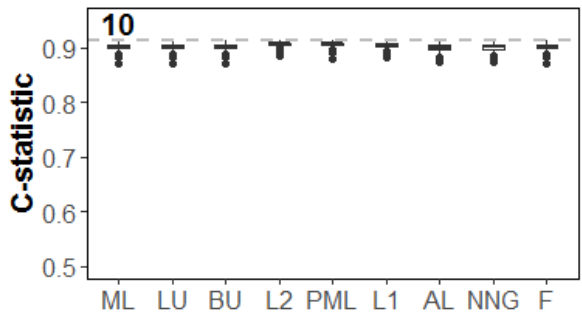
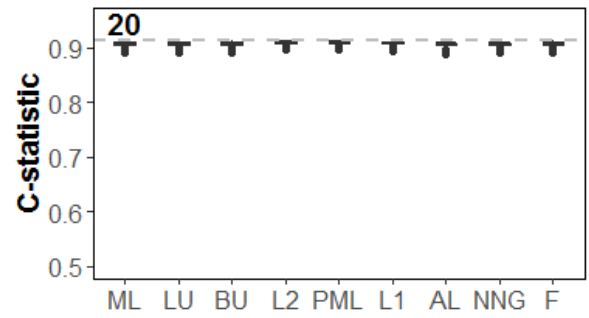
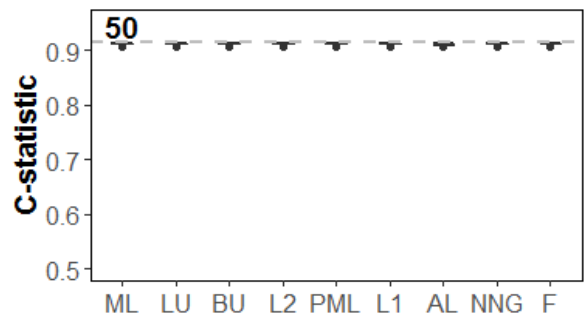

Figure S7. Scatter plots of calibration slopes with shrinkage vs calibration slopes without shrinkage (maximum likelihood). The green lines indicate a slope of 1, which is the target value. The blue line is the diagonal, points on the diagonal had the same calibration slope with and without shrinkage. Red points refer to runs where maximum likelihood gave a slope >1. ML, maximum likelihood; LU, uniform shrinkage based on likelihood; BU, uniform shrinkage based on bootstrap; L2, ridge (L2 penalty); PML, penalized maximum likelihood; L1, LASSO (L1 penalty); AL, adaptive LASSO; NNG, non-negative garrote; F, Firth's correction.

I. 5 true predictors, 0 correlation, 10% event rate, 3 events per variable (EPV)

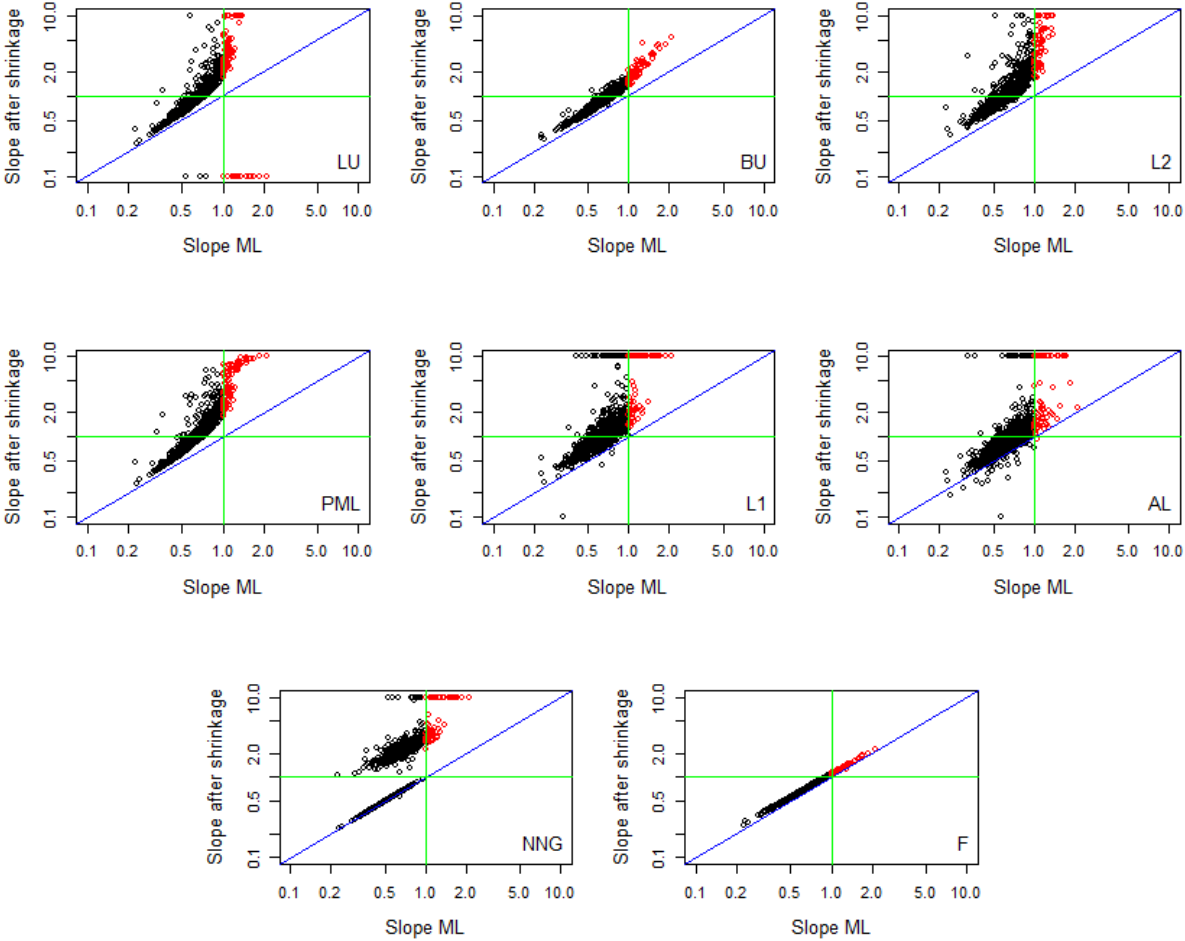

II. 5 true predictors, 0 correlation, 10% event rate, 5 EPV

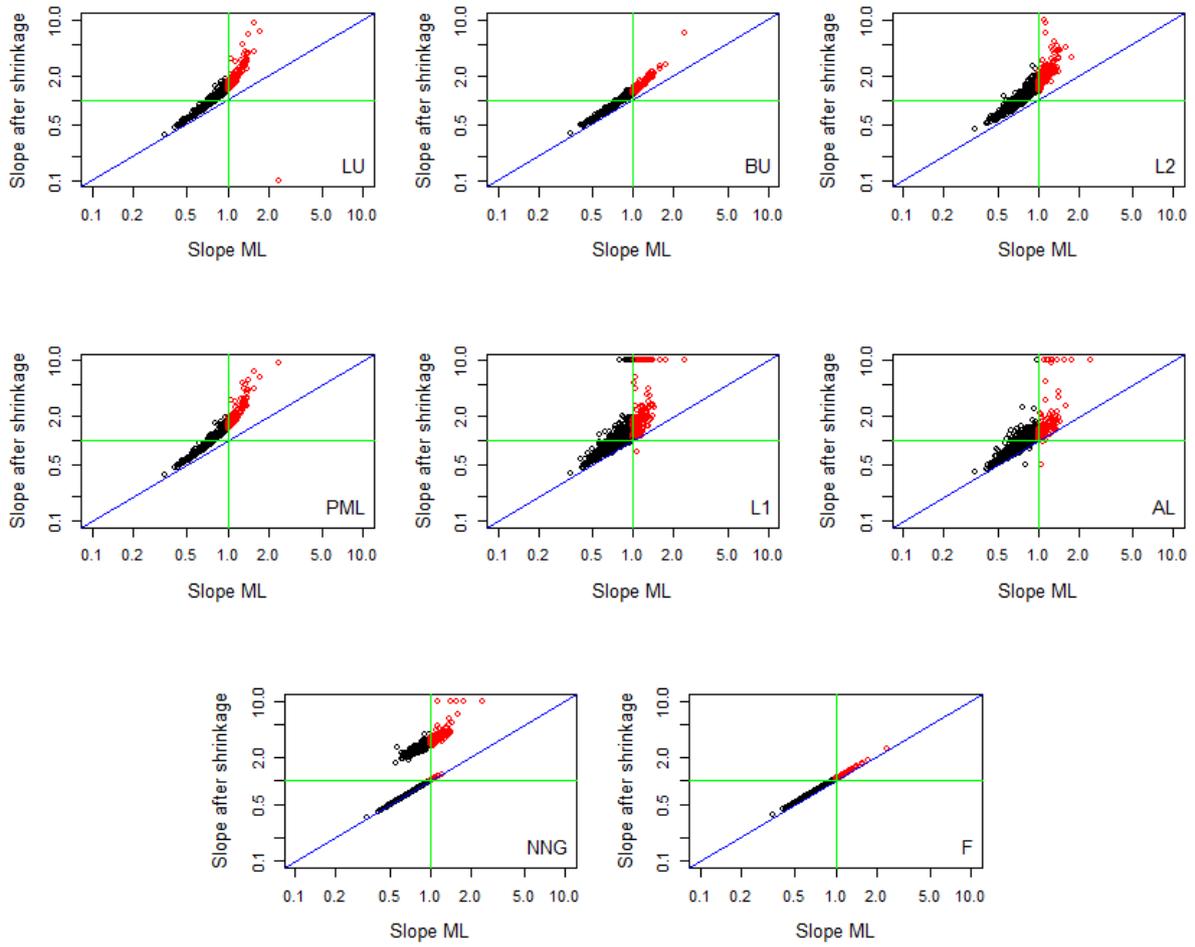

III. 5 true predictors, 0 correlation, 10% event rate, 10 EPV

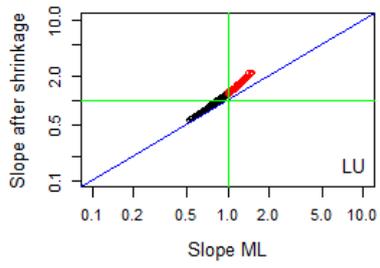 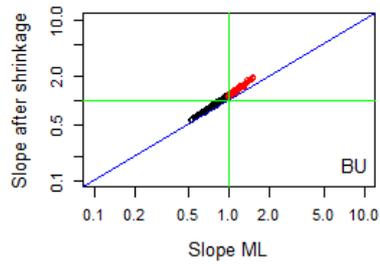 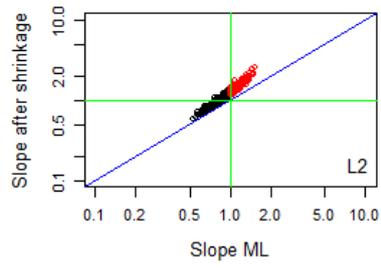
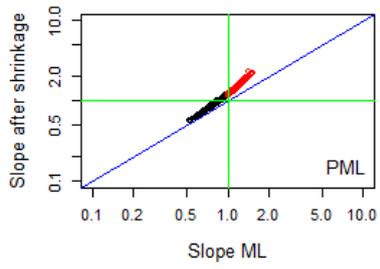 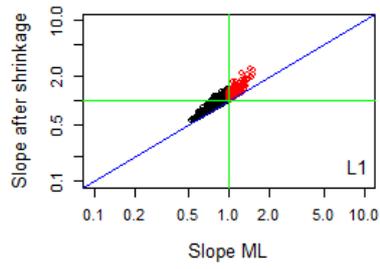 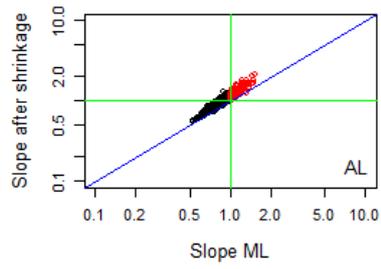
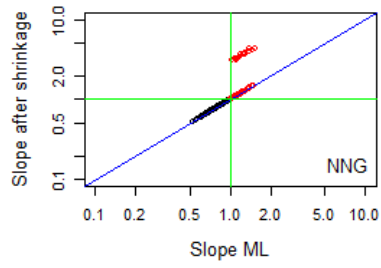 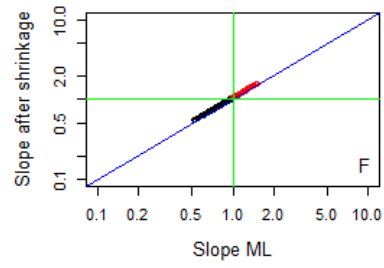

IV.     5 true predictors, 0 correlation, 10% event rate, 20 EPV

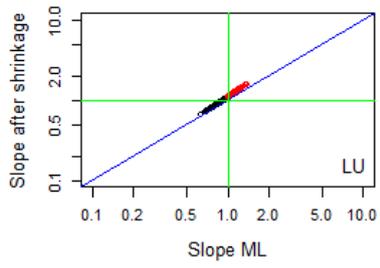 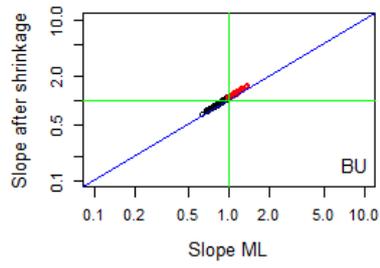 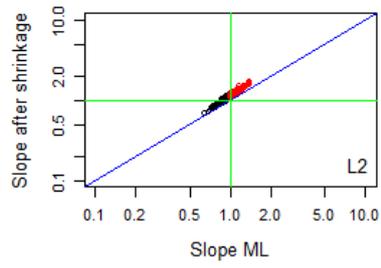
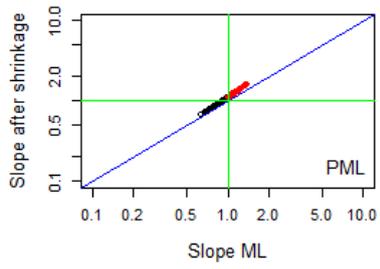 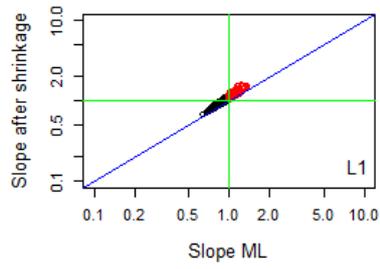 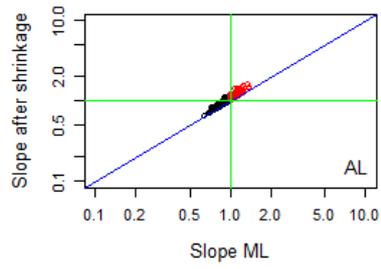
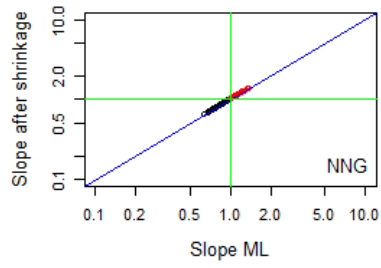 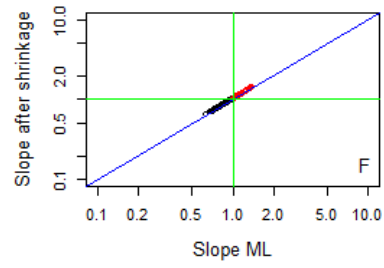

V. 5 true predictors, 0 correlation, 10% event rate, 50 EPV

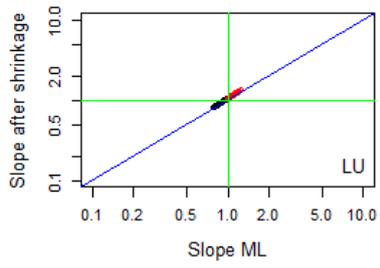 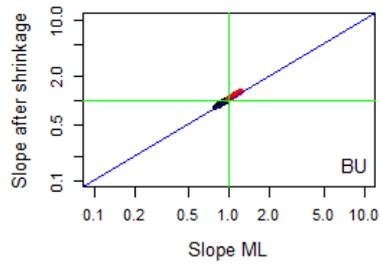 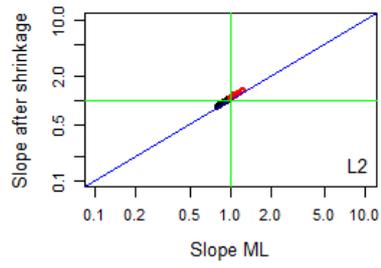
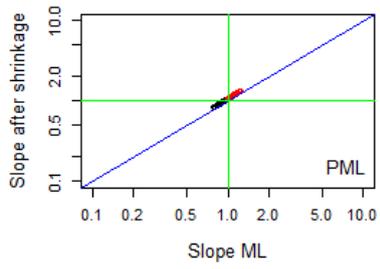 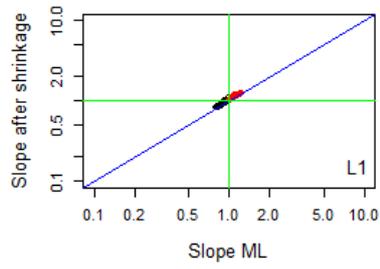 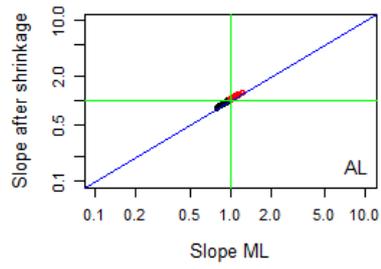
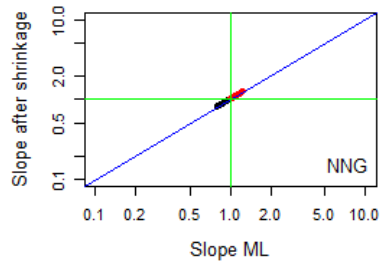 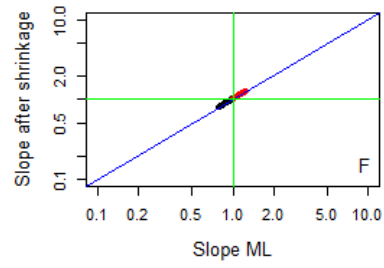

VI.  5 true predictors, 0.5 correlation, 10% event rate, 3 EPV

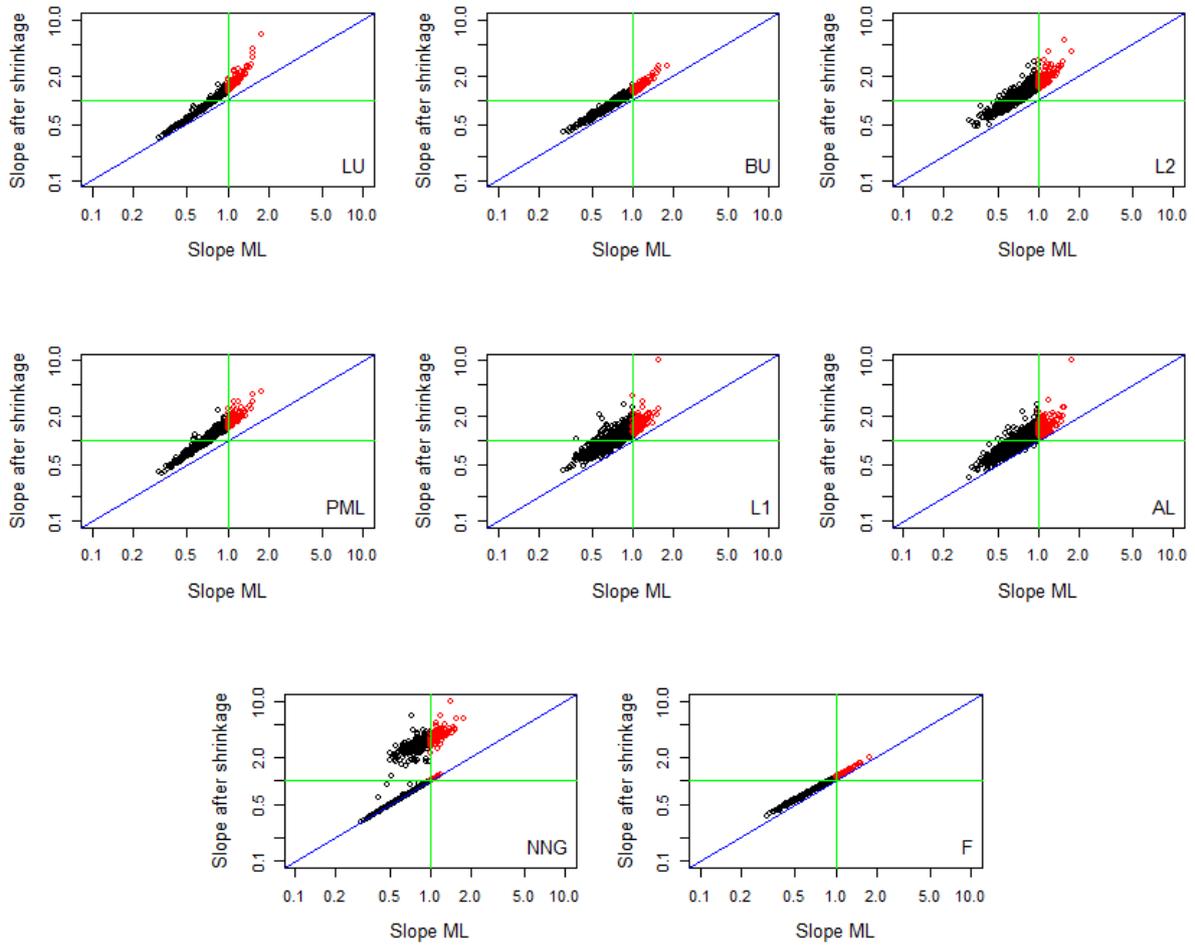

VII. 5 true predictors, 0.5 correlation, 10% event rate, 5 EPV

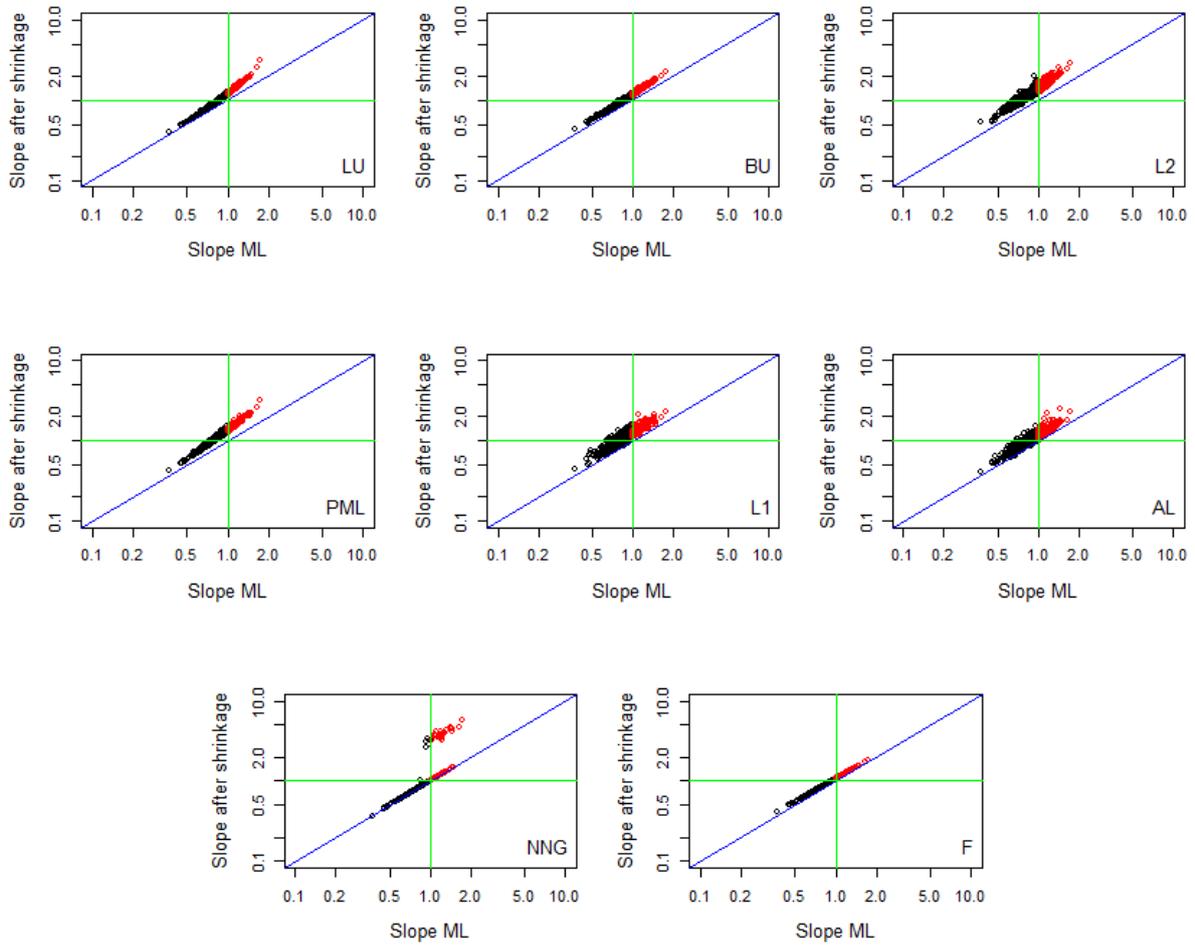

VIII.     5 true predictors, 0.5 correlation, 10% event rate, 10 EPV

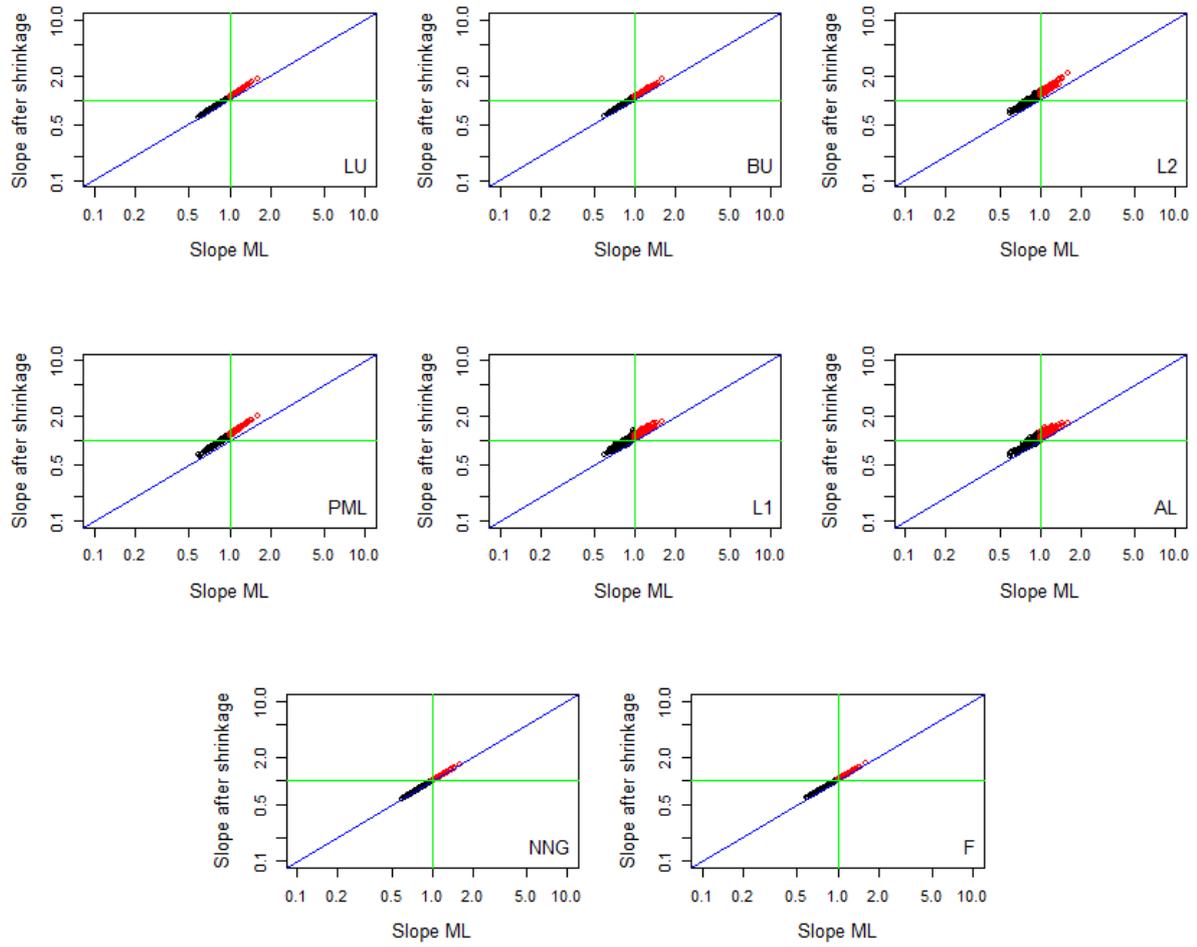

IX. 5 true predictors, 0.5 correlation, 10% event rate, 20 EPV

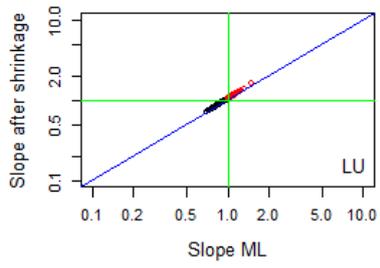
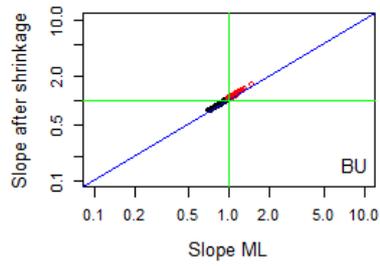
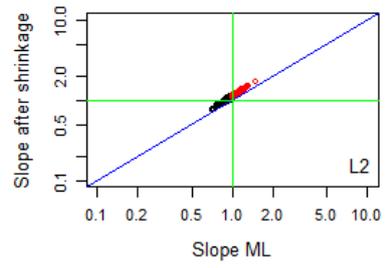
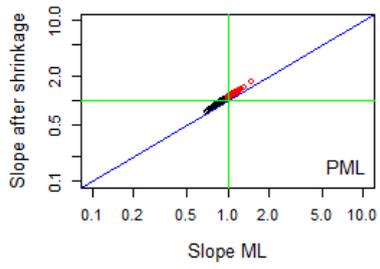
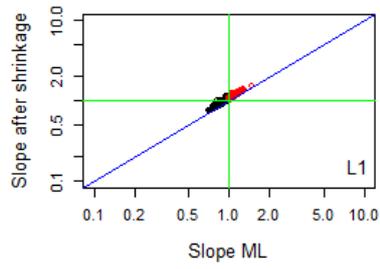
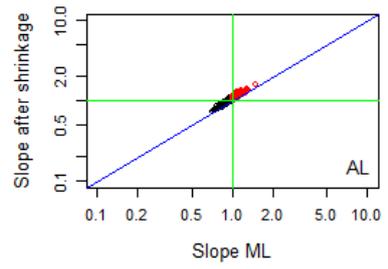
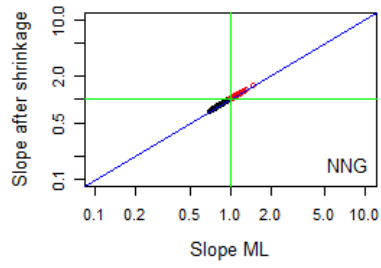
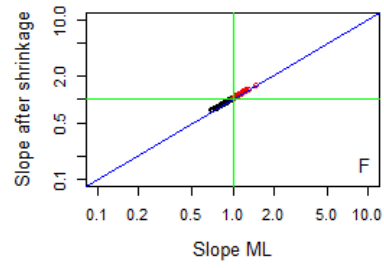

X.   5 true predictors, 0.5 correlation, 10% event rate, 50 EPV

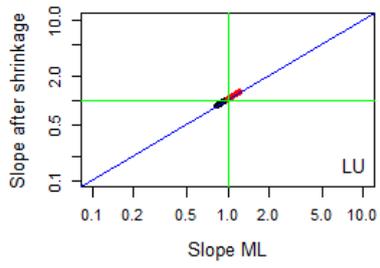
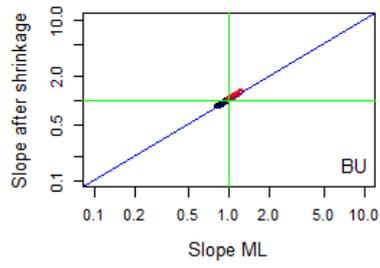
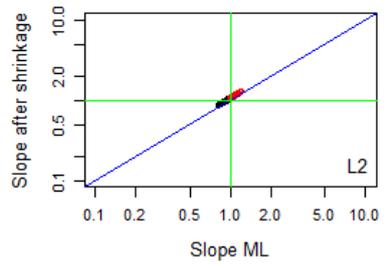
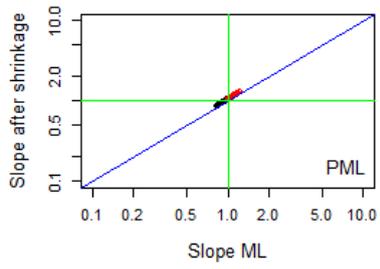
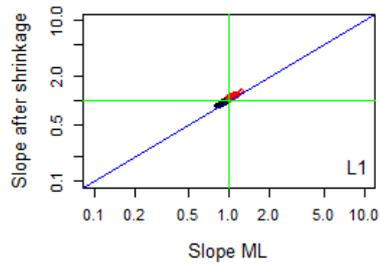
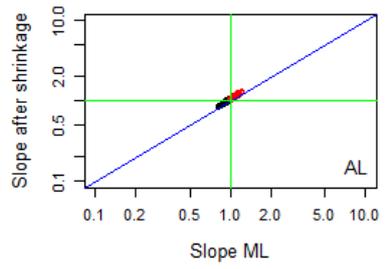
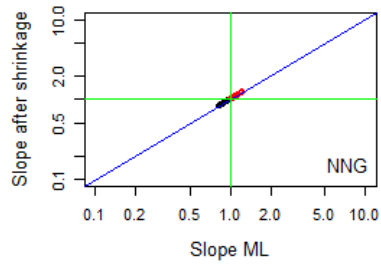
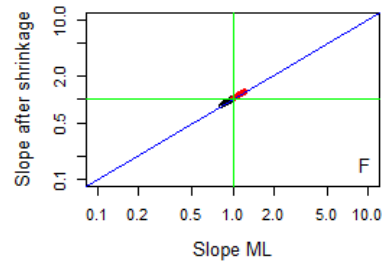

XI.     5 true predictors, 0 correlation, 50% event rate, 3 EPV

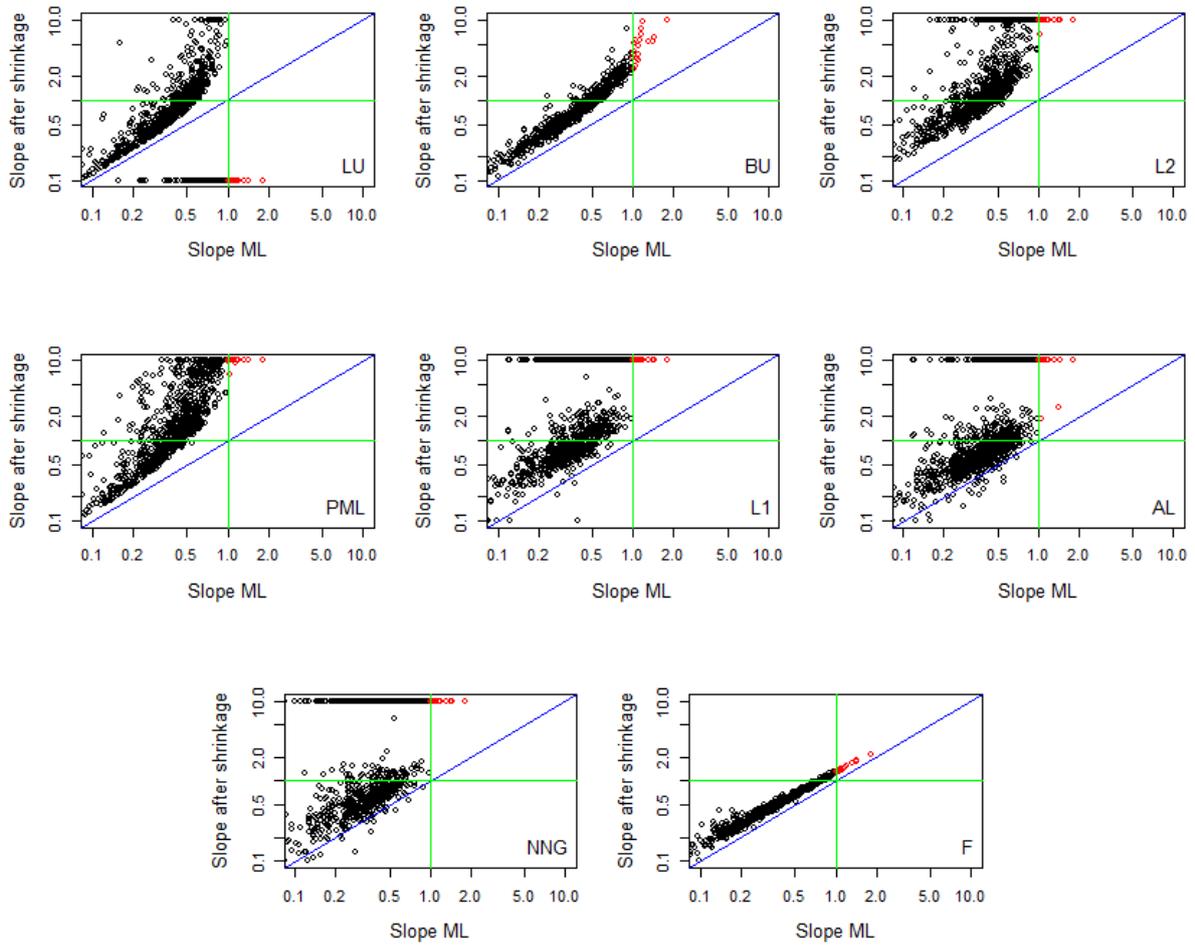

XII. 5 true predictors, 0 correlation, 50% event rate, 5 EPV

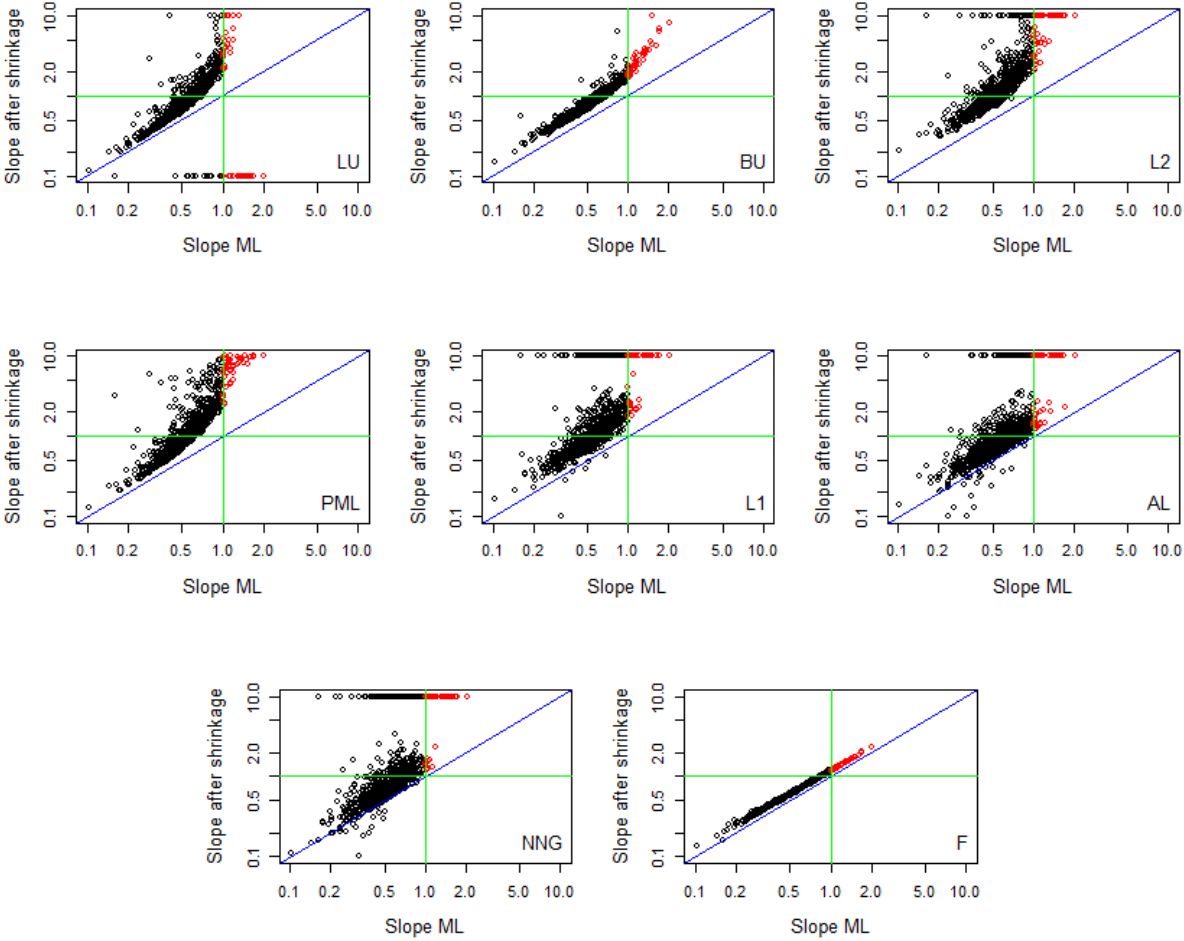

XIII. 5 true predictors, 0 correlation, 50% event rate, 10 EPV

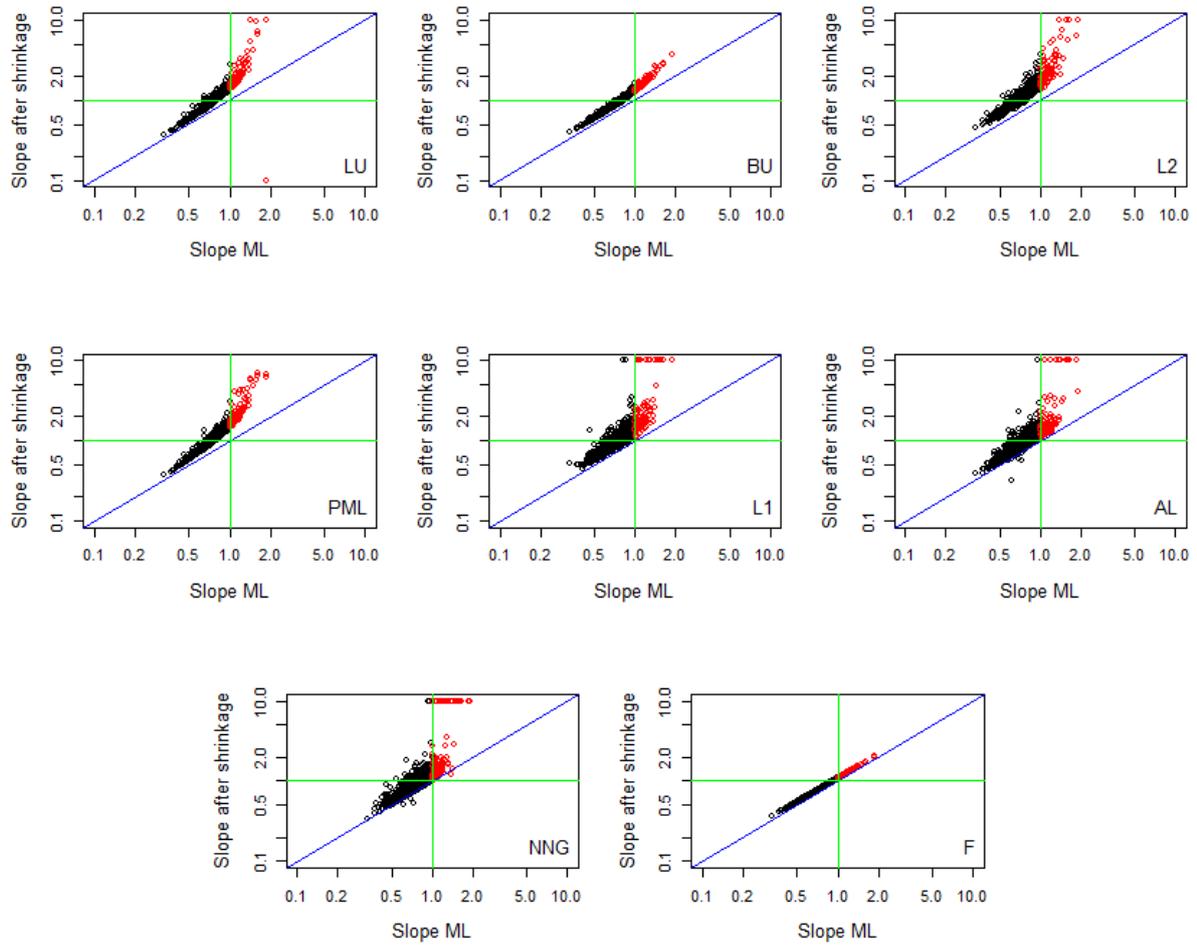

XIV. 5 true predictors, 0 correlation, 50% event rate, 20 EPV

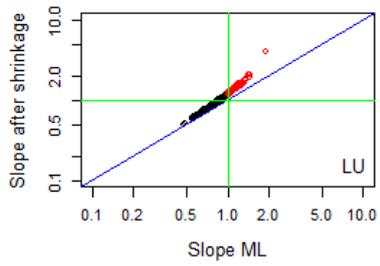
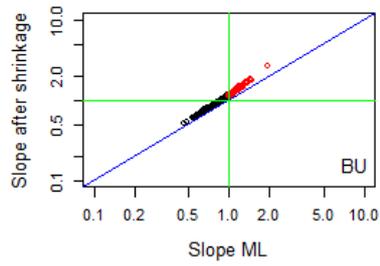
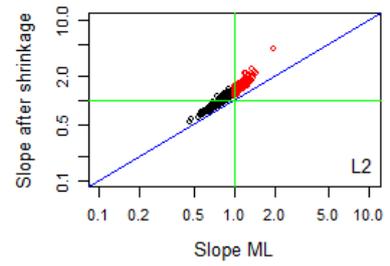
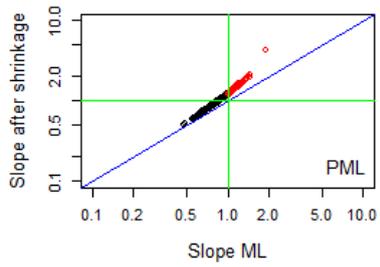
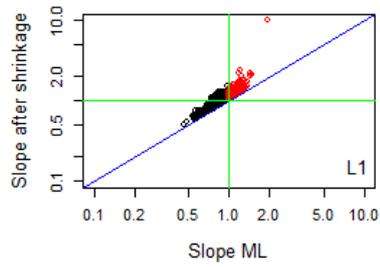
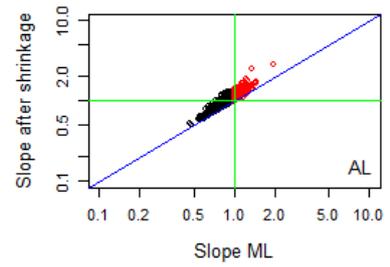
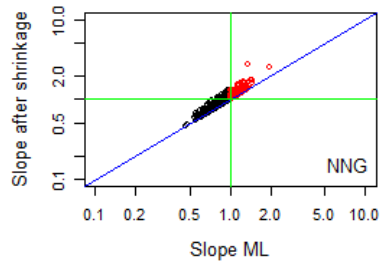
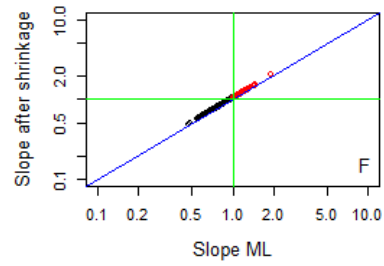

XV. 5 true predictors, 0 correlation, 50% event rate, 50 EPV

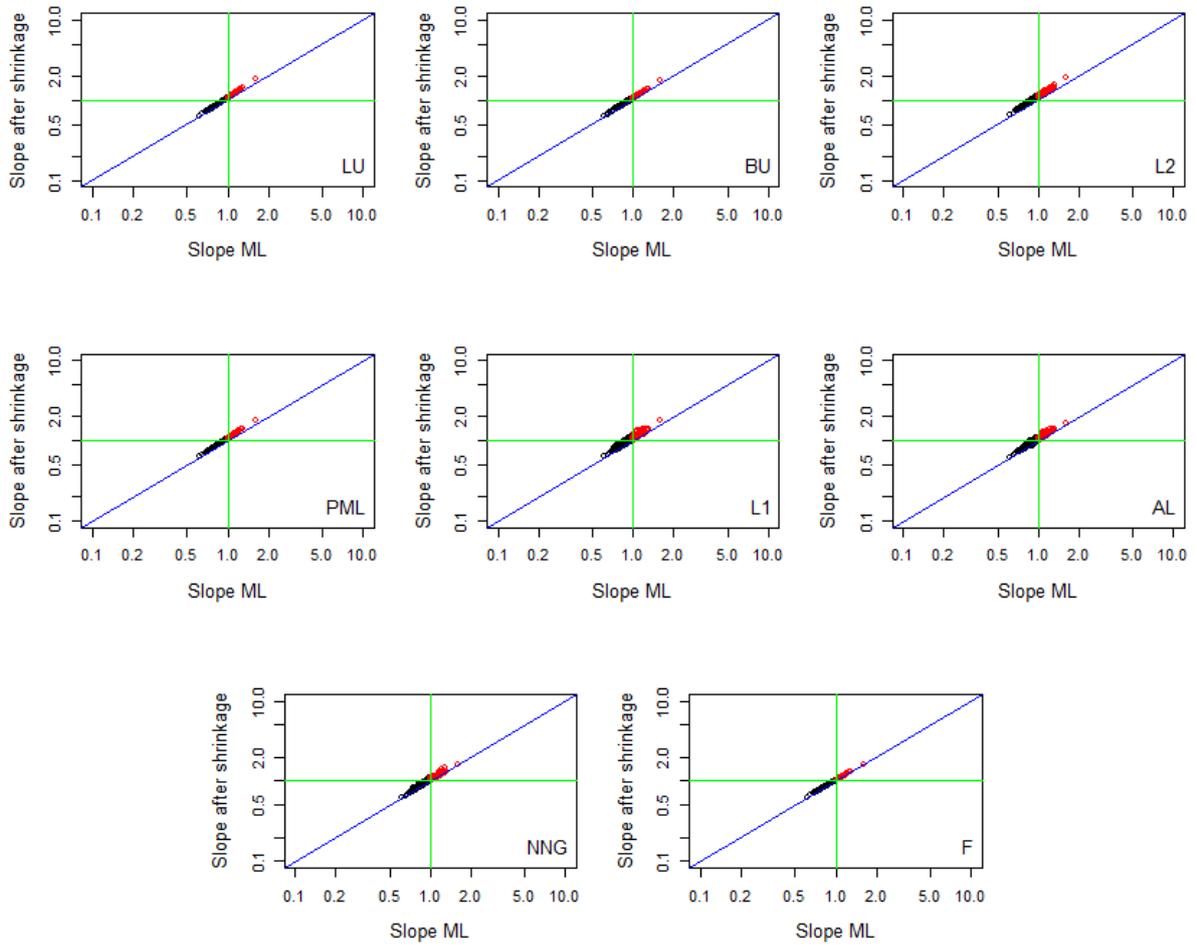

XVI. 5 true predictors, 0.5 correlation, 50% event rate, 3 EPV

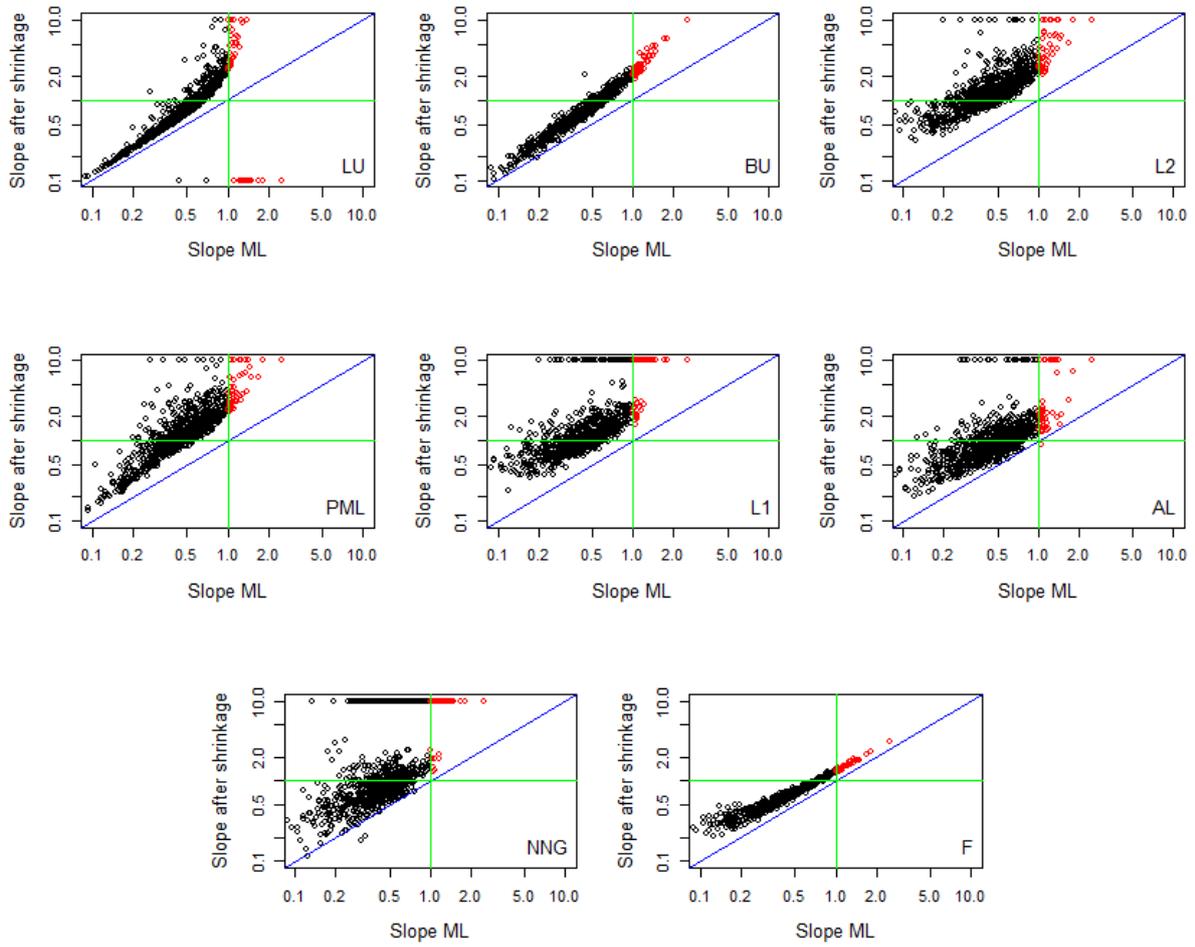

XVII. 5 true predictors, 0.5 correlation, 50% event rate, 5 EPV

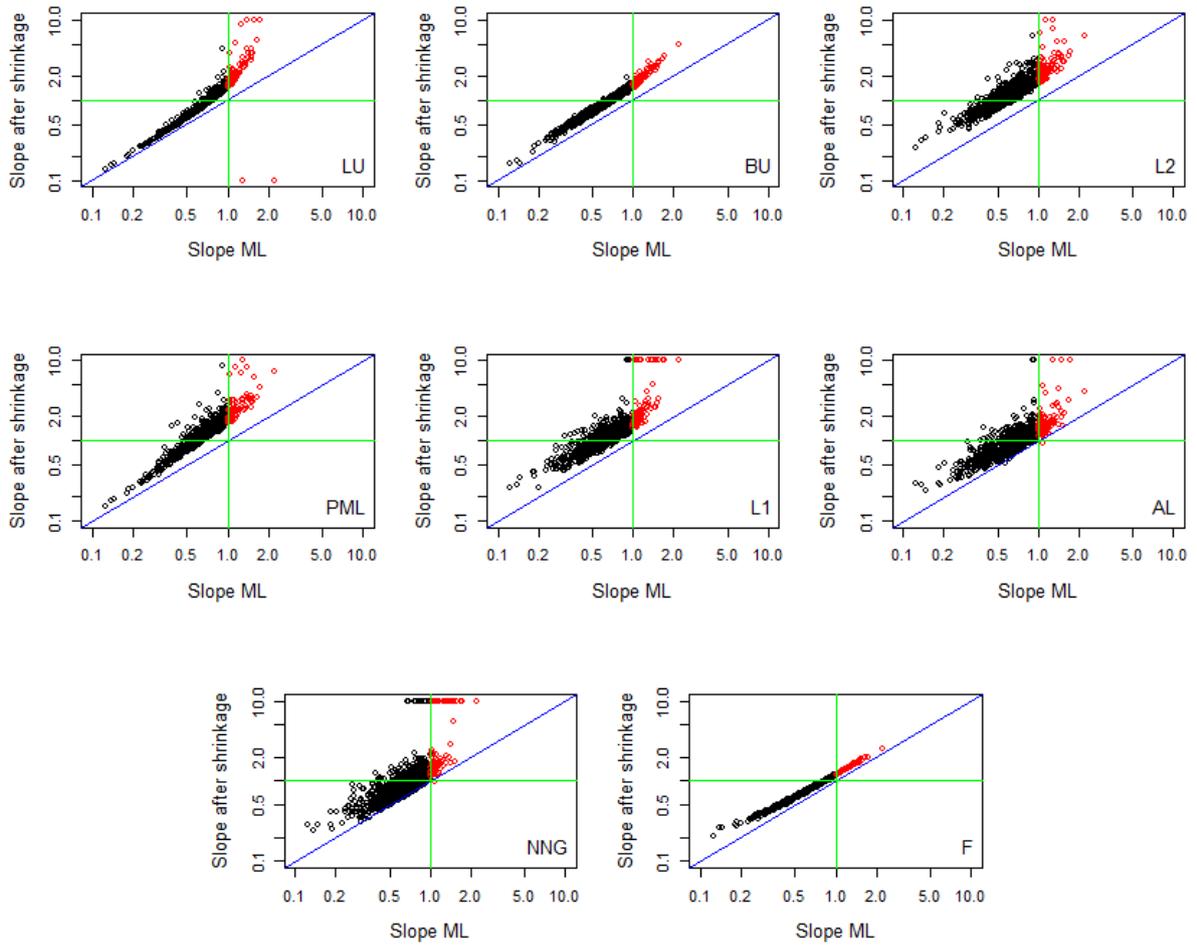

XVIII. 5 true predictors, 0.5 correlation, 50% event rate, 10 EPV

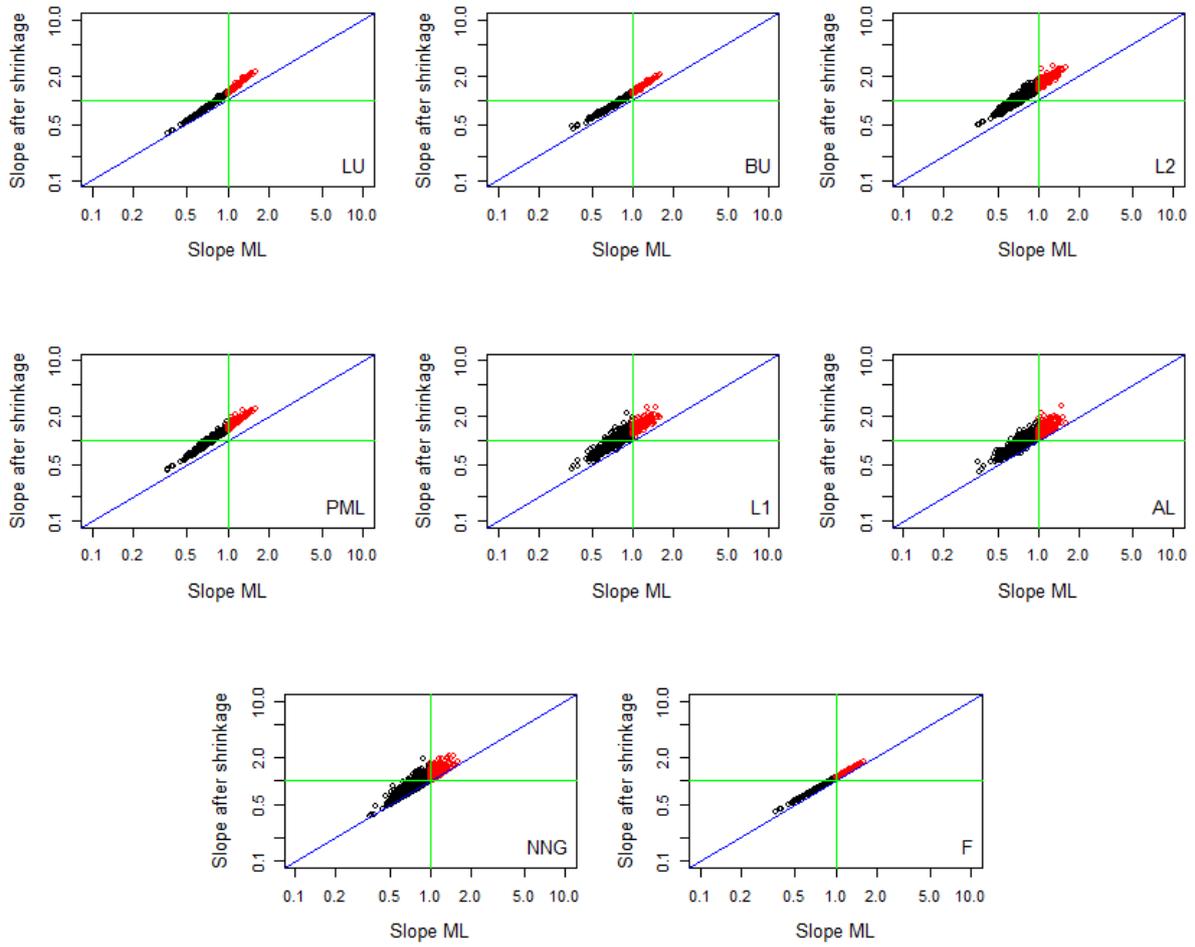

XIX.     5 true predictors, 0.5 correlation, 50% event rate, 20 EPV

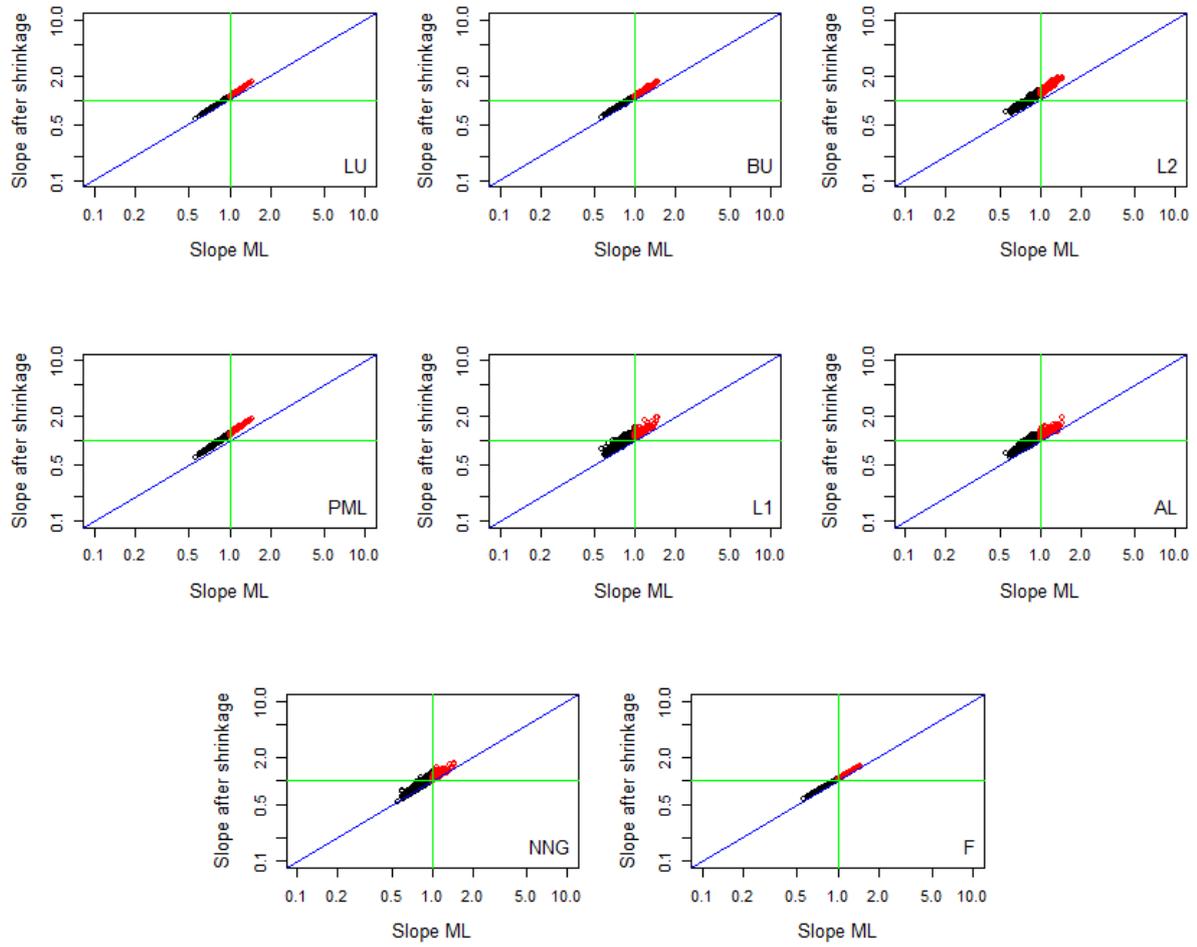

XX. 5 true predictors, 0.5 correlation, 50% event rate, 50 EPV

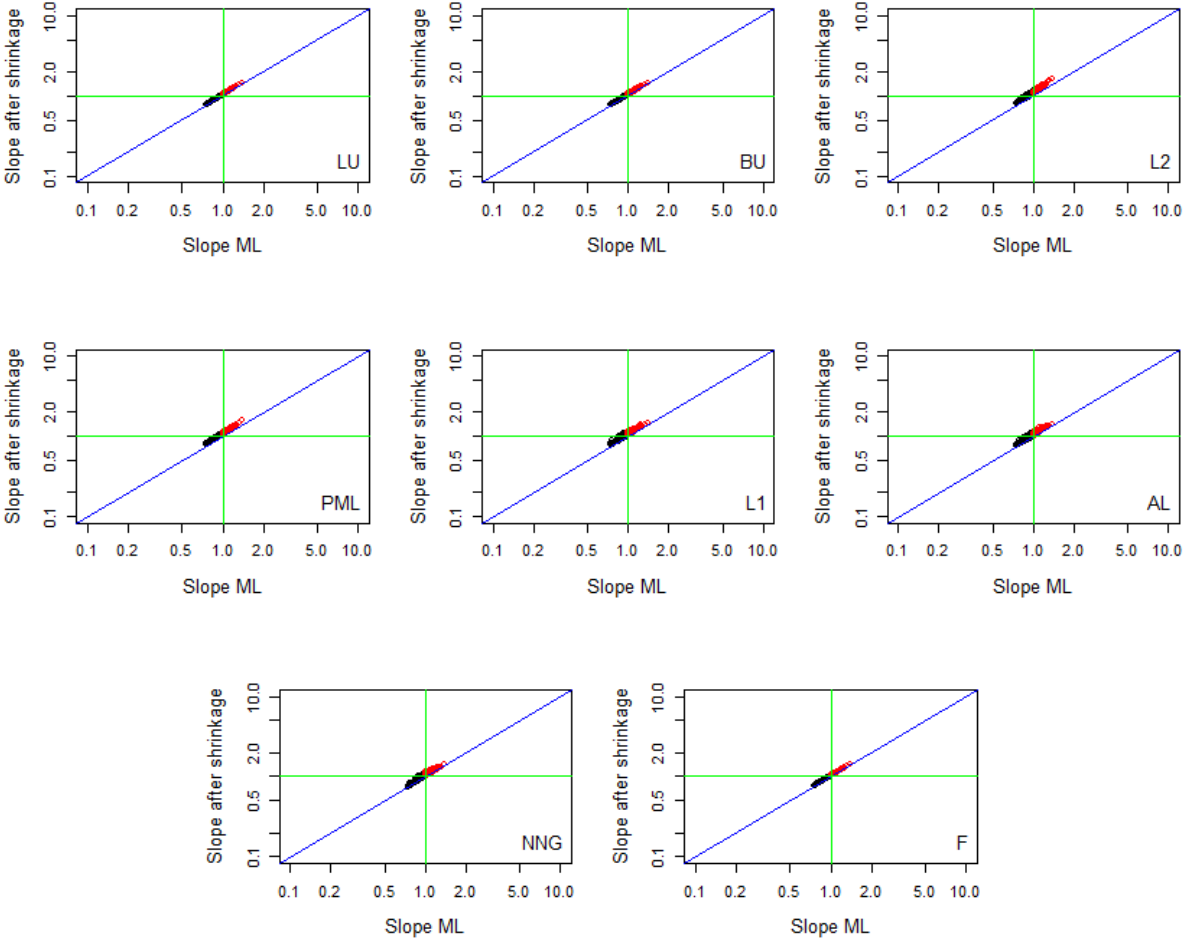

XXI. 5 true and 5 noise predictors, 0 correlation, 10% event rate, 3 EPV

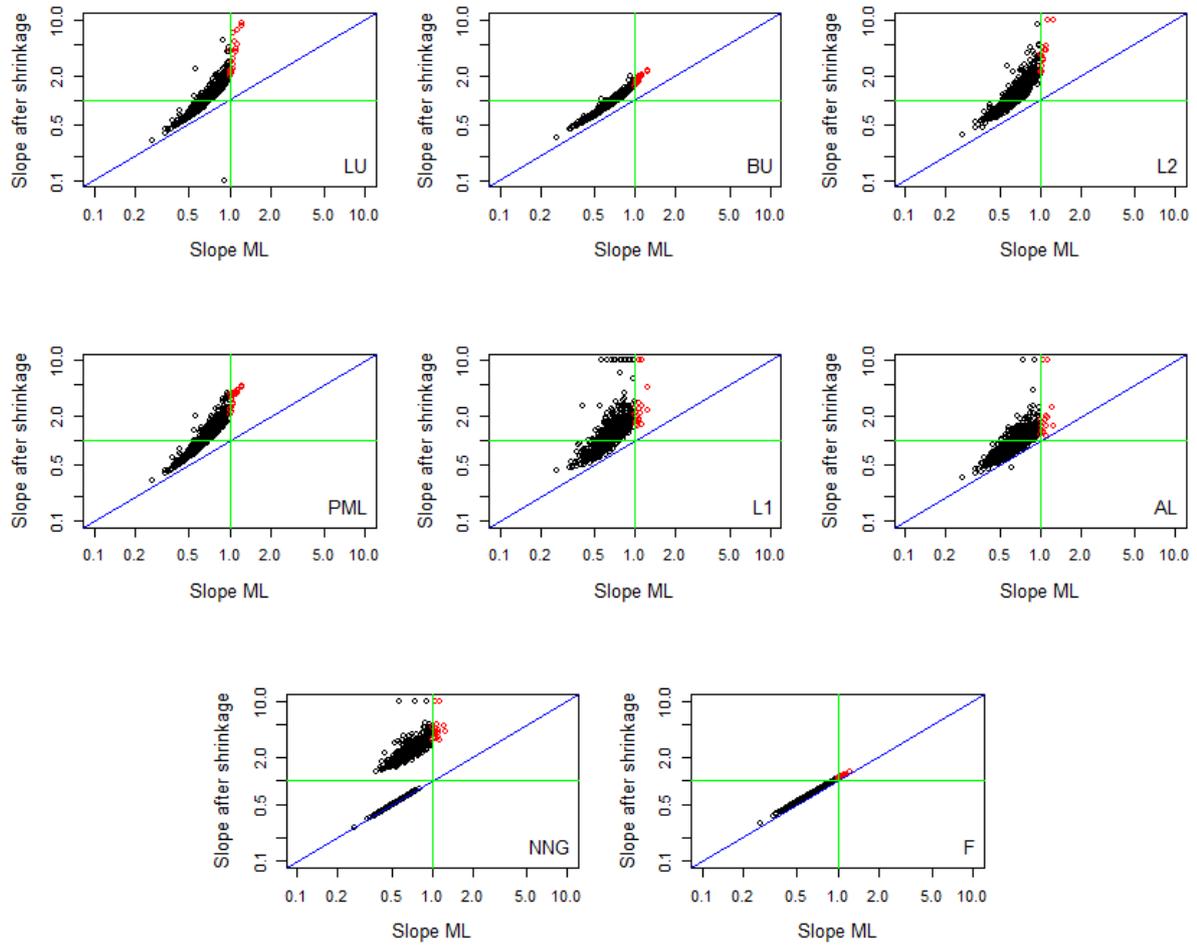

XXII. 5 true and 5 noise predictors, 0 correlation, 10% event rate, 5 EPV

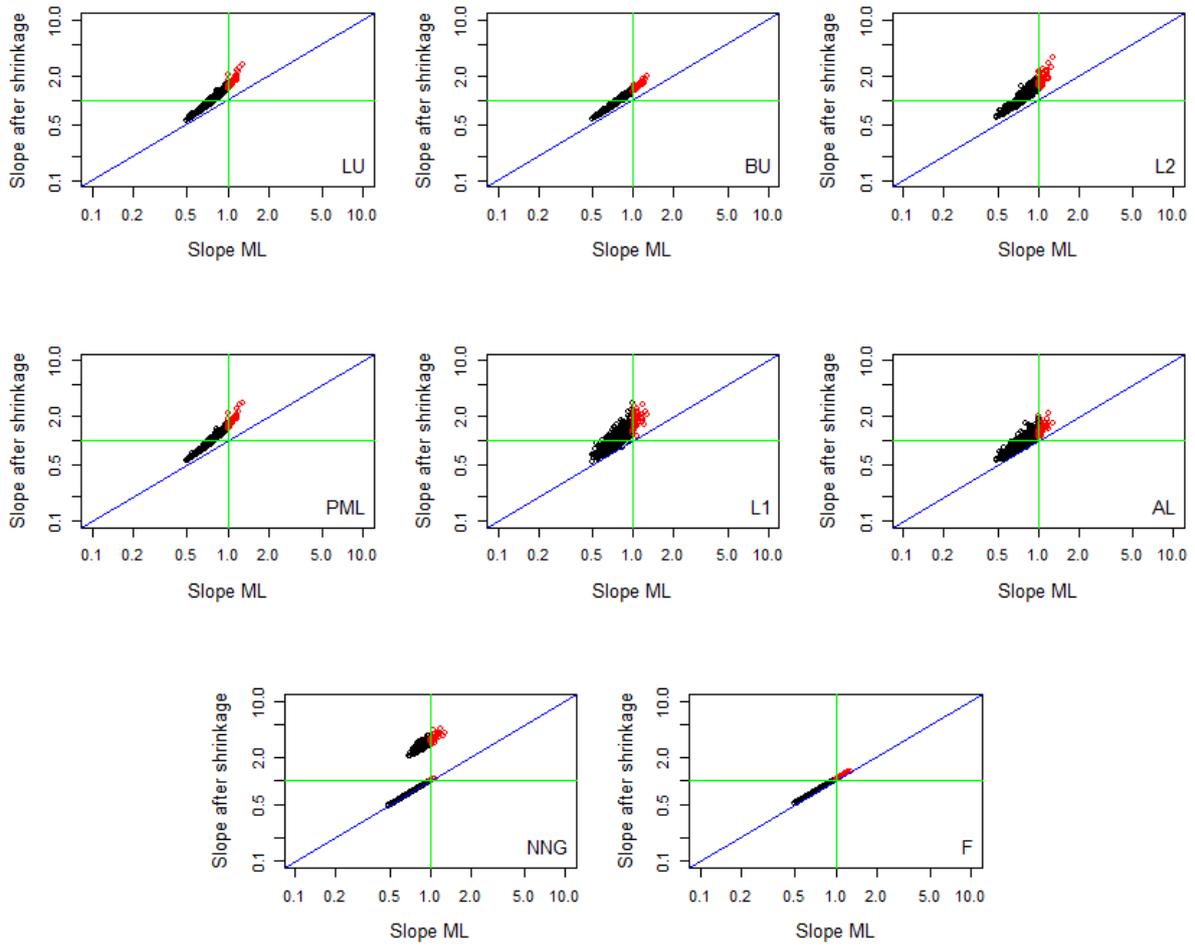

XXIII. 5 true and 5 noise predictors, 0 correlation, 10% event rate, 10 EPV

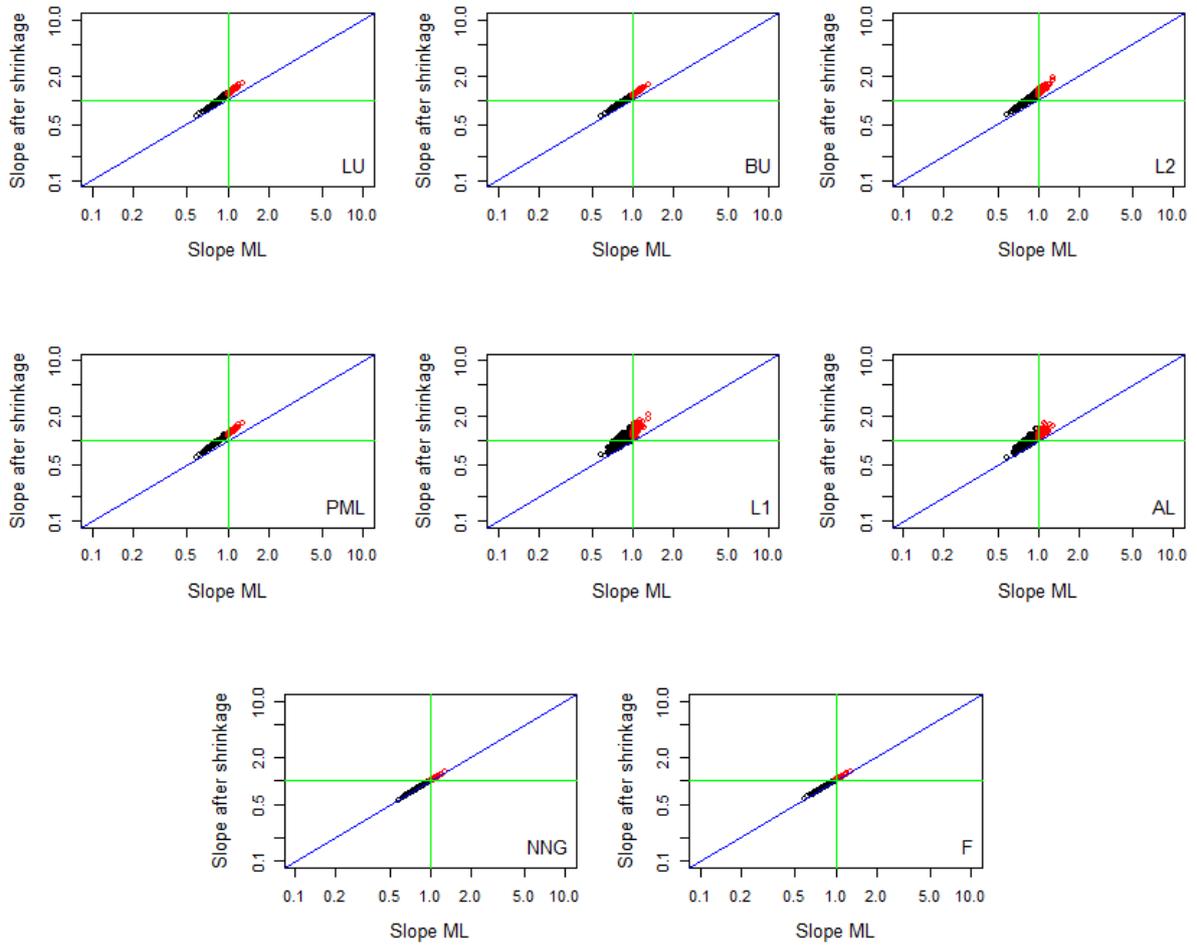

XXIV. 5 true and 5 noise predictors, 0 correlation, 10% event rate, 20 EPV

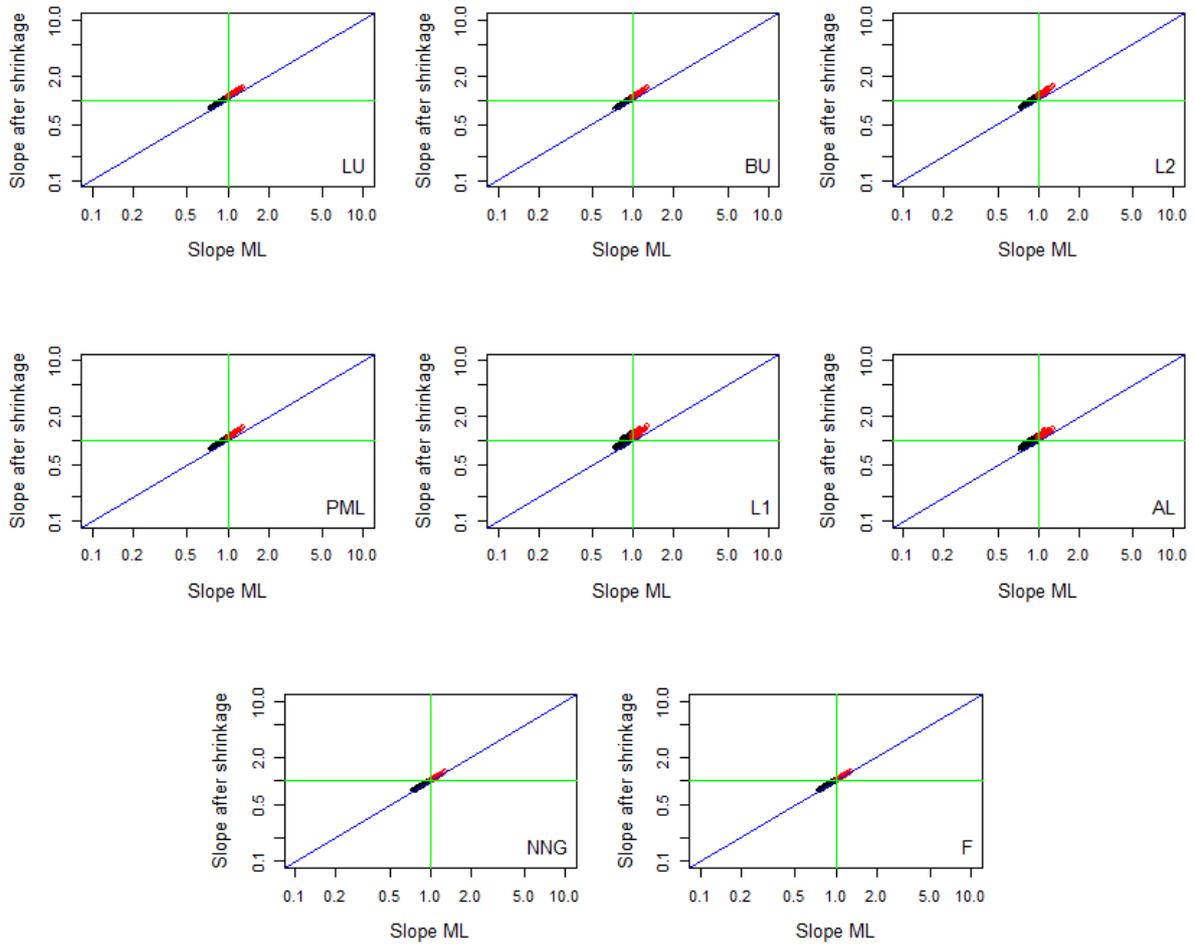

XXV.   5 true and 5 noise predictors, 0 correlation, 10% event rate, 50 EPV

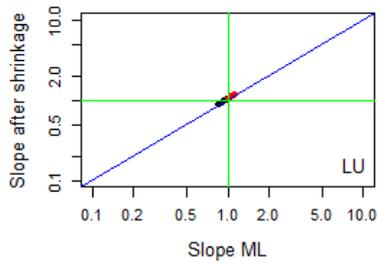
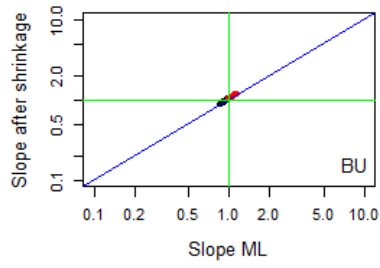
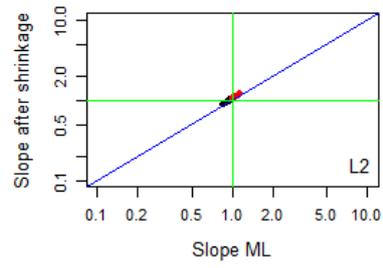
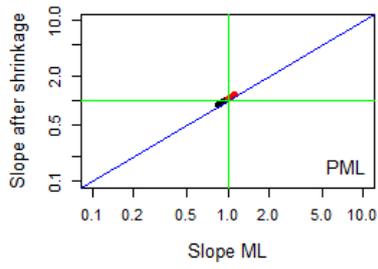
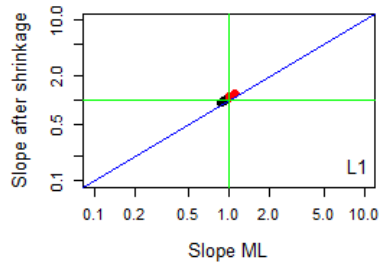
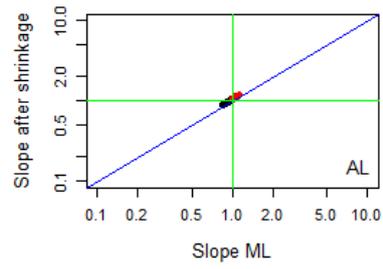
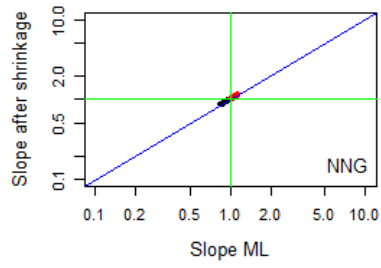
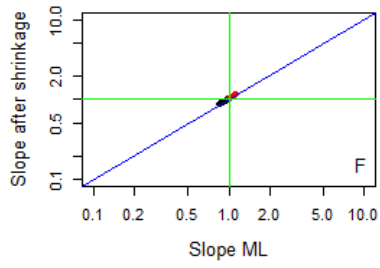

XXVI. 5 true and 5 noise predictors, 0.5 correlation, 10% event rate, 3 EPV

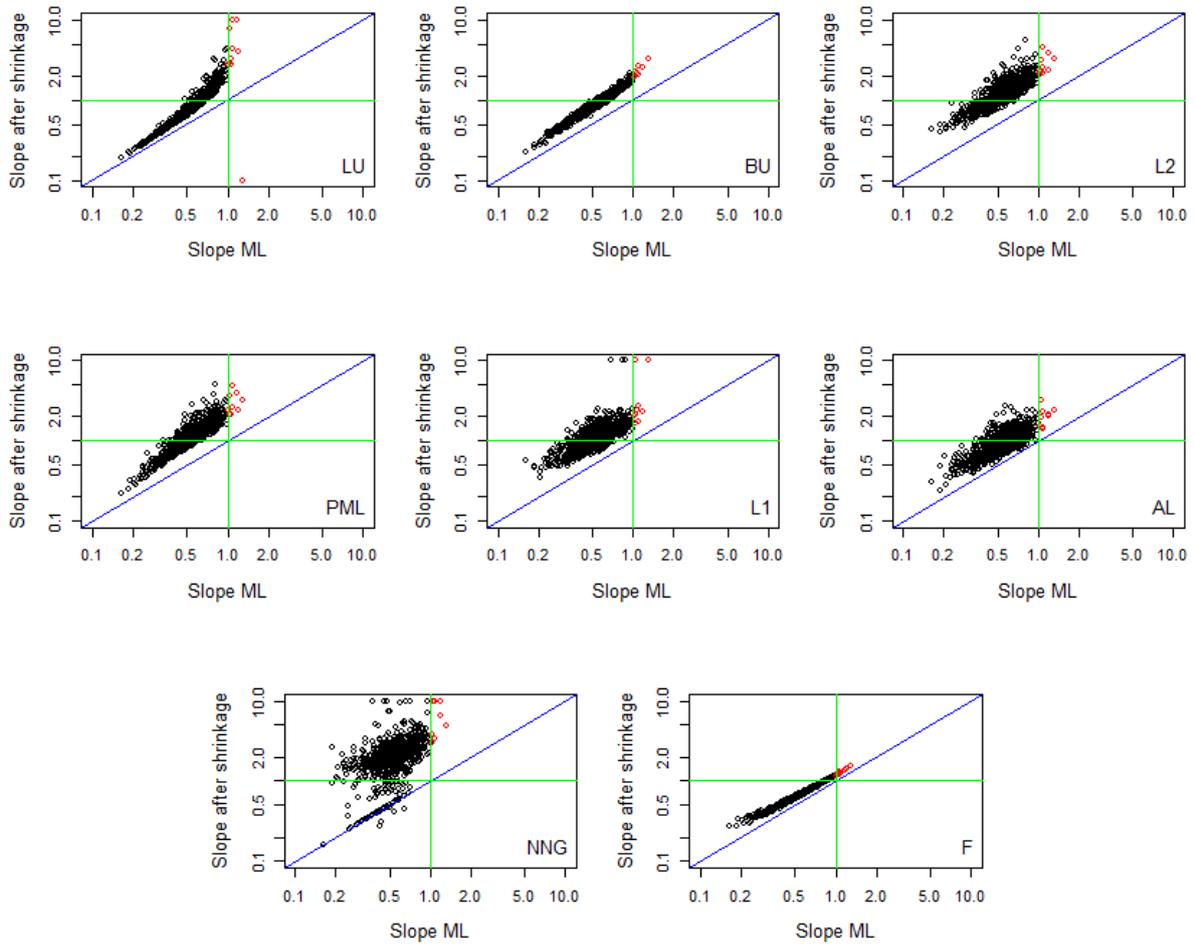

XXVII. 5 true and 5 noise predictors, 0.5 correlation, 10% event rate, 5 EPV

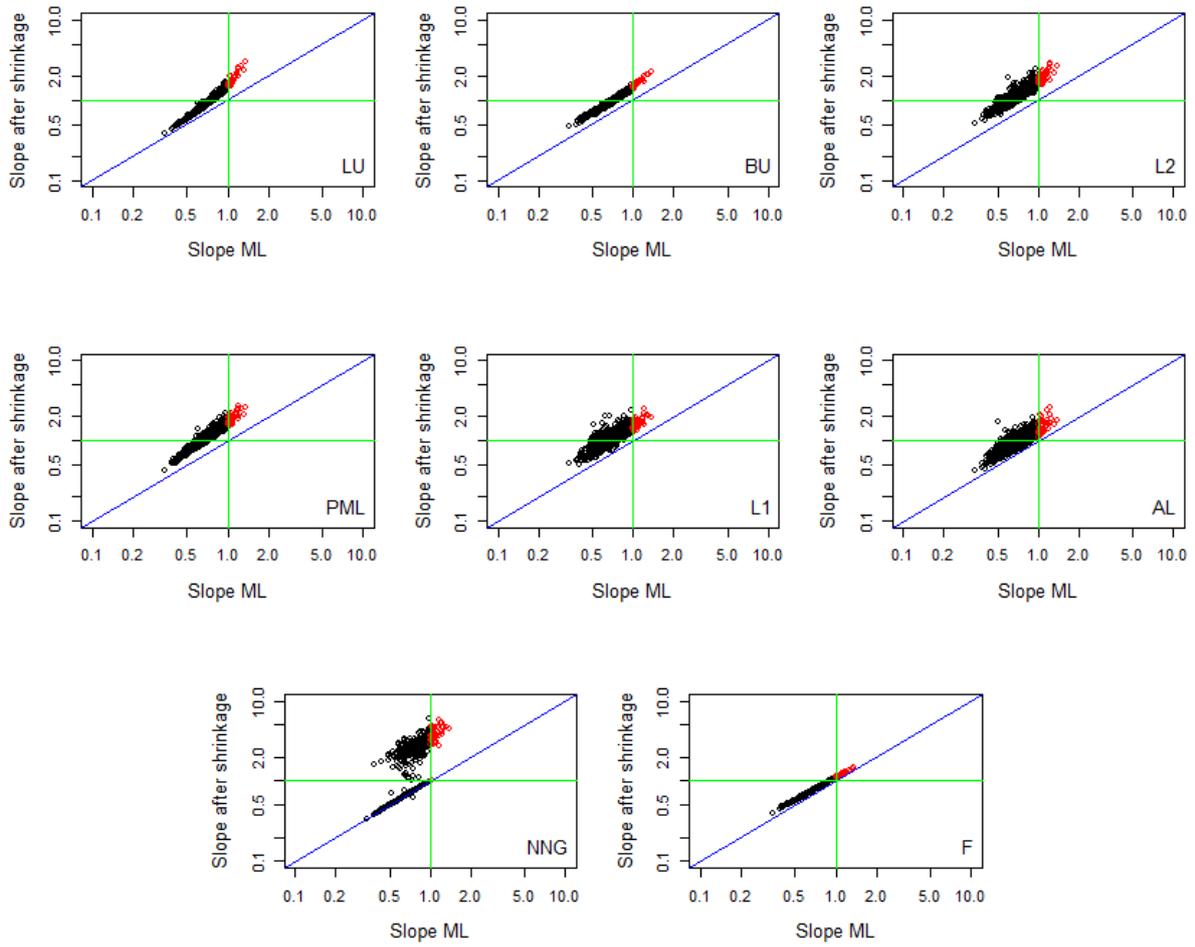

XXVIII. 5 true and 5 noise predictors, 0.5 correlation, 10% event rate, 10 EPV

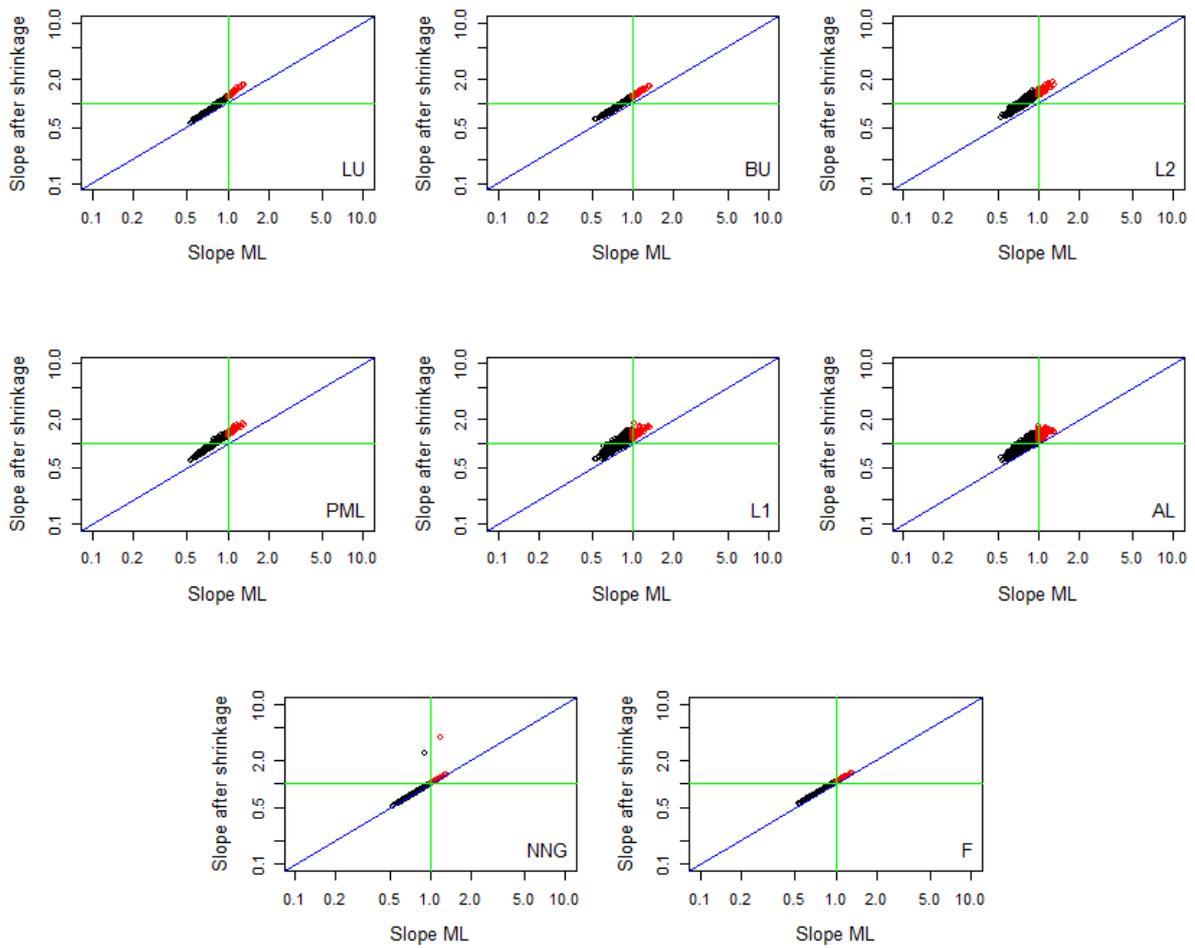

XXIX.    5 true and 5 noise predictors, 0.5 correlation, 10% event rate, 20 EPV

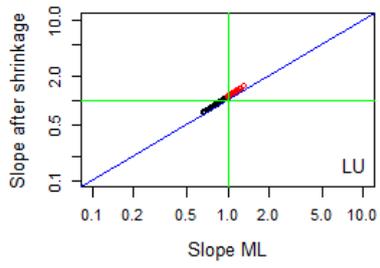
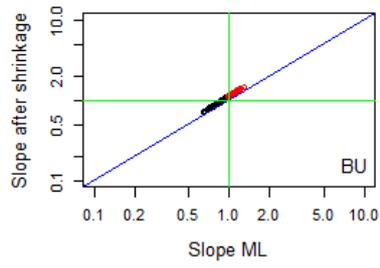
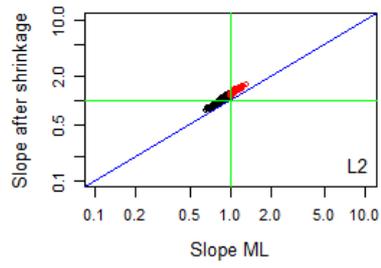
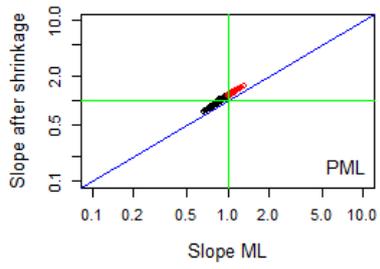
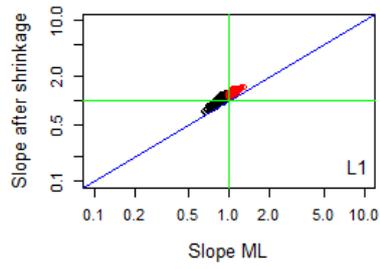
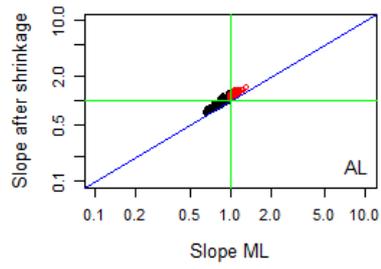
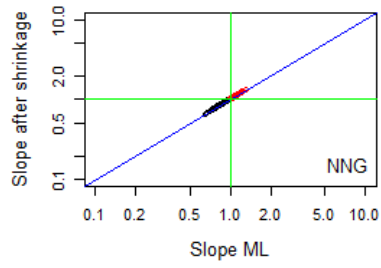
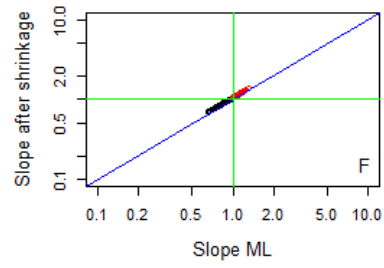

XXX. 5 true and 5 noise predictors, 0.5 correlation, 10% event rate, 50 EPV

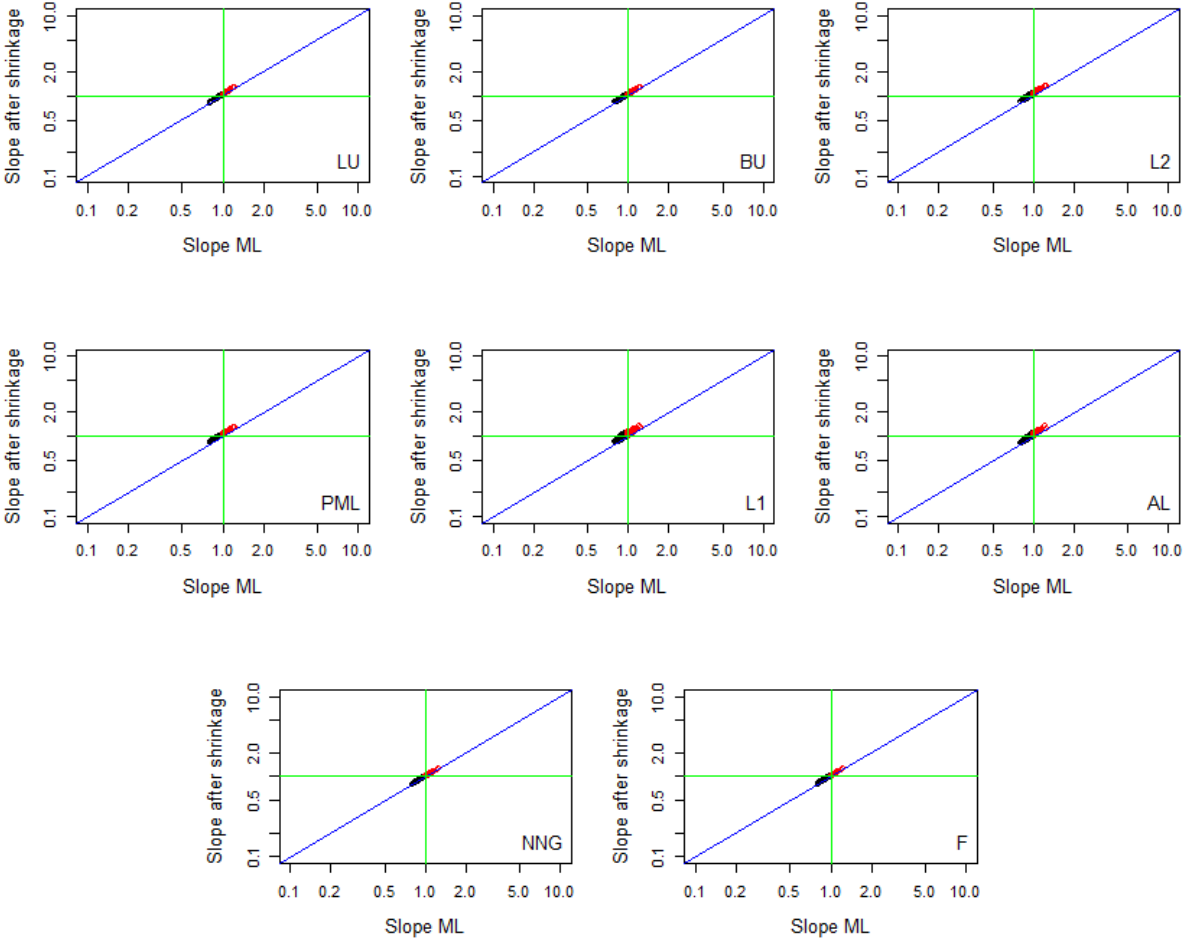

XXXI. 5 true and 5 noise predictors, 0 correlation, 50% event rate, 3 EPV

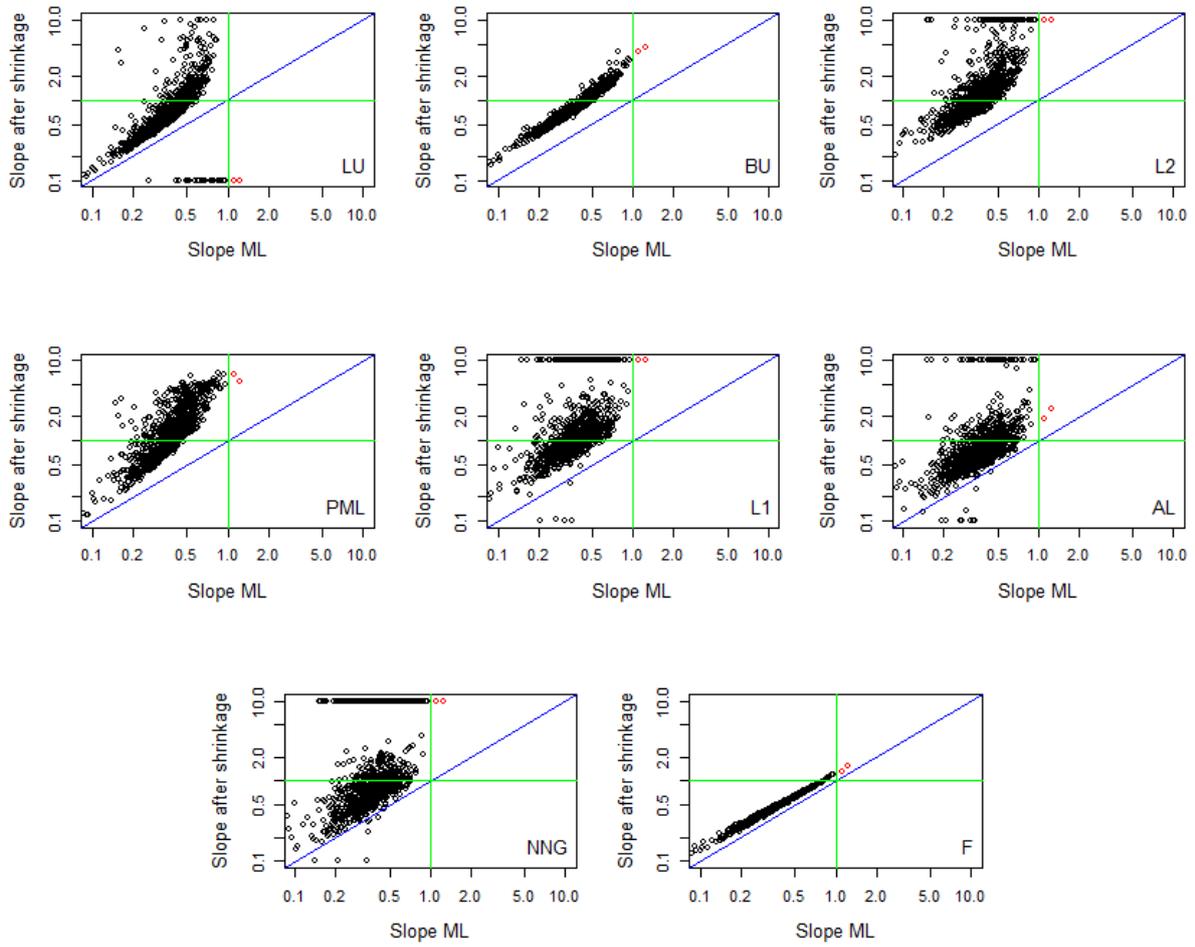

XXXII. 5 true and 5 noise predictors, 0 correlation, 50% event rate, 5 EPV

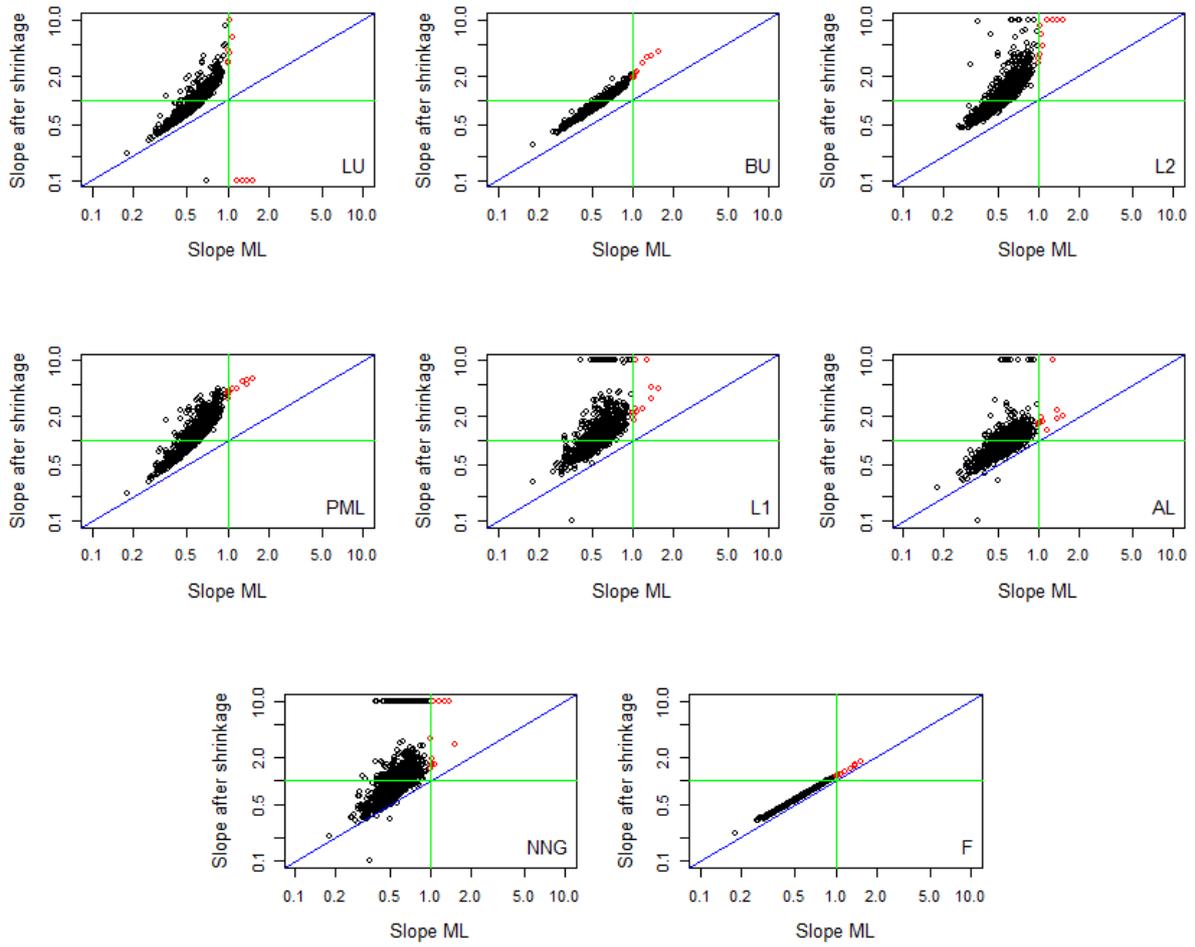

XXXIII. 5 true and 5 noise predictors, 0 correlation, 50% event rate, 10 EPV

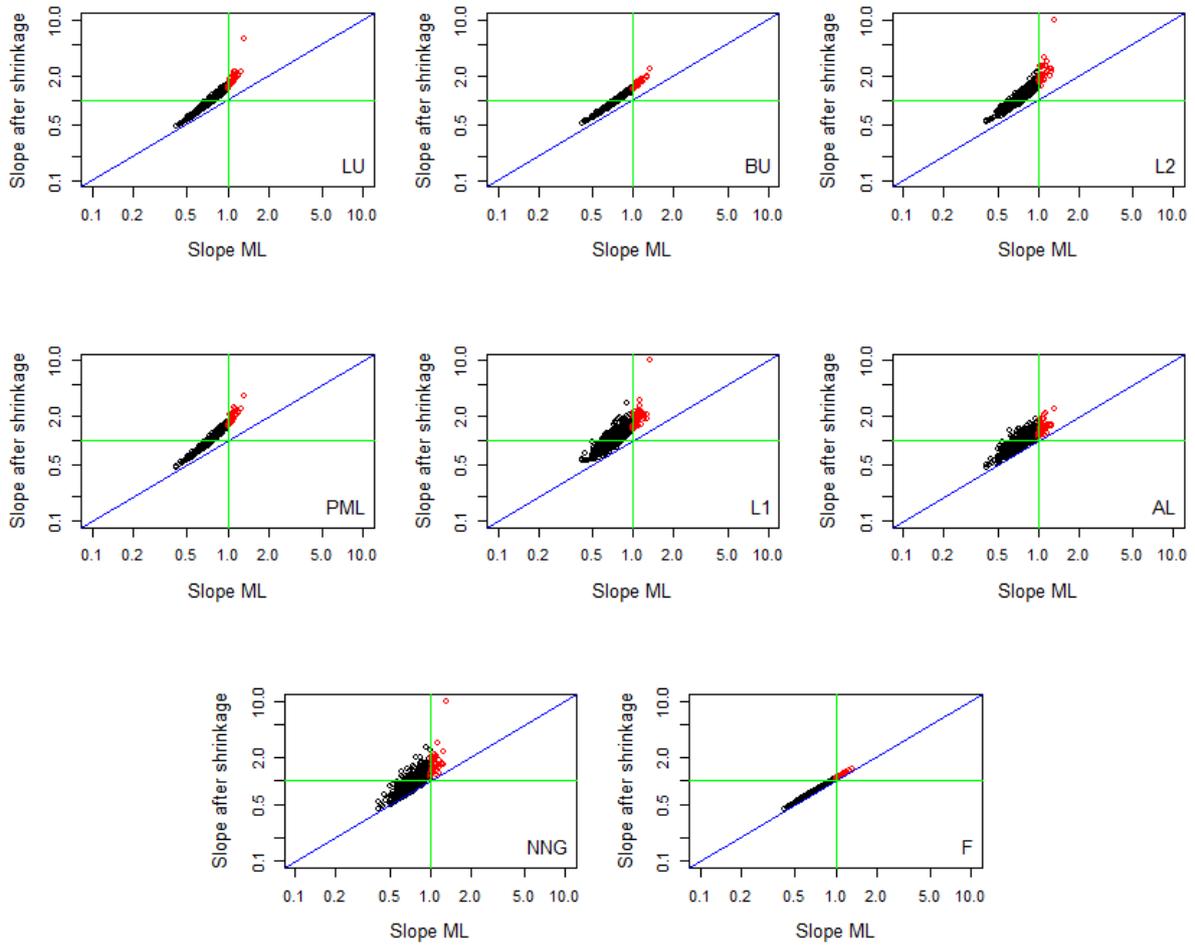

XXXIV. 5 true and 5 noise predictors, 0 correlation, 50% event rate, 20 EPV

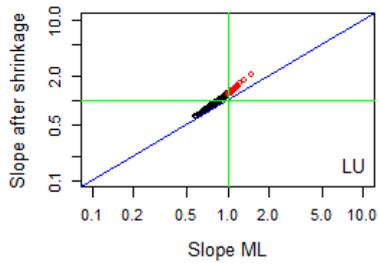
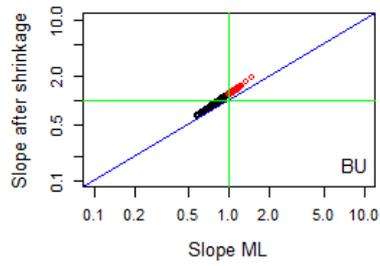
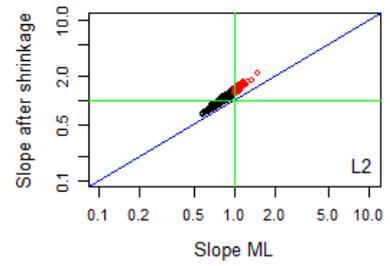
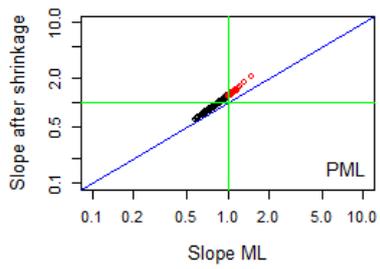
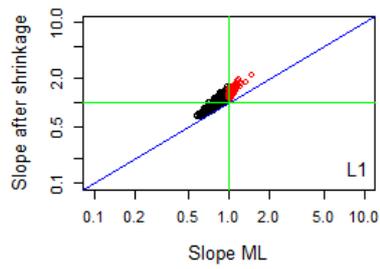
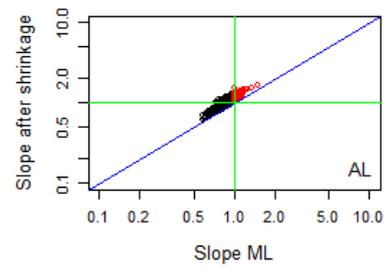
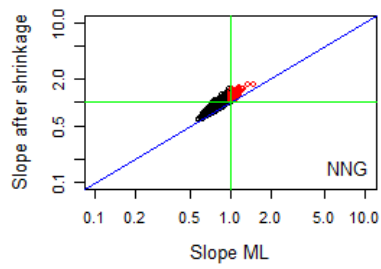
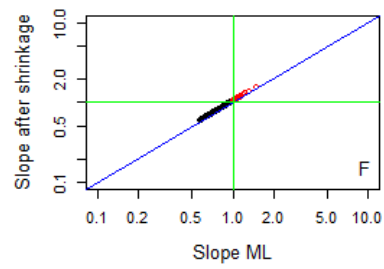

XXXV.     5 true and 5 noise predictors, 0 correlation, 50% event rate, 50 EPV

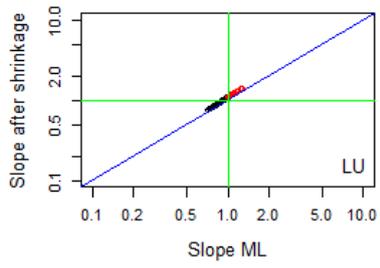
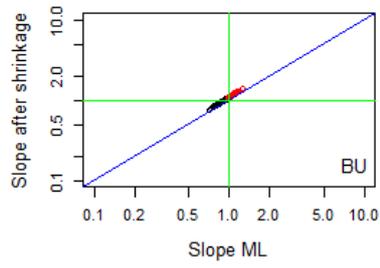
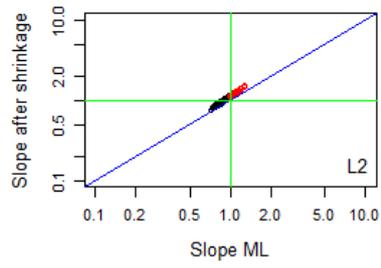
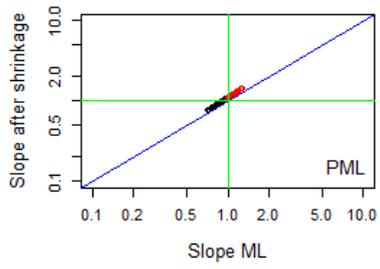
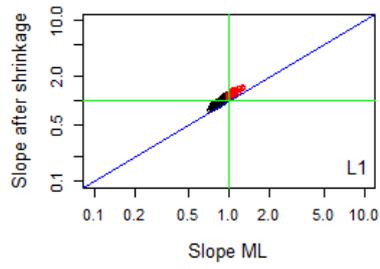
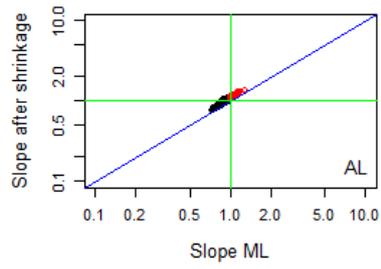
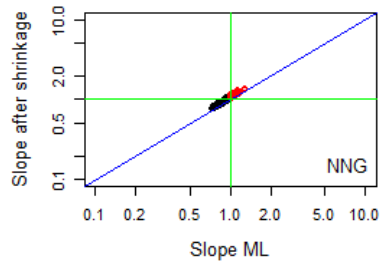
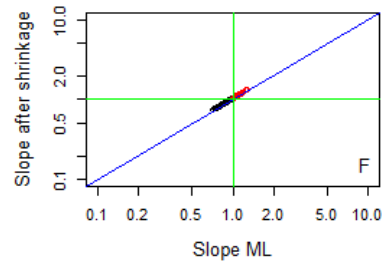

XXXVI. 5 true and 5 noise predictors, 0.5 correlation, 50% event rate, 3 EPV

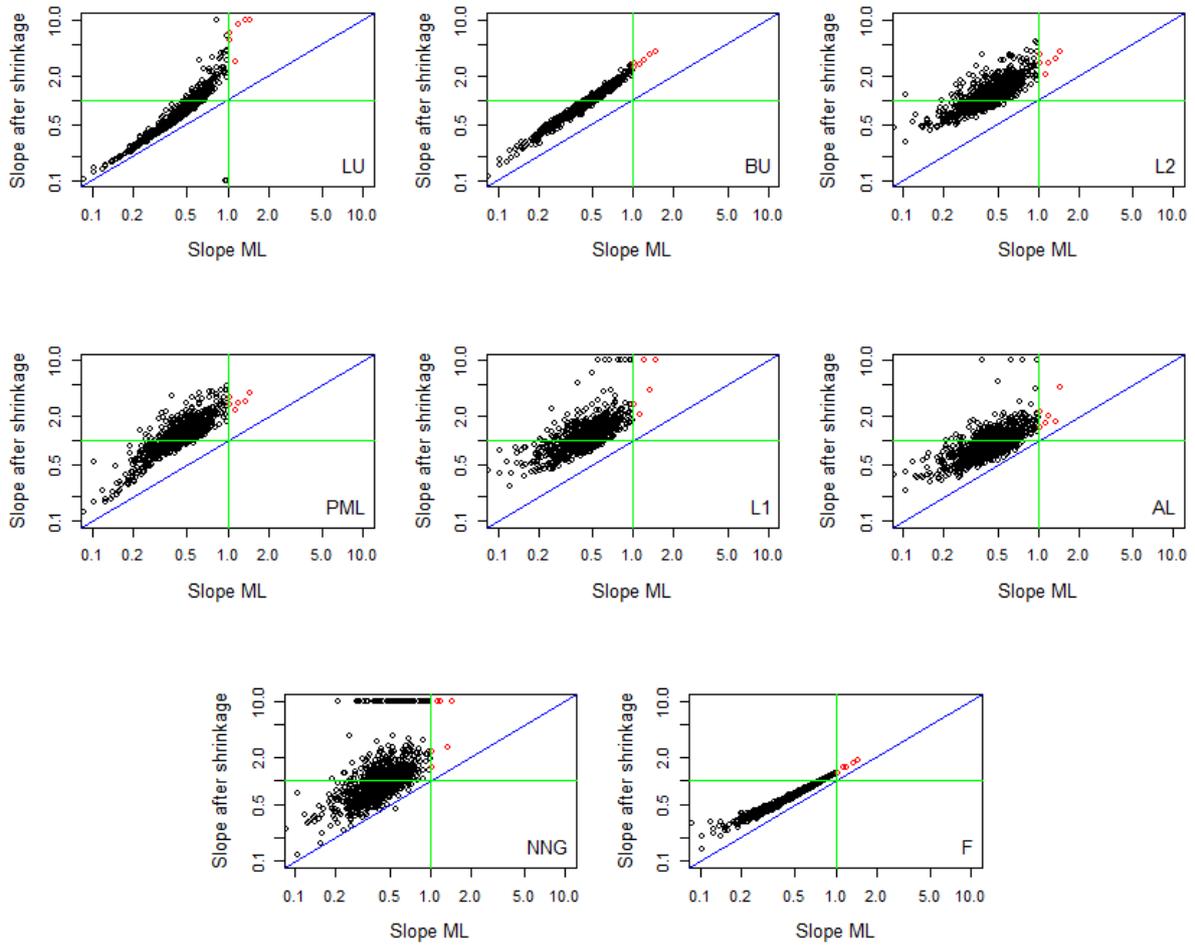

XXXVII. 5 true and 5 noise predictors, 0.5 correlation, 50% event rate, 5 EPV

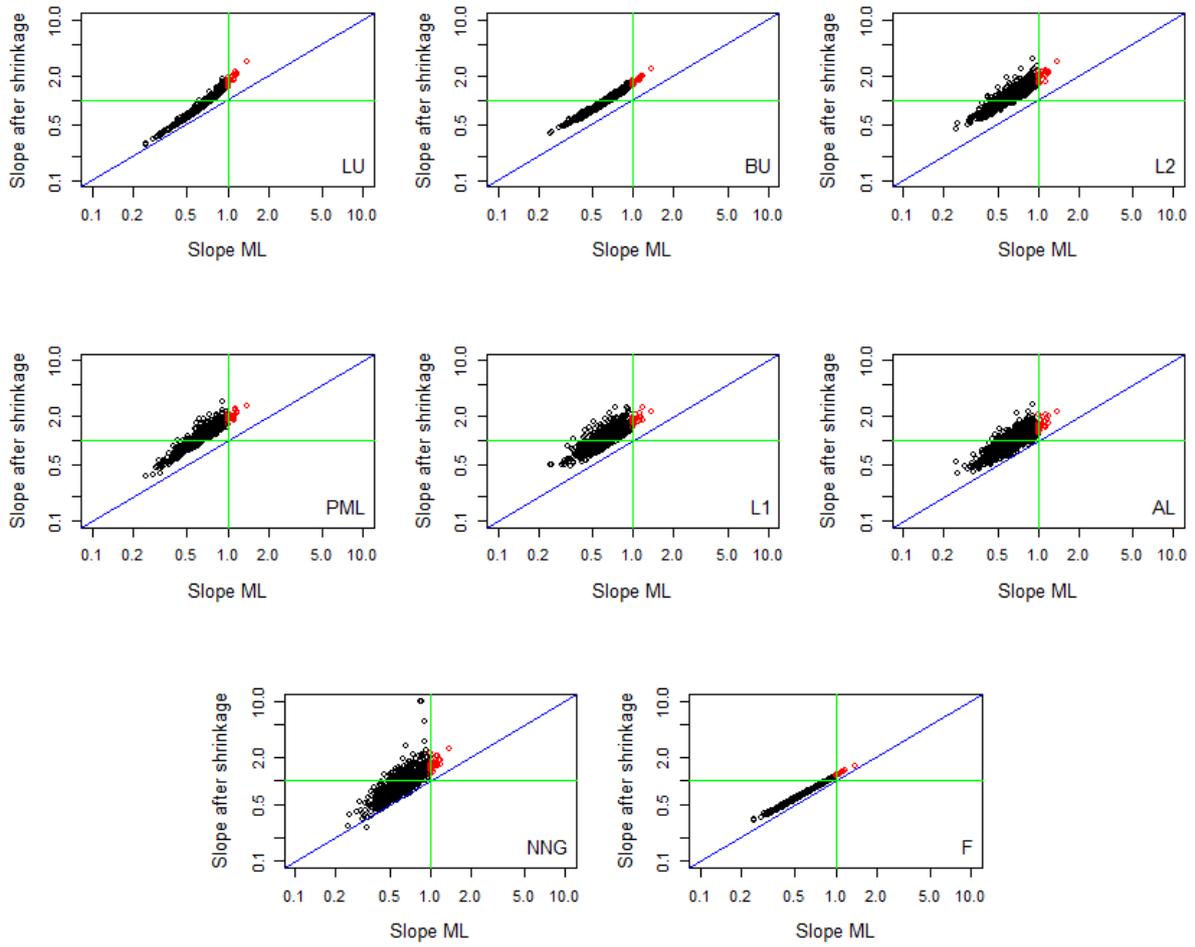

XXXVIII. 5 true and 5 noise predictors, 0.5 correlation, 50% event rate, 10 EPV

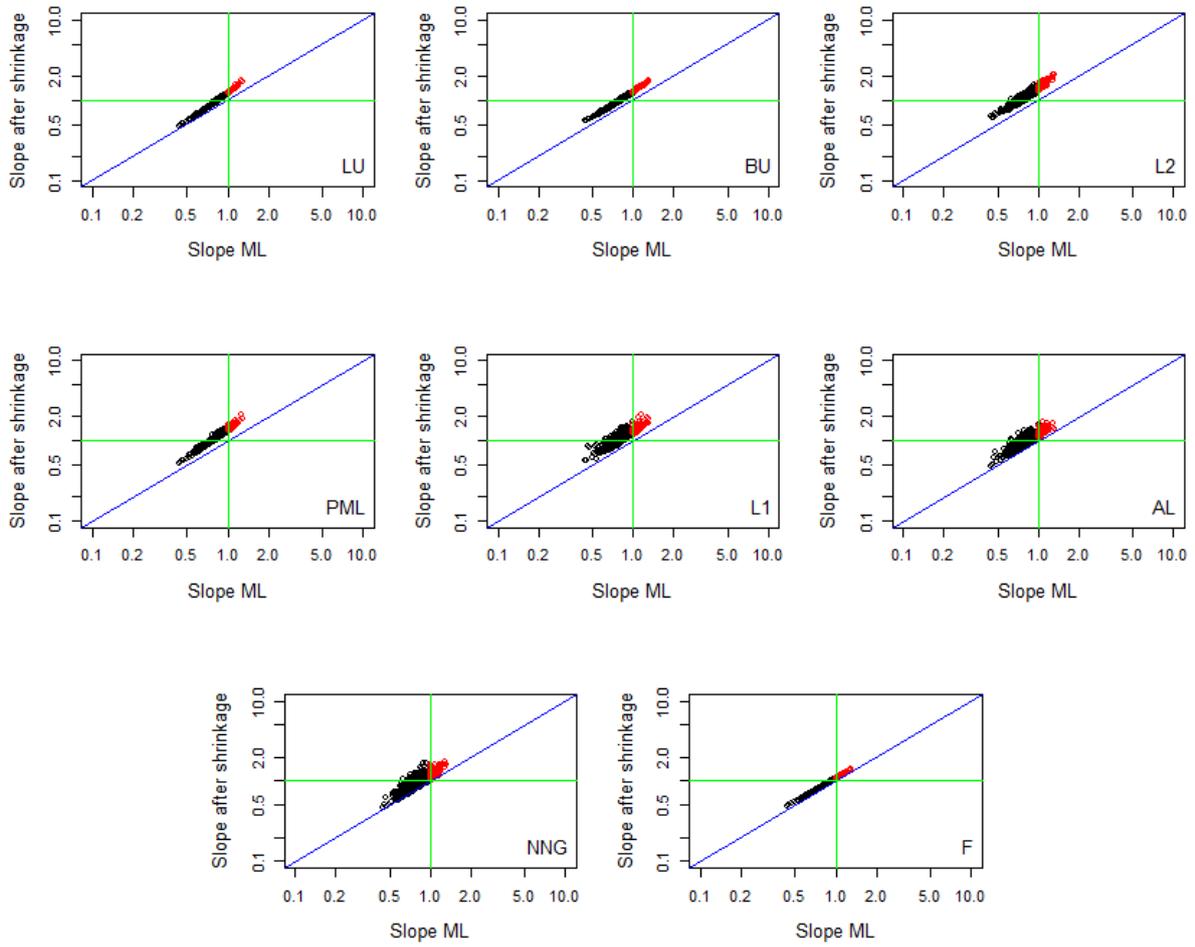

XXXIX. 5 true and 5 noise predictors, 0.5 correlation, 50% event rate, 20 EPV

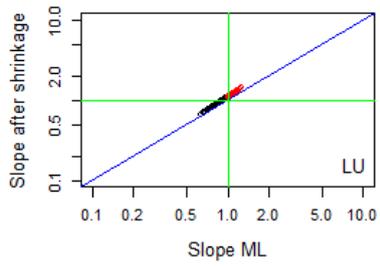
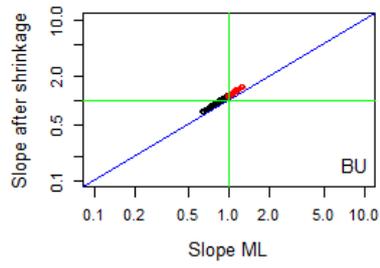
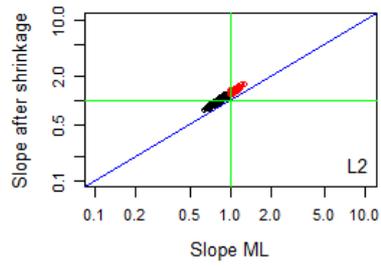
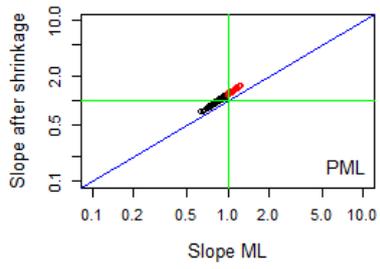
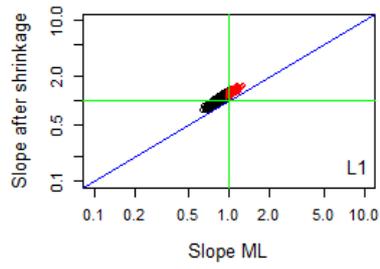
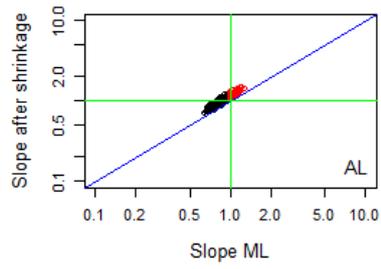
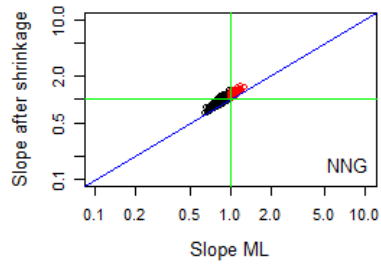
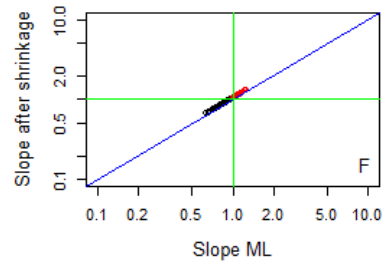

XL. 5 true and 5 noise predictors, 0.5 correlation, 50% event rate, 50 EPV

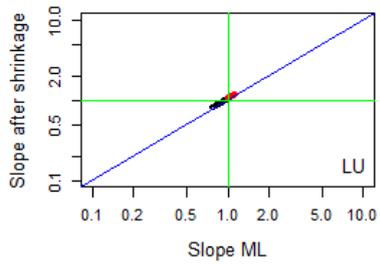
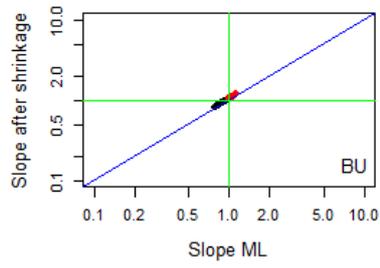
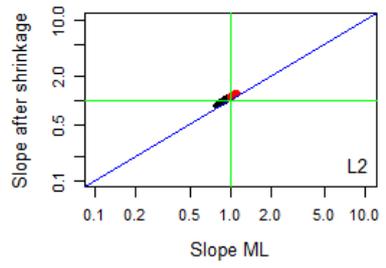
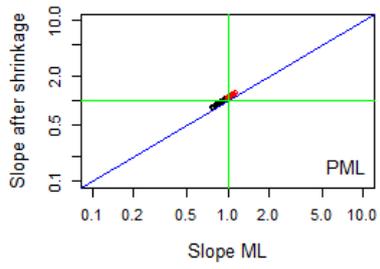
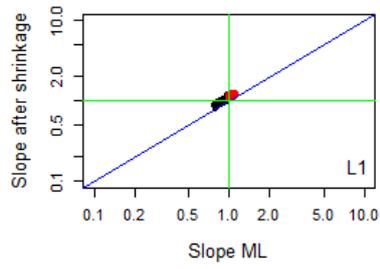
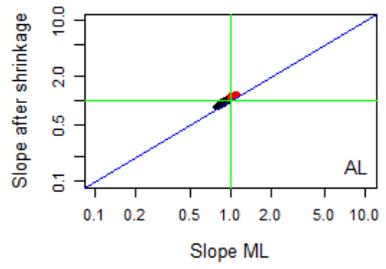
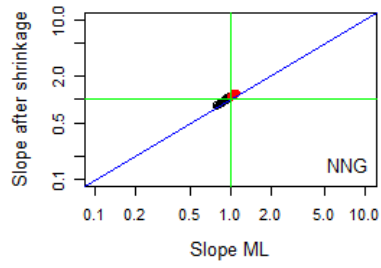
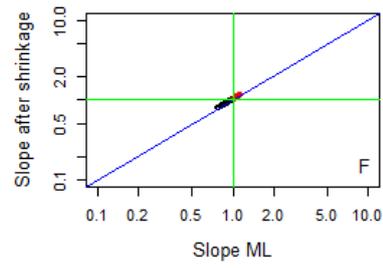

XLI.    10 true predictors, 0 correlation, 10% event rate, 3 EPV

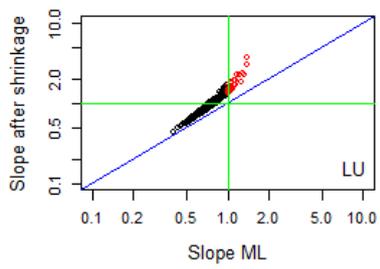 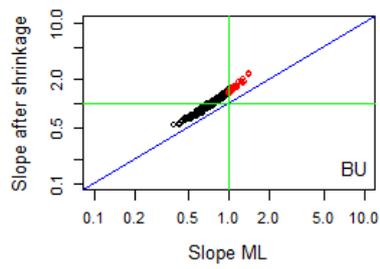 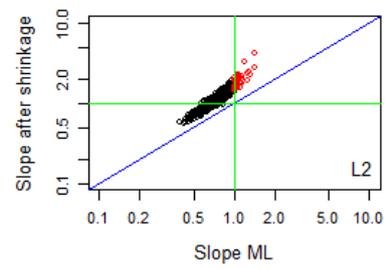
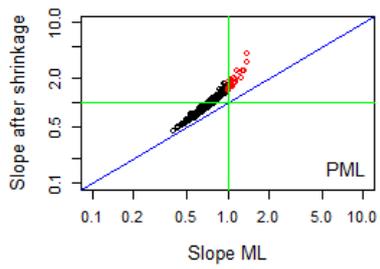 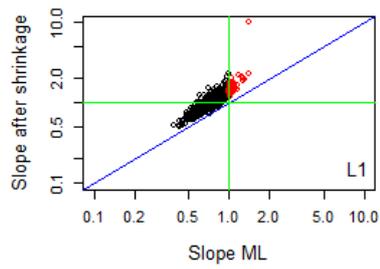 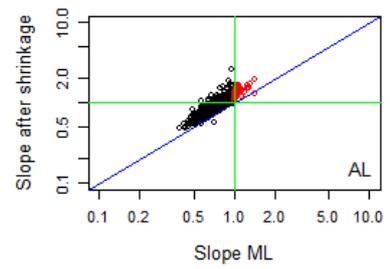
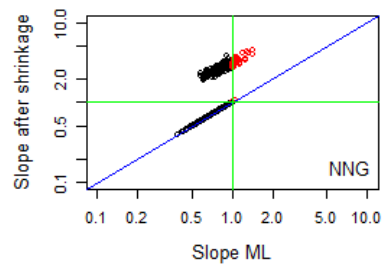 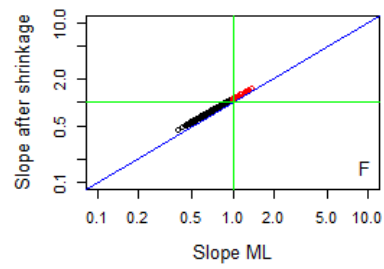

XLII. 10 true predictors, 0 correlation, 10% event rate, 5 EPV

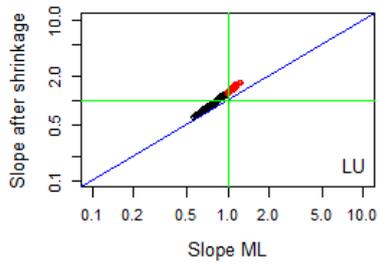
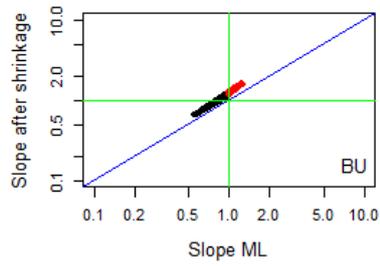
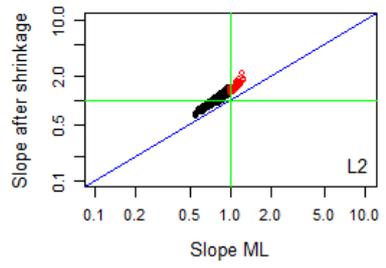
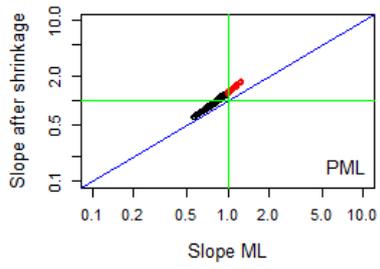
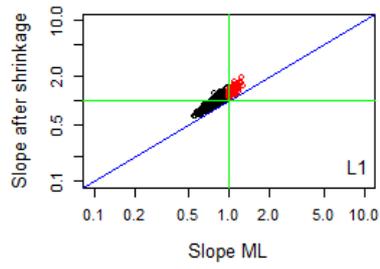
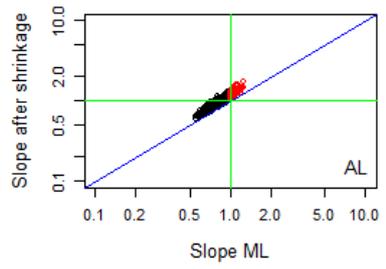
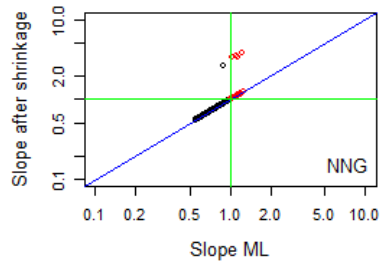
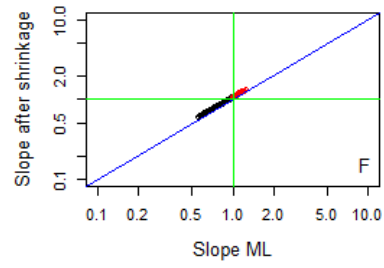

XLIII. 10 true predictors, 0 correlation, 10% event rate, 10 EPV

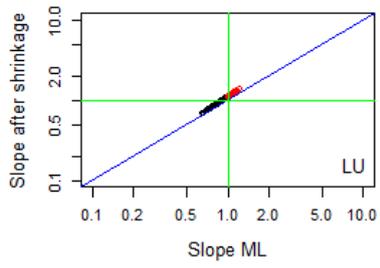 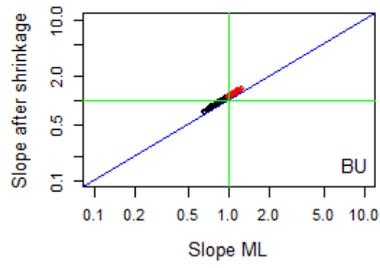 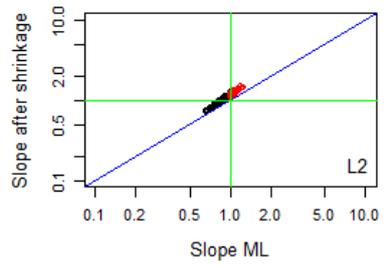
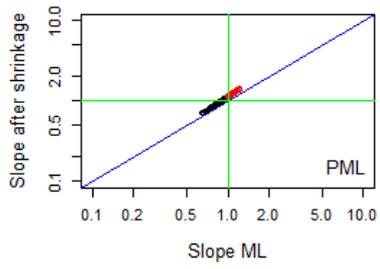 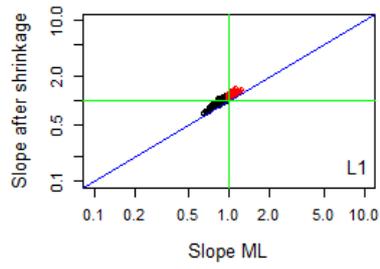 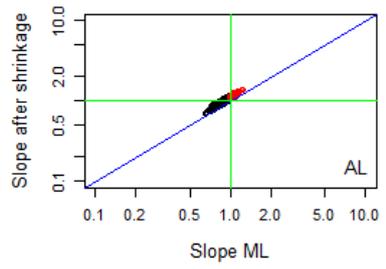
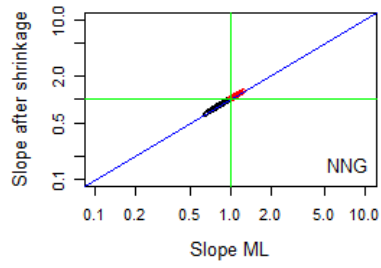 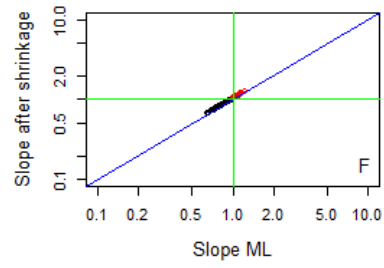

XLIV. 10 true predictors, 0 correlation, 10% event rate, 20 EPV

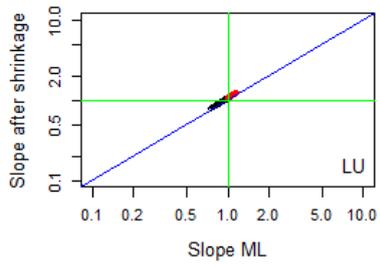
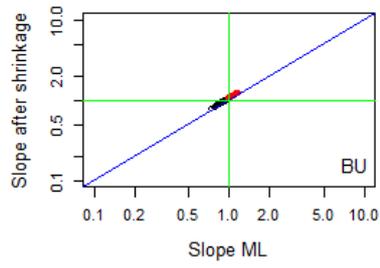
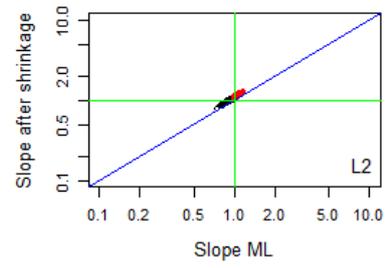
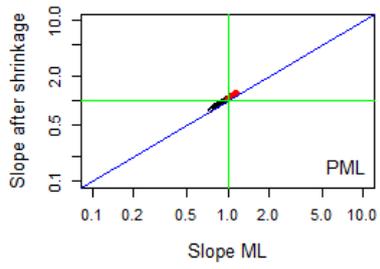
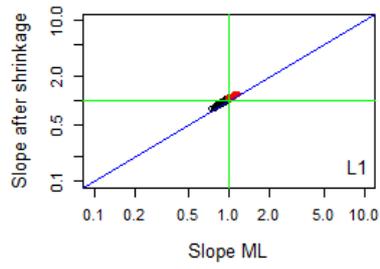
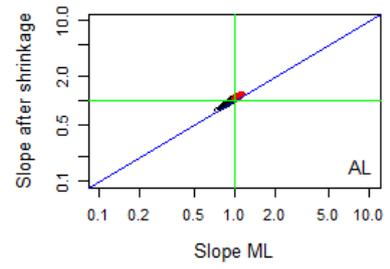
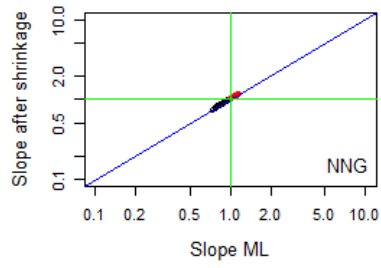
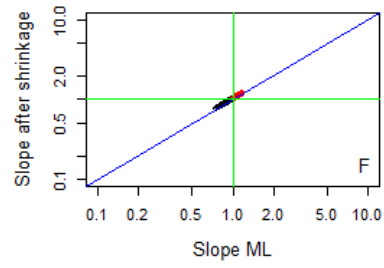

XLV. 10 true predictors, 0 correlation, 10% event rate, 50 EPV

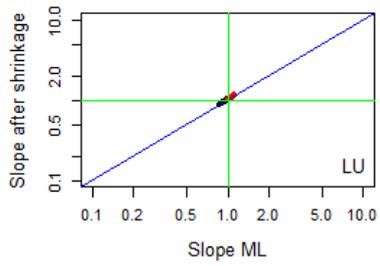
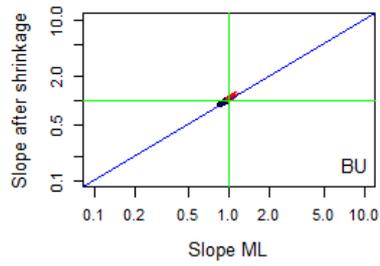
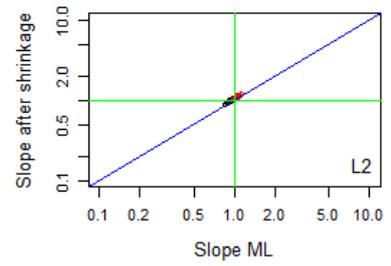
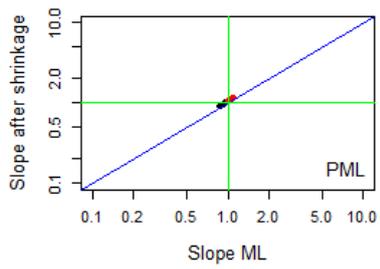
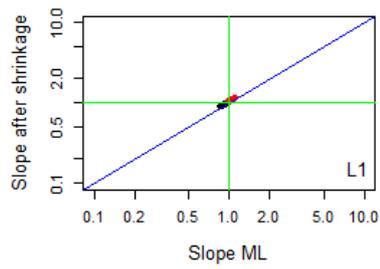
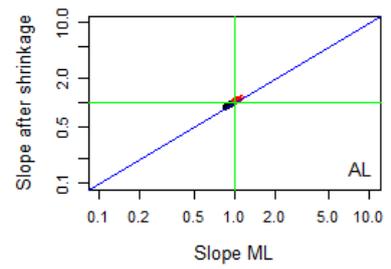
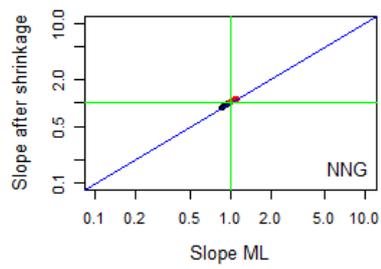
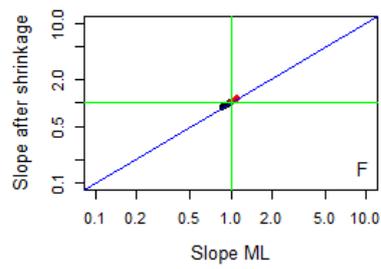

XLVI. 10 true predictors, 0.5 correlation, 10% event rate, 3 EPV

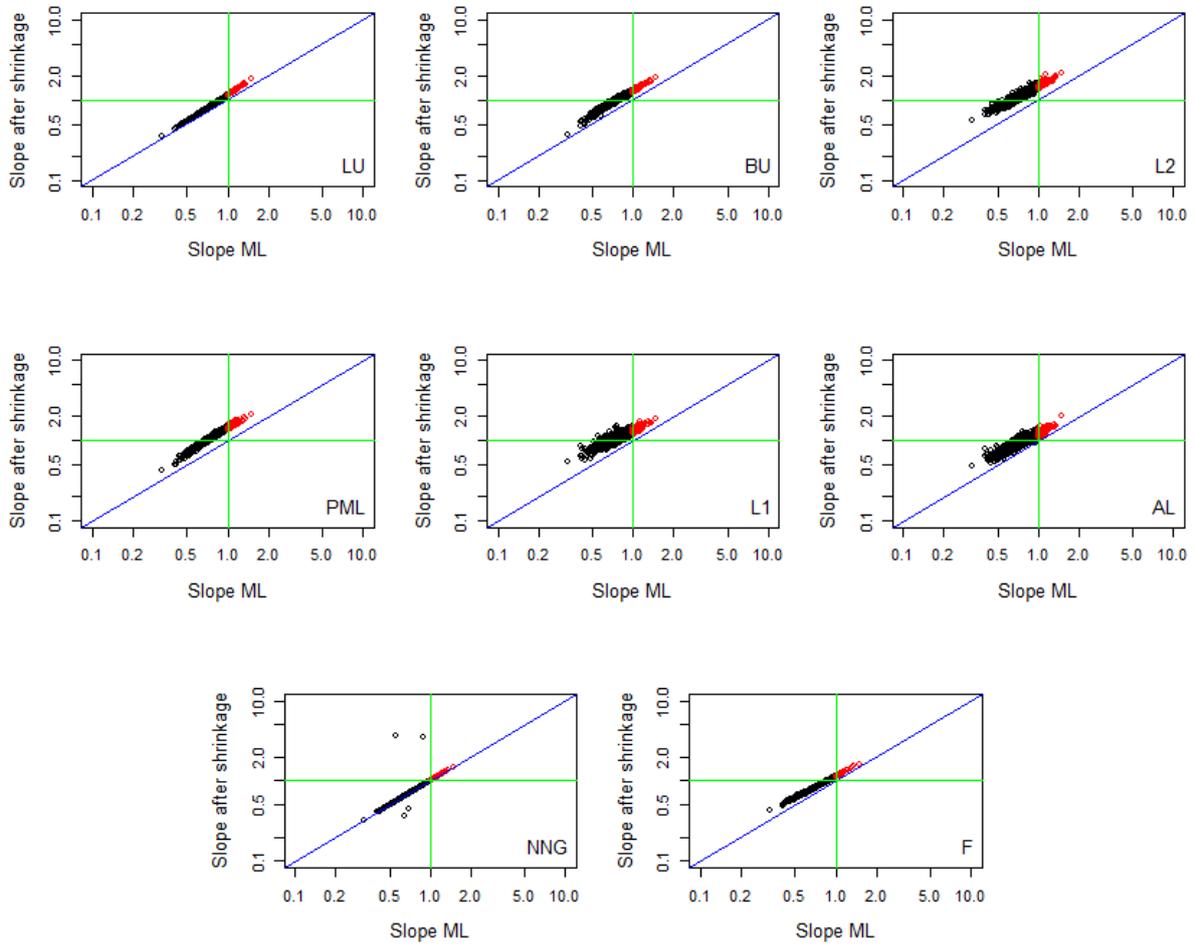

XLVII. 10 true predictors, 0.5 correlation, 10% event rate, 5 EPV

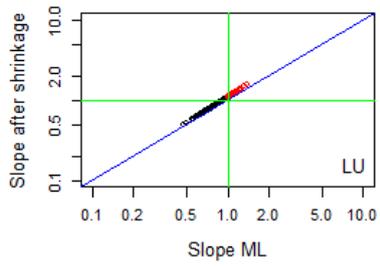
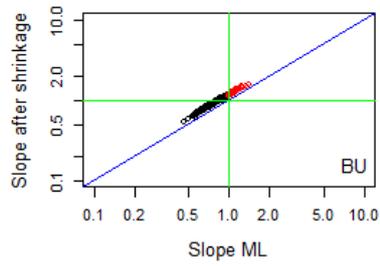
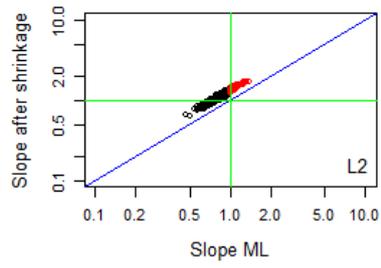
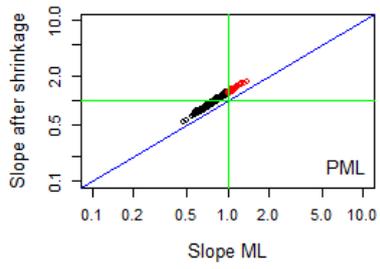
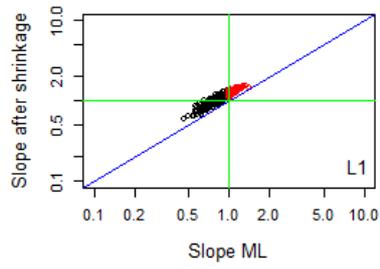
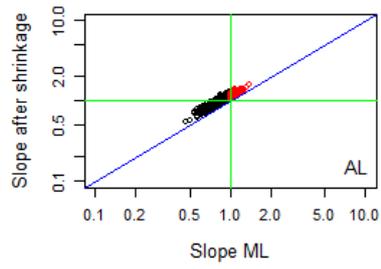
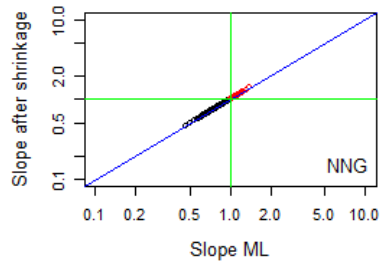
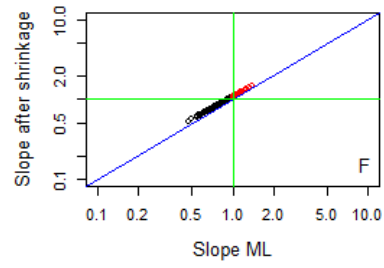

XLVIII. 10 true predictors, 0.5 correlation, 10% event rate, 10 EPV

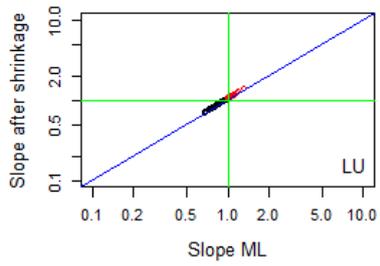
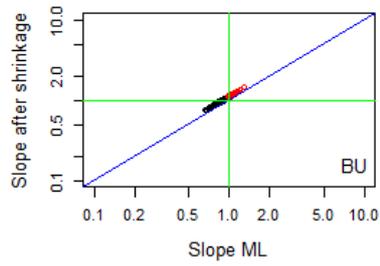
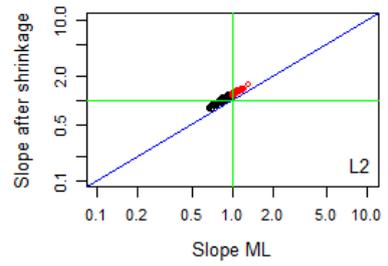
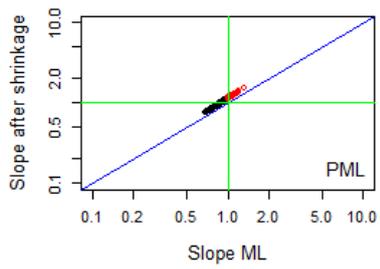
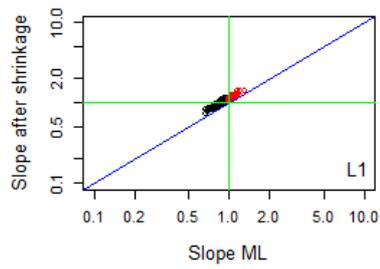
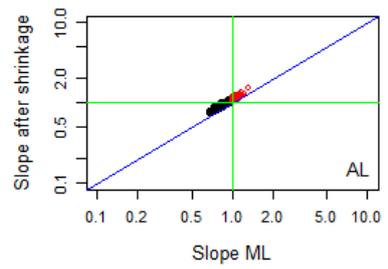
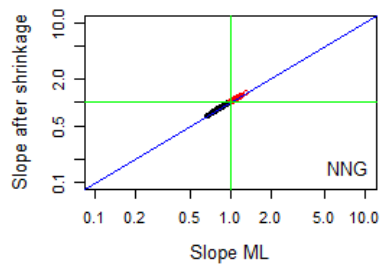
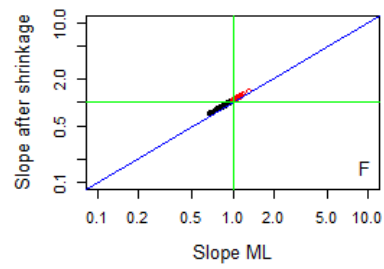

XLIX. 10 true predictors, 0.5 correlation, 10% event rate, 20 EPV

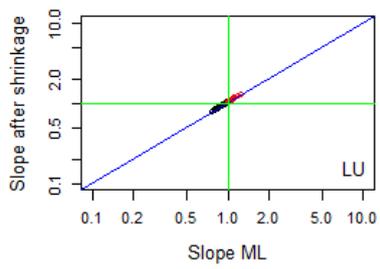
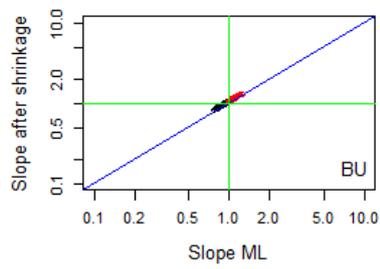
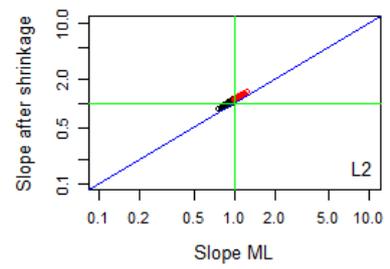
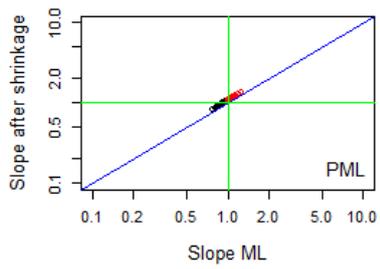
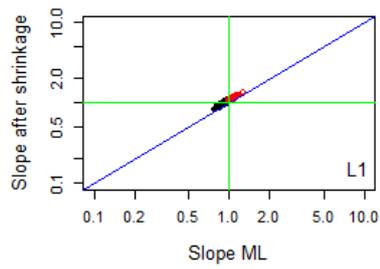
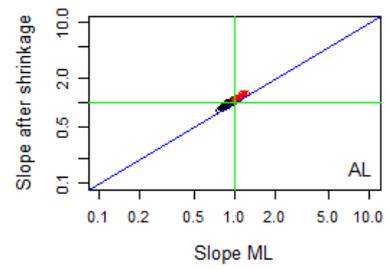
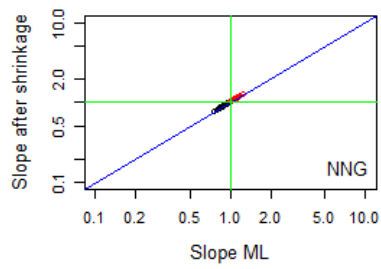
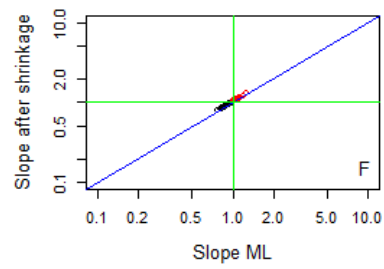

L. 10 true predictors, 0.5 correlation, 10% event rate, 50 EPV

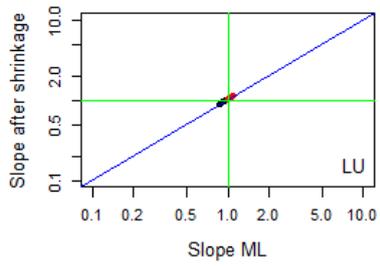
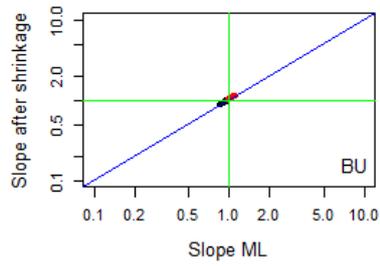
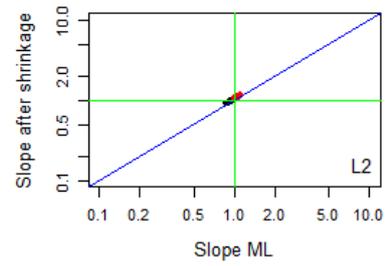
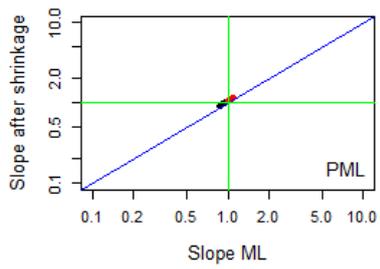
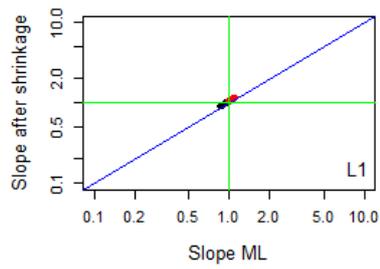
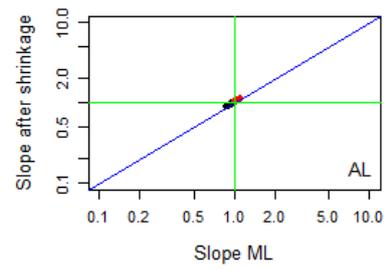
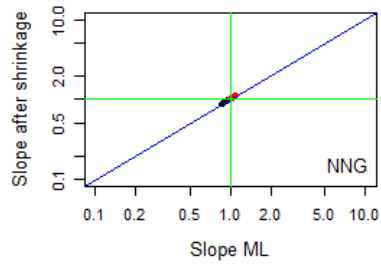
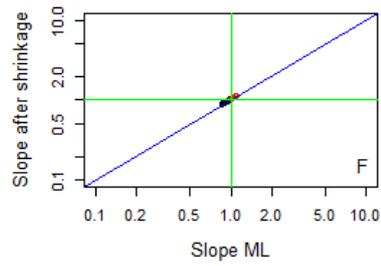

LI. 10 true predictors, 0 correlation, 50% event rate, 3 EPV

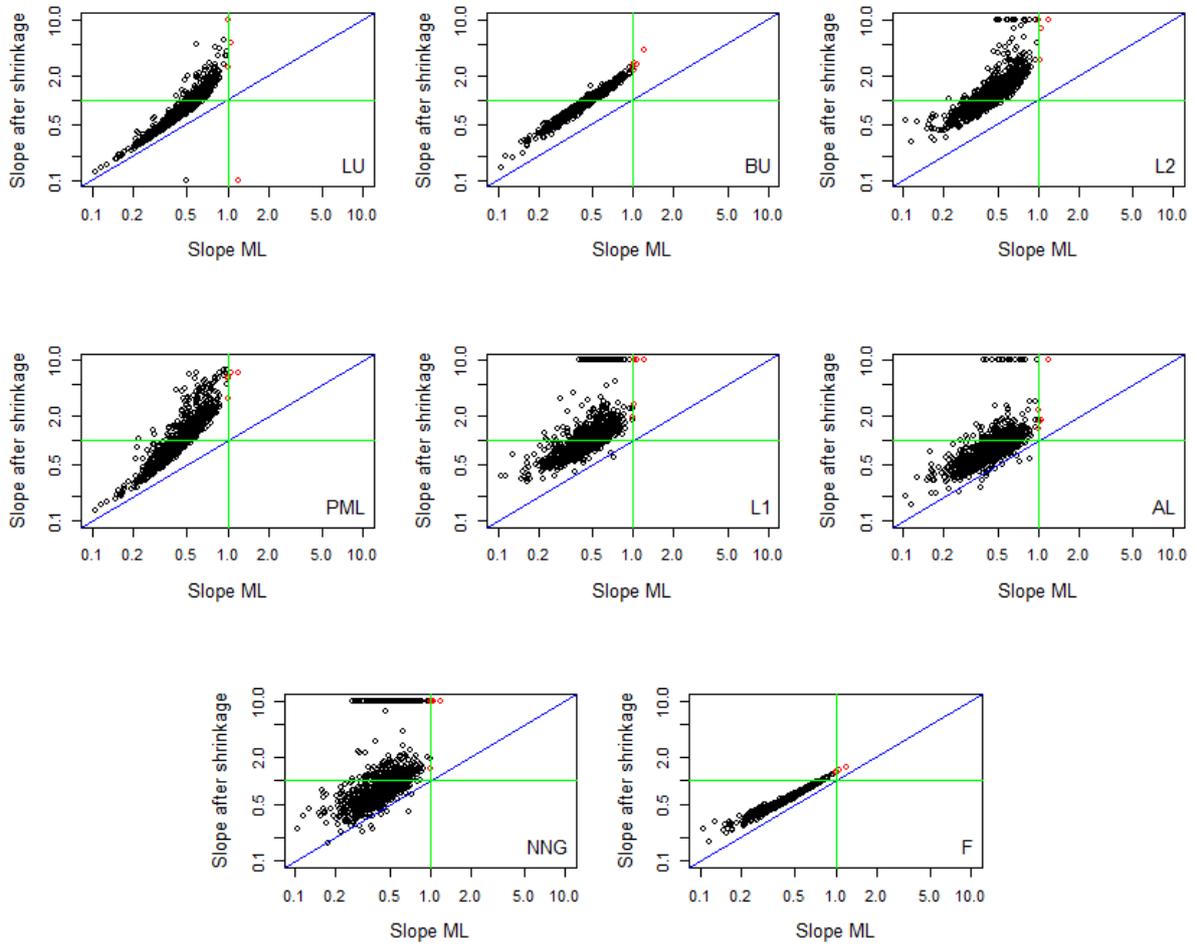

LII. 10 true predictors, 0 correlation, 50% event rate, 5 EPV

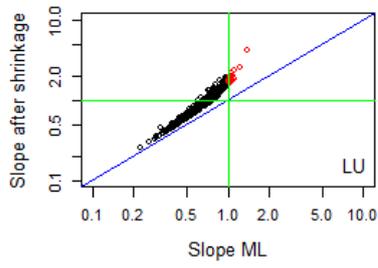
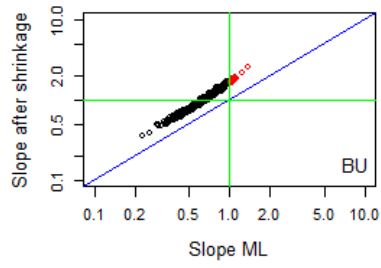
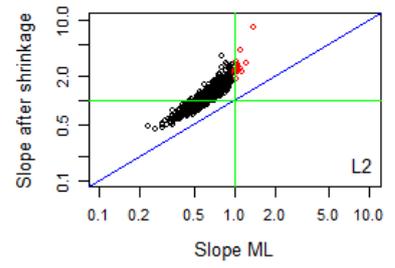
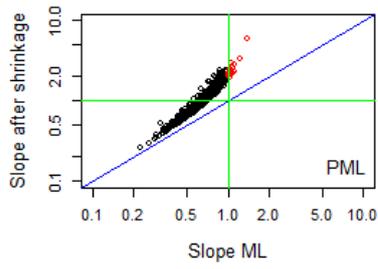
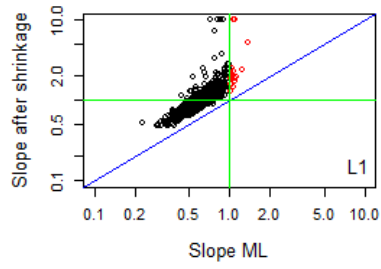
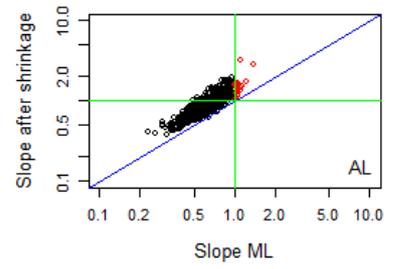
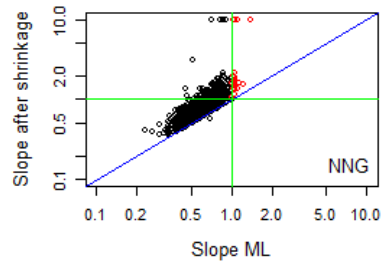
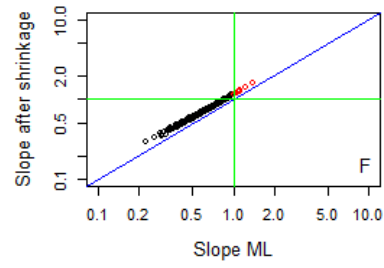

LIII. 10 true predictors, 0 correlation, 50% event rate, 10 EPV

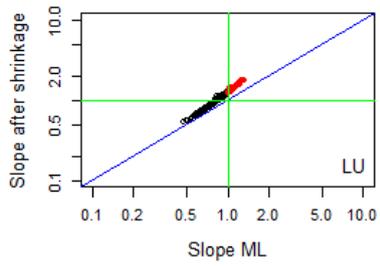 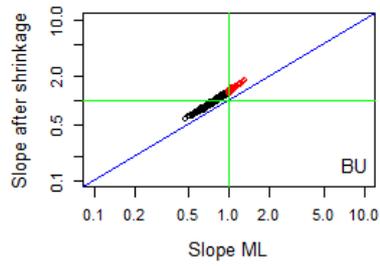 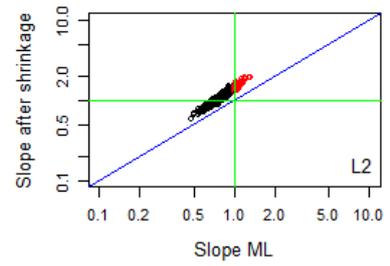
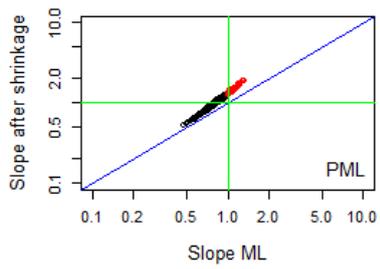 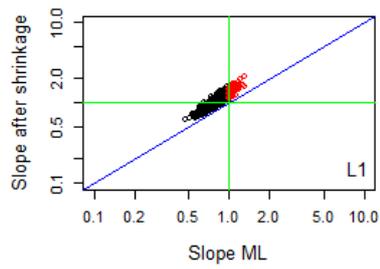 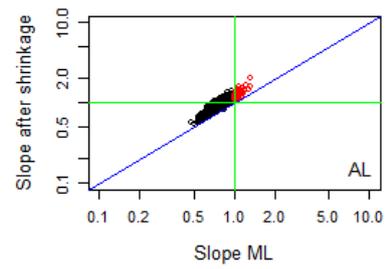
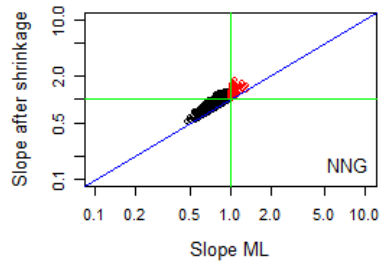 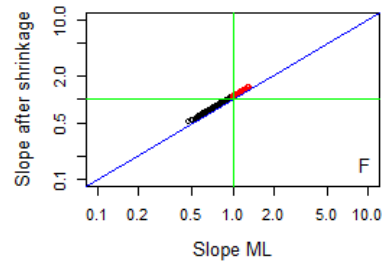

LIV.  10 true predictors, 0 correlation, 50% event rate, 20 EPV

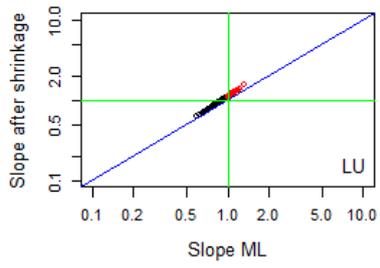 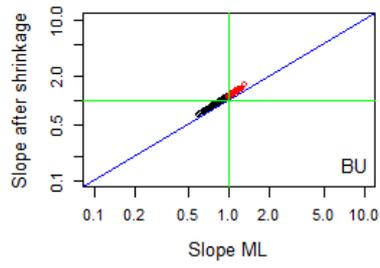 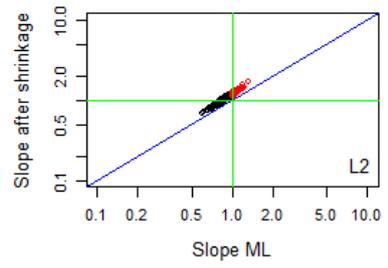
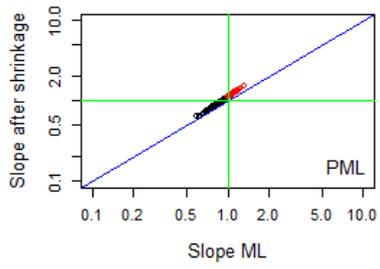 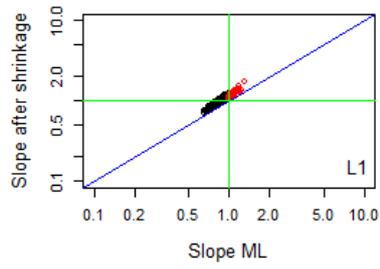 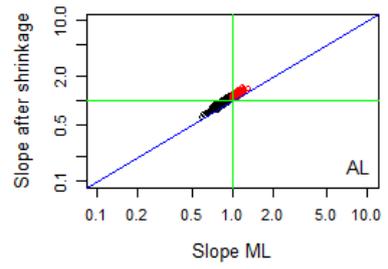
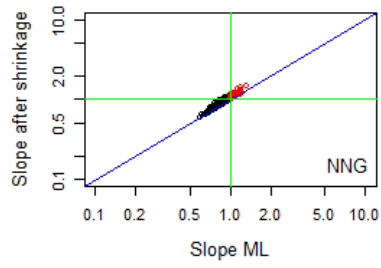 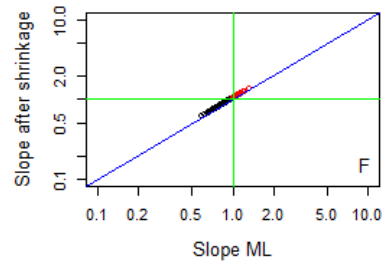

LV.     10 true predictors, 0 correlation, 50% event rate, 50 EPV

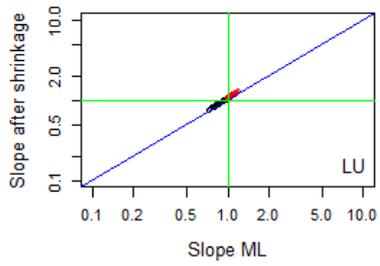 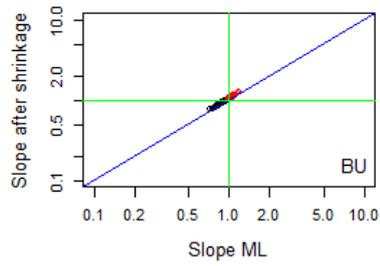 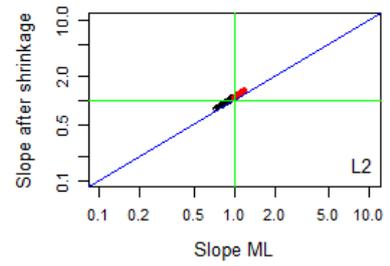

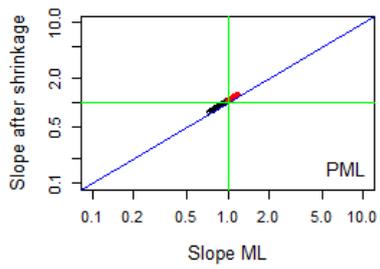 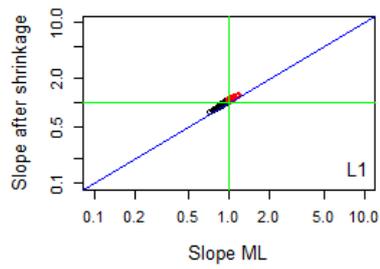 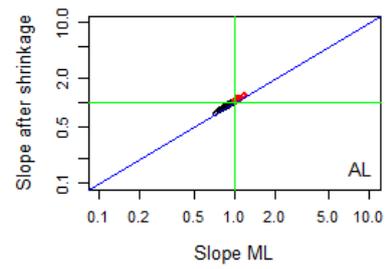

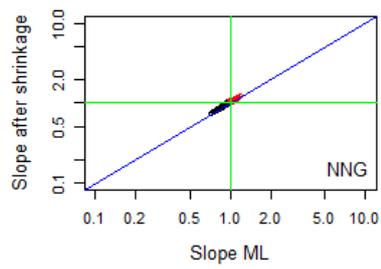 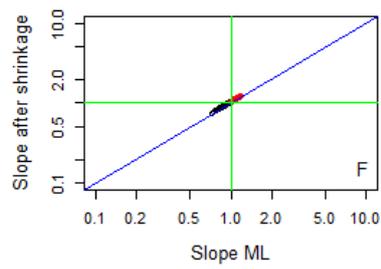

LVI. 10 true predictors, 0.5 correlation, 50% event rate, 3 EPV

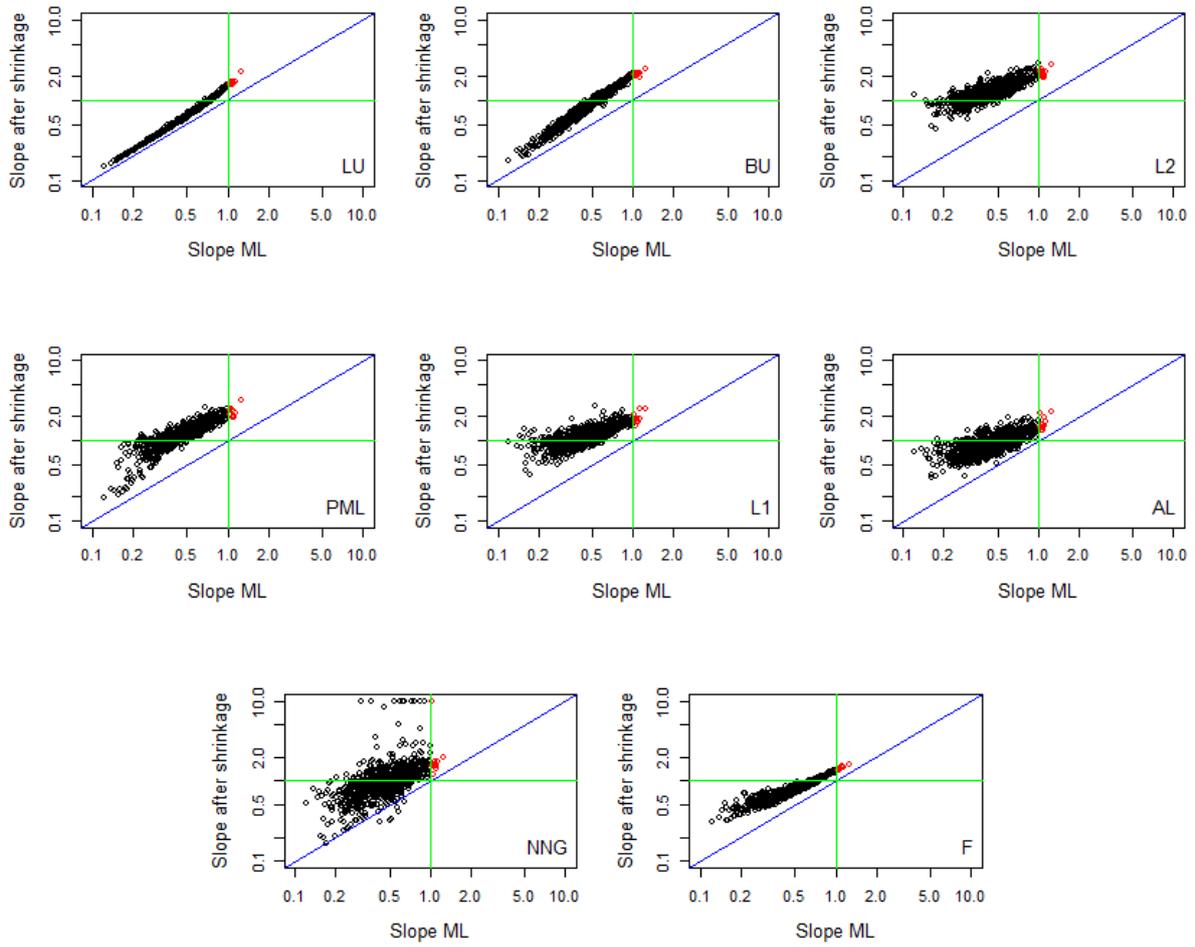

LVII. 10 true predictors, 0.5 correlation, 50% event rate, 5 EPV

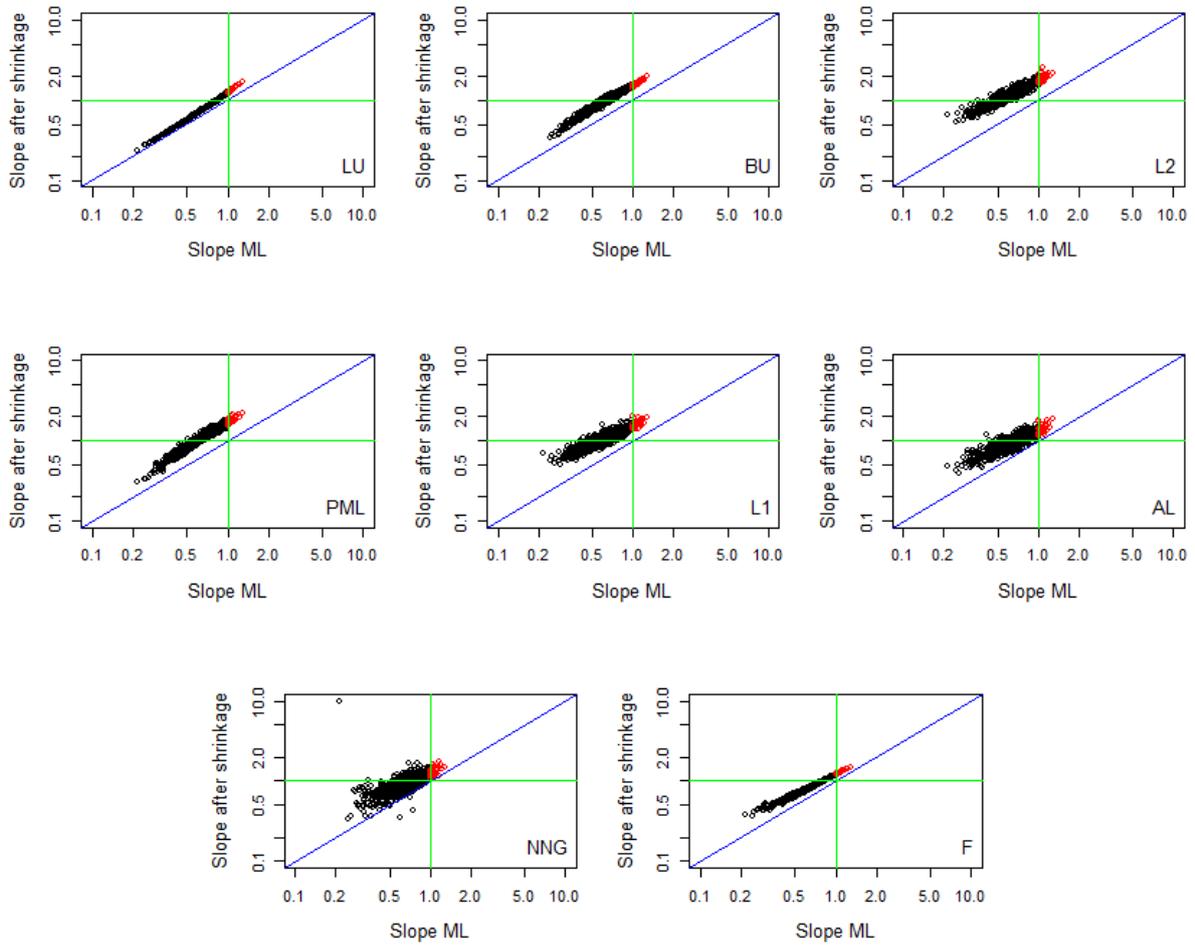

LVIII. 10 true predictors, 0.5 correlation, 50% event rate, 10 EPV

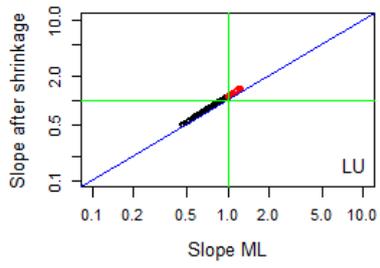
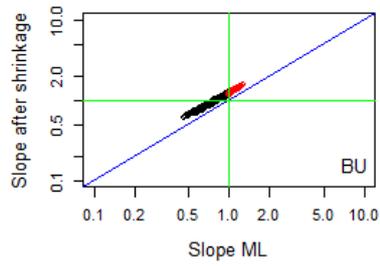
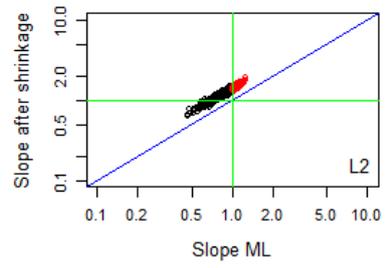
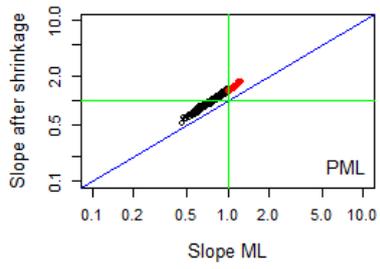
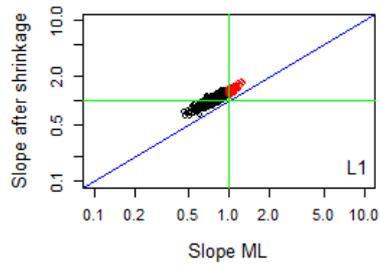
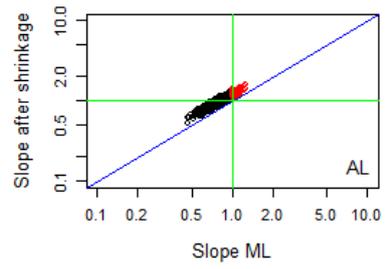
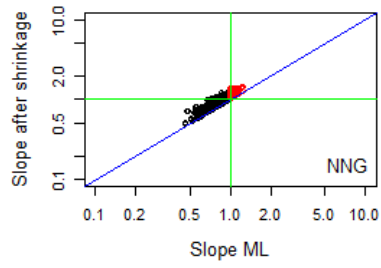
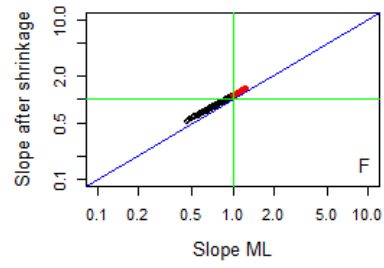

LIX. 10 true predictors, 0.5 correlation, 50% event rate, 20 EPV

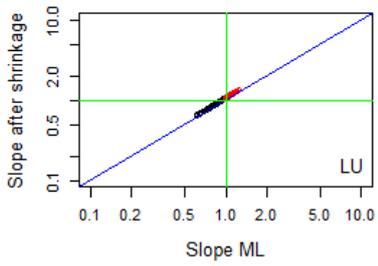 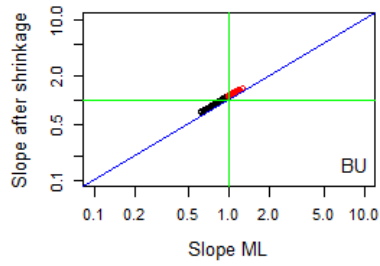 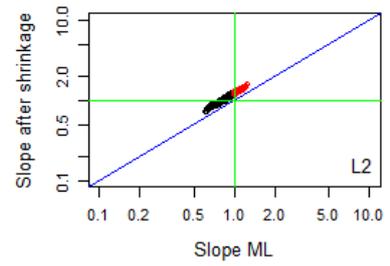
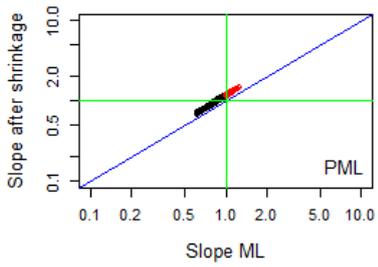 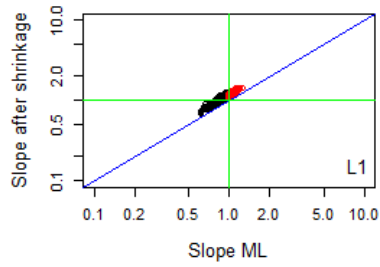 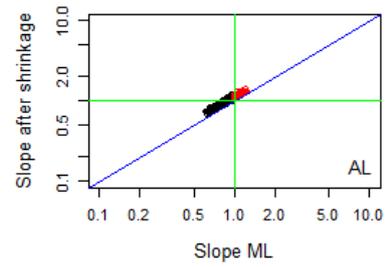
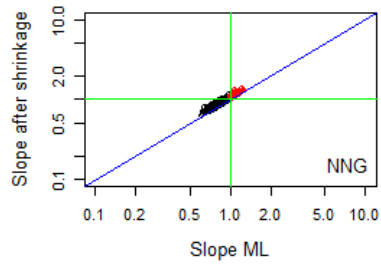 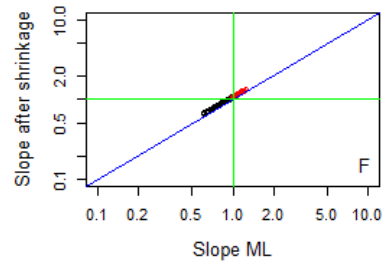

LX. 10 true predictors, 0.5 correlation, 50% event rate, 50 EPV

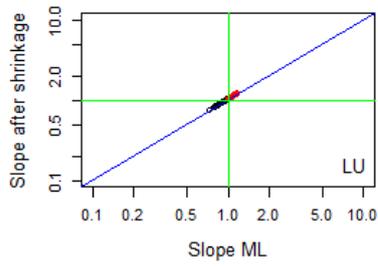
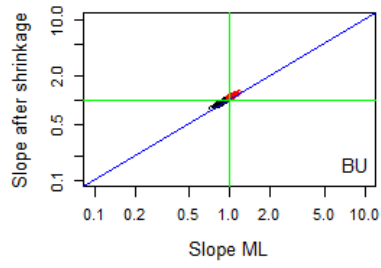
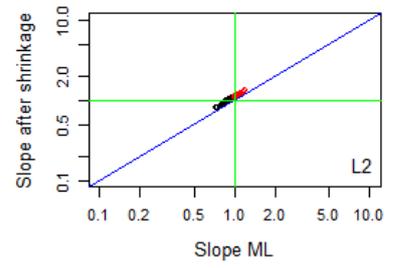
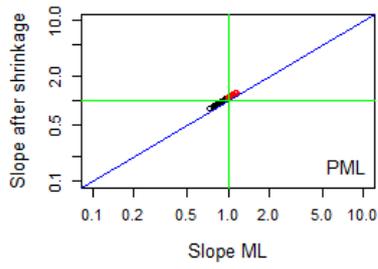
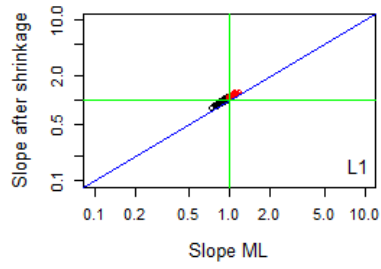
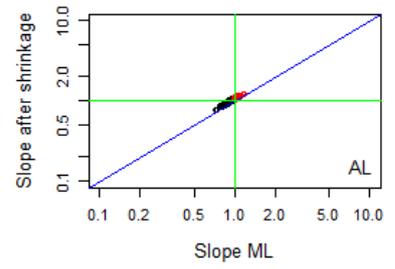
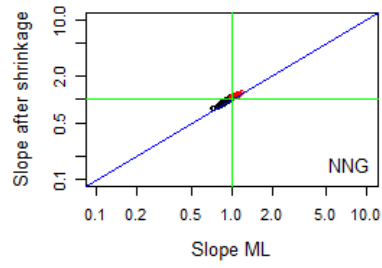
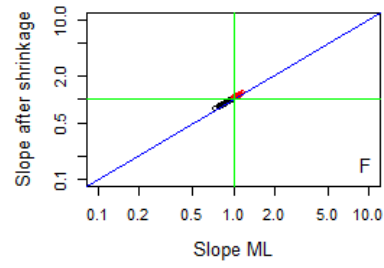

Figure S8. Spearman correlation between estimated and optimal shrinkage by scenario and method.
ML, maximum likelihood; LU, uniform shrinkage based on likelihood; BU, uniform shrinkage based on bootstrap; L2, ridge (L2 penalty); PML, penalized maximum likelihood; L1, LASSO (L1 penalty); AL, adaptive LASSO; NNG, non-negative garrote; F, Firth's correction.

A. Scenarios with 5 true predictors

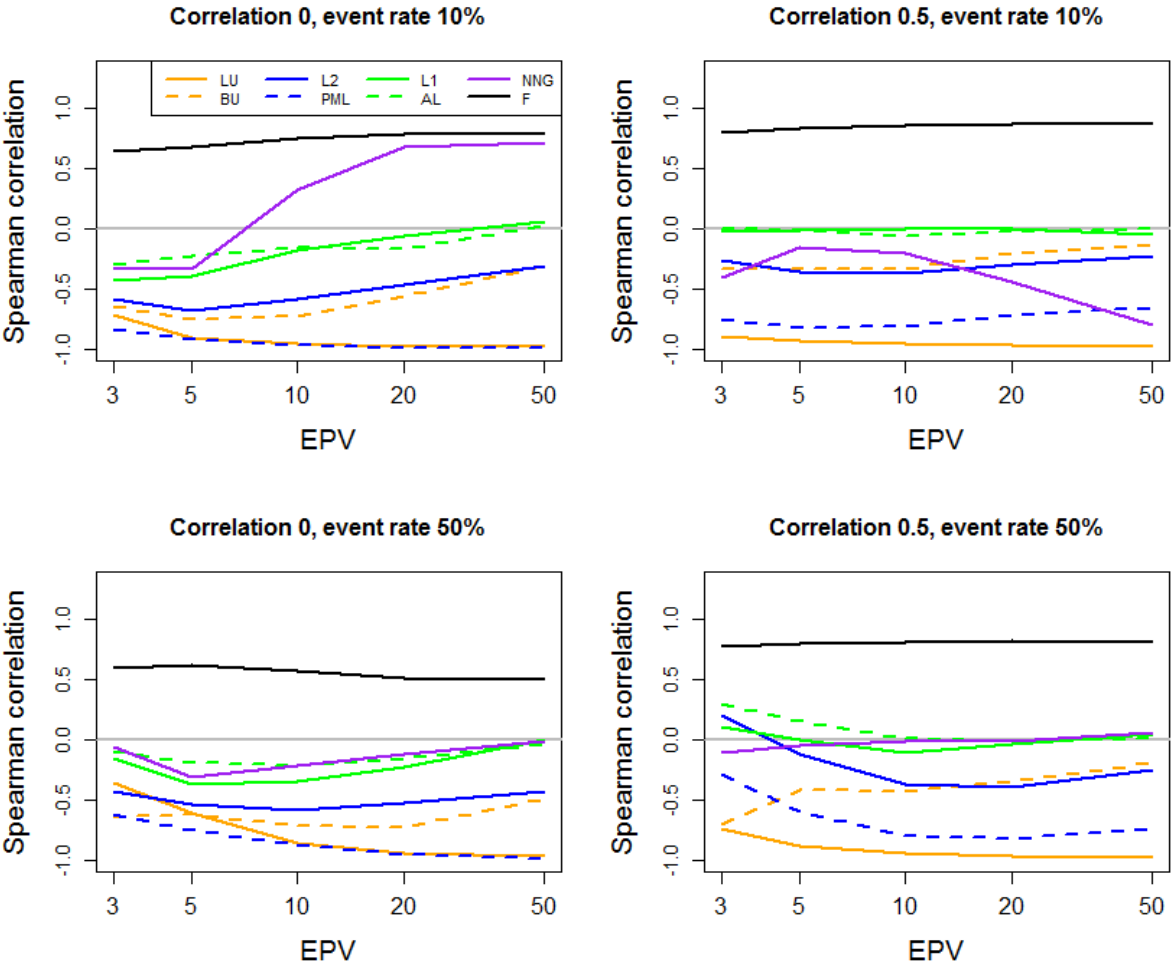

B. Scenarios with 5 true and 5 noise predictors

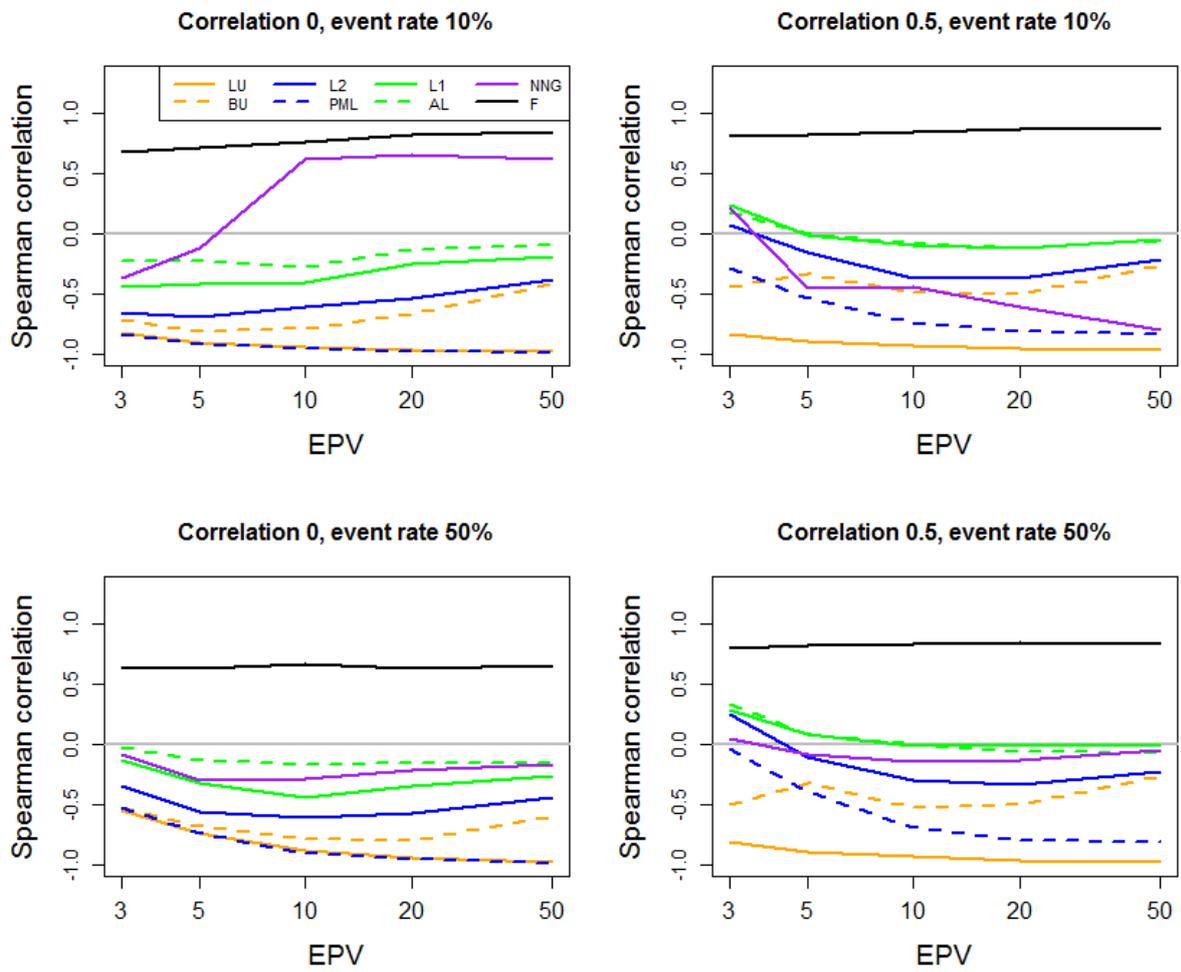

C. Scenarios with 10 true predictors

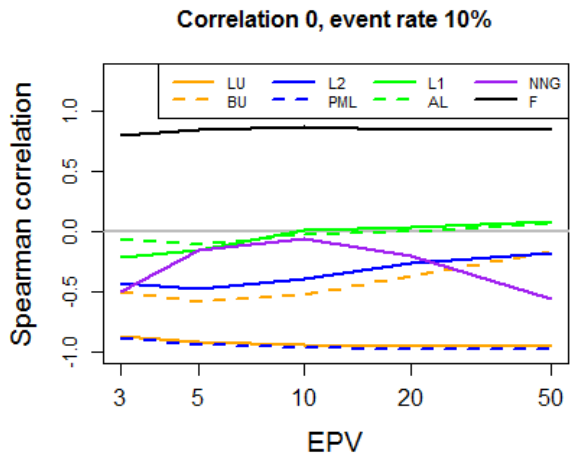
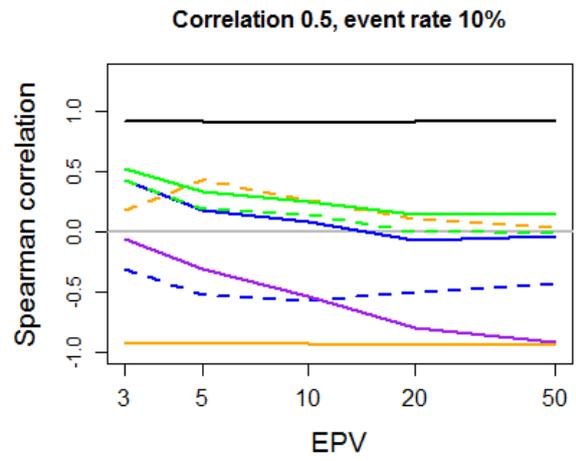
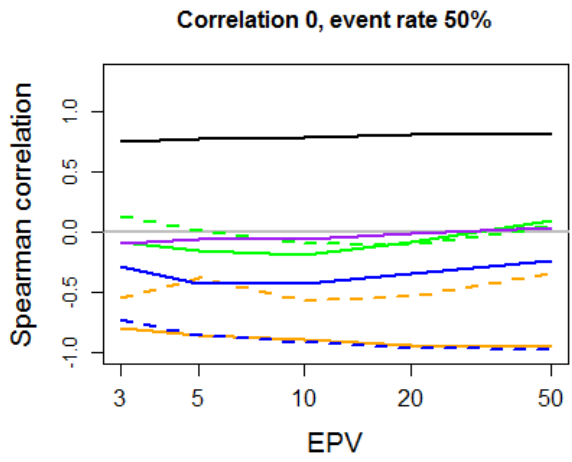
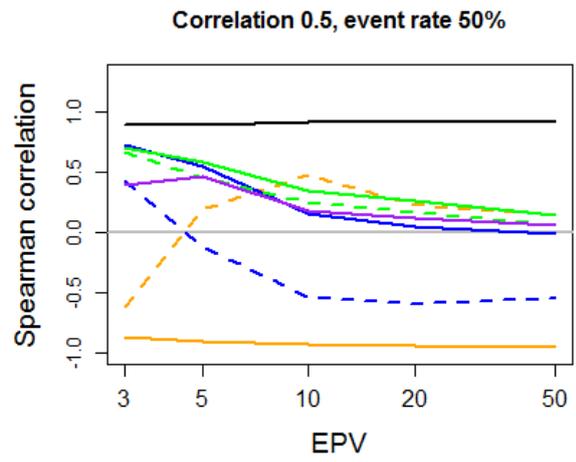

Figure S9. Mean bias in true coefficients per scenario and method. ML, maximum likelihood; LU, uniform shrinkage based on likelihood; BU, uniform shrinkage based on bootstrap; L2, ridge (L2 penalty); PML, penalized maximum likelihood; L1, LASSO (L1 penalty); AL, adaptive LASSO; NNG, non-negative garrote; F, Firth's correction.

A. Scenarios with 5 true predictors

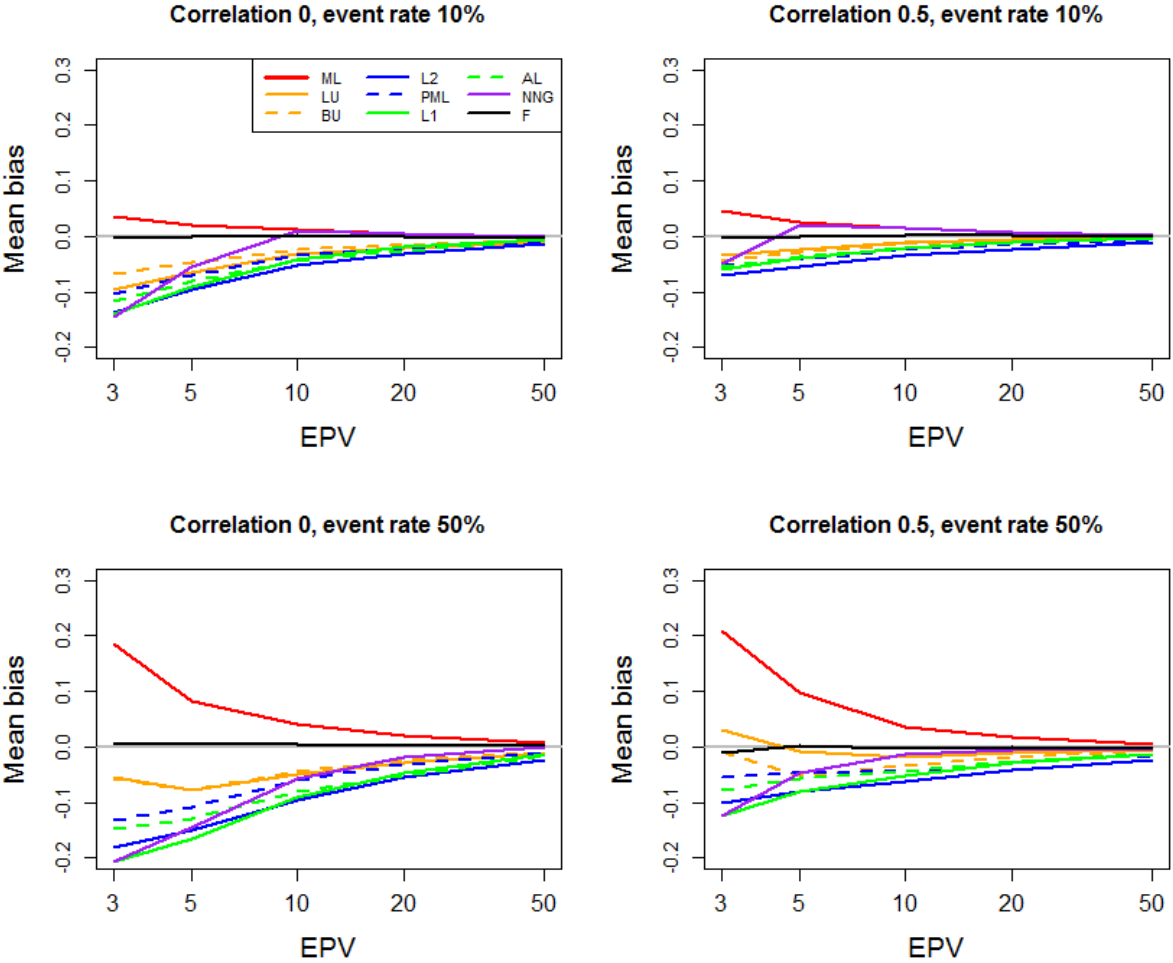

B. Scenarios with 5 true and 5 noise predictors

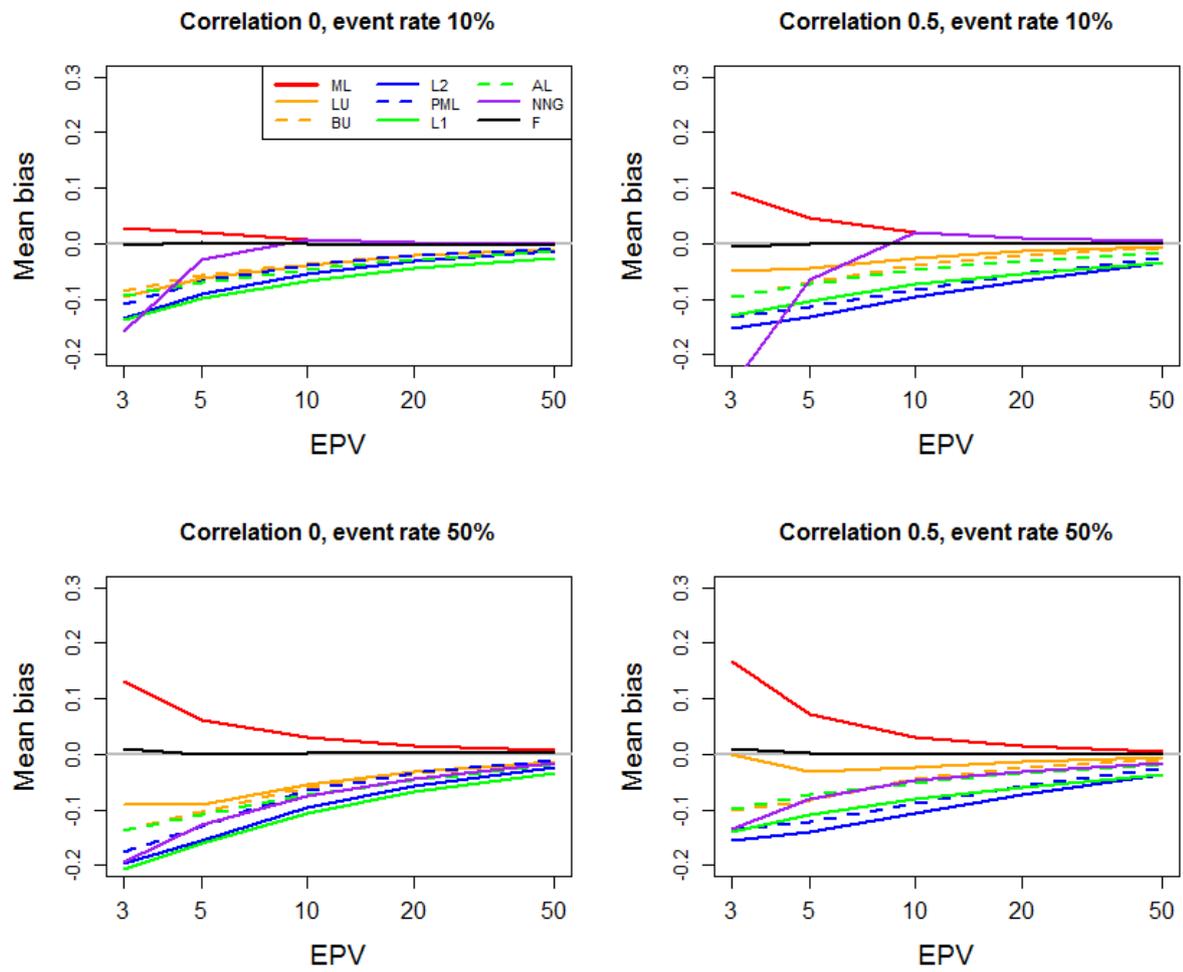

C. Scenarios with 10 true predictors

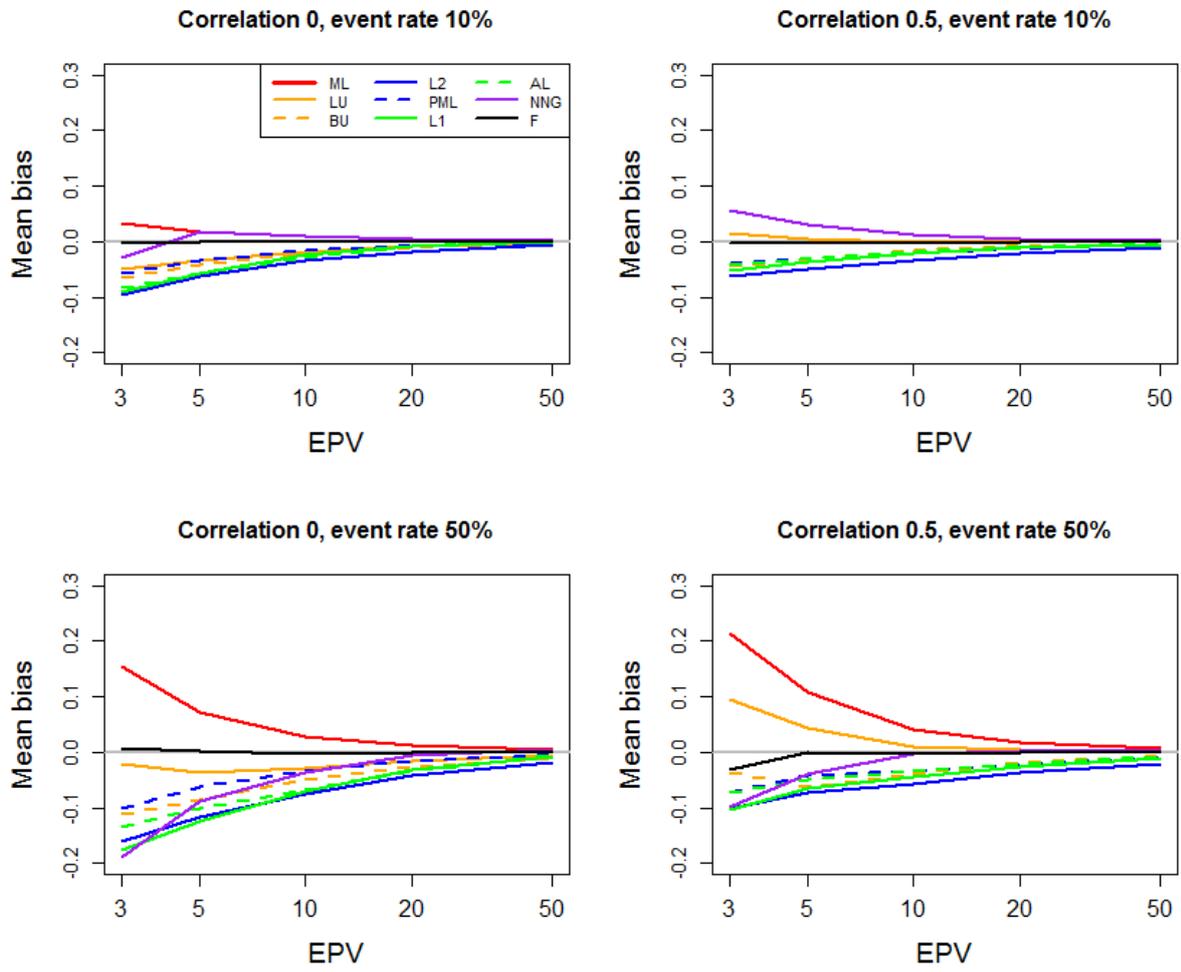

Figure S10. Mean bias in noise coefficients per method, in scenarios with 5 true and 5 noise coefficients. ML, maximum likelihood; LU, uniform shrinkage based on likelihood; BU, uniform shrinkage based on bootstrap; L2, ridge (L2 penalty); PML, penalized maximum likelihood; L1, LASSO (L1 penalty); AL, adaptive LASSO; NNG, non-negative garrote; F, Firth's correction.

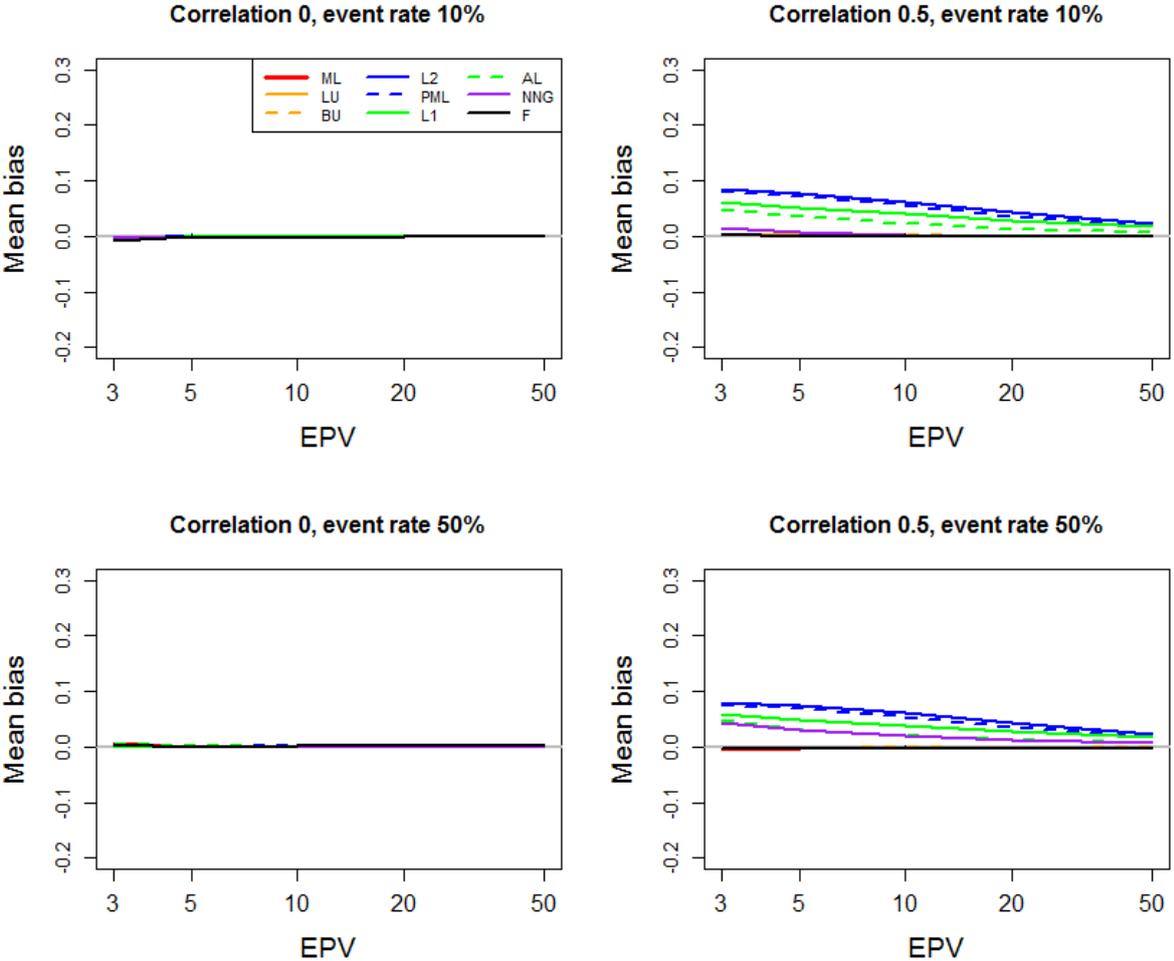

Figure S11. Mean number of selected variables per scenario, for methods based on LASSO procedures. L1, LASSO (L1 penalty); AL, adaptive LASSO; NNG, non-negative garrote.

A. Scenarios with 5 true predictors

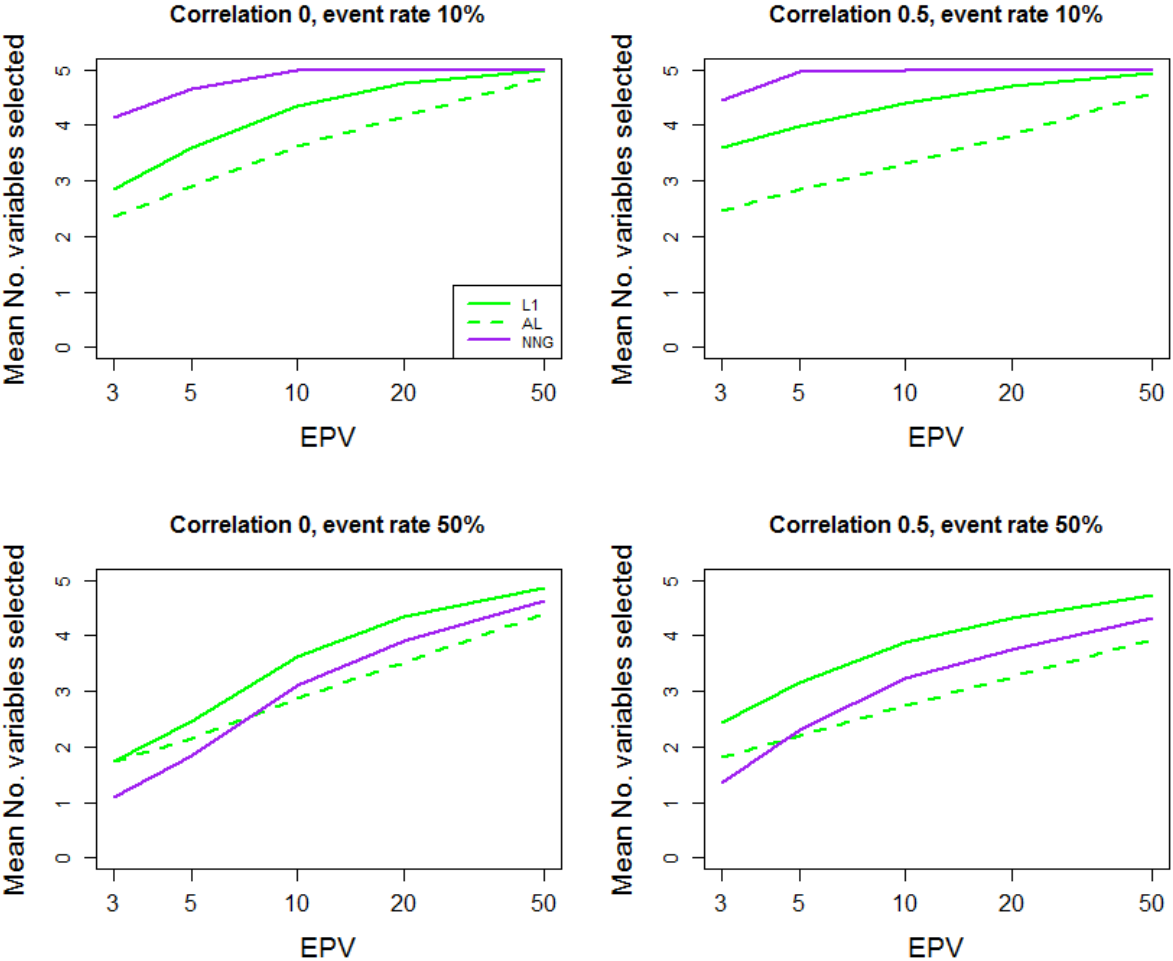

B. Scenarios with 5 true and 5 noise predictors

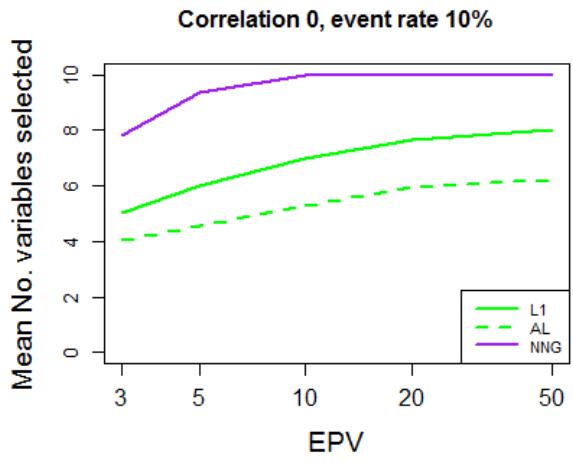
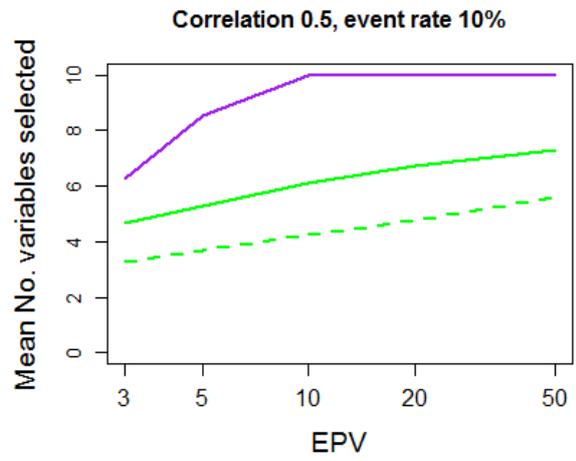
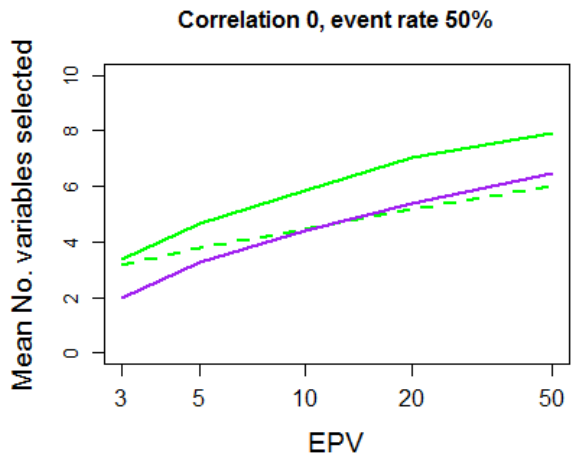
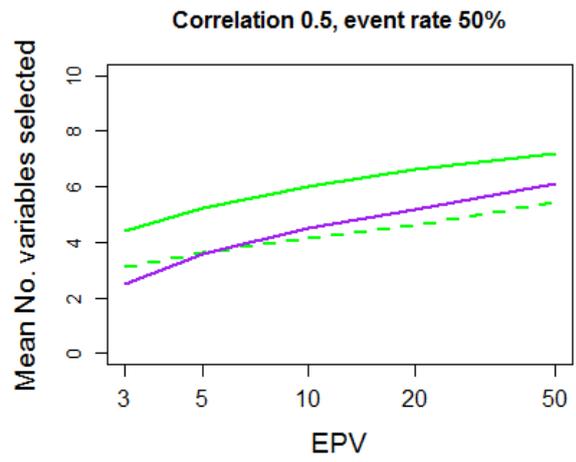

C. Scenarios with 10 true predictors

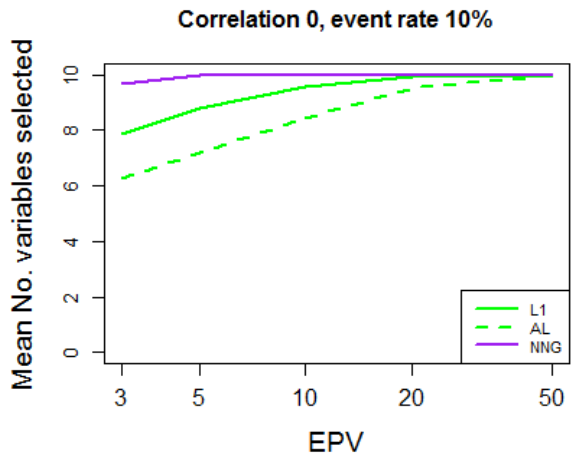
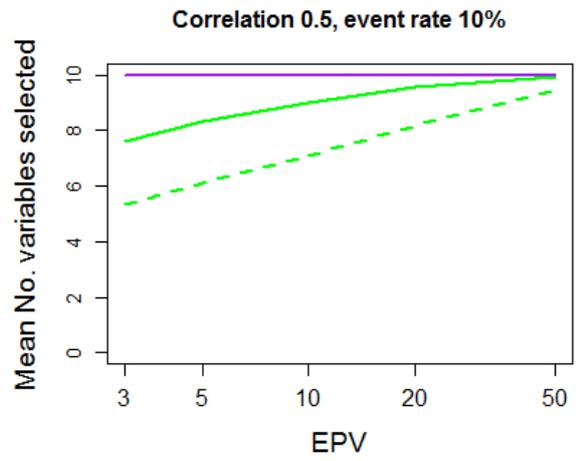
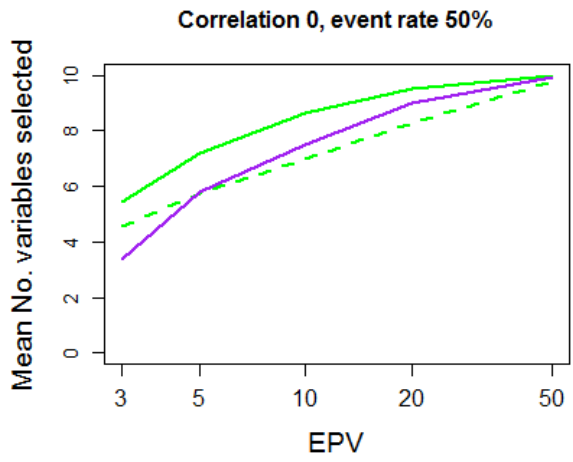
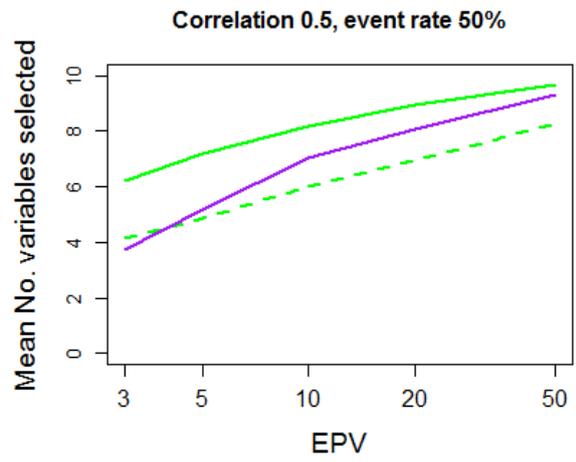

Figure S12. Mean number of selected noise coefficients per method, in scenarios with 5 true and 5 noise coefficients. L1, LASSO (L1 penalty); AL, adaptive LASSO; NNG, non-negative garrote.

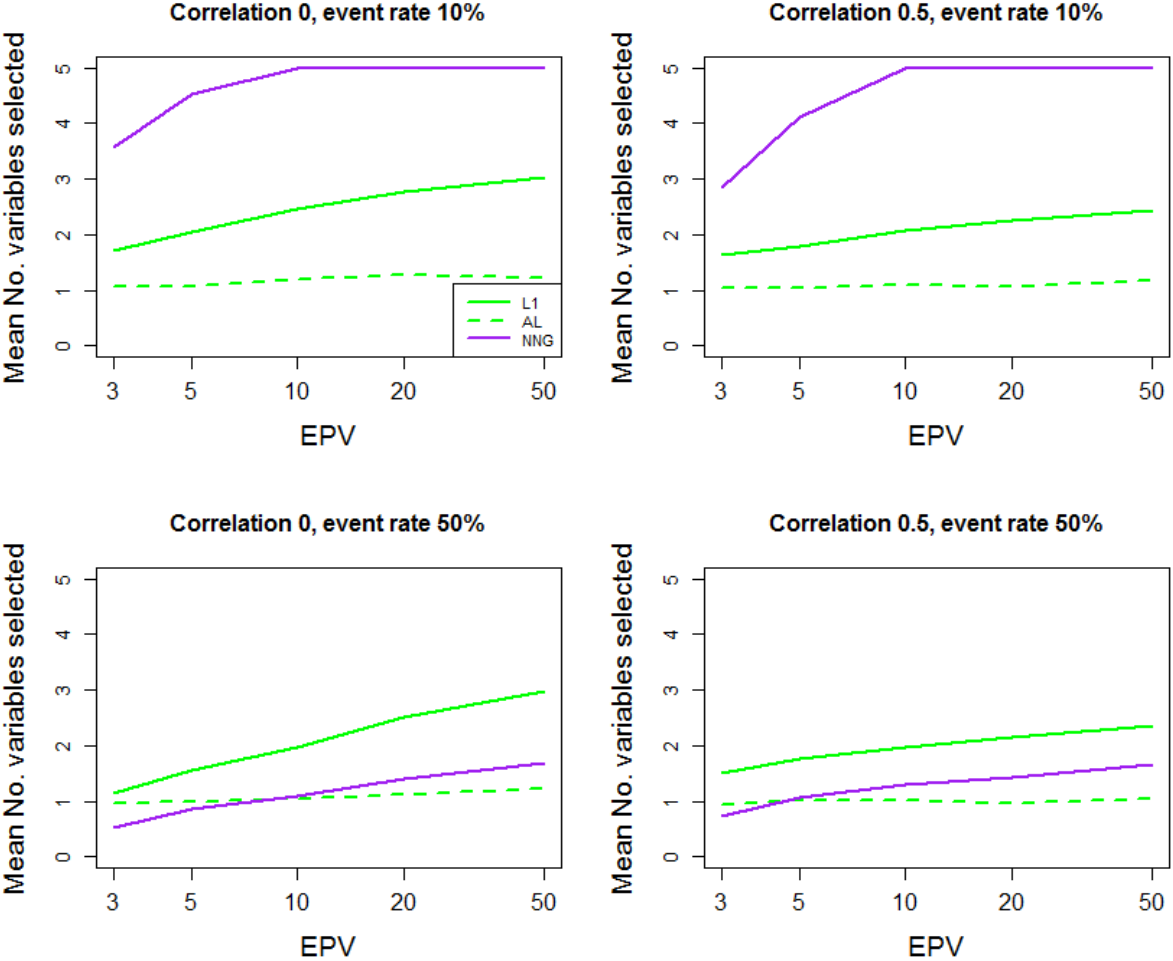